\documentclass[final]{cvpr}
\usepackage{multirow}
\usepackage{array}
\usepackage{graphicx}
\usepackage{booktabs} 
\usepackage{enumerate}
\usepackage{times}
\usepackage{epsfig}
\usepackage{graphicx}
\usepackage{amsmath}
\usepackage{amssymb}
\usepackage{soul}
\usepackage{multirow}
\usepackage{booktabs}
\usepackage{wrapfig}
\usepackage{footnote}
\usepackage{subcaption}
\usepackage[pagebackref=true,breaklinks=true,colorlinks,bookmarks=false]{hyperref}

\newcommand{\joint}{\ensuremath{\mathbf{j}}}
\newcommand{\rot}{\ensuremath{R}}
\newcommand{\trans}{\ensuremath{T}}
\newcommand{\dbox}{\ensuremath{B}}

\newcommand{\embed}{\ensuremath{E}}
\newcommand{\loss}{\ensuremath{\mathcal{L}}}
\newcommand{\targets}{\ensuremath{\mathcal{T}}}

\newcommand{\mov}{\ensuremath{m}}
\newcommand{\base}{\ensuremath{b}}

\newcommand{\src}{\ensuremath{s}}
\newcommand{\tgt}{\ensuremath{t}}
\newcommand{\srctotgt}{\ensuremath{{\src \rightarrow \tgt}}}

\newcommand{\disp}{\ensuremath{d}}
\newcommand{\ang}{\ensuremath{\theta}}

\newcommand{\Joint}{\ensuremath{J}}
\newcommand{\LocalAlign}{\ensuremath{L}}
\newcommand{\GlobalAlign}{\ensuremath{G}}

\newcommand{\Motions}{\ensuremath{\mathcal{M}}}
\newcommand{\potentialmotion}{\ensuremath{\mathbf{\tilde{m}}}}
\newcommand{\bestmotion}{\ensuremath{\mathbf{m^*}}}
\newcommand{\confidence}{\ensuremath{C}}
\newcommand{\shapepart}{\ensuremath{\mathbf{p}}}

\newcommand{\real}{\mathbb{R}}

\newenvironment{packed_itemize}
{\begin{itemize}
    \vspace{-\topsep}
    \setlength{\itemsep}{1pt}
    \setlength{\parskip}{0pt}
    \setlength{\parsep}{0pt}
}{\end{itemize}}

\newcommand{\nolistbottomspace}{\vspace{-\topsep}}

\newcommand\numberthis{\addtocounter{equation}{1}\tag{\theequation}}

\usepackage{enumitem}
\setlist[itemize]{leftmargin=*}
\setlist[enumerate]{leftmargin=*}

\begin{document}

\title{Unsupervised Kinematic Motion Detection for\\Part-segmented 3D Shape Collections}

\author{Xianghao Xu\\
Brown University\\
\and
Yifan Ruan\\
Brown University\\
\and
Srinath Sridhar\\
Brown University\\
\and
Daniel Ritchie\\
Brown University\\
}

\twocolumn[{%
\renewcommand\twocolumn[1]{#1}%
\maketitle
\begin{center}
    \centering
    \includegraphics[width=0.9\linewidth]{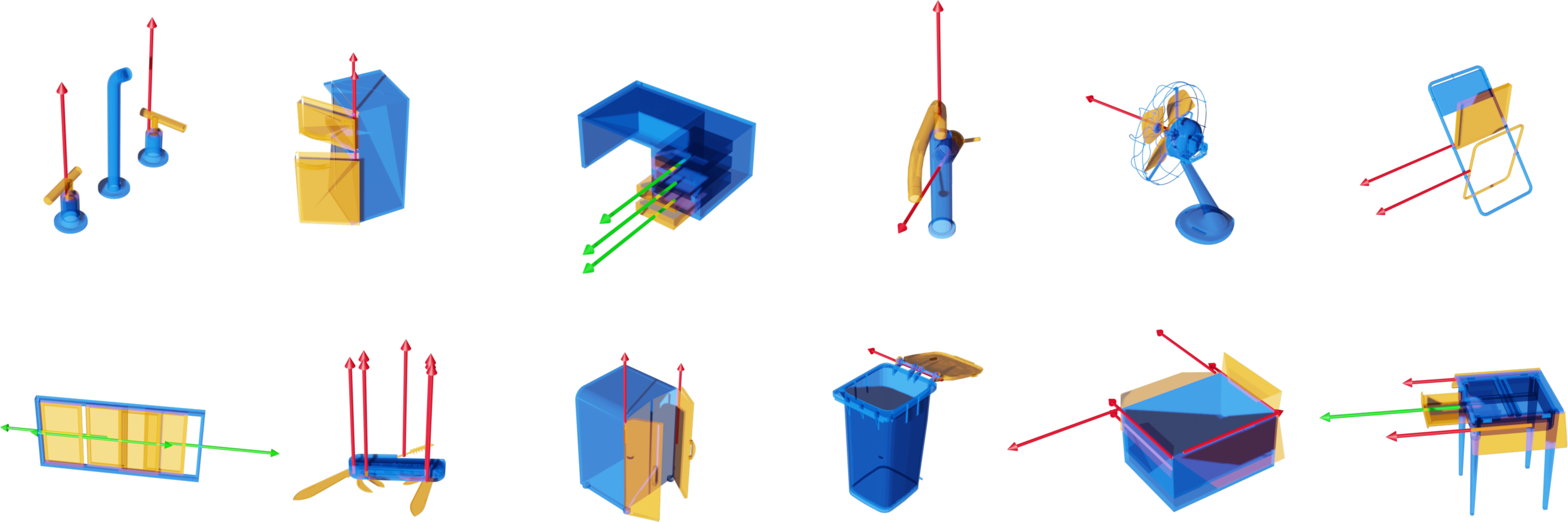}
    \captionof{figure}{Our method discovered these kinematic motion axes (and their ranges of motion) without any human supervision.
    It works by finding motion parameters such that one shape can transform into another from the same category.
    Moving parts are orange; static parts are blue; translation axes are green; rotation axes are red.
    }
    \label{fig:teaser}
\end{center}%
}]

\begin{abstract}

3D models of manufactured objects are important for populating virtual worlds and for synthetic data generation for vision and robotics.
To be most useful, such objects should be \emph{articulated}: their parts should move when interacted with.
While articulated object datasets exist, creating them is labor-intensive.
Learning-based prediction of part motions can help, but all existing methods require annotated training data.
In this paper, we present an unsupervised approach for discovering articulated motions in a part-segmented 3D shape collection.
Our approach is based on a concept we call \emph{category closure}: any valid articulation of an object's parts should keep the object in the same semantic category (e.g. a chair stays a chair).
We operationalize this concept with an algorithm that optimizes a shape's part motion parameters such that it can transform into other shapes in the collection.
We evaluate our approach by using it to re-discover part motions from the PartNet-Mobility dataset.
For almost all shape categories, our method's predicted motion parameters have low error with respect to ground truth annotations, outperforming two supervised motion prediction methods.
\end{abstract}

\section{Introduction}

3D models of manufactured objects are important for many applications: populating virtual worlds for games, AR/VR experiences, animation, interior design, and architectural visualization; creating synthetic training data for data-hungry computer vision models~\cite{PlayingForData,RenderingSUNCG}; simulated training for robots to learn to navigate or to detect and manipulate objects before being deployed in the real world~\cite{EmbodiedQA,AI2Thor,Habitat,MINOS,CHALET,xiang2020sapien}.
Ideally, such 3D objects should be \emph{articulated}: each part should specify how it moves (e.g. cabinet drawers slide open).
Recognizing the value of such data, researchers have created datasets of articulated 3D object models~\cite{FirstPartMobilityPrediction,Shape2Motion,xiang2020sapien}.
However, annotating 3D objects with kinematic motions requires human time and effort.
One alternative approach is to use machine learning to predict kinematic part motions for a shape, automating some manual annotation effort.
While such methods have been proposed, all rely on supervised learning with 3D shapes that already have motion annotations.

In this paper, we present an unsupervised method that discovers kinematic motions in a consistently part-segmented 3D shape collection.
What makes our method possible is an insight we call \emph{category closure}: given an object of category $C$ (e.g. chairs), all valid articulations of its parts will produce a shape that is still in category $C$.
While appealing in theory, this insight is challenging to apply in practice, as it is non-trivial to determine without supervision whether an articulated shape belongs to a category.
We address this challenge via the following observation: given a collection of shapes of the same category, an articulation of one shape's parts definitely remains in the same category if that articulation can transform the shape into other shapes from the collection.
Implicit in this observation is the assumption that the shape collection contains some degree of part pose variation.

Based on these insights, we design an alternating optimization scheme for discovering part articulations in a collection of shapes.
In one optimization phase, the system learns an embedding space in which shapes which can transform to one another via articulation are close.
The learning signal is based on feedback from another phase, in which groups of nearby shapes in the embedding space are selected and articulation parameters are optimized to try to transform one shape in the group into the others.
As this optimization problem is underconstrained, the system uses commonsense and physically-inspired priors to avoid finding implausible part motions.
Our approach predicts the type of motion (rotational, translation, or static) as well as the motion parameters (axes of motion, centers of rotation, ranges of motion).

We evaluate our approach by predicting articulations for shapes in PartNet-Mobility, a dataset of consistently part-segmented objects which have ground-truth kinematic motion annotations with which we can compare~\cite{xiang2020sapien}.
Our approach discovers motion parameters which exhibit low error with respect to the ground truth, outperforming two supervised motion prediction approaches on almost all shape categories.
In summary, our contributions are:
\begin{packed_itemize}
    \item The concept of category closure as self-supervision for discovering valid kinematic part motions.
    \item An alternating optimization scheme which implements this concept by finding motion parameters which transform objects into other objects of the same category.
\nolistbottomspace
\end{packed_itemize}
Code and data for this paper are at \url{https://github.com/xxh43/UKMD}
\section{Related Work}

\paragraph{Articulated object datasets}
Researchers have built datasets of part-segmented shapes with kinematic motions.
One includes an unreleased dataset of 368 moving joints~\cite{FirstPartMobilityPrediction}, manually annotated from ShapeNet~\cite{chang2015shapenet}.
The Shape2Motion dataset~\cite{Shape2Motion} (no longer available) contained 2,240 3D objects from 45 categories, sourced from ShapeNet and the 3D Warehouse~\cite{3DWarehouse} and manually annotated with kinematic motions.
PartNet-Mobility~\cite{xiang2020sapien} consists of over 2,000 objects in 47 categories, also from ShapeNet.
All these datasets were manually annotated, which is labor intensive.
Other prior work proposes a machine-learning-assisted interface for rapidly writing simple programs to annotate shapes with kinematic motions~\cite{MotionAnnotationPrograms}.
This system reduces human labeling effort but does not eliminate it.
We seek a method that require no human labeling.

\paragraph{Predicting part mobilities}
Early work on automatic mobility prediction includes illustrating the motions of mechanical assemblies~\cite{mitra2010illustrating}, analyzing multiple instances of an object in a scene~\cite{sharf2014mobility}, slippage analysis for deformable mesh models~\cite{xu2009joint}, and inferring kinematic chains based on motion trajectories~\cite{yan2006automatic}.
These methods rely on having high-fidelity, physically accurate joint geometry or access to multiple observations of the same shape in different poses; in contrast, large shape collections have widely varying geometric quality and only contain a single observation of each shape.
The problem has also been studied in robotics for manipulating unknown articulating objects~\cite{pillai2015learning, sturm2011probabilistic, hausman2015active}.
In computer vision, machine learning has been applied to unstructured point clouds to jointly segment them into parts and predict their motions~\cite{Shape2Motion, RPMNet, DeepPartInduction, hu2017learning}.
Closet to our work is the system of Hu et al.~\cite{FirstPartMobilityPrediction}, which also assumes consistently-segmented manufactured object meshes.
Given a new object, it retrieves the best-matching example and transfers its motion to the input.
These methods require labeled examples; in contrast, our approach leverages the principle of category closure as form of self-supervision.
In concurrent work, Kawana et al. \cite{Kawana2021} jointly predict part segmentation and part articulations via a neural network trained with adversarial self-supervision.
Training this network requires many pose variations of each training shape.
In contrast, our method works with only \emph{one} observation of each unique training object and requires fewer unique training objects.

\paragraph{Estimating articulation from images}
Estimating 3D articulation from images and depth maps has been widely studied for humans~\cite{huang2020arch, shotton2011real, mehta2017vnect, ballan2012motion, alldieck2018detailed, kanazawa2018end, mueller2018ganerated, joo2018total} and more recently for articulating objects~\cite{li2020category, zhang2021strobenet, abbatematteo2019learning}, assuming a known kinematic structure.
When the structure is unknown, a recent method~\cite{mu2021sdf} has proposed to disentangle shape and appearance using a neural network to estimate parts, joints, and joint angles.
Unlike these methods, ours requires no kinematic structure or other supervision..
\section{Approach}
\label{sec:approach}

\begin{figure*}[t!]
    \centering
    \includegraphics[width=\linewidth]{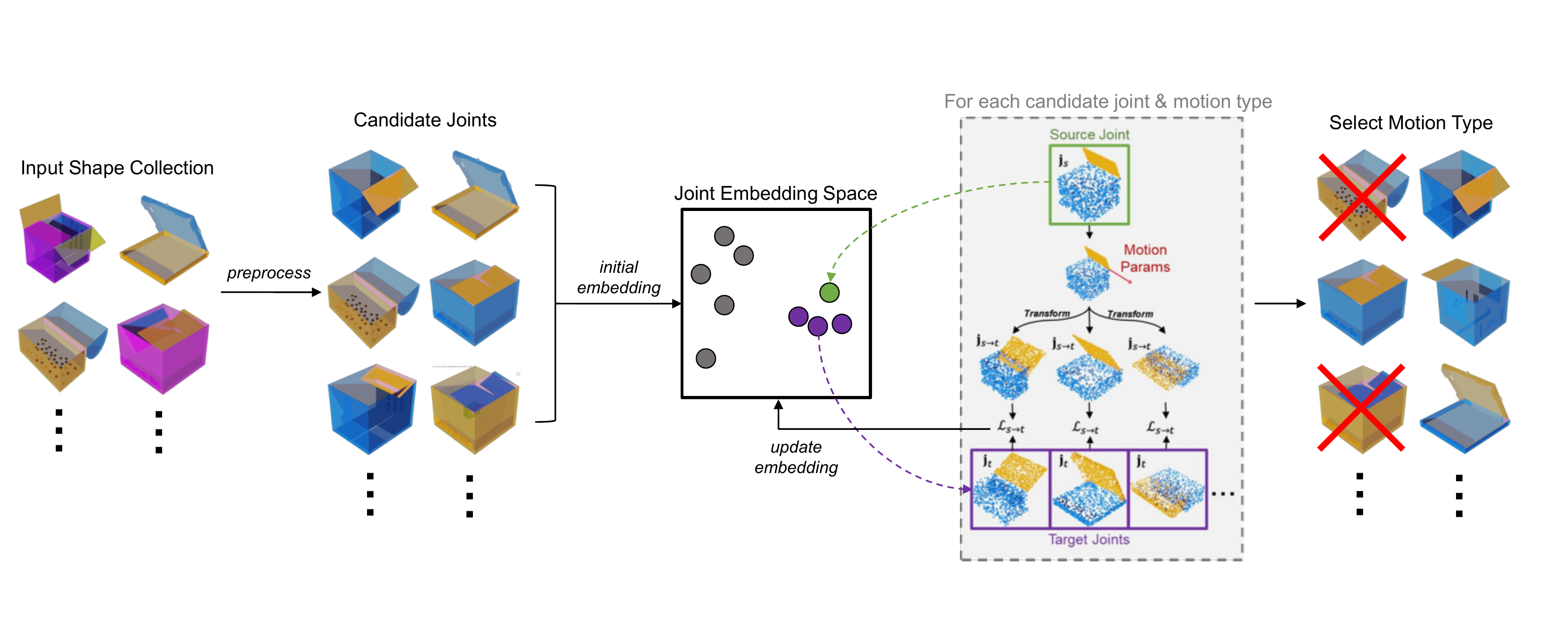}
    \vspace{-2.5em}
    \caption{
    Given a collection of consistently part-segmented shapes from the same semantic category, our system first extracts a set of candidate \emph{joints} consisting of connected parts, one of which may move about the other.
    It then initializes an embedding space in which two joints should be nearby if one can be transformed into the other through a valid hinge or prismatic motion.
    For each candidate joint $\joint_\src$, it samples nearby joints $\joint_\tgt$ and optimize for motion parameters which approximately transform $\joint_\src$ to each $\joint_\tgt$ (producing $\joint_\srctotgt$).
    The reconstruction error is used as feedback to improve the embedding space, and the process repeats.
    This results in multiple possible motions for each joint, so we heuristically select which motion (if any) is best.
    Best viewed zoomed.
    }
    \label{figure:approach}
\end{figure*}

Figure~\ref{figure:approach} shows an overview of our system, which takes as input a set of part-segmented shapes.
We assume the segmentations are consistent (i.e. shapes are segmented at the same granularity, though some shapes may have parts that others do not) but do not require part labels.
Such data can be scalably produced from online 3D model repositories~\cite{Turbosquid,CGTrader,AdobeStock3D} using e.g. machine-learning-assisted segmentation tools~\cite{ScalableActiveSegmentation}.
We also assume that input shapes exhibit some part pose variations (i.e. parts are not modeled in exactly the same pose across every shape in the dataset).
We examine the impact of input pose variation on our method's performance in Section~\ref{sec:results}.

From the input shapes, we create a set of candidate \emph{joints}: each joint $\joint$ consists of a moving part $\joint^\mov$ and a base part $\joint^\base$, i.e. the fundamental unit of kinematic motion.
For each part in each shape, we create one joint in which that part is $\joint^\mov$ and the largest other part to which it is connected is $\joint^\base$.
Part connectivity is automatically inferred from geometry; see supplemental.
For each joint, our goal is to determine what type of motion (if any) applies to it, and what its motion parameters are.
We consider two types of joints: hinge (rotational) joints, parameterized by an axis, a center of rotation, and a range of angles; and prismatic (translational) joints, parameterized by an axis and a range of displacements~\cite{murray2017mathematical}.

To solve this problem, we observe that a valid joint motion is one that can transform the joint into other joints from the same collection (assuming parts occur in different poses throughout the collection);
this is the concept of \emph{category closure}.
Our solution is a two phase, alternating optimization scheme.
The first phase, for a given joint, identifies `target joints' to which it should be transformed.
Not all joints are good targets, e.g. we do not want to transform a cabinet door into a cabinet drawer.
Our system identifies good targets by building an embedding space in which joints are nearby if they are good transformation targets for one another.
The more other joints into which we can transform one joint through a motion, the more confident we can be that this motion is correct.

To learn this embedding space, the system relies on feedback from the second phase.
Here, given a source joint, a set of target joints, and a candidate motion type, the system optimizes for motion parameters that transform the source into the target.
This problem is underconstrained and can produce implausible motions; thus, we introduce commonsense and physically-inspired priors to steer the system toward good solutions.

These two phases are iterated: feedback about how well a source joint can be transformed to its targets is used to improve the embedding space; the improved embedding space leads to new target joints which help the system optimize for better motions.
This iterative process produces multiple possible motions for each part.
Thus, the system uses a heuristic final phase to determine which type of motion (or no motion) is most plausible.

\section{Identifying Transformation Target Joints}
\label{sec:clustering}

Given a candidate joint, the goal of this phase is to construct a set of `target joints' to which that joint should be transformed via a kinematic motion.
The more other joints into which we can transform one joint through a motion, the more confident we can be in that motion.
To solve this problem, we construct an embedding space in which two joints are close by if one is a good target for the other.

\paragraph{Initial embedding}
Initially, the system has no information about which joints can transform into other joints via valid motions.
Thus, we construct an initial embedding based on which joints can transform into others through \emph{any} affine transformation.
For every pair of joints $(\joint_1, \joint_2)$, we optimize for a rotation, translation, and scale (where we penalize the anistropy of the scale)
for both $\joint_1^\mov$ and $\joint_1^\base$ to bring them as close as possible to $\joint_2^\mov$ and $\joint_2^\base$ by minimizing bidirectional chamfer distance (assuming all objects are consistently upright-oriented, $\joint_1^\base$'s rotation reduces to a single rotation about the up axis).

We then use the optimization residuals to produce a $N \times N$ similarity matrix for a collection of $N$ joints ($N \in [100, 200]$, in our experiments). We set the similarities between joints that have a different number of connected components in either their moving part or base part to zero, to prevent these structurally-different joints from being grouped as source-target pairs.
An embedding can be constructed from this matrix, but we do not need to do so---for our purposes, it suffices to select, for a source joint $\joint_\src$ the $16$ most similar joints as its set of potential target joints.

\paragraph{Iterative improvement}
On each iteration of the system, for each source joint and its target joints, we run the motion optimization procedure described in Section~\ref{sec:motionopt}.
This produces a transformation reconstruction loss $\loss^\text{recon}_\srctotgt$ for each pair of source and target joints $(\joint_\src, \joint_\tgt)$.
The system uses these losses to learn a new embedding space, where the distance between two joint embeddings should be proportional to their loss:
\begin{equation}
    \loss^{\text{embed}} = \frac{1}{N}\sum_{\src=1}^{N}\frac{1}{k}\sum_{\tgt \in \targets_\src}|\loss^\text{recon}_\srctotgt - \alpha ||\embed(\joint_\src) - \embed(\joint_\tgt)||_{2} |
\end{equation}
where $N$ is the total number of joints, $k=5$ is the number of target joints per source joint, $\targets_\src$ is the set of $k$ targets for source joint $\src$, and $E$ is a PointNet encoder~\cite{qi2017pointnet} whose parameters (and $\alpha$) are the variables of optimization.
A joint is fed to the encoder as a point cloud with a per-point one-hot indicator of whether the point belongs to the moving part or base part.
We minimize this loss using Adam~\cite{Kingma2014AdamAM}.
We then select $k$ new target joints for each source joint $\joint_\src$ by sampling $\joint_\tgt \sim \text{exp}(-||\embed(\joint_\src) - \embed(\joint_\tgt) ||_2)$.
The system then moves to the next iteration.

\section{Optimizing for Joint Motion Parameters}
\label{sec:motionopt}

Given a source joint $\joint_\src$, a set of target joints $\{ \joint_\tgt \}$, and a motion type (hinge or prismatic), the goal of this phase is to optimize for motion parameters that can transform the source joint to each of the targets.
As a pre-process, we first optimize for rotations $\theta_{\tgt \rightarrow \src}$ about the up axis that bring each target joint $\joint_\tgt$ into closest alignment with $\joint_\src$ (via bidirectional chamfer distance).

\subsection{Parameterized transformation model}

We first define the parametric function by which one joint $\joint_\src$ is transformed into another joint $\joint_\tgt$:

\paragraph{Kinematic motion}
We use $\trans^\Joint_\src$ to denote a prismatic (translational) joint transformer for joint $\src$ (implicitly parameterized by a translation direction vector).
We use $\trans^\Joint_\src(\joint_\src^\mov, \disp)$ to denote articulating the source joint $\joint_\src$'s moving part $\joint_\src^\mov$ with translational displacement $\disp$.
Similarly, we use $\rot^\Joint_\src$ to denote a hinge (rotational) joint transformer for joint $\src$ (implicitly parameterized by an axis and center of rotation).
We use $\rot^\Joint_\src(\joint_\src^\mov, \ang)$ to denote articulating the source joint $\joint_\src$'s moving part $\joint_\src^\mov$ with rotation angle $\ang$.

\paragraph{Additional pose transformations}
In addition to kinematic motion, we may need additional pose transformations to align the source and target joint.
We use $\trans^\GlobalAlign_\srctotgt$ and $\rot^\GlobalAlign_\srctotgt$ to denote a translation and a rotation about world-up that are applied to the entire joint $\joint_\src$ to help globally align it with the target joint $\joint_\tgt$.
The moving part sometimes also needs additional degrees of freedom relative to the base part.
For example, to transform the bottom drawer in a cabinet into the top drawer, we need an additional upward translation.
For this, we also define a local alignment translation $\trans^\LocalAlign_\srctotgt$.
For translational joints, to ensure that this local alignment translation cannot become redundant with joint motion, it is projected into the plane perpendicular to the axis of translation.

\paragraph{Geometric deformers}
To transform a joint to a geometrically different joint, we must also permit some deformation of joint geometries, in addition to pose variation.
$\dbox_\srctotgt$ denotes a \emph{box deformer}, whose degrees of freedom are the scales of the 6 faces of a part's bounding box, which allows adjusting the bulk shape of a part.
This box is aligned with the part's local coordinate frame, so its deformations are independent of the part's pose.

\begin{figure}[t!]
    \centering
    \includegraphics[width=\linewidth]{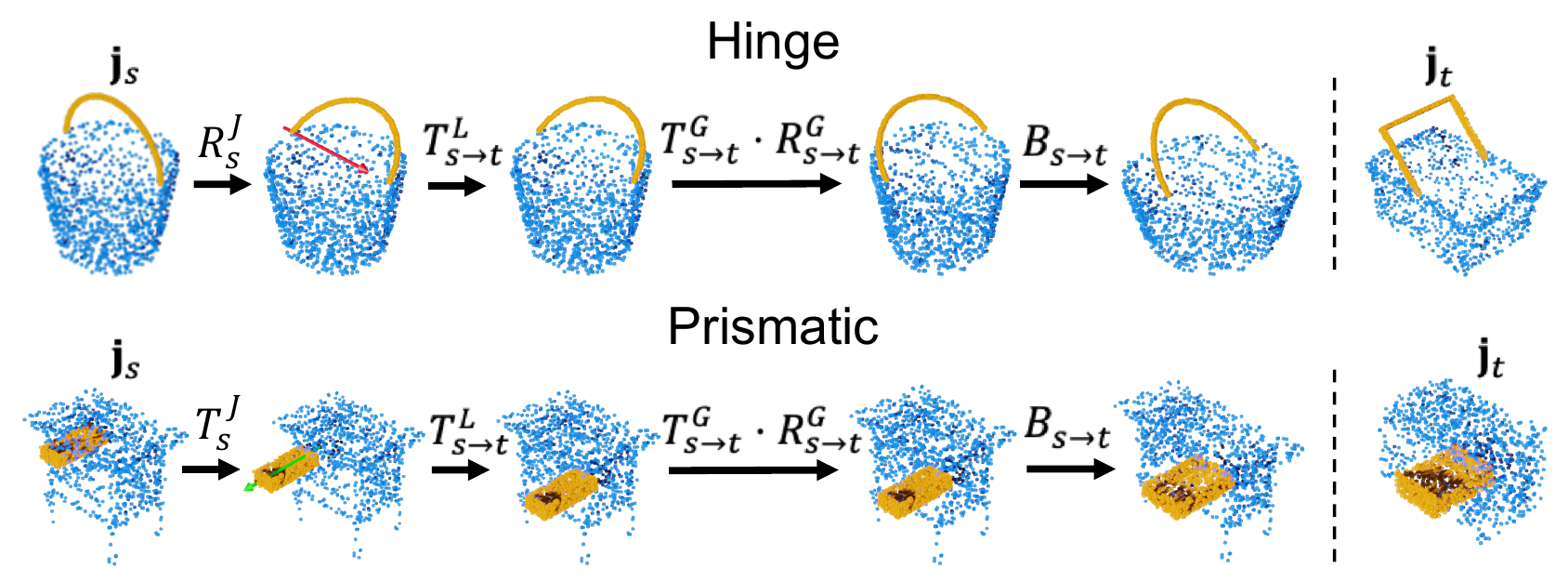}
    \vspace{-1.25em}
    \caption{
    How our transformation models for hinge and prismatic motions transform one joint $\joint_\src$ to another $\joint_\tgt$.
    }
    \label{fig:motionmodels}
\end{figure}

\paragraph{Final transformation model}
Given the functions defined above, we can now define the complete, optimizable transformation function that takes the moving part of a joint $\src$ to that of another joint $\tgt$ via a prismatic motion:
\begin{equation}
    \joint^\mov_\srctotgt =
        \dbox^\mov_\srctotgt(
            \{\trans^\GlobalAlign \cdot \rot^\GlobalAlign \cdot \trans^\LocalAlign\}_\srctotgt \cdot
                \trans^\Joint_\src(\joint^\mov_\src, \disp_\srctotgt)
        )
\end{equation}

Similarly, for a hinge motion:
\begin{equation}
    \joint^\mov_\srctotgt =
        \dbox^\mov_\srctotgt(
            \{\trans^\GlobalAlign \cdot \rot^\GlobalAlign \cdot \trans^\LocalAlign\}_\srctotgt \cdot
                \rot^\Joint_\src(\joint^\mov_\src, \ang_\srctotgt)
        )
\end{equation}

For both types of motion, the base part transforms as:
\begin{equation}
    \joint^\base_\srctotgt =
        \dbox^\base_\srctotgt(
            \{\trans^\GlobalAlign \cdot \rot^\GlobalAlign\}_\srctotgt \cdot
                \joint^\base_\src
        )
\end{equation}

Figure~\ref{fig:motionmodels} illustrates these transformation sequences.

\subsection{Loss functions}
\label{sec:losses}

To produce plausible transformations from a source joint $\joint_\src$ to a target $\joint_\tgt$, we optimize the parameters of the above transformation model with respect to several loss functions:

\paragraph{Reconstruction loss}
First and foremost, the transformed source joint must approximately reconstruct the target joint:
\begin{equation}
    \loss^\text{recon}_\srctotgt = \frac{D_\text{chamfer}(\joint^\mov_\srctotgt, \joint^\mov_\tgt)}{\text{diag}(\joint^\mov_\srctotgt)} + 
    \frac{D_\text{chamfer}(\joint^\base_\srctotgt, \joint^\base_\tgt)}{\text{diag}(\joint^\base_\srctotgt)}
\end{equation}
where $D_\text{chamfer}$ denotes bidirectional chamfer distance between two point-sampled parts, and $\text{diag}(\cdot)$ is the diagonal length of a part (to normalize these distances). 

Many settings of the transformation model parameters will give low reconstruction loss, most of which do not correspond to valid kinematic motions.
Thus, we design priors to encourage the optimization to find plausible motions:

\paragraph{Small joint motion penalty}
The optimization can find spurious motions by setting the amount of motion (i.e. the displacement $\disp$ or rotation angle $\ang$) to a small value.
Thus, we propose to penalizes motions smaller than a threshold $\tau$.
For prismatic joints,
\begin{equation}
    \loss^\text{joint}_\srctotgt = w_\text{joint} \cdot  \max(\tau_\disp - | \disp_\srctotgt |, 0)
\end{equation}
and for hinge joints,
\begin{equation}
    \loss^\text{joint}_\srctotgt = w_\text{joint} \cdot  \max(\tau_\ang - | \ang_\srctotgt |, 0)
\end{equation}

This term is particularly important for finding correct motions for parts whose geometry does not visibly change over the course of articulation (e.g. a spinning wheel).

\paragraph{Large alignment transform penalties}
While the global alignment transforms $\trans^\GlobalAlign_\srctotgt$, $\rot^\GlobalAlign_\srctotgt$ and local alignment transform $\trans^\LocalAlign_\srctotgt$ are often necessary, the optimization should try to perform as much of the transformation as possible using the joint motion.
To this end, we introduce loss terms to penalize the magnitude of the alignment transforms:
\begin{equation}
    \loss^\text{align}_\srctotgt = w^\GlobalAlign_\text{align}| \rot^\GlobalAlign_\srctotgt| + w^\LocalAlign_\text{align}|\trans^\LocalAlign_\srctotgt|
\end{equation}

\paragraph{Large deformation penalty}
Similarly, we must also restrict the box deformers $\dbox_\srctotgt$ from being responsible for more of the transformation than is necessary:
\begin{equation}
    \loss^\text{deform}_\srctotgt = w_\text{deform} ( |\dbox^\mov_\srctotgt| + |\dbox^\base_\srctotgt| )
\end{equation}
where $|B|$ is the sum of all box face absolute displacements.

\paragraph{Collision penalty}
For a joint motion to be physically valid, the moving part must not collide with the base part.
We define a collision penalty loss $\loss^\text{collide}_\srctotgt$ which enforces this property.
We sample $n$ equally-spaced values along the interval $[0,\disp_\srctotgt]$ for prismatic joints ($[0,\theta_\srctotgt]$ for hinge joints) and transform the moving part to that pose.
The collision penalty for each pose is the mean penetration distance of each point in the base part point cloud to the moving part.
The overall collision penalty is then the mean of these per-timestep penalties.

For a hinge joint:

\begin{equation}
    \loss^\text{collide}_\srctotgt = \frac{w_\text{collide}}{n|\joint^\base_\src|} \sum_{i=1}^n \sum_{\mathbf{x} \in \joint^\base_\src} \max(0, d_0-\text{sdf}(\mathbf{x}, \rot^\Joint_\src(\joint^\mov_\src, i \cdot \ang_\srctotgt)) )
\end{equation}

where $\text{sdf}(\mathbf{x}, \joint^\mov)$ is the signed distance from the point $\mathbf{x}$ to the minimum volume bounding box of the moving part $\joint^\mov$ (negative signed distance $\rightarrow$ the point is inside the box) and $d_0$ is the largest penetration distance of any point $\mathbf{x} \in \joint^\base_\src$ at $i=0$ (i.e. some joints may initially have a small degree of interpenetration, which we should not penalize).
The loss for a prismatic joint is defined analogously (replace $\rot^\Joint_\src(\joint^\mov_\src, i \cdot \ang_\srctotgt)$ with $\trans^\Joint_\src(\joint^\mov_\src, i \cdot \disp_\srctotgt)$).

\paragraph{Detachment penalty}
In addition to not colliding with the base part, a moving part should also not \emph{detach} from its base part over the course of its motion.
For hinge joints, we minimize the distance between the nearest 10 original contact points in the rotated moving part $\rot^\Joint_\src(\joint^\mov_\src, i \cdot \ang_\srctotgt)$ (denoted as $\mathcal{S}_{\joint^\mov_\src}$) and the nearest 10 original contact points in $\joint^\base_\src$ (denoted as $\mathcal{S}_{\joint^\base_\src}$):
\begin{equation}
    \loss^\text{detach}_\srctotgt = \frac{w_\text{detach}}{n} \sum_{i=1}^n \max(0,\sum_{\mathbf{x, y} \in {\mathcal{S}_{\joint^\mov_\src}, \mathcal{S}_{\joint^\base_\src}}} ||\mathbf{x} - \mathbf{y}||_2 - r)
\end{equation}
where $r=0.01$ is a distance penalty threshold.
In addition to the above term, we also add a term which penalizes the center of rotation from non-physically falling outside of the moving part's bounding box.

For prismatic joints, we penalize the distance between the moving part $\joint^\mov_\src$ and the $50$ points on the base part $\joint^\base_\src$ which are closest to it (denoted as $\mathcal{N}_{\joint_\src}$):
\begin{equation}
    \loss^\text{detach}_\srctotgt = \frac{w_\text{detach}}{n|\mathcal{N}_{\joint_\src}|} \sum_{i=1}^n \sum_{\mathbf{x} \in \mathcal{N}_{\joint_\src}} \max(0, \text{sdf}(\mathbf{x}, \rot^\Joint_\src(\joint^\mov_\src, i \cdot \ang_\srctotgt)) )
\end{equation}

\paragraph{Hyperparameters}
We empirically define two sets of values for the various loss weights $w$: one set for optimizing hinge joints; one set for optimizing prismatic joints.
These weights are kept constant across all shape categories in all of our experiments.
Values for weights and other hyperparameters can be found in supplemental.

\subsection{Optimization procedure}

To optimize the parameters of the transformation model, we combine all the above losses together into one:
\begin{align*}
    \loss_\srctotgt &= \loss^\text{recon}_\srctotgt + \loss^\text{joint}_\srctotgt + \loss^\text{align}_\srctotgt + \loss^\text{deform}_\srctotgt + \loss^\text{collide}_\srctotgt + \loss^\text{detach}_\srctotgt
    \\
    \loss &= \frac{1}{kN} \sum_{s=1}^N \sum_{\tgt \in \targets_\src} \loss_\srctotgt \numberthis
\end{align*}
We minimize $\loss$ using the Adam optimizer.

\paragraph{Multiple initializations}
As this optimization problem is non-convex, we solve it multiple times with different initializations to avoid local minima.
For hinge joints, we use the 3 axes of the minimum volume bounding box of the moving part as the initial rotation axes.
We use the centroid of the moving part and the centers of 4/6 of its bounding box faces as the initial rotation centers (the four with the largest distance to the part centroid).
For prismatic joints, we use the longest 2 axes of the minimum volume bounding box of the moving part as the initial axes.
This results in 2 initializations for prismatic joints and 15 (3 $\times$ 5) initializations for hinge joints; we choose the one which gives the smallest $\loss_\srctotgt$.

\paragraph{Axis post-processing}
The optimization often gives motion axes that noisily oscillate around a good solution.
Thus, we use a post-processing step to `snap' the axes.
We check if the axis is close to any of the three world axes or the three principal axes of the moving part or base part.
If the dot product of the optimized axis and any of these axes is $> 0.975$, we snap the axis to it.

\paragraph{Determining range of motion}
Finally, given optimized motion parameters for a joint $\joint_\src$, we estimate its range of motion.
For this, we sample $16$ nearby target joints from the embedding space (using the sampling procedure from Section~\ref{sec:clustering}) and optimize for a transformation from $\joint_\src$ to each of these targets, holding motion parameters fixed and only optimizing pose, deformation, and alignment transforms.
We estimate the joint's motion range as the range of poses for all of these target joints whose post-optimization $\loss^\text{recon}$ is less than a threshold (see supplemental).
We call these joints the `valid' targets for a motion.
This results in a motion range relative to the initial pose of $\joint_\src^\mov$ (i.e. for a hinge joint, $\ang = 0$ is the initial pose).

\section{Determining Joint Motion Type}

After multiple iterations of optimization, we are left with multiple potential hinge and prismatic motions $\potentialmotion$ for each candidate joint $\joint$.
In this section, we describe our procedure for determining (a) which of these motions is the best for each joint and (b) whether a part moves at all or should instead be labeled as static.

\paragraph{Selecting the best candidate motion}
We start by considering the set of `valid' target joints for each potential motion $\potentialmotion$, as described above.
Intuitively, a motion is more likely to be correct if (a) it allows the source joint to reach more valid targets, and (b) the targets exhibit a wider range of poses.
Let $N^\potentialmotion_\text{valid}$ be the number of valid target joints for a motion, which addresses (a).
For (b), we discretize the predicted range of motion into a set of equally-sized bins and let $N^\potentialmotion_\text{bin}$ be the number of these bins which contain the pose of at least one of the valid target joints.
We then define our confidence in this potential motion as:
\begin{equation}
    \confidence^\potentialmotion = \lambda_1 N^\potentialmotion_\text{valid} + \lambda_2 N^\potentialmotion_\text{bin}
\end{equation}
We select whichever potential motion $\potentialmotion$  has the highest confidence as the best motion $\bestmotion$.
See supplemental for the values of $\lambda_1, \lambda_2$.

\paragraph{Distinguishing moving vs. static parts}

To identify whether a part $\shapepart$ should be movable or static, we look at the number of candidate motions $\potentialmotion$ in which it is is used as a base or moving part, as well as our confidence in those motions.
Intuitively, a part used as a moving part in many high-confidence motions is more likely to be movable; a part used as a base part in many high-confidence motions is more likely to be static.
Let $\Motions^\shapepart_\text{mov}$ and $\Motions^\shapepart_\text{base}$ be the set of candidate motions in which part $\shapepart$ is used as a moving or base part, respectively.
Our confidences that this part is movable or static are:

\begin{align*}
    \confidence^\shapepart_\text{mov} &= \frac{\lambda_3}{|\Motions^\shapepart_\text{mov}|} \sum_{\potentialmotion \in \Motions^\shapepart_\text{mov}}  \confidence^\potentialmotion + \lambda_4 |\Motions^\shapepart_\text{mov}|
    \\
    \confidence^\shapepart_\text{static} &= \frac{\lambda_3}{|\Motions^\shapepart_\text{base}|} \sum_{\potentialmotion \in \Motions^\shapepart_\text{base}}  \confidence^\potentialmotion  + \lambda_4 |\Motions^\shapepart_\text{base}| \numberthis
\end{align*}

If $\confidence^\shapepart_\text{static} / \confidence^\shapepart_\text{mov}$ is greater than a threshold, the part is labeled static.
We also always label the largest part of every shape as static.
See supplemental for $\lambda_3, \lambda_4,$ and threshold values.

\section{Results}
\label{sec:results}

Here we evaluate our system's ability to discover accurate kinematic motions without supervision.

\paragraph{Dataset}
We evaluate our method on PartNet-Mobility, a dataset of part-segmented 3D shapes annotated with ground-truth kinematic articulations~\cite{xiang2020sapien}.
We run experiments on 18 categories of objects:
Box, Bucket, Clock, Door, Fan, Faucet, Folding Chair, Knife, Laptop, Pliers, Refrigerator, Scissors, Stapler, StorageFurniture, Table, Trash Can, USB, and Window.
These categories were chosen to give good coverage of the different types of motions which occur in PartNet-Mobility.
We perform various filtering steps on this data: joints with invalid motion ranges, moving parts that are extremely small relative to the overall shape (which are not well-represented in point cloud from), etc. See supplemental for details.
Our final evaluation dataset contains 753 shapes with 1939 parts.

Our method assumes that the input shapes exhibit pose variations.
Some shape categories in PartNet-Mobility have this property; others have all shapes in a neutral pose.
We normalize pose variation across categories by randomly sampling a pose from within each movable part's range of motion.
In some of our experiments, we examine the impact that the amount of pose variation has on our method.
Also, all shapes in a PartNet-Mobility category are aligned to a common coordinate frame, but not all in-the-wild shape collections exhibit this property.
We randomly rotate each shape about its up vector.

\begin{table}[t!]
    \centering
    \scriptsize
    \caption{Comparing the performance of BaseNet (with and without pre-alignment) vs. Our method on predicting the motion attributes of shapes in PartNet-Mobility.
    }
    \resizebox{\columnwidth}{!}{\begin{tabular}{lcccc}
        \toprule
        \textbf{Method} & \textbf{Type Acc$\uparrow$} & \textbf{Axis Err ($^\circ$)$\downarrow$} & \textbf{Center Err (\%)$\downarrow$} & \textbf{Range IoU$\uparrow$}
        \\
        \midrule
        BaseNet & 0.87 & 30.71 & 24.28 & 0.44
        \\
        BaseNet + align & 0.93 & 13.24 & 22.18 & 0.45
        \\
        Ours & 0.84 & \textbf{6.09} & \textbf{9.12} & 0.46
        \\
        \bottomrule
    \end{tabular}}
    \label{tab:comparison_basenet_summary}
\end{table}

\begin{table}[t!]
    \centering
    \scriptsize
    \caption{Comparing the performance of Shape2Motion (with and without pre-alignment) vs. Our method on predicting the motion attributes of shapes in PartNet-Mobility. S2M does not handle static motion type and does not predict motion range.
    }
    \resizebox{\columnwidth}{!}{\begin{tabular}{lccc}
        \toprule
        \textbf{Method} & \textbf{Type(w/o static) Acc$\uparrow$} & \textbf{Axis Err ($^\circ$)$\downarrow$} & \textbf{Center Err (\%)$\downarrow$} 
        \\
        \midrule
        S2M & 0.91 & 33.65 & 25.58
        \\
        S2M + align & 0.92 & 15.80 & 16.67
        \\
        Ours & 0.80 & \textbf{6.09} & \textbf{9.12}
        \\
        \bottomrule
    \end{tabular}}
    \label{tab:comparison_s2m_summary}
\end{table}

\begin{figure*}
    \centering
    \small
    \setlength{\tabcolsep}{1pt}
    \begin{tabular}{rcccccccc}
        \raisebox{2.5em}{BaseNet} & 
        \includegraphics[width=0.1\linewidth]{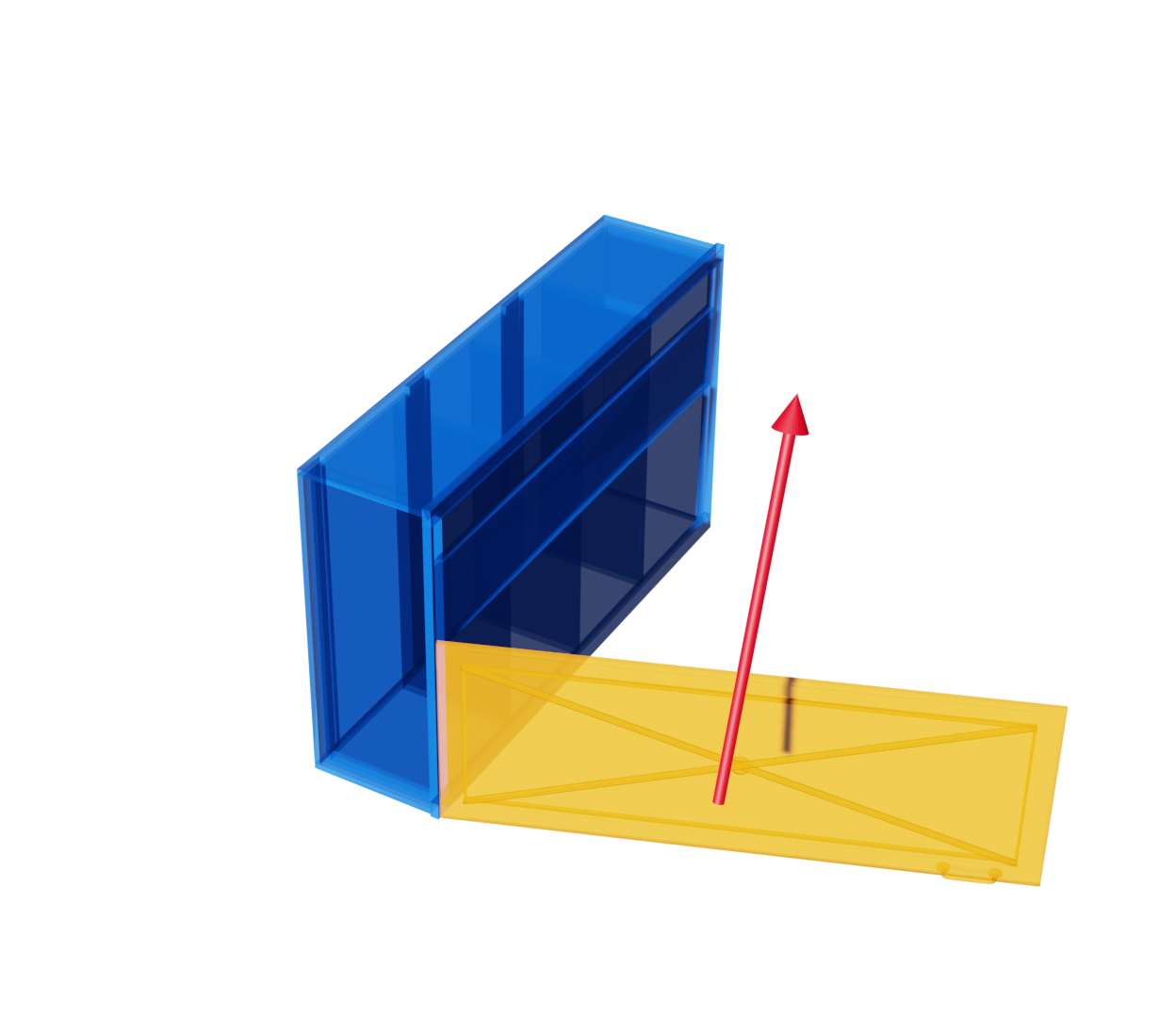} &
        \includegraphics[width=0.1\linewidth]{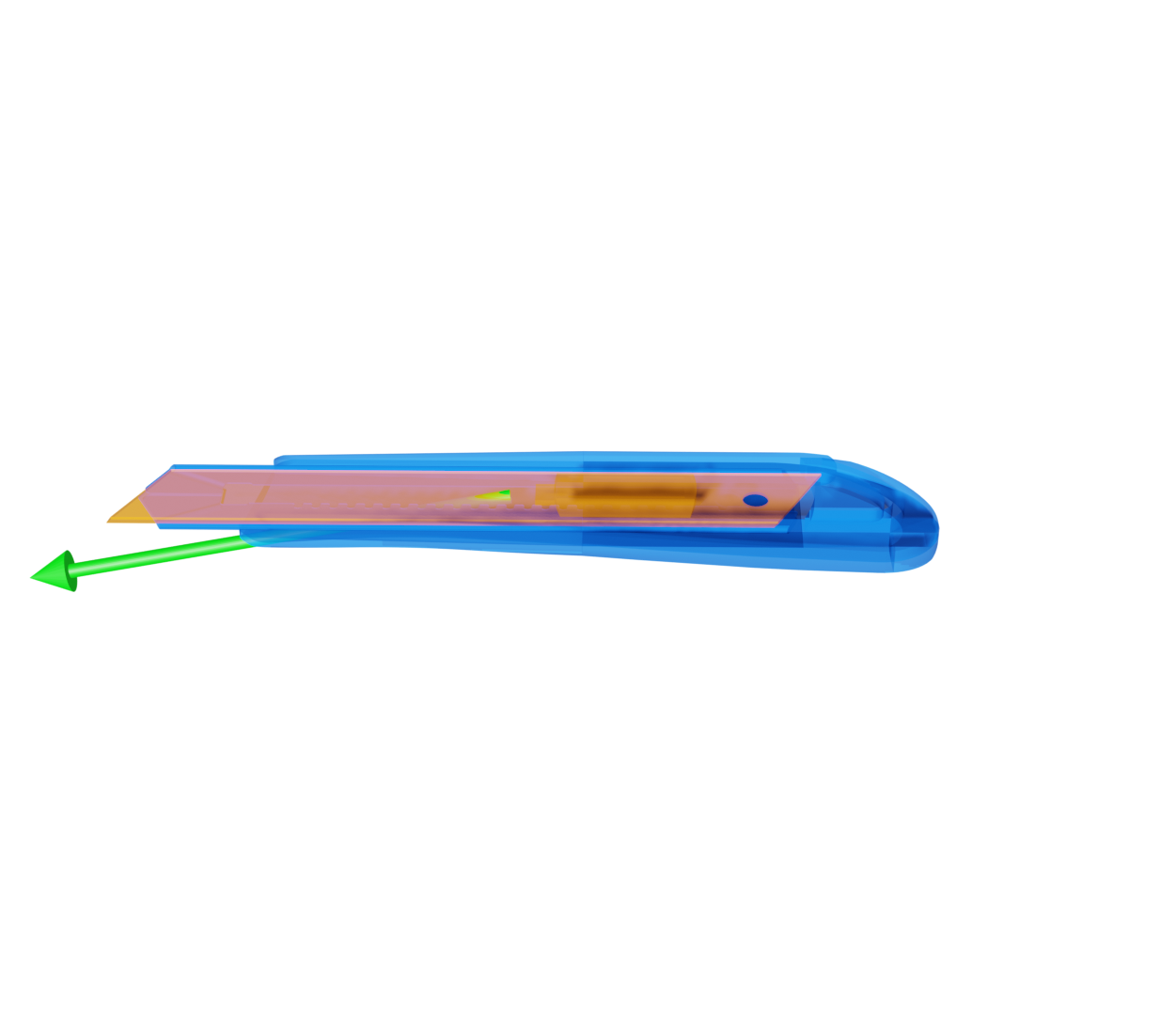} &
        \includegraphics[width=0.1\linewidth]{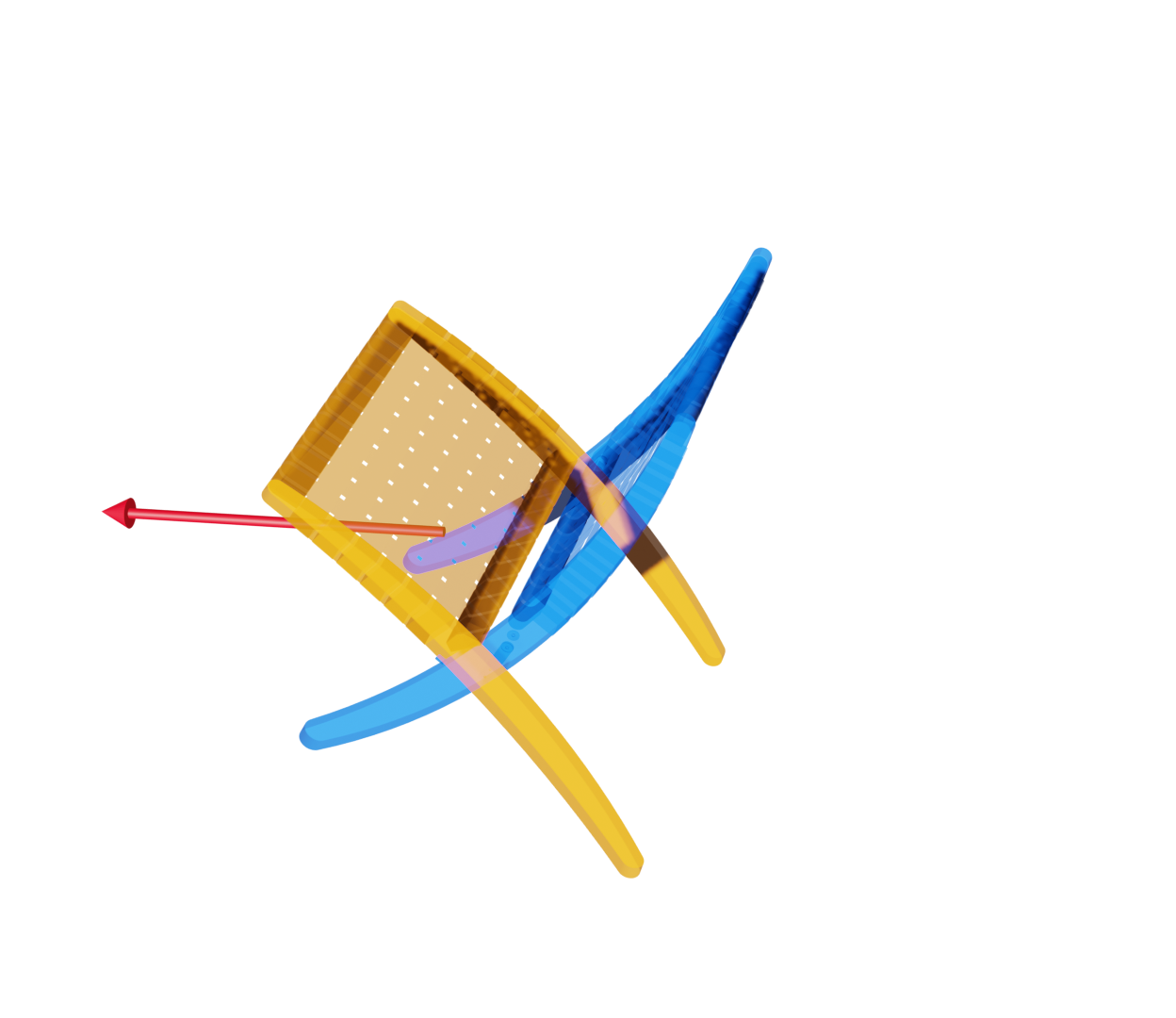} &
        \includegraphics[width=0.1\linewidth]{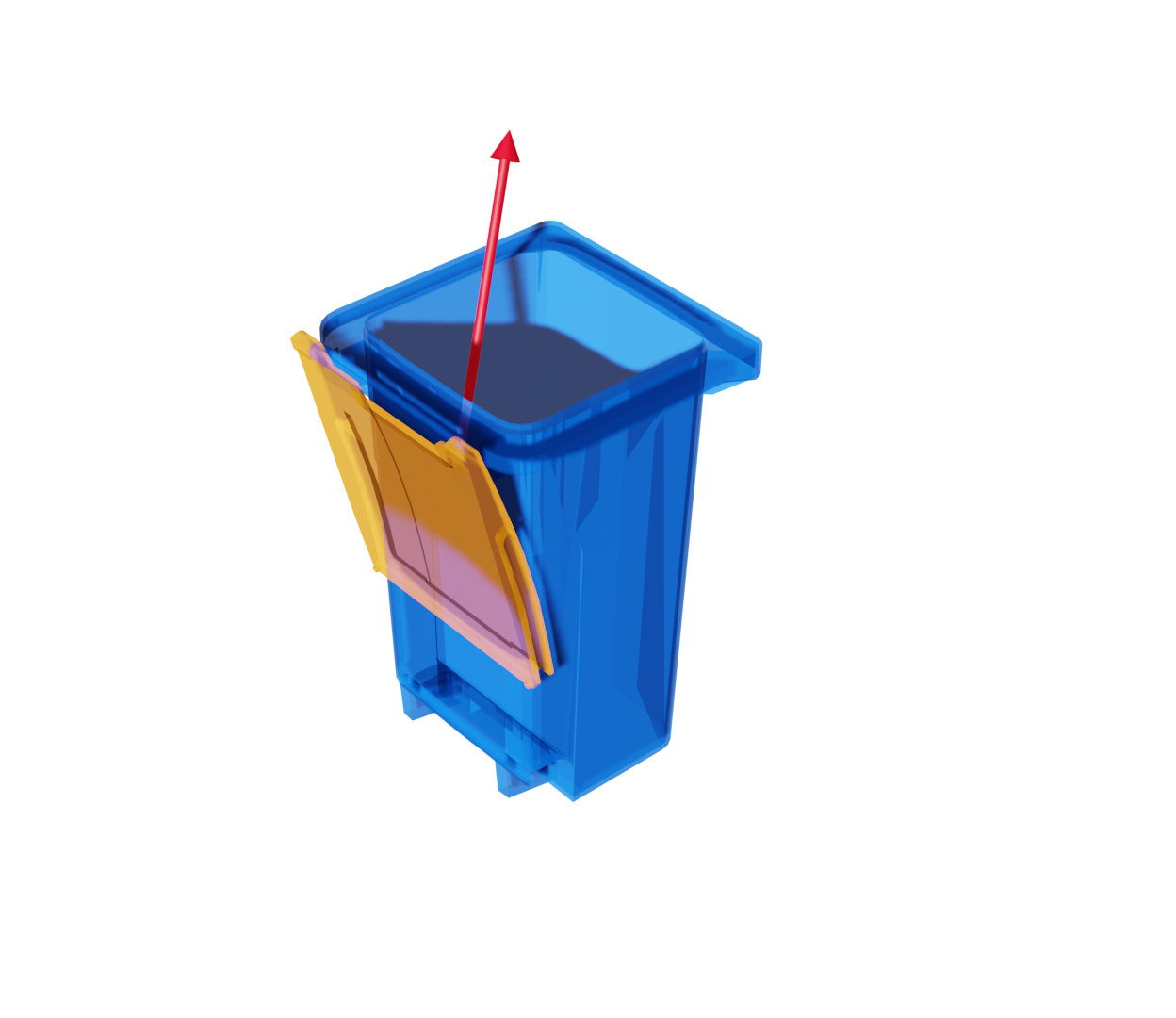} &
        \includegraphics[width=0.1\linewidth]{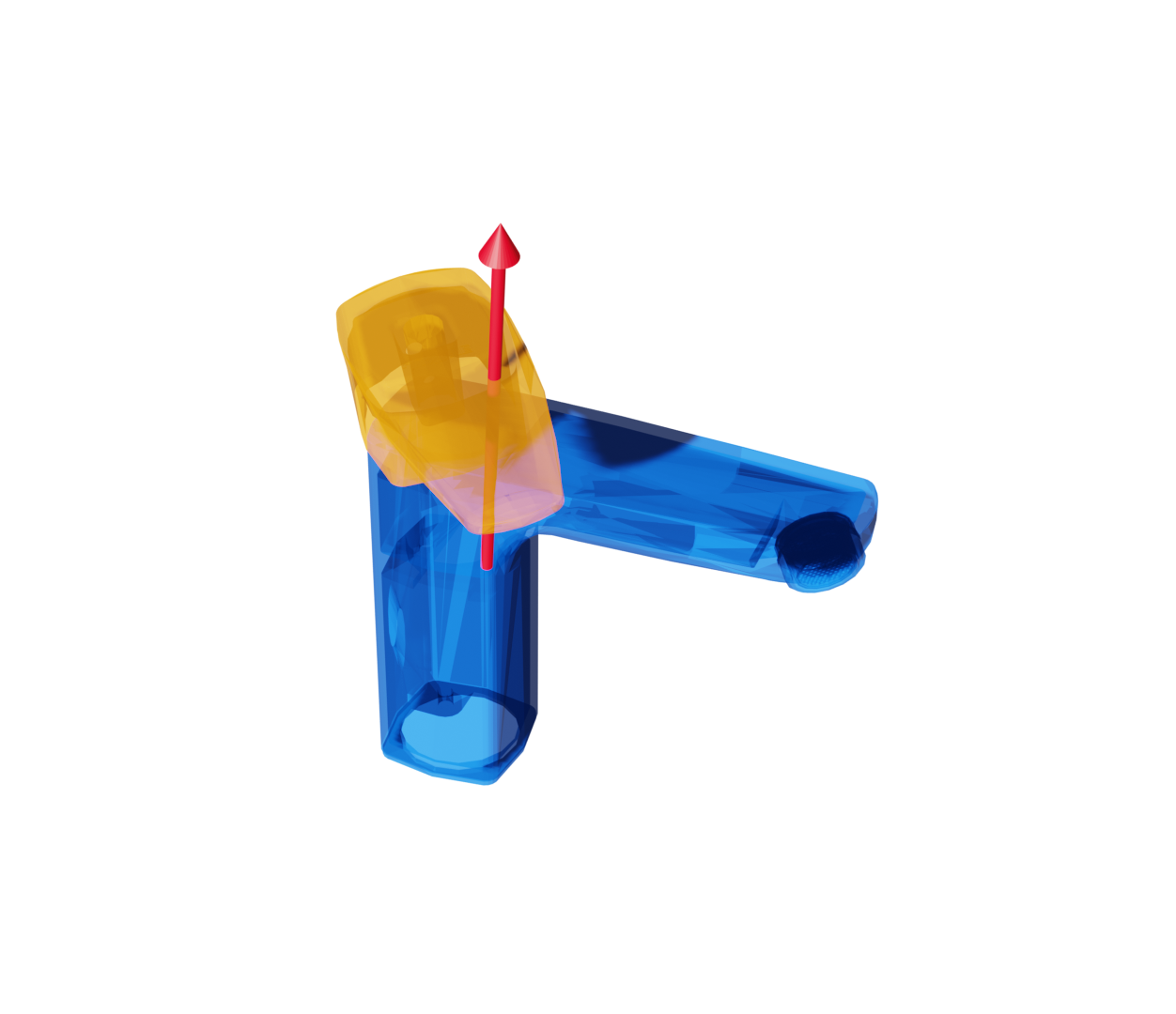} &
        \includegraphics[width=0.1\linewidth]{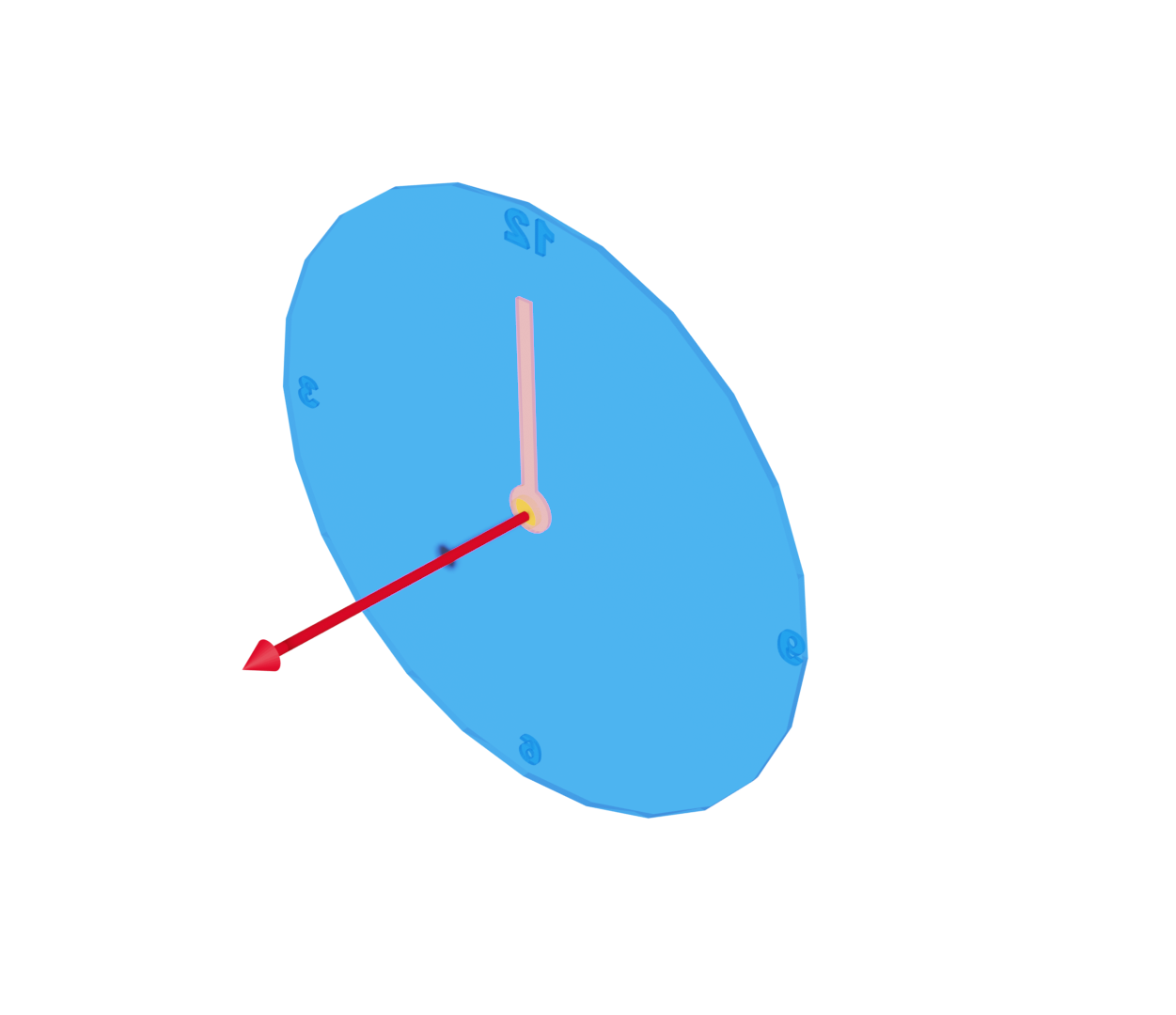} &
        \includegraphics[width=0.1\linewidth]{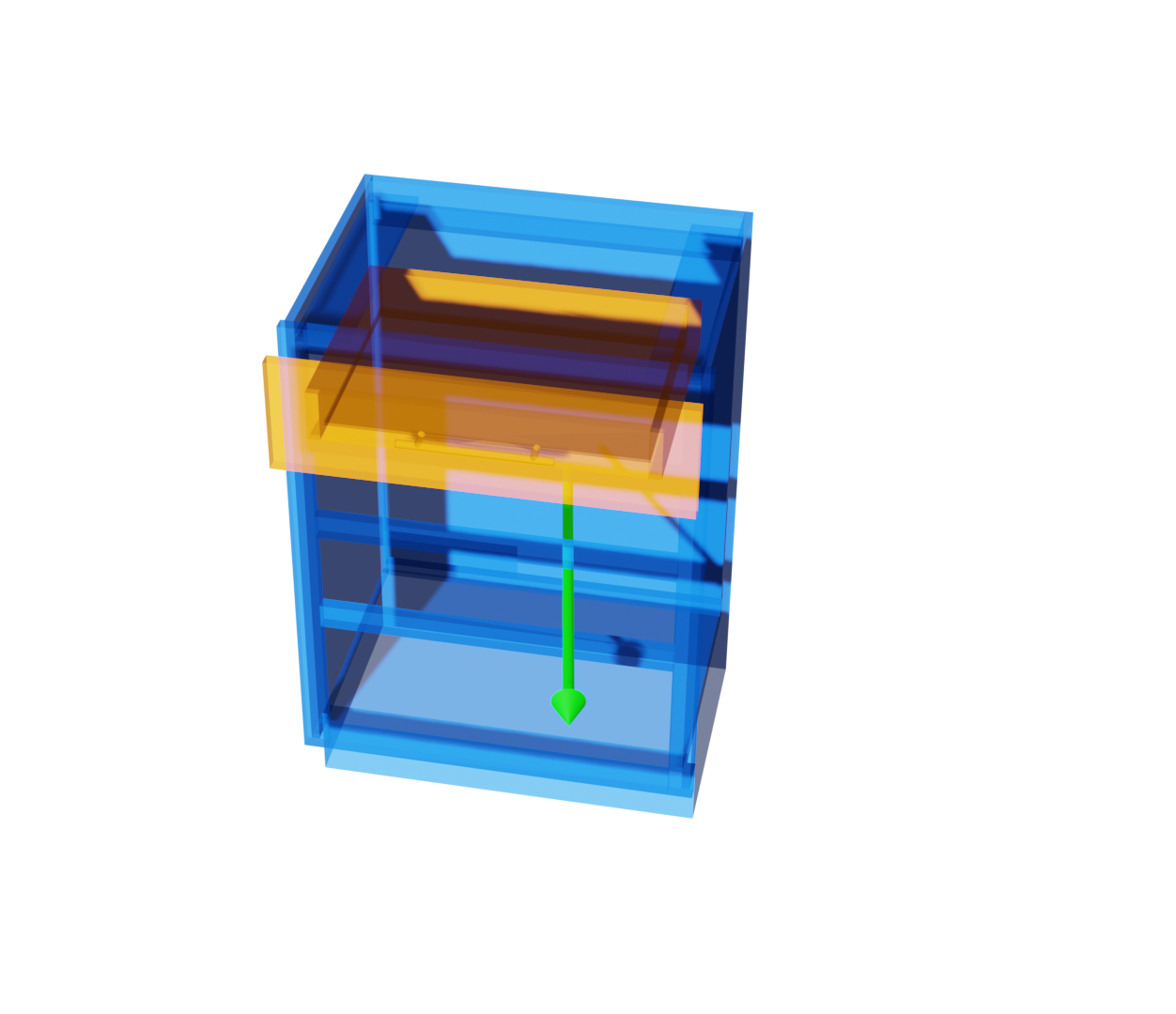} &
        \includegraphics[width=0.1\linewidth]{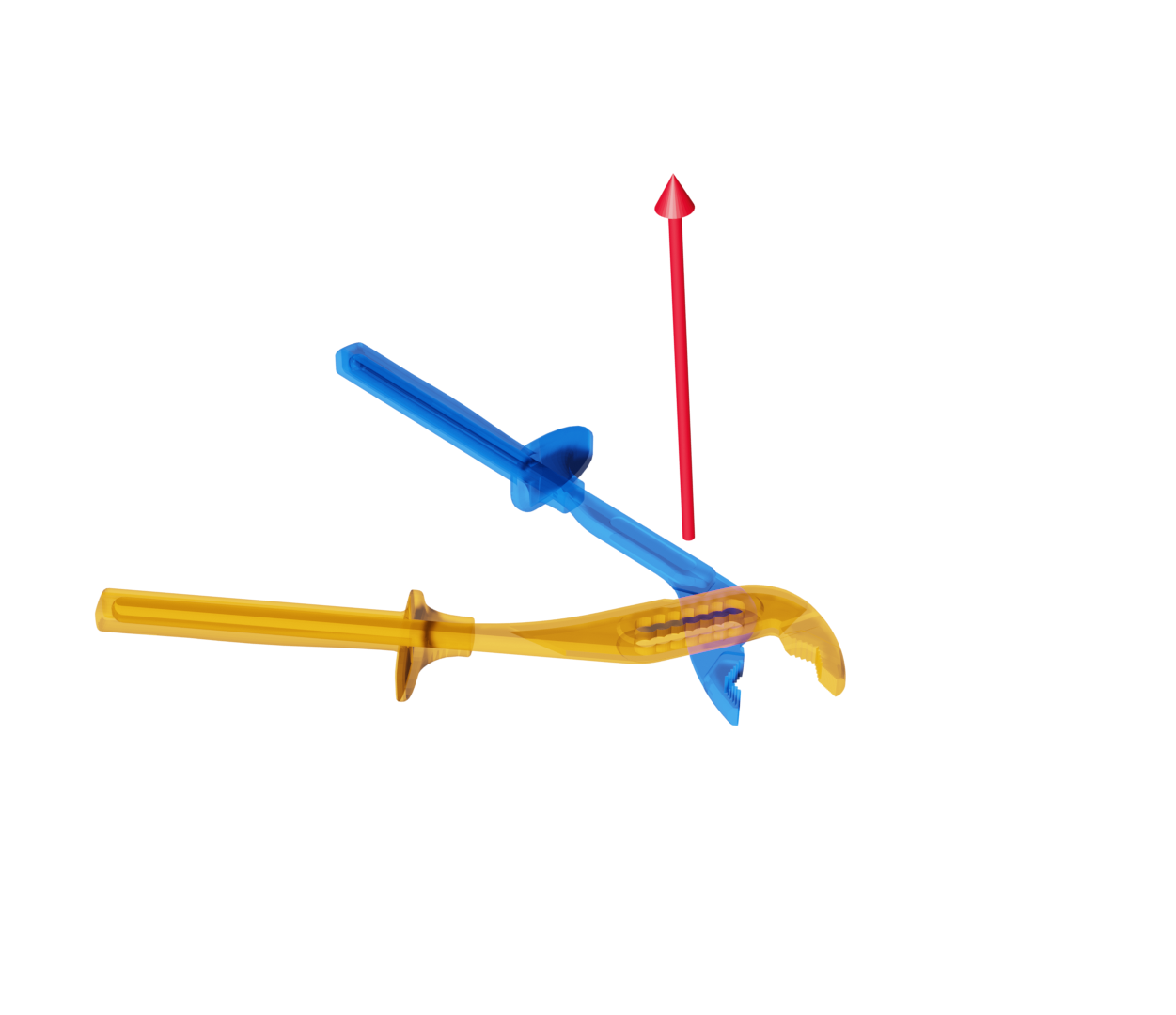}
        \\
        \raisebox{2.5em}{Ours} & 
        \includegraphics[width=0.1\linewidth]{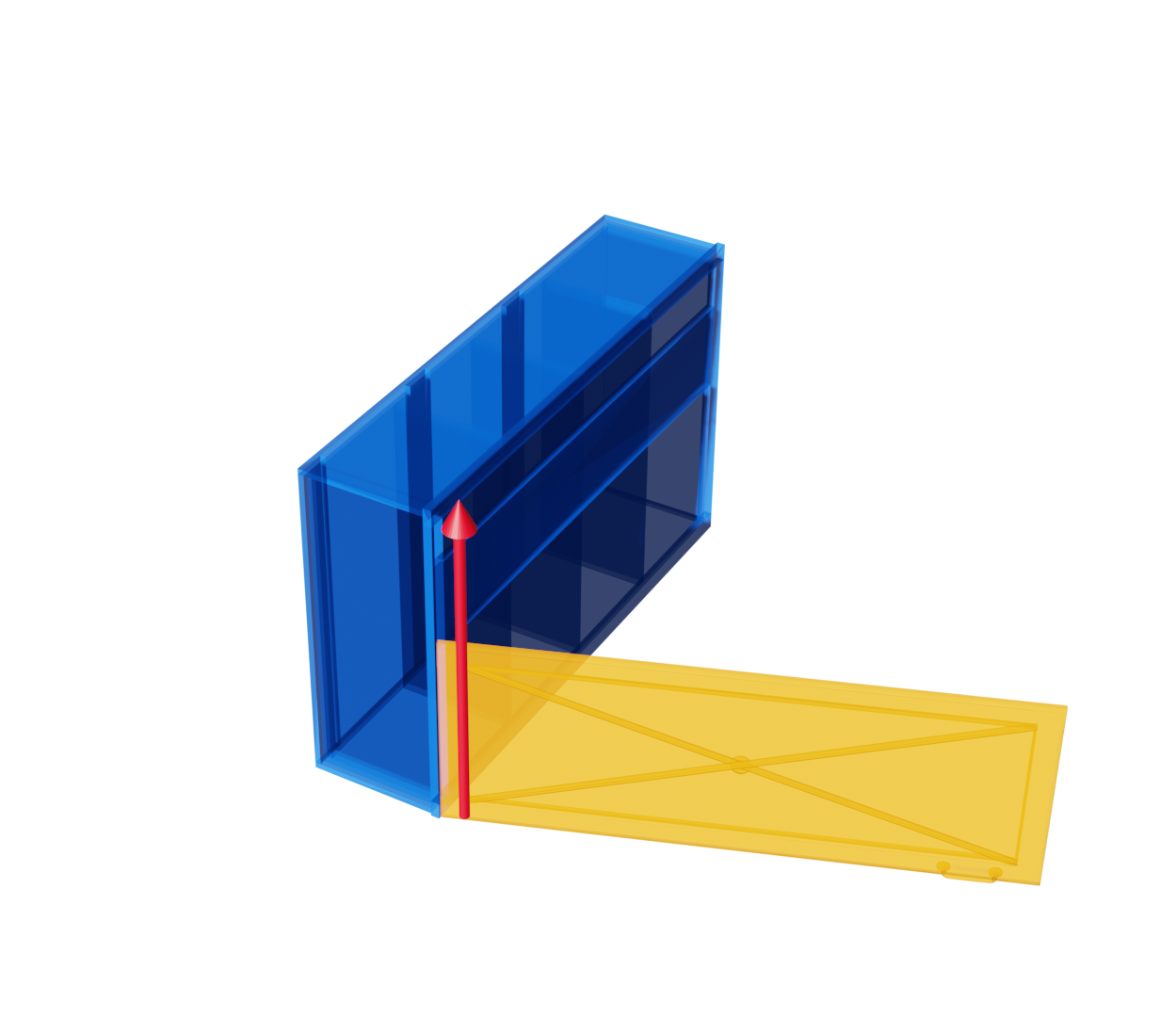} &
        \includegraphics[width=0.1\linewidth]{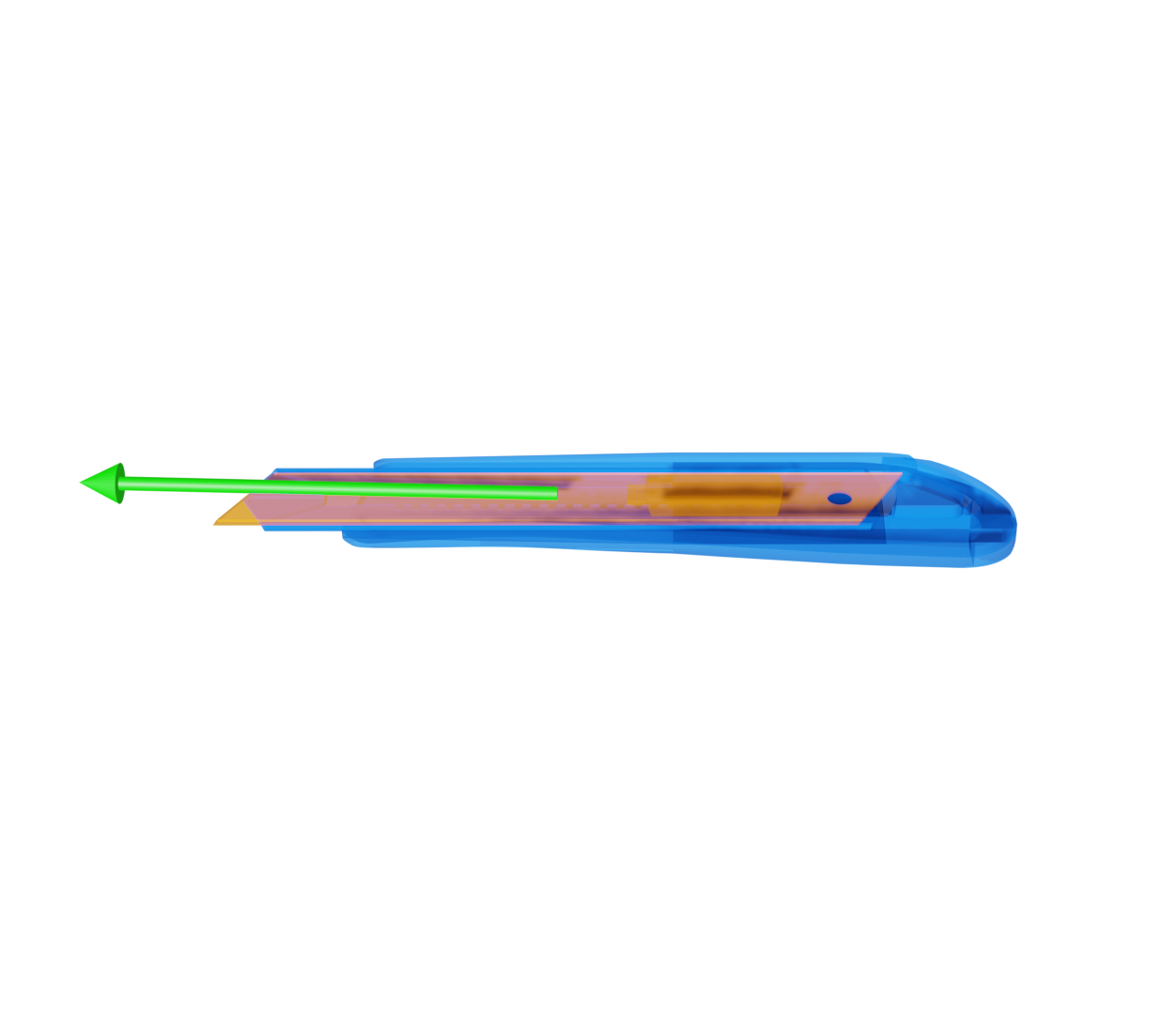} &
        \includegraphics[width=0.1\linewidth]{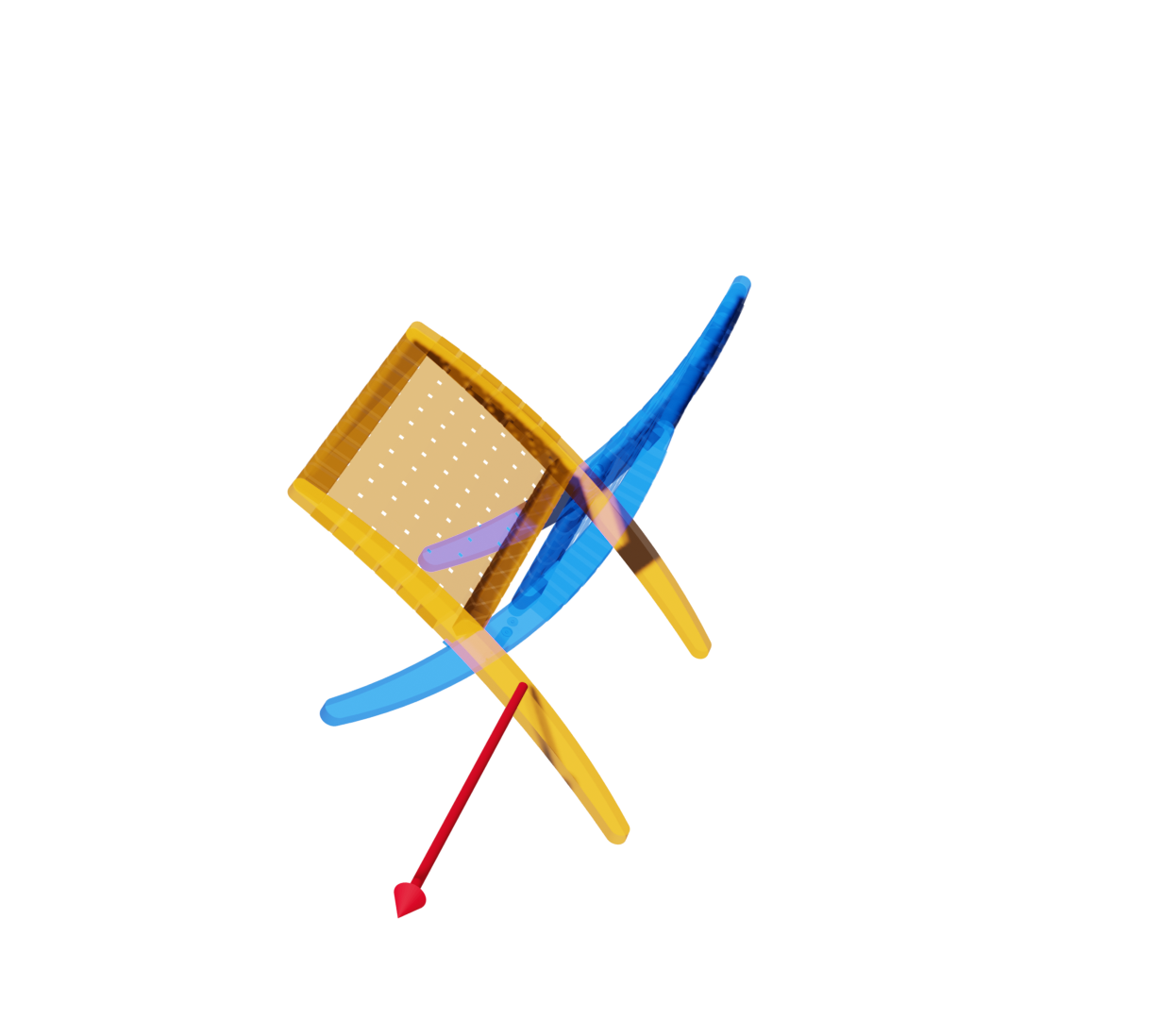} &
        \includegraphics[width=0.1\linewidth]{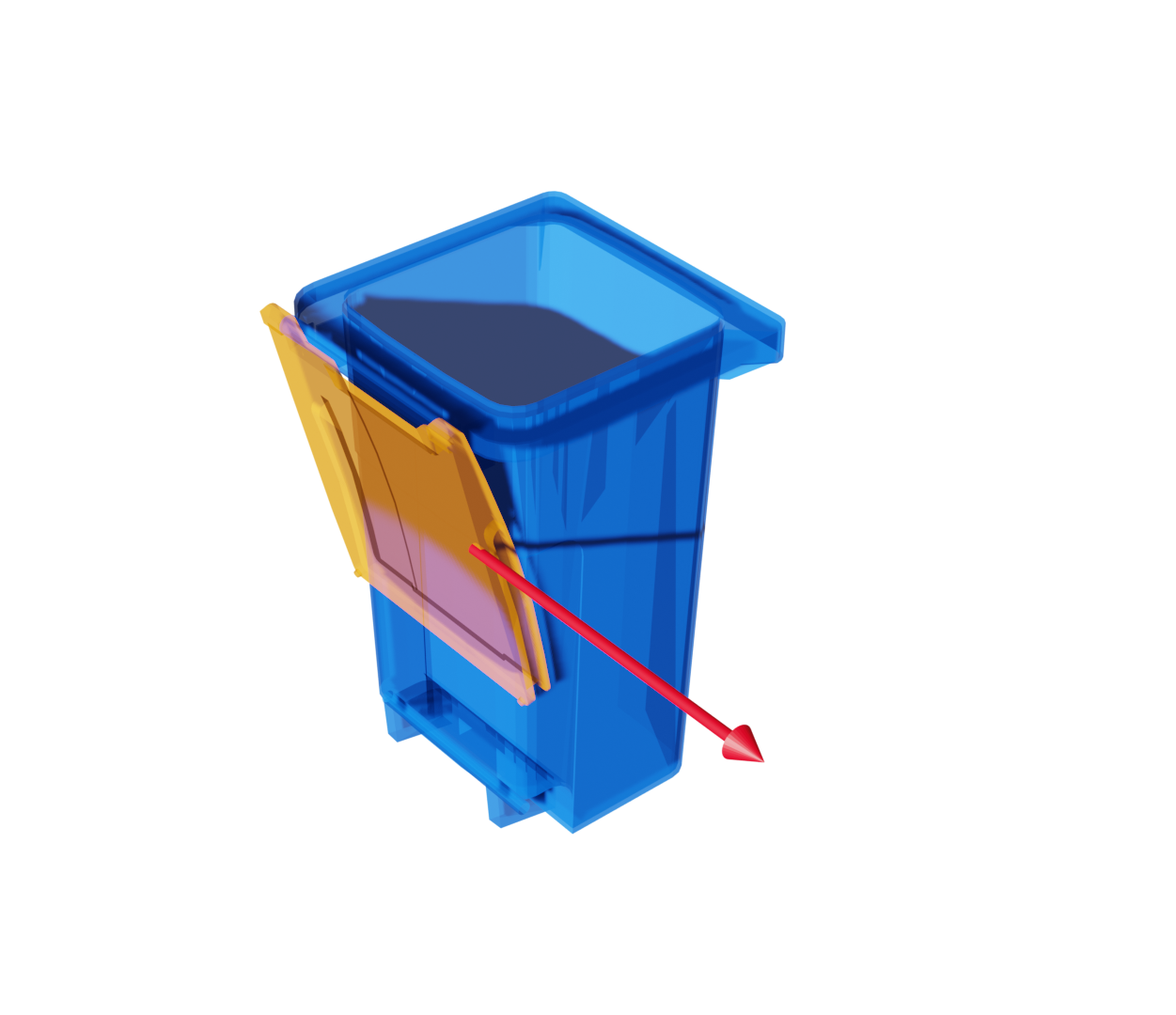} &
        \includegraphics[width=0.1\linewidth]{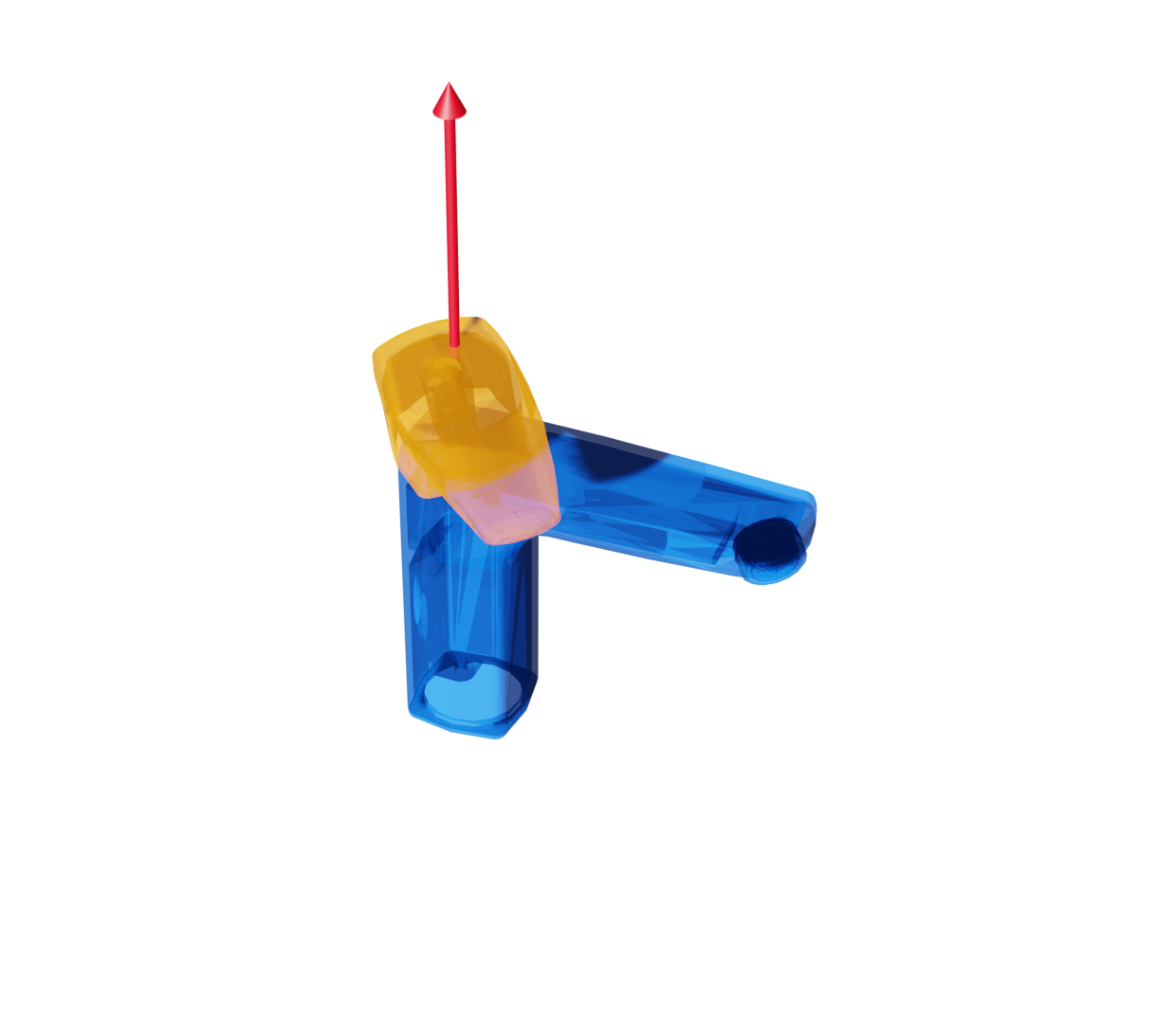} &
        \includegraphics[width=0.1\linewidth]{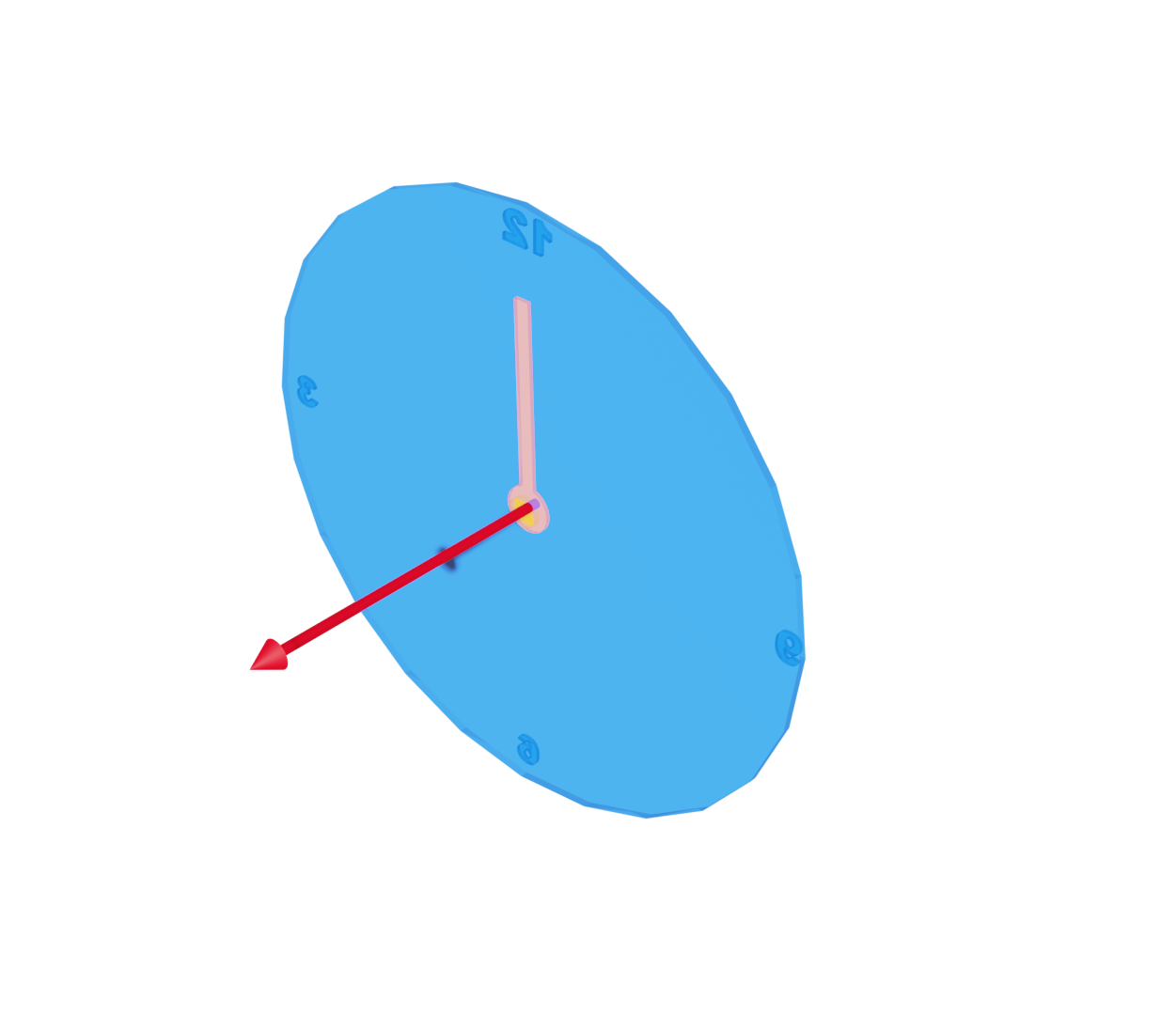} &
        \includegraphics[width=0.1\linewidth]{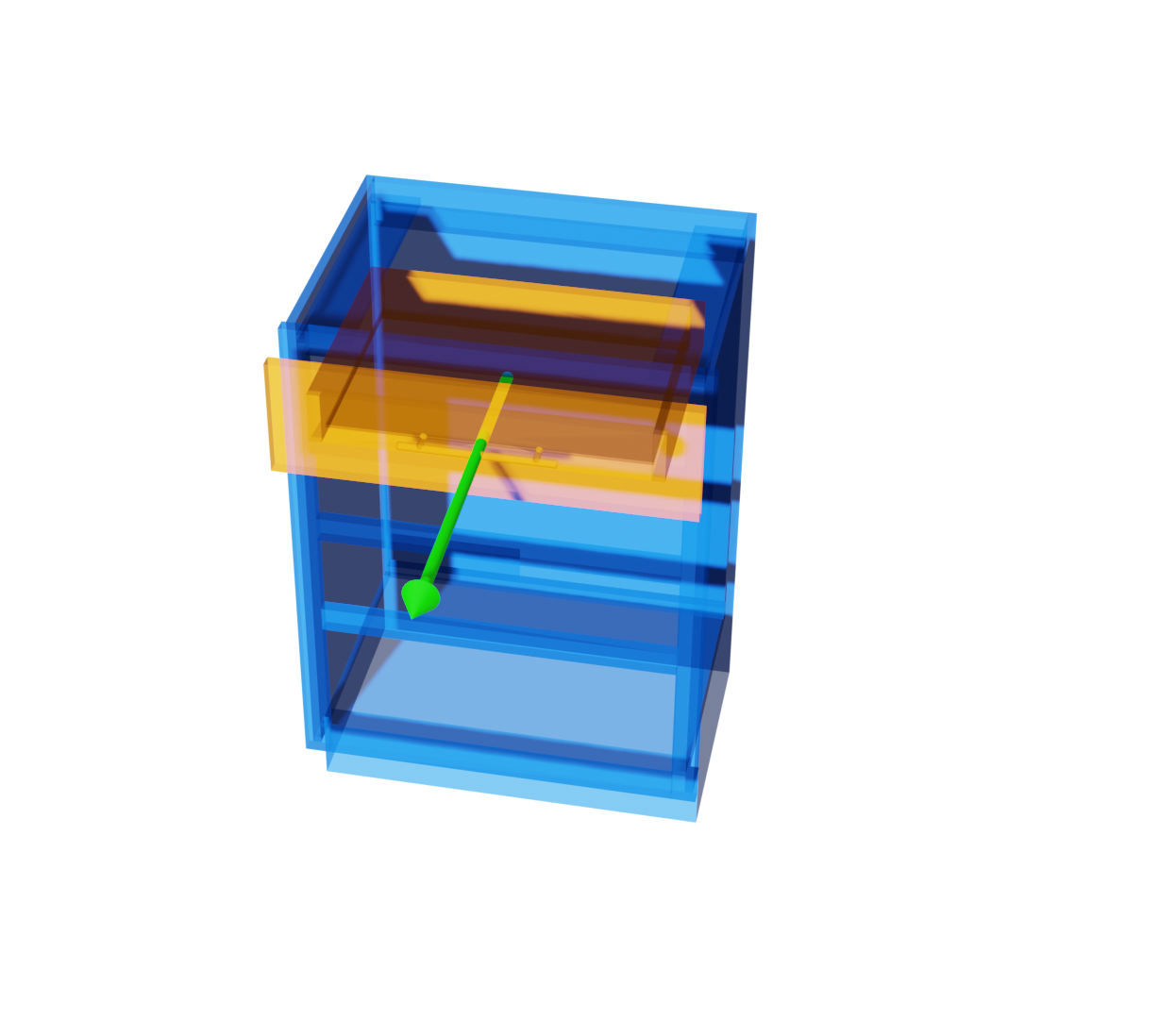} &
        \includegraphics[width=0.1\linewidth]{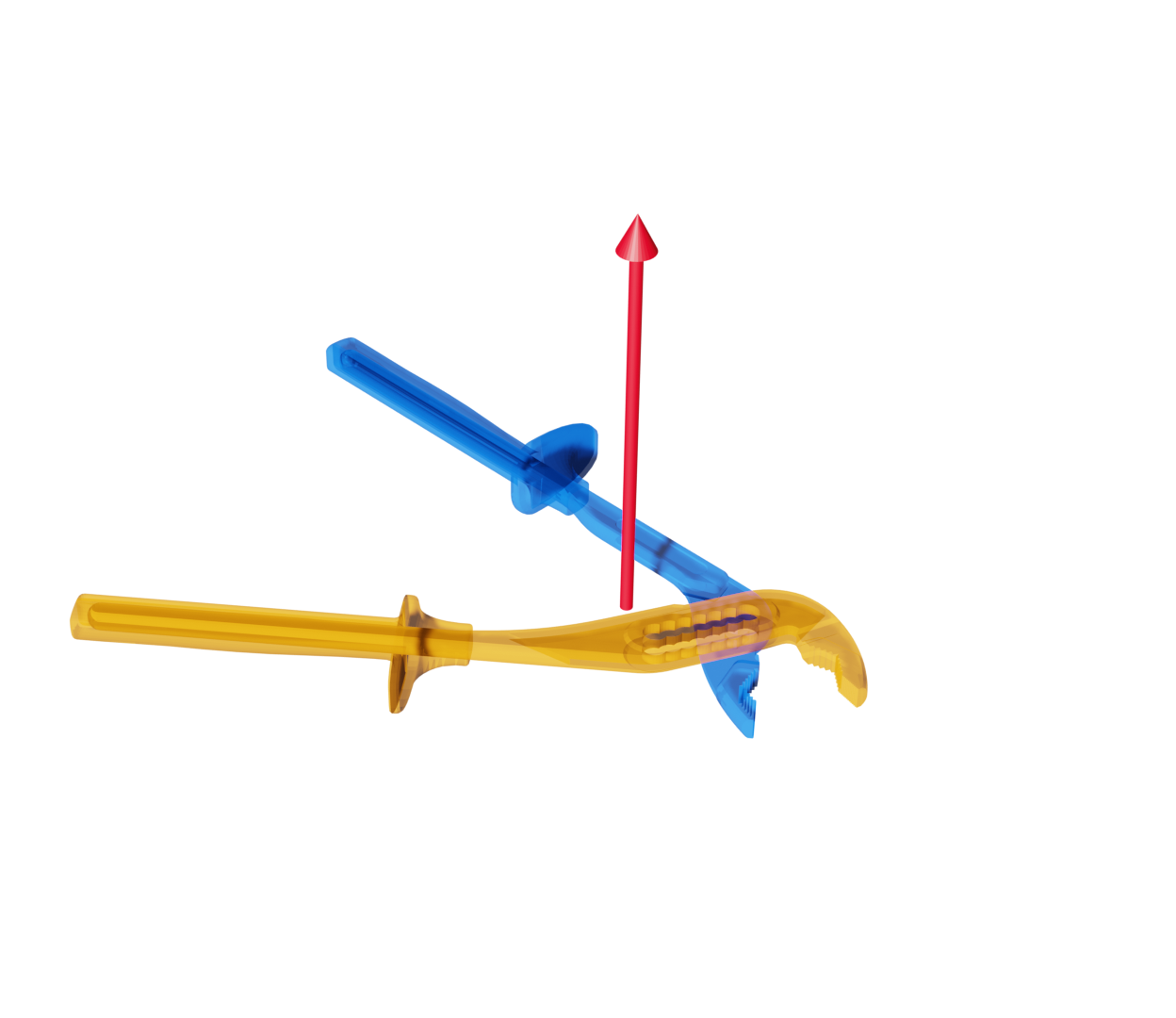}
        \\
        \raisebox{2.5em}{Ground Truth} & 
        \includegraphics[width=0.1\linewidth]{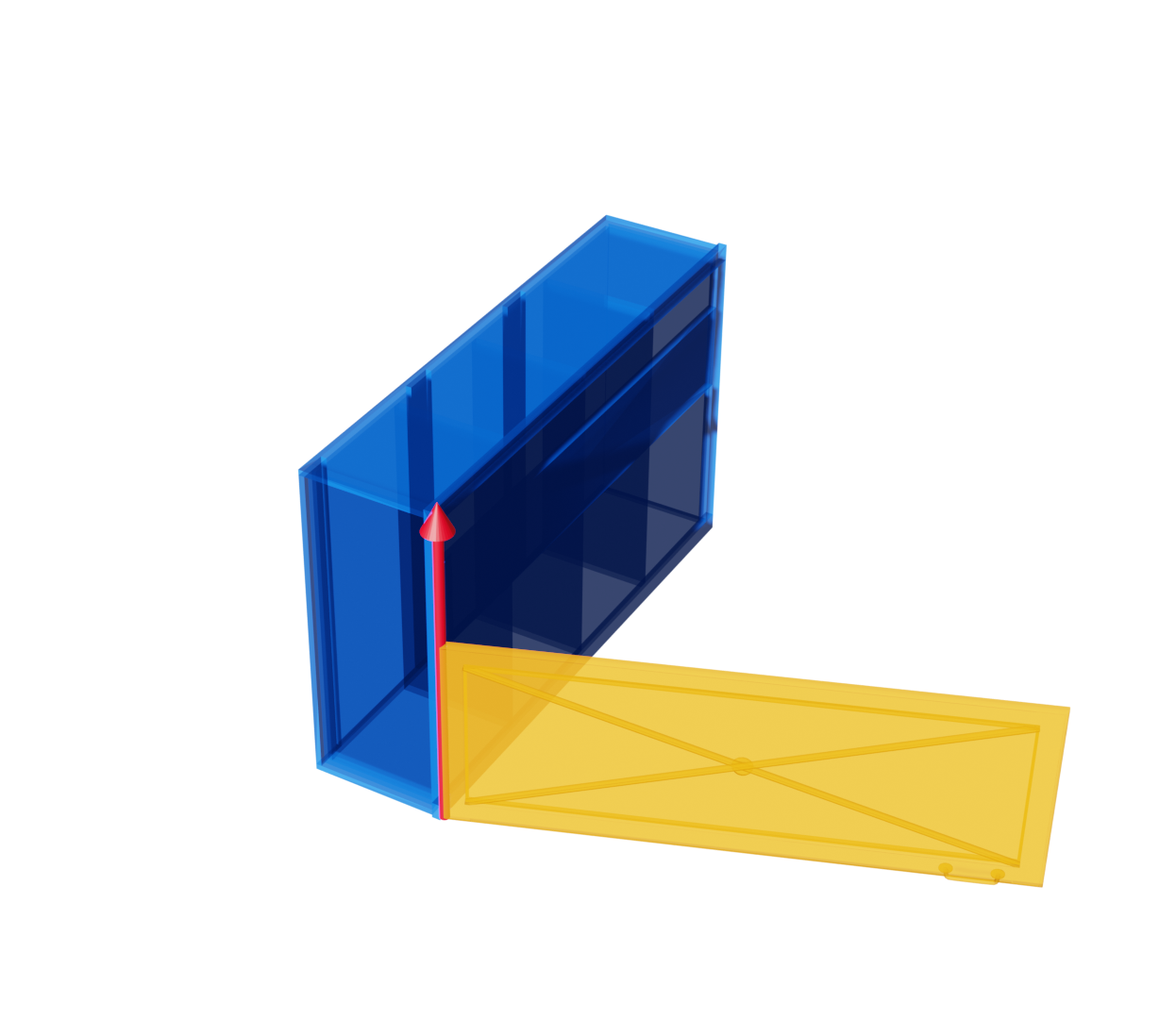} &
        \includegraphics[width=0.1\linewidth]{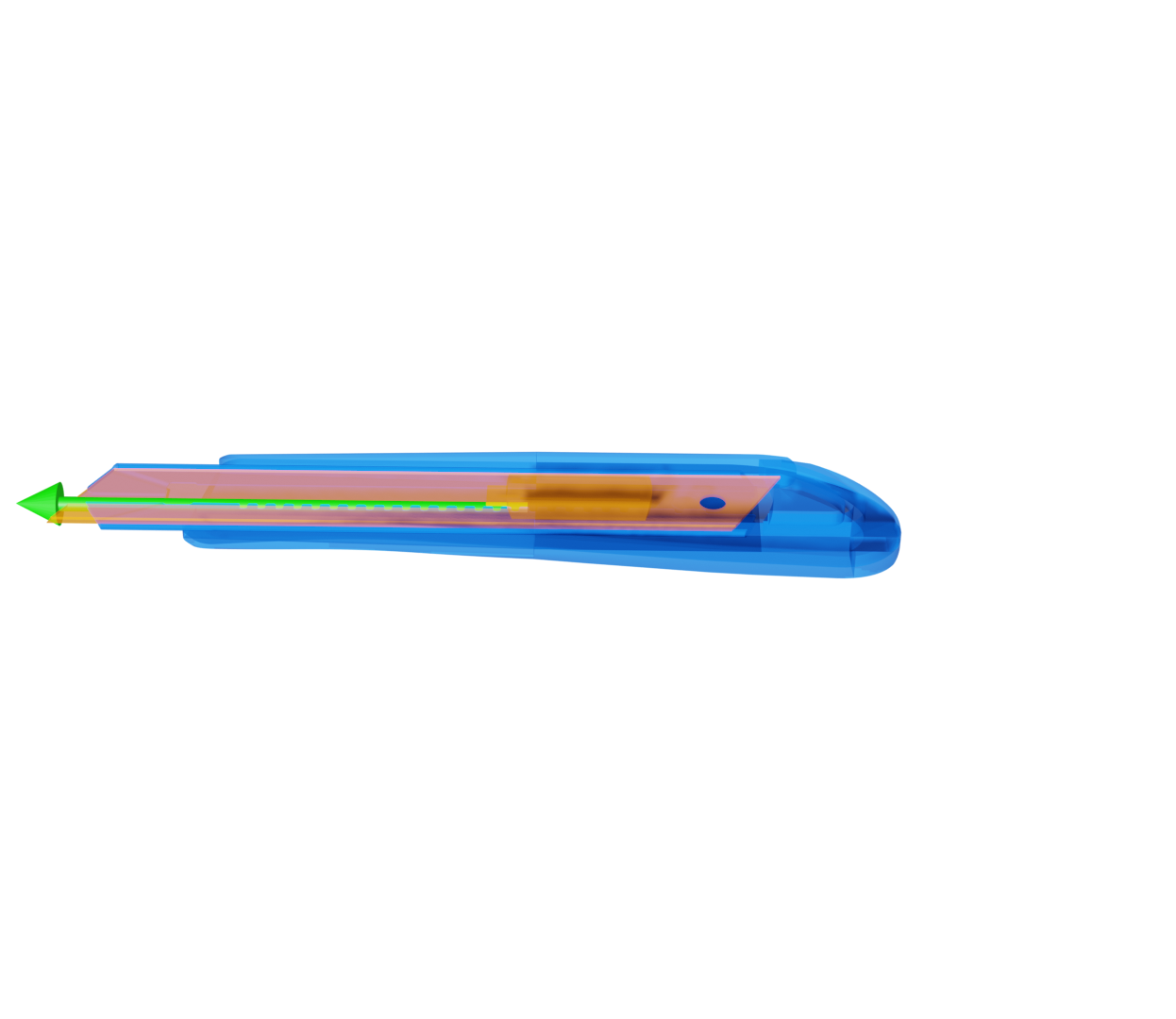} &
        \includegraphics[width=0.1\linewidth]{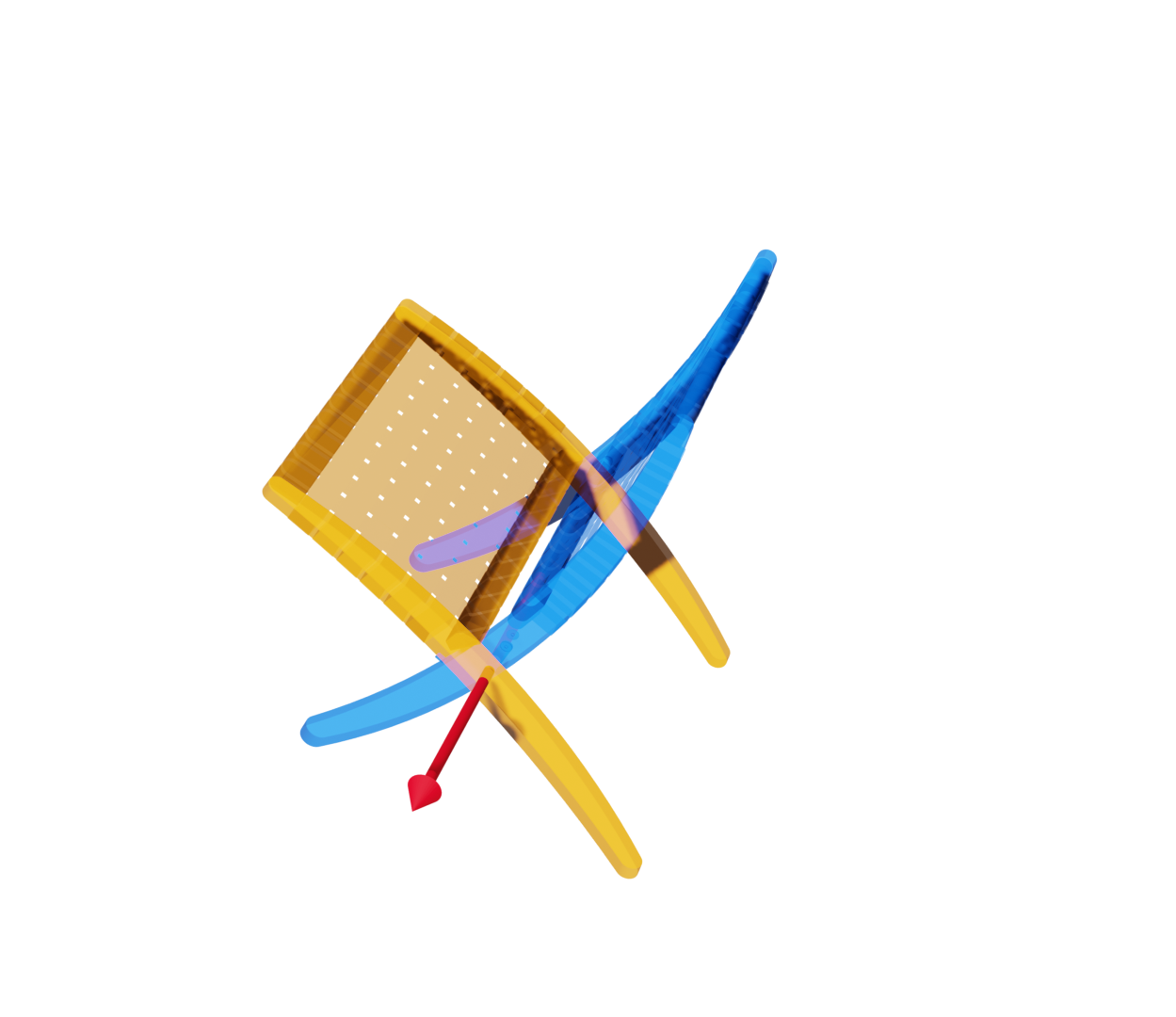} &
        \includegraphics[width=0.1\linewidth]{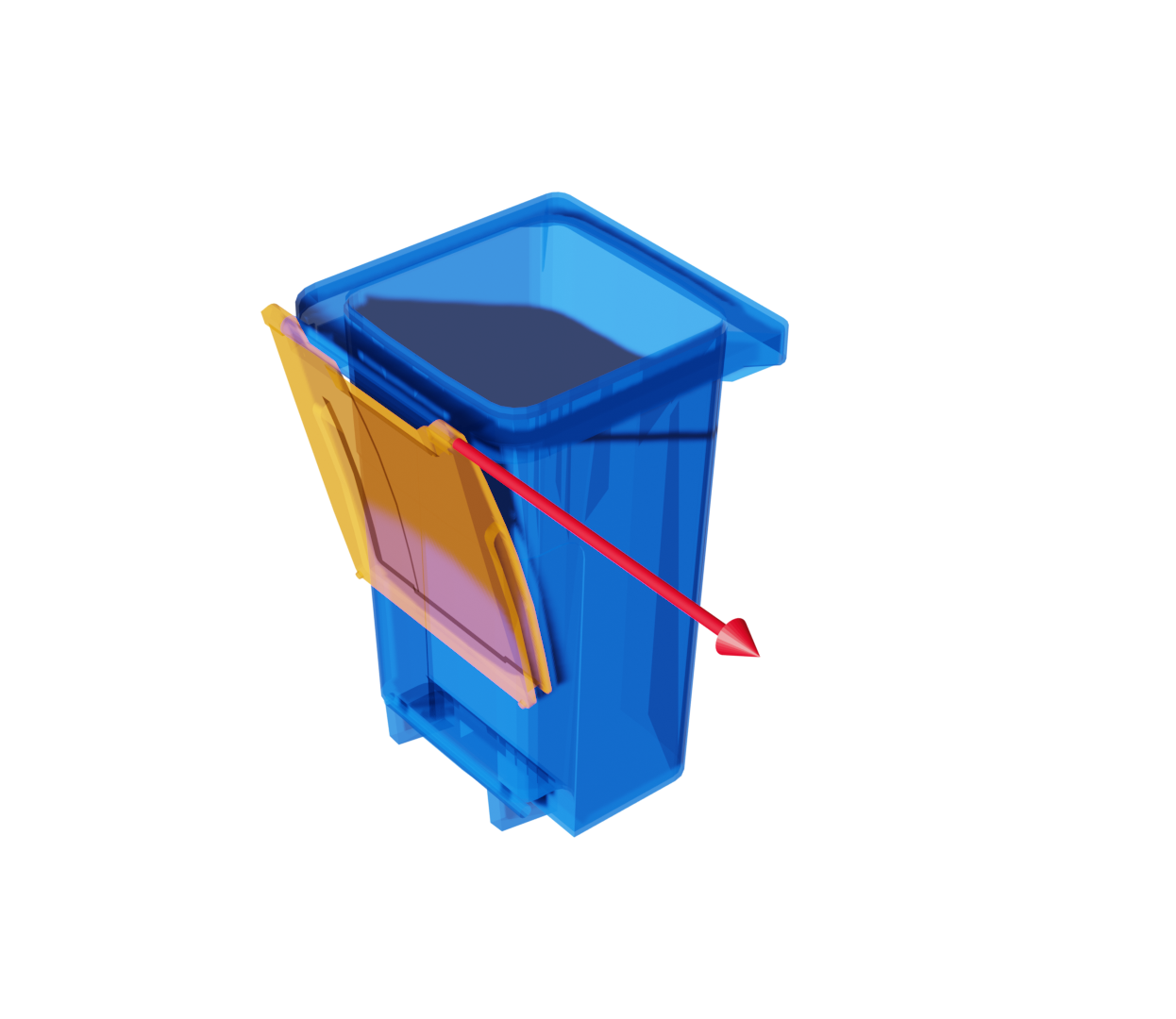} &
        \includegraphics[width=0.1\linewidth]{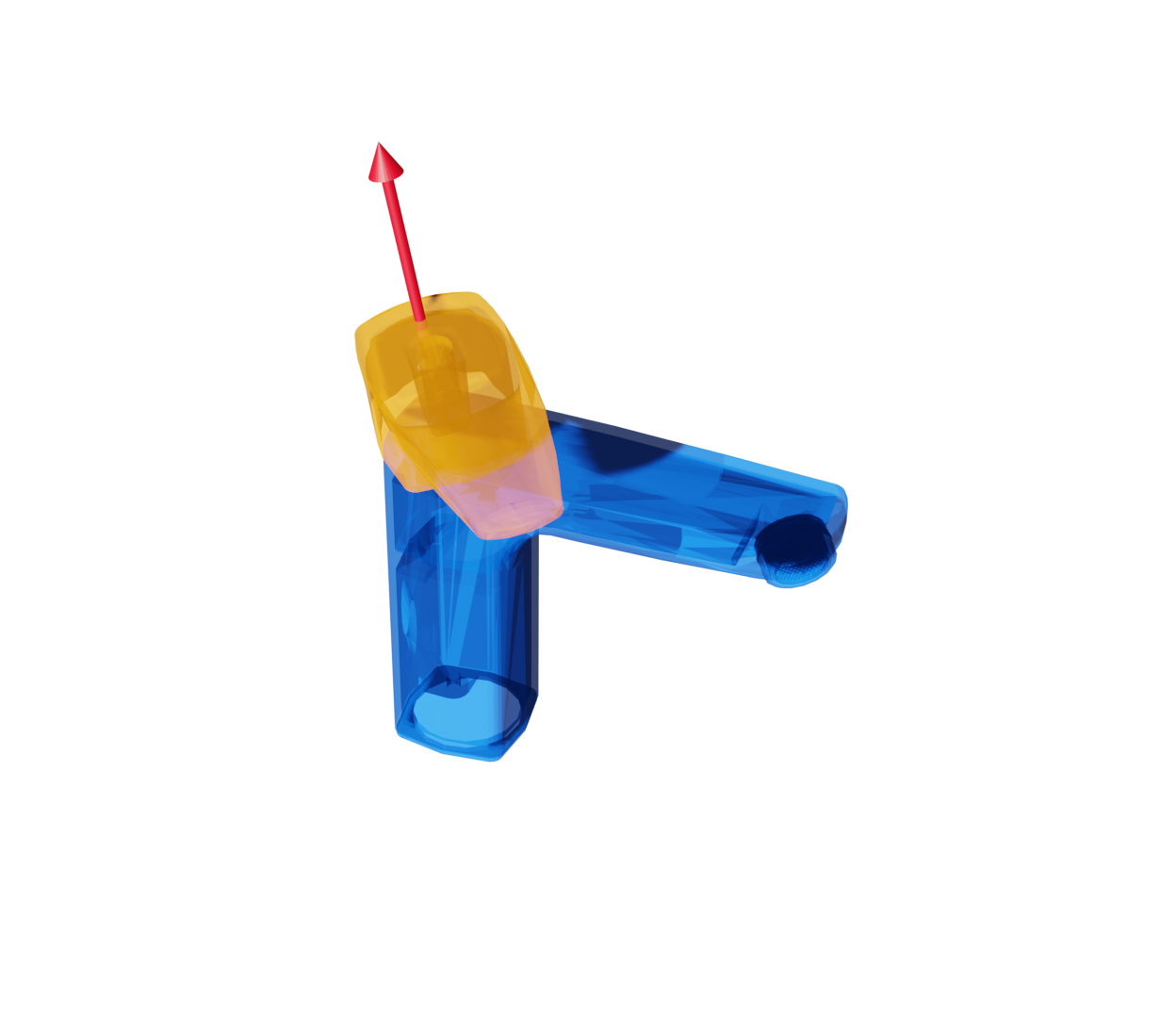} &
        \includegraphics[width=0.1\linewidth]{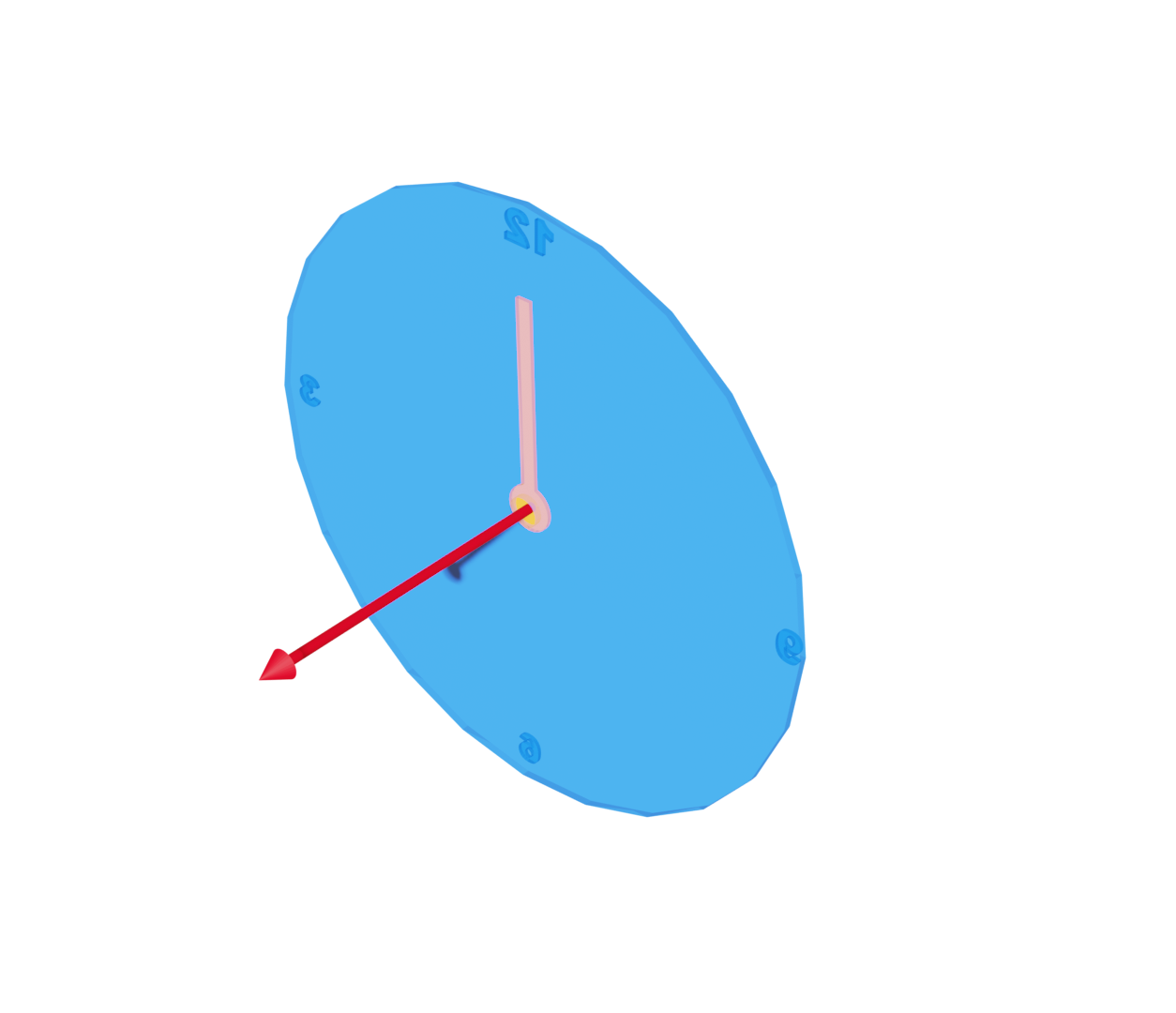} &
        \includegraphics[width=0.1\linewidth]{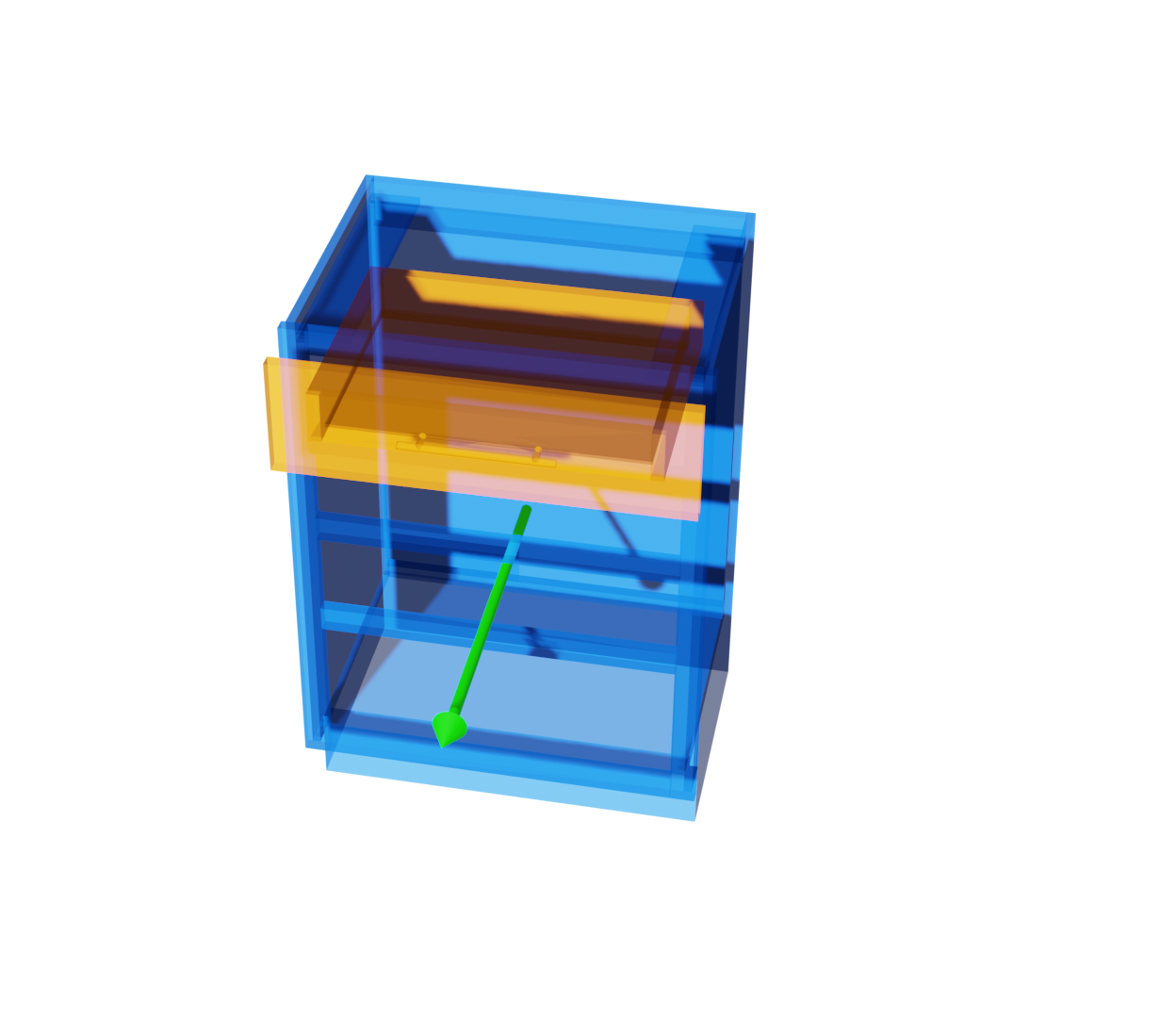} &
        \includegraphics[width=0.1\linewidth]{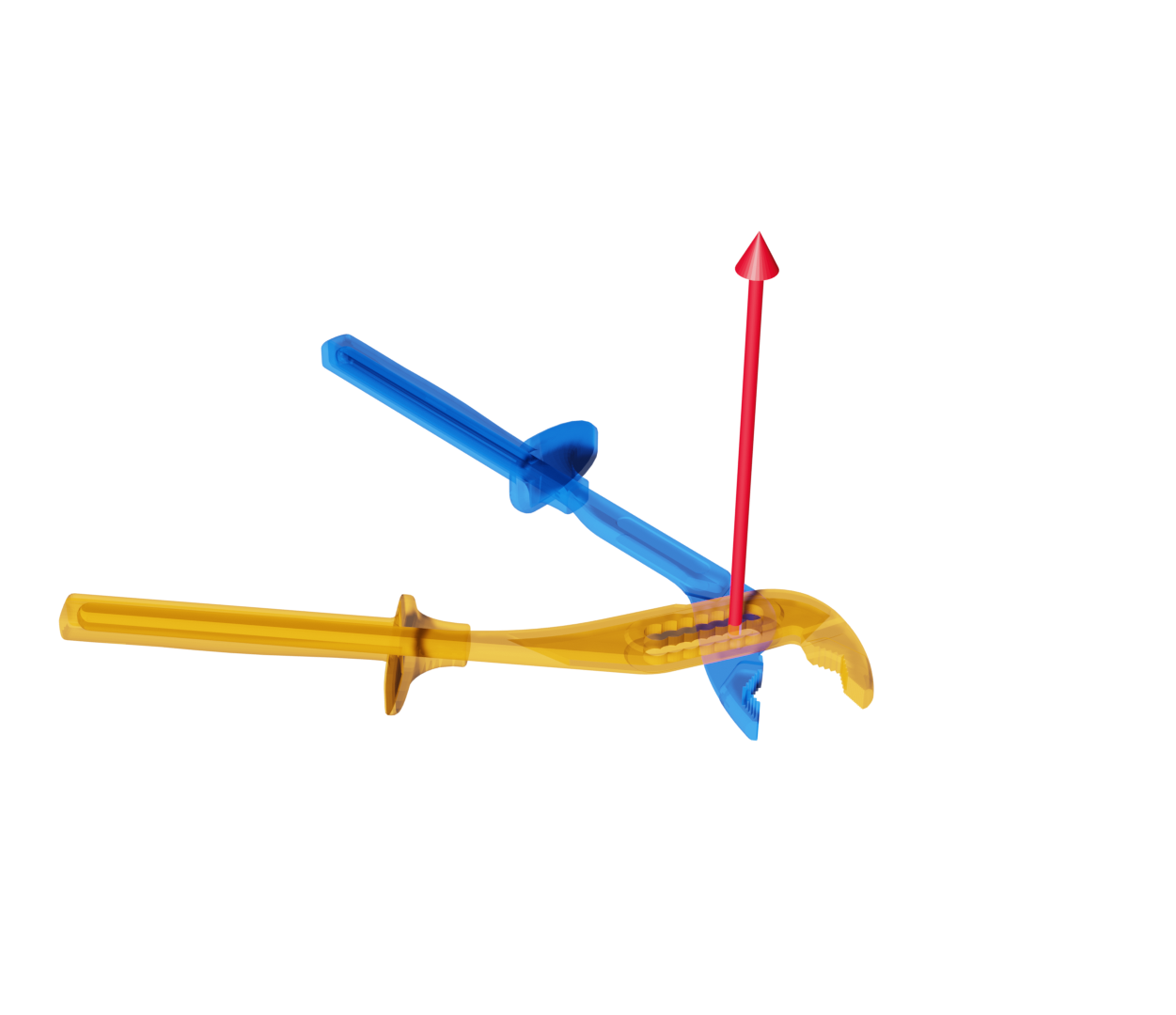}
    \end{tabular}
    \caption{Qualitative examples of our method vs. BaseNet on predicting motion parameters for PartNet-Mobility.}
    \label{fig:comparison}
\end{figure*}

\paragraph{Comparison to baselines}
There is no existing prior work which performs unsupervised kinematic motion detection; thus, we compare to existing supervised approaches.
We stress that these approaches require training data, whereas ours can be applied on categories of shapes that have never been seen before.
We compare our method to two supervised baselines:
\begin{packed_itemize}
    \item \emph{BaseNet:} This network takes in a point cloud of a shape, where each point carries an extra one-hot dimension indicating whether it is a member of the part for which motion should be predicted. This point cloud is passed through a PointNet encoder, the output of which is fed into four separate fully connected branches which predict motion type (hinge, prismatic, static), axis ($\real^3$), center of rotation ($\real^3$), and range ([min, max] $\in \real^2$). The entire network is trained jointly, with cross entropy loss for the motion type branch and MSE loss for the other branches.
    \item \emph{Shape2Motion (S2M):} A network which jointly predicts part segmentations and part motions for point clouds~\cite{Shape2Motion}. For fair comparison with our method, we modify the network to take ground truth part segmentations and only predict motions.
\end{packed_itemize}
For both supervised methods, we evaluate them with and without a pre-alignment step, in which objects are rotated above the up axis to roughly align them via chamfer distance.

See supplemental for more details on these baselines.
We split our filtered collection of joints 60\%/40\% into train/test sets.
The supervised baselines are trained on the train set, our method is run on all joints, and all methods are evaluated on the test set.

We evaluate test set motion predictions with these metrics:
\begin{packed_itemize}
    \item \emph{Motion type accuracy (Type Acc)}: percentage of joints whose motion type (static, prismatic, hinge) is correctly predicted. Since Shape2Motion does not handle `static' parts, we omit that label for its training and evaluation.
    \item \emph{Axis angular error (Axis Err)}: mean difference (in degrees) between predicted axis directions and their ground-truth values.
    \item \emph{Rotation center error (Center Err)}: mean distance (in percentage of the part's bounding box diagonal length) between predicted centers of rotation and ground truth rotation axes.
    \item \emph{Range of motion accuracy (Range IoU)}: mean intersection over union between predicted and ground truth ranges of motion.
\nolistbottomspace
\end{packed_itemize}
Table~\ref{tab:comparison_basenet_summary} and Table~\ref{tab:comparison_s2m_summary} show quantitative results of this comparison; a breakdown by category is in supplemental.
Our method has complementary strengths to BaseNet and Shape2Motion: These two supervised methods are better at predicting motion type; ours is better at motion parameters. BaseNet's and Shape2Motion's higher type accuracy is not surprising: ternary motion type is the easiest quantity for a network to learn to predict (given the limited training data, it is harder to learn to regress continuous motion parameters). Motion type is also the easiest information for a human annotator to provide. A hybrid approach might be best in practice: combining weakly-supervised motion type labeling with our approach for predicting motion parameters. Shape2Motion doesn't outperform BaseNet by a large margin, which we hypothesize is due to some combination of (1) the network being complicated and requiring more training data, (2) training on ground truth segmentations rather than jointly inferring them resulting in a weaker learned representation. 

\begin{figure}[t!]
    \centering
    \setlength{\tabcolsep}{1pt}
    \begin{tabular}{rccc}
         \raisebox{2.5em}{Ours} & 
         \includegraphics[width=0.18\linewidth]{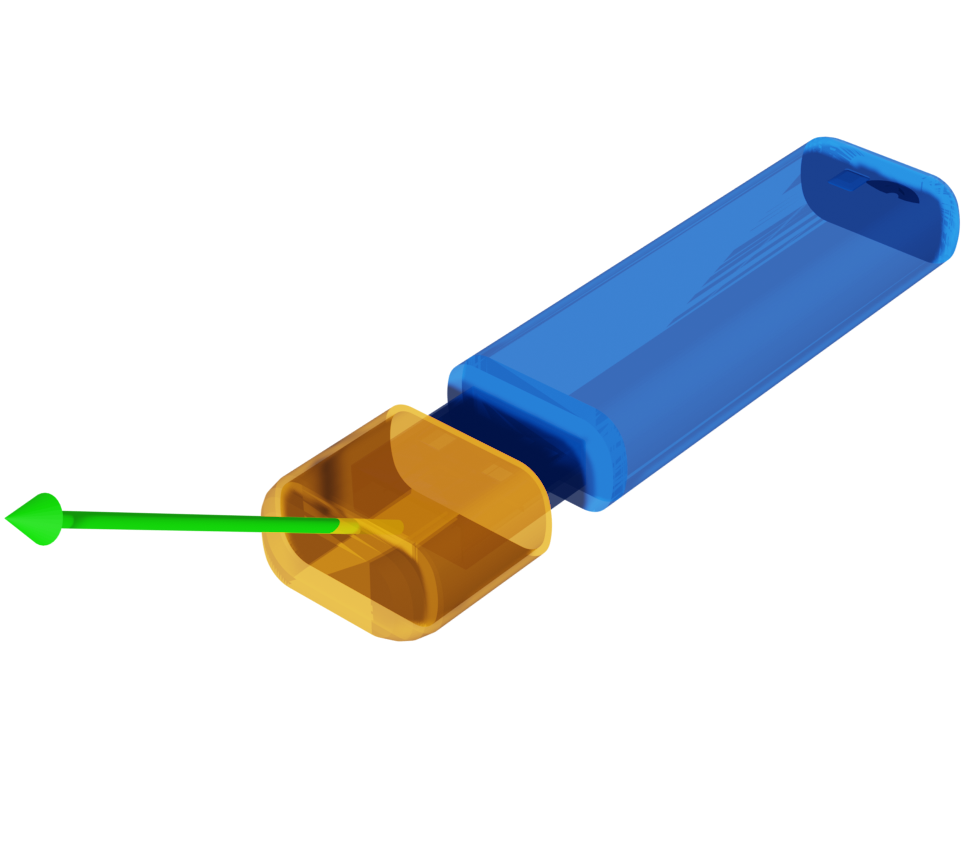} &
         \includegraphics[width=0.18\linewidth]{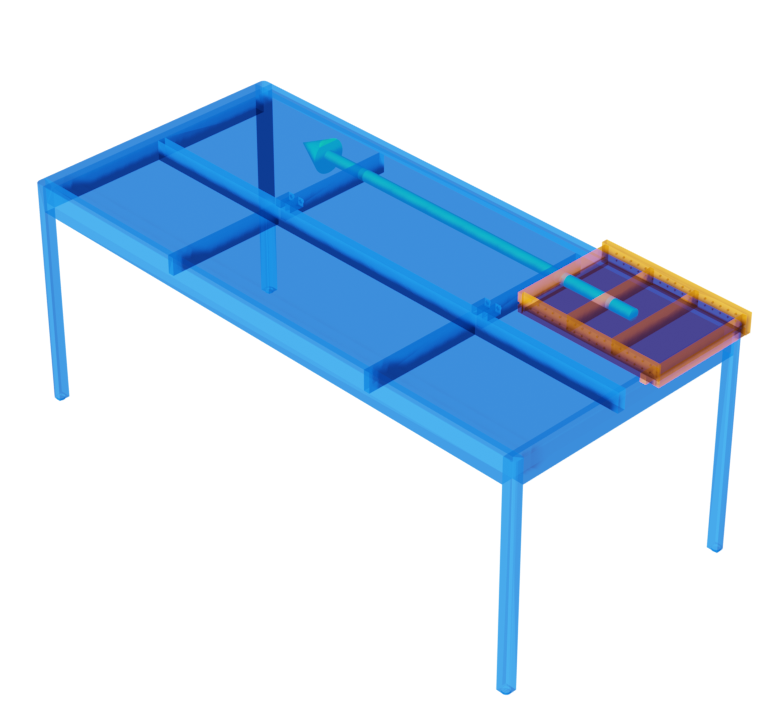} &
         \includegraphics[width=0.18\linewidth]{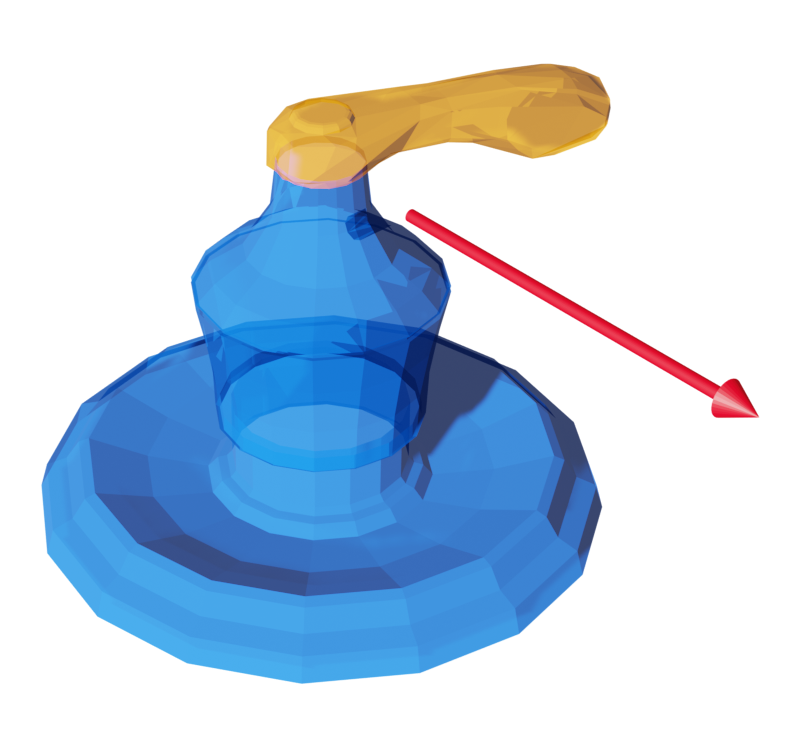}
         \\
         \raisebox{2.5em}{GT} & 
         \includegraphics[width=0.18\linewidth]{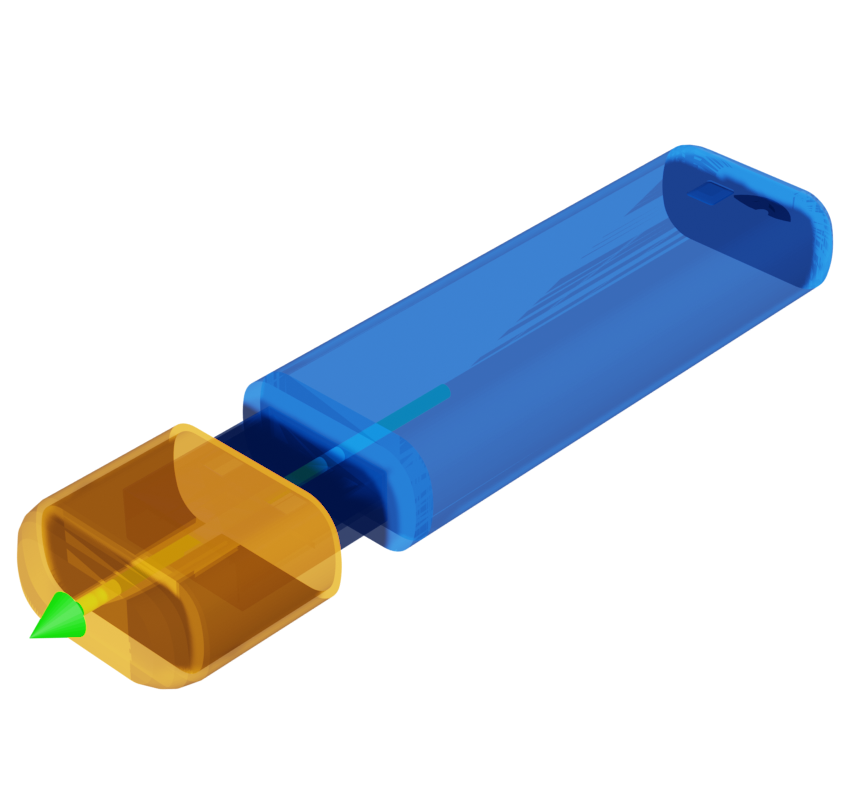} &
         \includegraphics[width=0.18\linewidth]{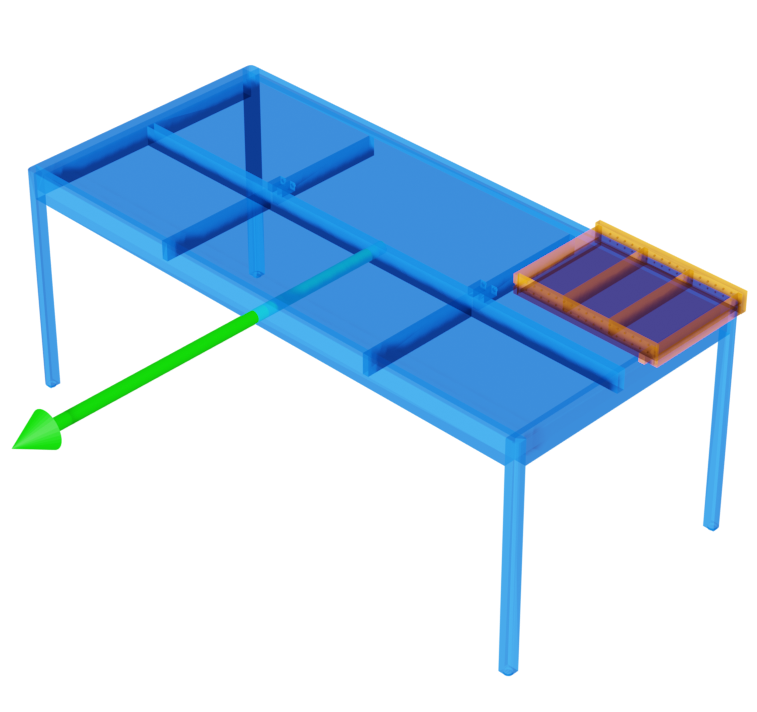} &
         \includegraphics[width=0.18\linewidth]{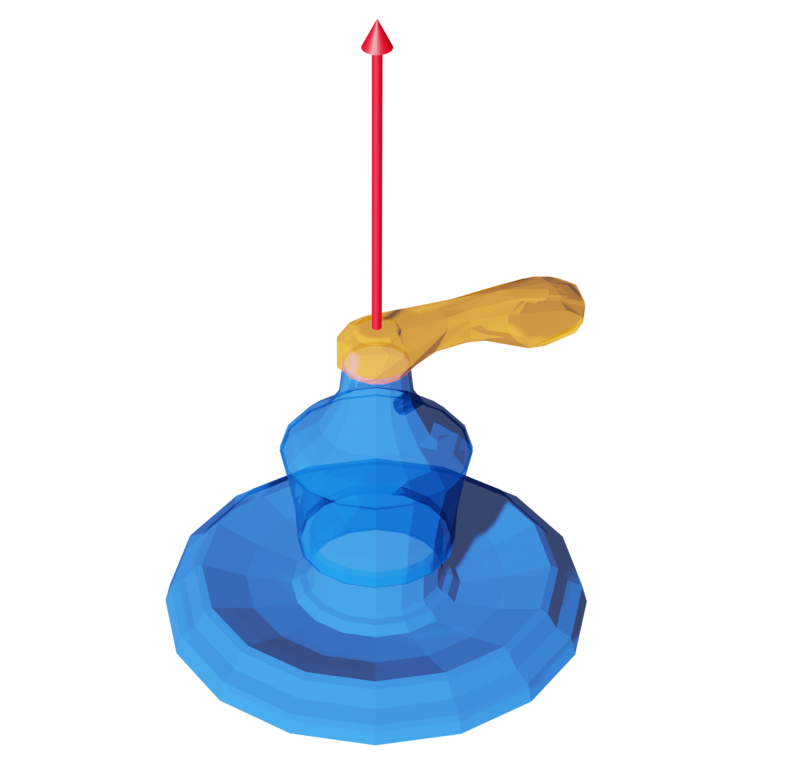}
    \end{tabular}
    \caption{Typical failure cases (USB, Table, Faucet).
    The Table and Faucet examples disagree with ground-truth annotations but are still plausible. 
    }
    \label{fig:failures}
\end{figure}

Figure~\ref{fig:comparison} qualitatively compares our method to BaseNet.
Our physically-inspired approach degrades more gracefully than the learning-based approach, which can produce nonsensical outputs due to insufficient training data.
Figure~\ref{fig:failures} additionally shows some of the ways our method gracefully fails.
For the table and faucet joints shown, the motions found are different from PartNet-Mobility's ground truth but are nonetheless plausible: the table drawer could slide side-to-side rather than front-to-back; the faucet handle could rotate vertically rather than / as well as laterally.
The USB failure may be due to incorrect grouping of joints, not enough similar joints to group, or suboptimal loss weight hyperparameters.

\paragraph{Sensitivity to amount of pose variation}
Table~\ref{tab:pose_variation} shows how our method performs on Doors when the amount of pose variation is increased (see supplemental for details)
Results are poor with no variation, but even a little increases performance dramatically.

\begin{table}[t!]
    \centering
    \scriptsize
    \caption{Analyzing how the performance of our method on Doors varies with amount of pose variation in the input shape collection. The number of target joints $k = 3$ for this experiment.}
    \resizebox{\columnwidth}{!}{\begin{tabular}{lcccc}
        \toprule
        \textbf{Pose variation level} & \textbf{Type Acc$\uparrow$} & \textbf{Axis Err ($^\circ$)$\downarrow$} & \textbf{Center Err (\%)$\downarrow$} & \textbf{Range IoU$\uparrow$}
        \\
        \midrule
        0 & 0.54 & 39.21 & 12.70 & 0.03 \\
        1 & 0.89 & 0.02 & 16.33 & 0.25 \\
        2 & 0.91 & 0.03 & 10.09 & 0.42 \\
        3 & 0.91 & 0.09 & 10.48 & 0.49 \\
        4 & 0.91 & 0.09 & 4.59 & 0.37 \\
        5 & 0.91 & 0.00 & 6.34 & 0.43 \\
        \bottomrule
    \end{tabular}}

    \label{tab:pose_variation}
\end{table}

\paragraph{Running time}
Our method iterates complex optimizations on multiple joint groups, so it can be time-consuming: one iteration on a set of 100 joints with target joint number $k = 5$ takes our PyTorch implementation about 1 hour on an 8-core Intel i9 machine with 32 GB RAM and a NVIDIA RTX 2080Ti GPU.
However, compute time can be small price to pay to avoid human annotation time.

\paragraph{Additional results}
The supplemental contains additional qualitative results, a study on sensitivity to number of target joints, an ablation study on model components, and a visualization of the learned joint embedding space.
\section{Conclusion}

In this paper, we presented an unsupervised approach for discovering part motions in part-segmented 3D shape collections.
Our approach is based on \emph{category closure}: a valid articulation of a shape's parts should not change the semantic category of that shape.
We operationalize this insight via an algorithm that finds motion parameters for a joints that transforms into other joints from the same category.
Our approach successfully rediscovers a large percentage of motions in the PartNet-Mobility dataset, often outperforming a supervised motion prediction network.

Our system has some limitations.
It cannot handle moving parts whose size is a small fraction of their base parts, due to point cloud resolution limits.
More fundamentally, our method assumes that the input shape collection contains similar joints in different poses.
This is often true (e.g. lamp arms, swivel chair bases), but some shapes are typically modeled in a canonical pose (e.g. cabinet doors are usually modeled as closed).
These shapes may pose a challenge for our method (or any unsupervised method).

Finally, our method assumes the input part segmentation is fairly consistent, e.g. it would not perform well on cabinets if each cabinet door was broken into a different number of segments.
In the future, we would like to extend our method to handle such data by developing a system for proposing ways to group different part fragments.
Combined with an automatic shape over-segmentation method, this would allow our method to discover shape parts as well as their motions without any supervision.

\section{Acknowledgements}
This work was funded in part by NSF Award \#1941808.
Daniel Ritchie is an advisor to Geopipe and owns equity in the company. Geopipe is a start-up that is developing 3D technology to build immersive virtual copies of the real world with applications in various fields, including games and architecture. Srinath Sridhar was supported by a Google Research Scholar Award.

{\small
\bibliographystyle{ieee_fullname}
\bibliography{main}
}

\clearpage
\setcounter{section}{0}
\renewcommand\thesection{\Alph{section}}
\renewcommand\thesubsection{\thesection.\arabic{subsection}}
\section{Supplemental Material}






\subsection{Data Preprocessing}

Here we describe how we convert a segmented shape into joints. Given a shape as a collection of part meshes, we first uniformly sample the surface of each part with points, and then for each pair of parts, we compute the nearest distance between any two points from each part. If this distance is below a threshold, these two parts will be determined as they are connected to each other. With this process, we can build a part graph over the shape by adding edges between parts. Once the graph is built, for each part, we will first check if it is a cut part(node) of the part graph (A cut node of a graph is the node once removed, will make the graph disconnected into multiple components). If it is not a cut part, we will use this part as the moving part and pick the largest connected part as the initial base part and the final point cloud for base part is acquired by sampling all geometry inside the bounding box of this initial base part. If it is a cut part, which means some other part may need to move with this part, in this case, we will run a Breadth First Search on the part graph to retrieve all parts that are left alone as the consequence of assuming this part is removed. (By alone we mean it becomes a floating part that is not connected to any boundary(such as the side faces of the shape bounding box). All these retrieved floating parts are combined together with the current part to form the moving part. Once the moving part and the base part are determined, the joint is formed naturally.

For the pre-alignment for the two supervised baselines, in addition to the global shape alignment, the GT axes projected(by reversing the axis and range, which defines an equivalent motion) to the hemisphere defined by vector [1, 1, 1] for easier axis regression.

\subsection{System implementation details}
\begin{figure}[t!]
    \centering
    \includegraphics[width=\linewidth]{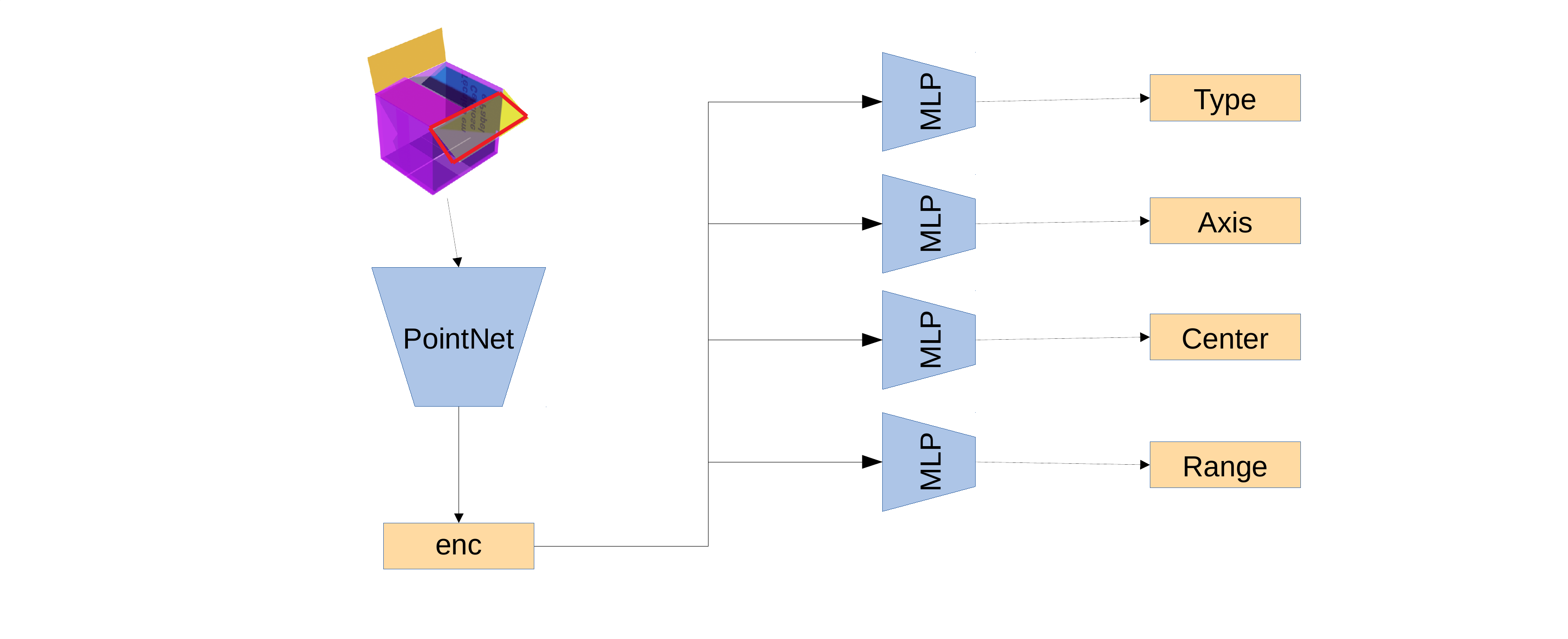}
    \caption{Architecture of the supervised BaseNet model}
    \label{fig:basenet}
\end{figure}

We also describe the learning model and hyper-parameters used throughout our experiments:
\begin{itemize}
    \item For model architecture of the joint encoder, Please see Table ~\ref{tab:encoder_pointnet}
    \item For model architecture of the supervised BaseNet network: Please see Table ~\ref{tab:basenet_pointnet} and Table ~\ref{tab:basenet_mlp} and fig ~\ref{fig:basenet}
\end{itemize}

\subsection{Hyper-parameters}

Joint encoding dimension: 48\\
Initial optimization encoder training epoch: 150\\
Initial optimization training learning rate: 0.04\\
Initial optimization inner group learning epoch: 700\\
Initial optimization joint group size: 16\\
Iterative optimization joint group size: 4 \\
Iterative optimization inner group learning rate: 0.008\\
Iterative optimization inner group learning epoch: 800\\
Iterative optimization iterations: 5 \\
Validate optimization group size: 16 \\

\noindent
Rotation small joint motion penalty weight: 0.001\\
Rotation large local alignment translation penalty weight: 0.0025\\
Rotation large alignment transform penalty weight: 0.00\\
Rotation collision penalty weight: 0.01\\
Rotation detachment penalty weight:0.01\\
Rotation center detachment penalty weight: 5.0\\
Rotation large deformation penalty weight: 0.001\\

\noindent
Translation small joint motion penalty weight: 0.001\\ 
Translation large local alignment translation penalty weight: 0.005\\
Translation large alignment transform penalty weight: 0.00\\
Translation collision penalty weight: 0.5\\
Translation detachment penalty weight: 0.5\\
Translation large deformation penalty weight: 0.001\\

\noindent
BaseNet learning rate: 0.0001\\
BaseNet training epoch: 150\\
BaseNet motion type loss weight 5.0\\
BaseNet motion direction loss weight 100.0\\
BaseNet motion center loss weight 20.0\\
BaseNet motion range loss weight 0.001\\

\begin{table}[t!]
    \centering
    \small
    \caption{
    Detailed architecture of the PointNet Joint Encoder we used in the project 
    }
    \begin{tabular}{@{}c@{}}
        \toprule
        \textbf{PointNet}
        \\
        \midrule
        \textbf{Conv1d} $\left(4, 64, 1 \right)$\\
        \textbf{Batchnorm1d}\\
        \textbf{LeakyRelu}\\
        \textbf{Conv1d} $\left(64, 128, 1 \right)$\\
        \textbf{Batchnorm1d}\\
        \textbf{LeakyRelu}\\
        \textbf{Conv1d} $\left(128, 512, 1 \right)$\\
        \textbf{MaxPool}\\
        \textbf{FC}$\left( 512 \times 256 \right)$\\
        \textbf{Batchnorm1d}\\
        \textbf{LeakyRelu}\\
        \textbf{FC}$\left( 256 \times enc\_dim \right)$\\
        \bottomrule
    \end{tabular}
    \vspace{0.1em}

    \label{tab:encoder_pointnet}
    \vspace{0.1mm}
\end{table}

\begin{table}[t!]
    \centering
    \small
    \caption{
    Detailed architecture of the PointNet model used in the BaseeNet method 
    }
    \begin{tabular}{@{}c@{}}
        \toprule
        \textbf{PointNet}
        \\
        \midrule
        \textbf{Conv1d} $\left(4, 64, 1 \right)$\\
        \textbf{Batchnorm1d}\\
        \textbf{LeakyRelu}\\
        \textbf{Conv1d} $\left(64, 128, 1 \right)$\\
        \textbf{Batchnorm1d}\\
        \textbf{LeakyRelu}\\
        \textbf{Conv1d} $\left(128, 512, 1 \right)$\\
        \textbf{MaxPool}\\
        \textbf{FC}$\left( 512 \times 256 \right)$\\
        \textbf{Batchnorm1d}\\
        \textbf{LeakyRelu}\\
        \textbf{FC}$\left( 256 \times 256 \right)$\\
        \bottomrule
    \end{tabular}
    \vspace{0.1em}

    \label{tab:basenet_pointnet}
    \vspace{0.1mm}
\end{table}

\begin{table}[t!]
    \centering
    \small
    \caption{
    Detailed architecture of the Single-Layer Perceptron Network used in the BaseNet method 
    }
    \begin{tabular}{@{}c@{}}
        \toprule
        \textbf{MLP}
        \\
        \midrule
        \textbf{FC}$\left( 256 \times motion\_parameter\_dim \right)$\\
        \bottomrule
    \end{tabular}
    \vspace{0.1em}

    \label{tab:basenet_mlp}
    \vspace{3mm}
\end{table}

\subsection{Shape2Motion Baseline}
Shape2Motion~\cite{Shape2Motion} jointly predicts part decomposition and associated kinematic motion parameters.
In order to have a fair comparison, we modified the Shape2Motion code to enable it to take ground segmentation information as the input.
Specifically, we replaced the predicted similarity matrix from stage one with the ground truth similarity matrix as input to stages two and three during both training and testing. We set the weight of segmentation loss training stage1 to be 0. We added BN layer and Activation Layer in the model of stage 2. We removed part proposal matching and part proposal merging logic, since we know the ground truth parts. 
To be comparable to our method, we also modified the network to predict the same set of motion types by removing the `rotation+translation' motion type.

\subsection{Hyper-parameters}

\noindent
Number of points per shape: 2048\\
stage1 training epoch: 80\\
stage1 learning rate: 0.001\\
stage2 training epoch: 140\\
stage2 learning rate: 0.0001\\
stage3 training epoch: 50\\
stage3 learning rate: 0.001\\

\subsection{Pose Variation Levels}
Given a pose variation level from 0 to 5, we set the pose of each part in the dataset by:\\
$min\_range = part.gt\_min\_range * 0.2 * motion\_level$\\
$max\_range = part.gt\_max\_range * 0.2 * motion\_level$\\
$pose = min\_range + random(0, 1) * (max\_range - min\_range)$\\
At motion level 0, every part has its pose set to the the minimum of its range of motion.
As the motion level increases, the sampled poses gradually spread out over the part's entire range of motion.

\subsection{Category Breakdown for Comparison Study}

Table~\ref{tab:comparison_basenet} shows the quantitative results for the experiment where we compare our method to supervised(BaseNet) baseline, broken down by individual shape category.

\begin{table}[t!]
    \centering
    \setlength{\tabcolsep}{1pt}
    \renewcommand{\arraystretch}{0.5}
    \scriptsize
    \caption{Comparing the performance of BaseNet(with and w/o alignment) vs. Our method(with target joints number k=5) learning approach on predicting the motion attributes of shapes in PartNet-Mobility.
    }
    \resizebox{\columnwidth}{!}{\begin{tabular}{llcccc}
        \toprule
        \textbf{Category} & \textbf{Method} & \textbf{Type Acc$\uparrow$} & \textbf{Axis Err ($^\circ$)$\downarrow$} & \textbf{Center Err (\%)$\downarrow$} & \textbf{Range IoU$\uparrow$}
        \\
        \midrule
        \multirow{2}{*}{\emph{Box}} & BaseNet & 0.95 & 32.72 & 35.72 & 0.41
        \\
        & BaseNet + align & 0.95 & 25.64 & 20.69 & 0.35
        \\
        & Ours & 0.95 & 0.02 & 5.40 & 0.45
        \\
        \midrule
        \multirow{2}{*}{\emph{Bucket}} & BaseNet & 1.0 & 28.68 & 12.27 & 0.50
        \\
        & BaseNet + align & 1.0 & 15.06 & 10.10 & 0.41
        \\
        & Ours & 1.0 & 2.73 & 4.76 & 0.57
        \\
                \midrule
        \multirow{2}{*}{\emph{Clock}} & BaseNet & 1.0 & 31.77 & 29.35 & 0.23
        \\
        & BaseNet + align & 1.0 & 4.32 & 17.94 & 0.19
        \\
        & Ours & 1.0 & 5.28 & 14.64 & 0.22
        \\
                \midrule
        \multirow{2}{*}{\emph{Door}} & BaseNet & 0.79 & 30.12 & 28.30 & 0.46
        \\
        & BaseNet + align & 0.89 & 5.97 & 13.03 & 0.45
        \\
        & Ours & 0.86 & 0.01 & 13.92 & 0.51
        \\
                \midrule
        \multirow{2}{*}{\emph{Fan}} & BaseNet & 1.0 & 24.70 & 11.54 & 0.40
        \\
        & BaseNet + align & 1.0 & 15.38 & 8.07 & 0.40
        \\
        & Ours & 0.74 & 13.90 & 11.17 & 0.09
        \\
                \midrule
        \multirow{2}{*}{\emph{Faucet}} & BaseNet & 0.94 & 32.68 & 26.98 & 0.43
        \\
        & BaseNet + align & 0.97 & 24.80 & 26.03 & 0.44
        \\
        & Ours & 0.72 & 26.96 & 15.65 & 0.30
        \\
                \midrule
        \multirow{2}{*}{\emph{FoldingChair}} & BaseNet & 0.83 & 28.23 & 14.52 & 0.48
        \\
        & BaseNet + align & 0.91 & 11.55 & 8.81 & 0.56
        \\
        & Ours & 1.0 & 0.05 & 5.83 & 0.64
        \\
                \midrule
        \multirow{2}{*}{\emph{Knife}} & BaseNet & 0.81 & 37.70 & 38.70 & 0.56
        \\
        & BaseNet + align & 0.97 & 9.48 & 66.60 & 0.55
        \\
        & Ours & 0.92 & 0.24 & 10.73 & 0.53
        \\
                \midrule
        \multirow{2}{*}{\emph{Laptop}} & BaseNet & 1.0 & 17.15 & 9.74 & 0.50
        \\
        & BaseNet + align & 1.0 & 4.50 & 5.40 & 0.64
        \\
        & Ours & 0.53 & 0.02 & 2.53 & 0.61
        \\
                \midrule
        \multirow{2}{*}{\emph{Pliers}} & BaseNet & 0.9 & 15.16 & 20.63 & 0.57
        \\
        & BaseNet + align & 0.85 & 9.65 & 15.05 & 0.57
        \\
        & Ours & 0.7 & 0.00 & 4.92 & 0.40
        \\
                \midrule
        \multirow{2}{*}{\emph{Refrigerator}} & BaseNet & 1.0 & 17.66 & 20.21 & 0.60 
        \\
        & BaseNet + align & 1.0 & 6.62 & 12.04 & 0.55
        \\
        & Ours & 0.94 & 0.13 & 3.69 & 0.62
        \\
                \midrule
        \multirow{2}{*}{\emph{Scissors}} & BaseNet & 0.44 & 39.77 & 16.48 & 0.41
        \\
        & BaseNet + align & 0.81 & 7.44 & 10.89 & 0.47
        \\
        & Ours & 0.55 & 0.52 & 4.92 & 0.53
        \\
        \midrule
        \multirow{2}{*}{\emph{Stapler}} & BaseNet & 0.93 & 43.76 & 35.45 & 0.50
        \\
        & BaseNet + align & 1.0 & 6.32 & 15.61 & 0.54
        \\
        & Ours & 0.97 & 1.78 & 15.55 & 0.60
        \\
        \midrule
        \multirow{2}{*}{\emph{StorageFurniture}} & BaseNet & 0.75 & 36.66 & 36.18 & 0.37
        \\
        & BaseNet + align & 0.72 & 16.87 & 31.13 & 0.46
        \\
        & Ours & 0.72 & 16.34 & 9.85 & 0.55
        \\
                \midrule
        \multirow{2}{*}{\emph{Table}} & BaseNet & 0.88 & 44.42 & 56.86 & 0.28
        \\
        & BaseNet + align & 0.83 & 15.28 & 56.33 & 0.27
        \\
        & Ours & 0.81 & 22.90 & 15.19 & 0.46
        \\
                        \midrule
        \multirow{2}{*}{\emph{TrashCan}} & BaseNet & 0.91 & 48.76 & 26.73 & 0.47
        \\
        & BaseNet + align & 0.95 & 35.47 & 30.11 & 0.43
        \\
        & Ours & 0.89 & 6.61 & 15.92 & 0.47
        \\
                \midrule
        \multirow{2}{*}{\emph{USB}} & BaseNet & 0.68 & 20.93 & 17.45 & 0.12
        \\
        & BaseNet + align & 0.86 & 14.86 & 13.74 & 0.13
        \\
        & Ours & 0.86 & 10.75 & 8.13 & 0.21
        \\
                \midrule
        \multirow{2}{*}{\emph{Window}} & BaseNet & 0.89 & 21.87 & 0.0 & 0.66
        \\
        & BaseNet + align & 0.98 & 9.10 & 37.64 & 0.75
        \\
        & Ours & 0.98 & 1.53 & 1.95 & 0.55
        \\
                \midrule
        \multirow{2}{*}{\textbf{\emph{Mean}}} & BaseNet & 0.87 & 30.71 & 24.28 & 0.44
        \\
        & BaseNet + align & 0.93 & 13.24 & 22.18 & 0.45
        \\
        & Ours & 0.84 & 6.09 & 9.12 & 0.46
        \\
        \bottomrule
    \end{tabular}}

    \label{tab:comparison_basenet}
\end{table}

Table ~\ref{tab:comparison_s2m} shows the quantitative results for the experiment where we compare our method to supervised(Shape2Motion) baseline, broken down by individual shape category.

\begin{table}[t!]
    \centering
    \setlength{\tabcolsep}{2pt}
    \renewcommand{\arraystretch}{0.5}
    \scriptsize
    \caption{Comparing the performance of Shape2Motion(with and w/o alignment) vs. Our method(with target joints number k=5) on predicting the motion attributes of shapes in PartNet-Mobility. S2M does not handle static motion type and motion range.}
    \label{tab:comparison_s2m}
    \resizebox{\columnwidth}{!}{\begin{tabular}{llccc}
        \toprule
        \textbf{Category} & \textbf{Method} & \textbf{Type Acc(w/o static)$\uparrow$} & \textbf{Axis Err ($^\circ$)$\downarrow$} & \textbf{Center Err (\%)$\downarrow$} 
        \\
        \midrule
        \multirow{2}{*}{\emph{Box}} 
        & S2M & 0.92 & 46.08 & 41.55
        \\
        & S2M + align & 0.92 & 33.10 & 17.05
        \\
        & Ours & 0.92 & 0.02 & 5.40
        \\
        \midrule
        \multirow{2}{*}{\emph{Bucket}} 
        & S2M & 1.0 & 47.36 & 28.42
        \\
        & S2M + align & 1.0 & 35.91 & 22.48
        \\
        & Ours & 1.0 & 2.73 & 4.76
        \\
        \midrule
        \multirow{2}{*}{\emph{Clock}} 
        & S2M & 1.0 & 54.80 & 59.30
        \\
        & S2M + align & 1.0 & 3.21 & 49.80
        \\
        & Ours & 1.0 & 5.28 & 14.64
        \\
        \midrule
        \multirow{2}{*}{\emph{Door}} 
        & S2M & 0.94 & 0.56 & 23.20
        \\
        & S2M + align & 0.94 & 0.50 & 15.09
        \\
        & Ours & 0.76 & 0.01 & 13.92
        \\
        \midrule
        \multirow{2}{*}{\emph{Fan}} 
        & S2M & 1.0 & 52.26 & 10.20
        \\
        & S2M + align & 1.0 & 34.15 & 11.79
        \\
        & Ours & 0.69 & 13.90 & 11.17
        \\
        \midrule
        \multirow{2}{*}{\emph{Faucet}} 
        & S2M & 0.96 & 35.75 & 19.52
        \\
        & S2M + align & 0.96 & 37.30 & 20.31
        \\
        & Ours & 0.75 & 26.96 & 15.65
        \\
        \midrule
        \multirow{2}{*}{\emph{FoldingChair}} 
        & S2M & 1.0 & 42.48 & 16.45
        \\
        & S2M + align & 1.0 & 1.29 & 9.83
        \\
        & Ours & 1.0 & 0.05 & 5.83
        \\
        \midrule
        \multirow{2}{*}{\emph{Knife}} 
        & S2M & 0.67 & 25.49 & 30.02
        \\
        & S2M + align & 0.71 & 1.85 & 32.68
        \\
        & Ours & 0.86 & 0.24 & 10.73 
        \\
        \midrule
        \multirow{2}{*}{\emph{Laptop}} 
        & S2M & 1.0 & 47.71 & 11.17
        \\
        & S2M + align & 1.0 & 4.15 & 3.33
        \\
        & Ours & 0.53 & 0.02 & 2.53
        \\
        \midrule
        \multirow{2}{*}{\emph{Pliers}} 
        & S2M & 1.0 & 0.69 & 16.21
        \\
        & S2M + align & 1.0 & 0.5 & 12.76
        \\
        & Ours & 0.7 & 0.00 & 4.92
        \\
        \midrule
        \multirow{2}{*}{\emph{Refrigerator}} 
        & S2M & 1.0 & 0.50 & 12.27
        \\
        & S2M + align & 1.0 & 0.52 & 10.15
        \\
        & Ours & 0.90 & 0.13 & 3.69
        \\
        \midrule
        \multirow{2}{*}{\emph{Scissors}} 
        & S2M & 1.0 & 0.62 & 26.06
        \\
        & S2M + align & 1.0 & 0.73 & 11.41
        \\
        & Ours & 0.56 & 0.52 & 4.92
        \\
        \midrule
        \multirow{2}{*}{\emph{Stapler}} 
        & S2M & 1.0 & 30.62 & 18.44
        \\
        & S2M + align & 1.0 & 2.03 & 17.37
        \\
        & Ours & 0.95 & 1.78 & 15.55
        \\
        \midrule
        \multirow{2}{*}{\emph{StorageFurniture}} 
        & S2M & 0.5 & 49.84 & 41.16
        \\
        & S2M + align & 0.65 & 30.74 & 8.94
        \\
        & Ours & 0.55 & 16.34 & 9.85
        \\
        \midrule
        \multirow{2}{*}{\emph{Table}} 
        & S2M & 0.71 & 35.88 & 64.48
        \\
        & S2M + align & 0.89 & 27.05 & 12.00
        \\
        & Ours & 0.73 & 22.90 & 15.19
        \\
        \midrule
        \multirow{2}{*}{\emph{TrashCan}} 
        & S2M & 0.83 & 56.26 & 28.52
        \\
        & S2M + align & 0.82 & 55.74 & 24.65
        \\
        & Ours & 0.83 & 6.61 & 15.92
        \\
        \midrule
        \multirow{2}{*}{\emph{USB}} 
        & S2M & 0.86 & 16.39 & 13.46
        \\
        & S2M + align & 0.79 & 3.28 & 20.44
        \\
        & Ours & 0.72 & 10.75 & 8.13
        \\
        \midrule
        \multirow{2}{*}{\emph{Window}} 
        & S2M & 0.95 & 62.39 & 0.0
        \\
        & S2M + align & 0.84 & 12.29 & 0.0
        \\
        & Ours & 0.97 & 1.53 & 1.95
        \\
        \midrule
        \multirow{2}{*}{\textbf{\emph{Mean}}}
        & S2M & 0.91 & 33.65 & 25.58
        \\
        & S2M + align & 0.92 & 15.80 & 16.67
        \\
        & Ours & 0.80 & 6.09 & 9.12
        \\
        \bottomrule
    \end{tabular}}

\end{table}

\subsection{Sensitivity to number of target joints}
Table~\ref{tab:num_target_joints} shows how our method performs on Buckets, TrashCans and Faucets when the number of target joints $k$ is varied.
More target joints tends to help find more accurate motion axes, particularly when the object category is geometrically simple (e.g. Buckets). 

\begin{table}[t!]
    \centering
    \scriptsize
    \caption{Analyzing how the performance of our model on Buckets, TrashCans and Tables varies with the number of target joints $k$ used during motion optimization.}
    \resizebox{\columnwidth}{!}{\begin{tabular}{llcccc}
        \toprule
        \textbf{Category} & $k$ & \textbf{Type Acc$\uparrow$} & \textbf{Axis Err ($^\circ$)$\downarrow$} & \textbf{Center Err (\%)$\downarrow$} & \textbf{Range IoU$\uparrow$}
        \\
        \midrule
        \emph{Bucket} & 2 & 1.0 & 15.10 & 7.51 & 0.43 \\
        & 3 & 1.0 & 4.43 & 6.92 & 0.51 \\
        & 4 & 1.0 & 3.41 & 7.66 & 0.51 \\
        & 5 & 1.0 & 0.18 & 5.35 & 0.54 \\
        \midrule
        \emph{TrashCan} & 2 & 0.86 & 16.60 & 14.45 & 0.44 \\
        & 3 & 0.875 & 14.58 & 13.22 & 0.47 \\
        & 4 & 0.89 & 5.76 & 15.39 & 0.47 \\
        & 5 & 0.89 & 6.41 & 16.26 & 0.46 \\
        \midrule
        \emph{Faucet} & 2 & 0.72 & 27.40 & 16.89 & 0.29 \\
        & 3 & 0.72 & 29.91 & 16.12 & 0.30 \\
        & 4 & 0.72 & 28.13 & 18.58 & 0.30 \\
        & 5 & 0.71 & 22.32 & 15.70 & 0.30 \\
        \bottomrule
    \end{tabular}}

    \label{tab:num_target_joints}
\end{table}

\subsection{Ablation Study}
We also explored the influence of several selected regularization terms used in our optimization loss function. Table ~\ref{tab:reg_term_ablation} shows the involvement of the additional transformation components and regularization terms leads to better results. For complex categories such as Fan, Faucet and Table, removing local alignment or deformation or physical constraints can lead to an increase of both axis and center error. For relatively simpler category such as Bucket, the small motion penalty helps with axis prediction and physical constraint helps with center prediction. The removal of local alignment and deformation makes less impact but does not mean they can improve the performance since both axis error and center error are already small.

\begin{table}[t!]
    \centering
    \setlength{\tabcolsep}{1pt}
    \scriptsize
    \caption{Results of our ablation study on categories: Bucket, Fan, Faucet, Table with target joints number k = 3. Each set corresponds to a shape category, each row in the set shows the motion attributes prediction performance without a specific feature enabled.}
    \resizebox{\columnwidth}{!}{\begin{tabular}{llcccc}
        \toprule
        \textbf{Category} & \textbf{Ablation} & \textbf{Type Acc$\uparrow$} & \textbf{Axis Err ($^\circ$)$\downarrow$} & \textbf{Center Err (\%)$\downarrow$} & \textbf{Range IoU$\uparrow$}
        \\
        \midrule
        \emph{Bucket} & all & 1.0 & 4.43 & 6.92 & 0.51\\
         & - Small motion penalty & 1.0 & 10.48 & 9.75 & 0.50 \\
         & - local alignment & 1.0 & 2.27 & 3.08 & 0.52\\
         & - deformation & 1.0 & 4.33 & 7.00 & 0.45\\
         & - collision and detach penalties & 1.0 & 1.38 & 10.35 & 0.53 \\
        \midrule
        \emph{Fan} & all & 0.77 & 11.19 & 13.47 & 0.08\\
         & - Small motion penalty & 0.79 & 15.72 & 18.21 & 0.07\\
         & - local alignment & 0.71 & 29.44 & 22.55 & 0.08\\
         & - deformation & 0.76 & 24.29 & 15.58 & 0.06\\
         & - collision and detach penalties & 0.69 & 15.48 & 23.69 & 0.08\\
         \midrule
        \emph{Faucet} & all & 0.74 & 28.84 & 15.00 & 0.30\\
         & - Small motion penalty & 0.74 & 27.42 & 18.15 & 0.27\\
         & - local alignment & 0.71 & 41.57 & 27.54 & 0.20\\
         & - deformation & 0.74 & 34.08 & 18.61 & 0.16\\
         & - collision and detach penalties & 0.72 & 30.01 & 53.44 & 0.29\\
         \midrule
        \emph{Table} & all & 0.79 & 24.52 & 6.89 & 0.43\\
         & - Small motion penalty & 0.80 & 19.23 & 15.46 & 0.46\\
          & - local alignment & 0.78 & 39.25 & 30.40 & 0.41\\
           & - deformation & 0.57 & 39.41 & 13.31 & 0.44\\
            & - collision and detach penalties & 0.97 & 50.07 & 48.00 & 0.40\\
        \bottomrule
    \end{tabular}}

    \label{tab:reg_term_ablation}
\end{table}

\subsection{Embedding space visualization}
\begin{figure}[t!]
    \centering
    \includegraphics[width=0.9\linewidth]{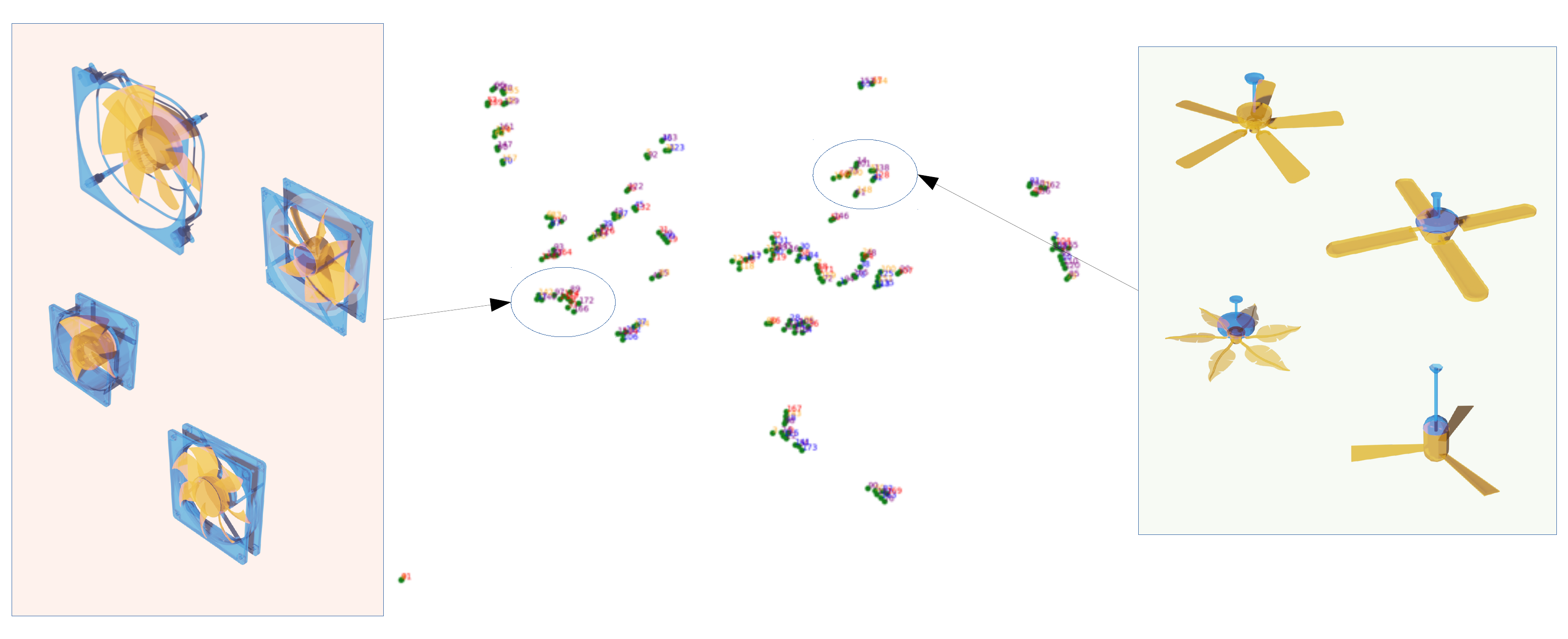}
    \caption{A 2D tSNE projection of the final joint embedding space for Fans.}
    \label{fig:embedding_viz}
\end{figure}

Figure~\ref{fig:embedding_viz} shows a 2D tSNE projection of the learned joint embeddings $\embed(\joint)$ for the Fan category.
The embedding space learns to separate different fan types (e.g. CPU fans vs. ceiling fans). 

\subsection{Additional Qualitative Results}

\begin{figure*}[t!]
    \centering
    \includegraphics[width=0.9\linewidth]{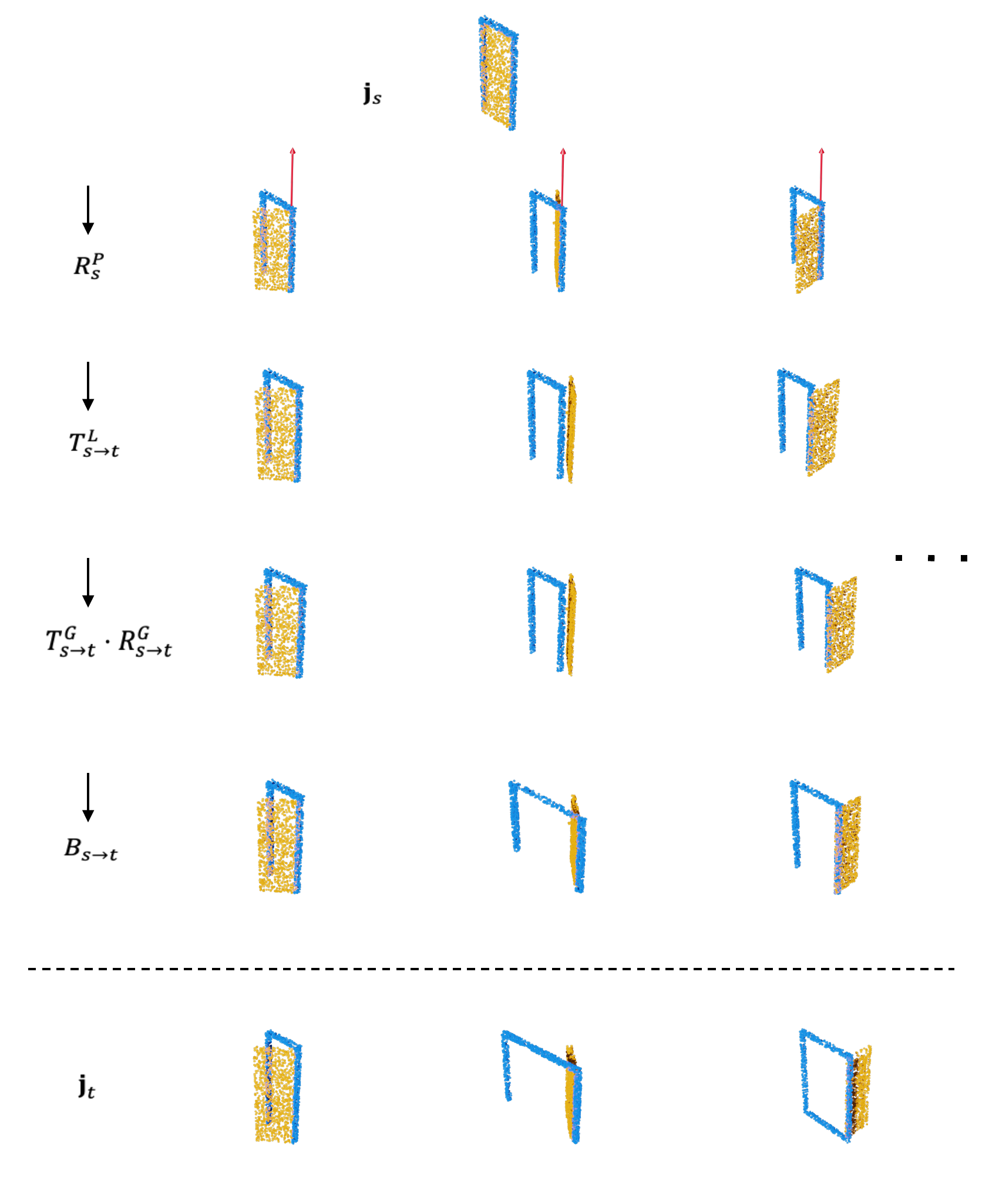}
    \caption{
    Transformation process for Door (shown only 3 joints in the group)
    }
    \label{figure:process_door}
\end{figure*}

\begin{figure*}[t!]
    \centering
    \includegraphics[width=0.9\linewidth]{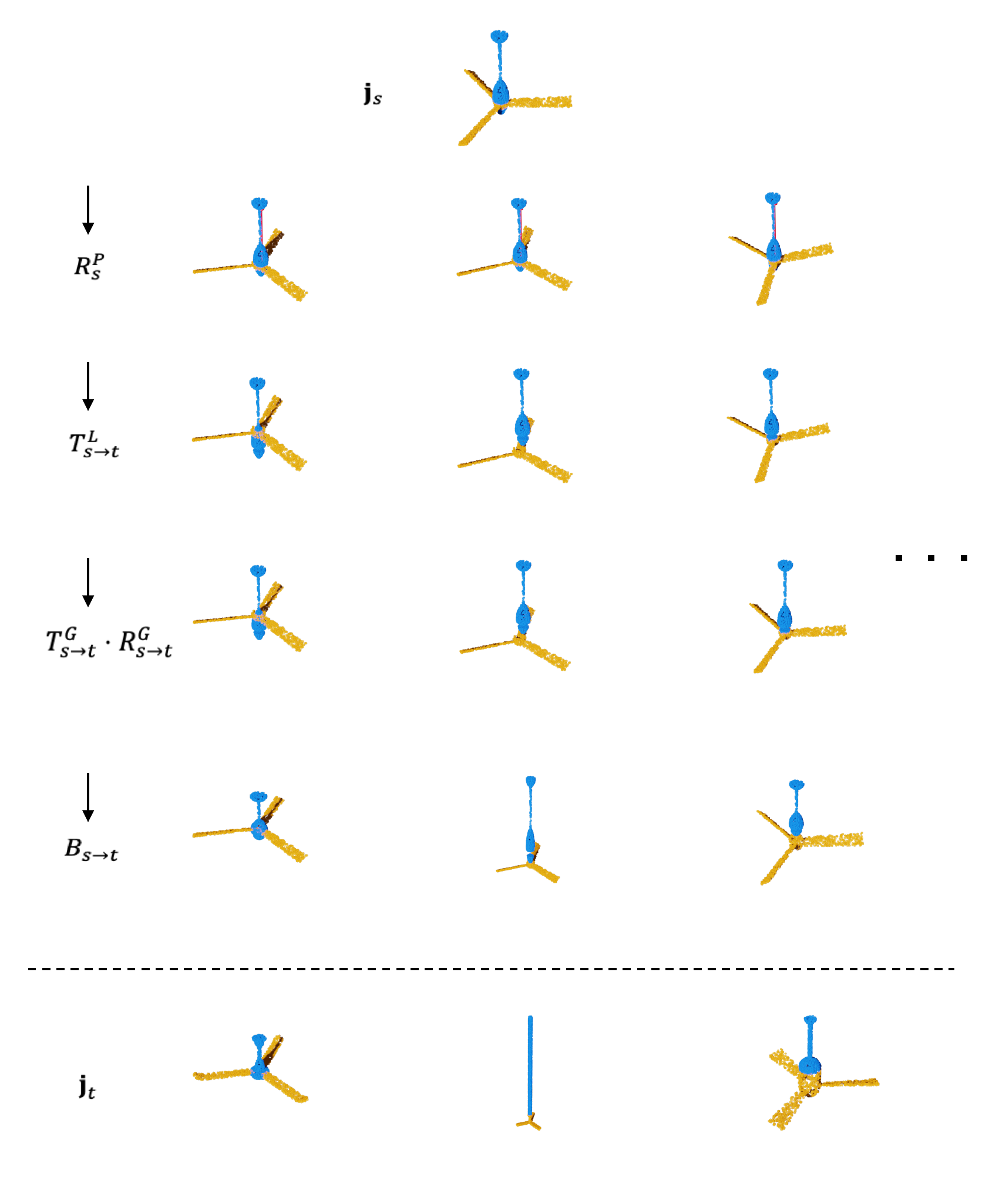}
    \caption{
    Transformation process for Fan (shown only 3 joints in the group)
    }
    \label{figure:process_fan}
\end{figure*}

\begin{figure*}[t!]
    \centering
    \includegraphics[width=0.9\linewidth]{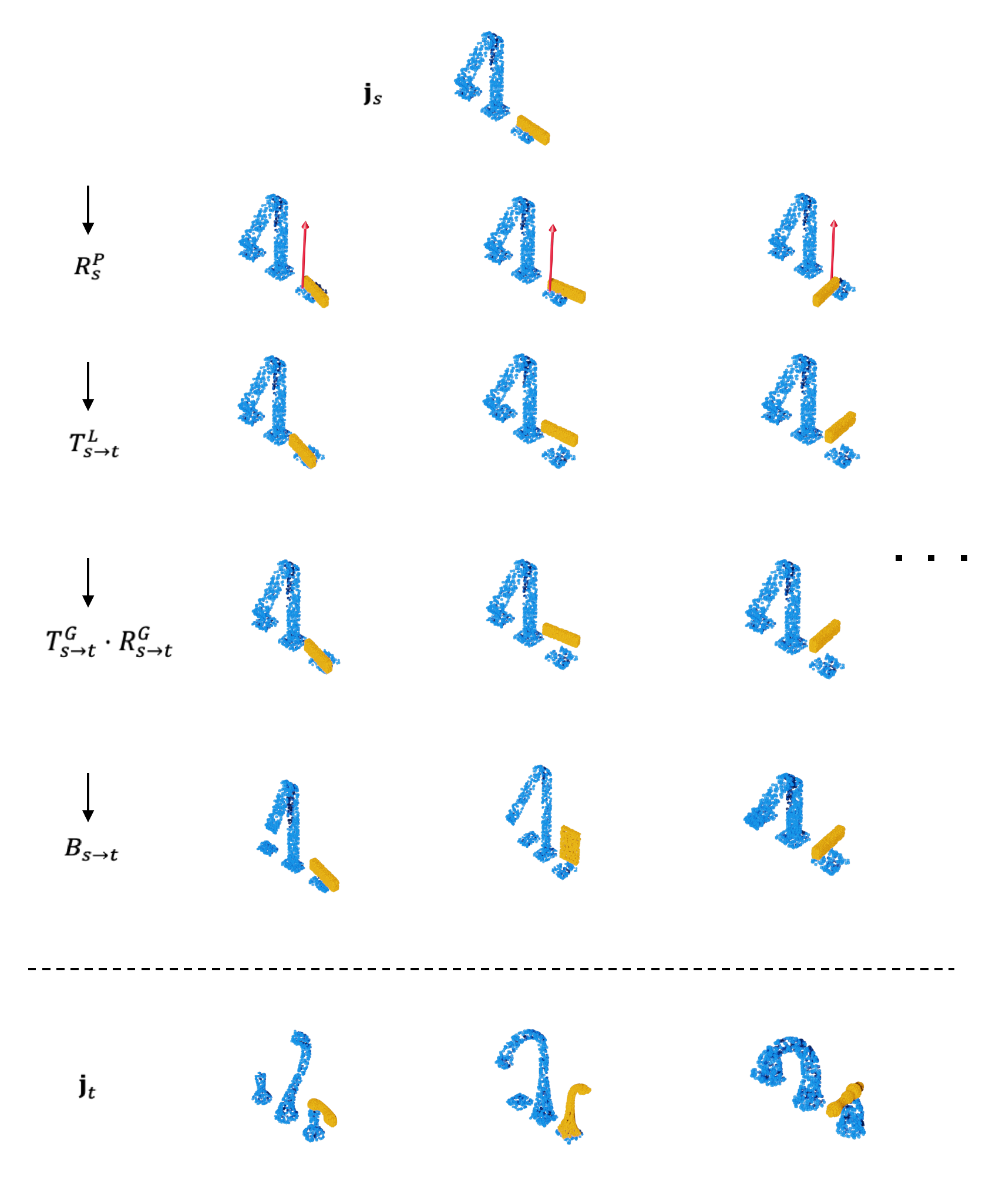}
    \caption{
    Transformation process for Faucet (shown only 3 joints in the group)
    }
    \label{figure:process_faucet}
\end{figure*}

\begin{figure*}[t!]
    \centering
    \includegraphics[width=0.9\linewidth]{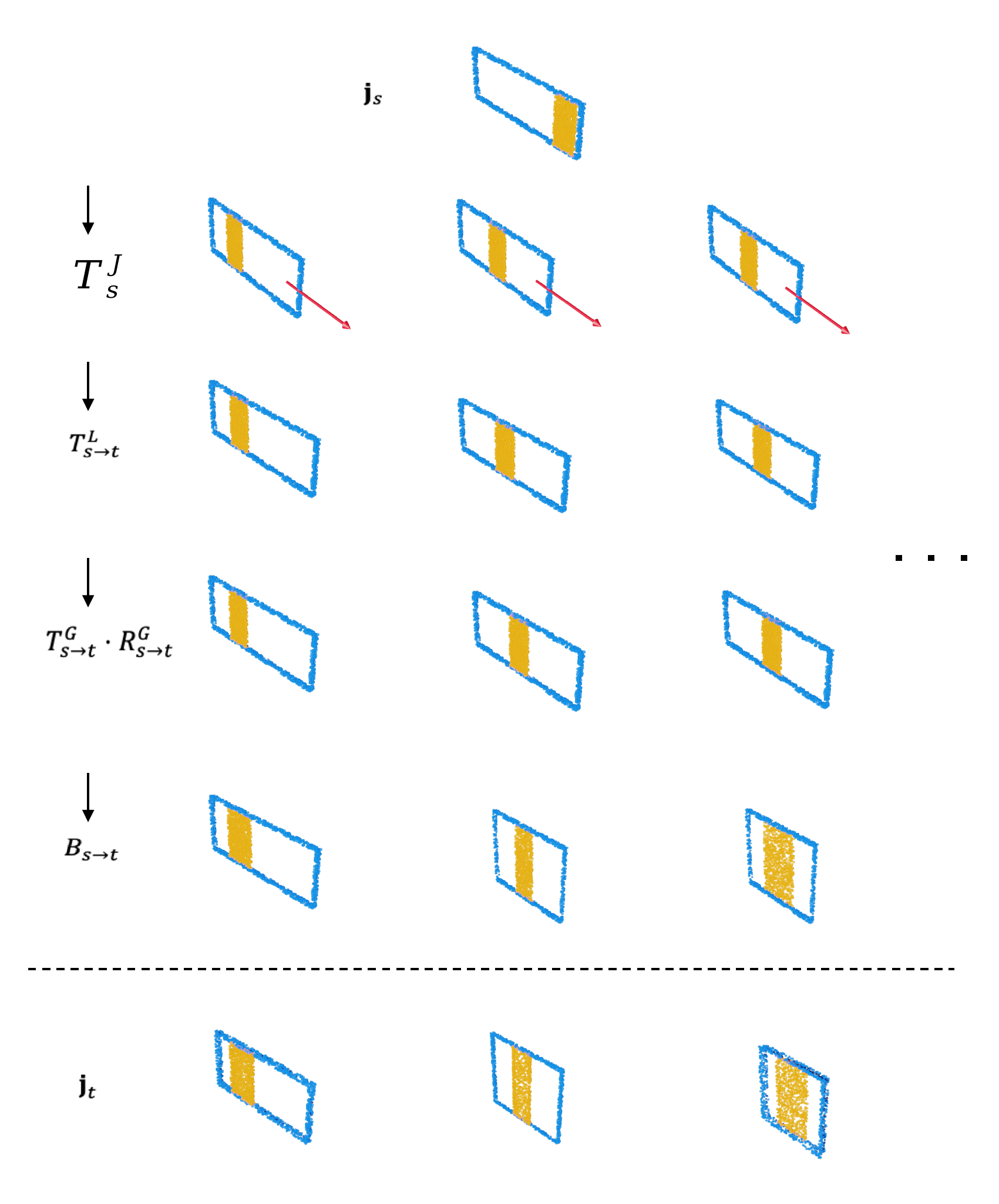}
    \caption{
    Transformation process for Window (shown only 3 joints in the group)
    }
    \label{figure:process_window}
\end{figure*}

Please see Figure~\ref{figure:process_door},~\ref{figure:process_fan},~\ref{figure:process_faucet},~\ref{figure:process_window}, for additional transformation process visualizations.

\begin{figure*}[ht!]
    \centering
    \setlength{\tabcolsep}{1pt}
    \begin{tabular}{ccc}
        
    \includegraphics[width=0.33\linewidth]{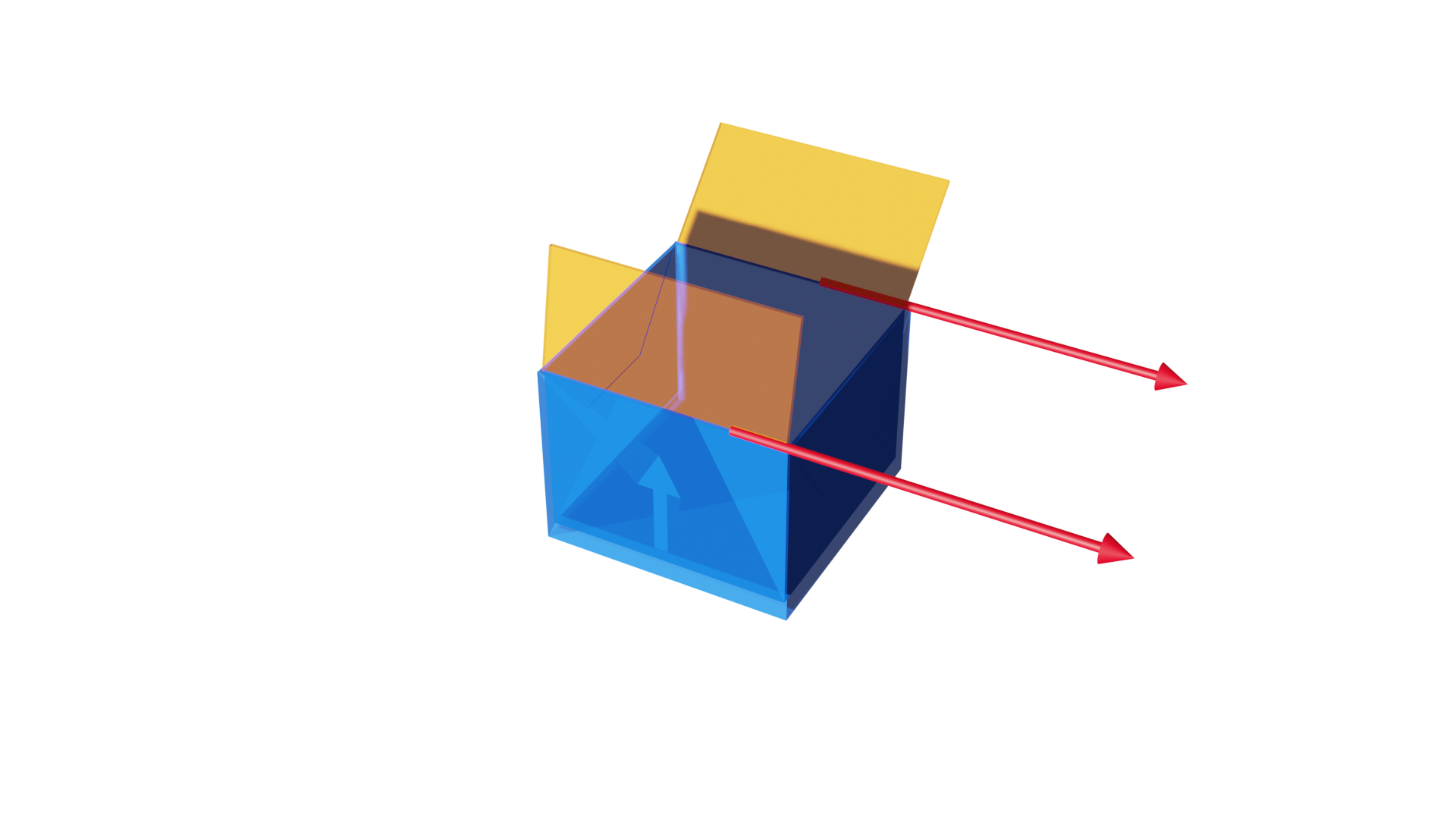}
    \includegraphics[width=0.33\linewidth]{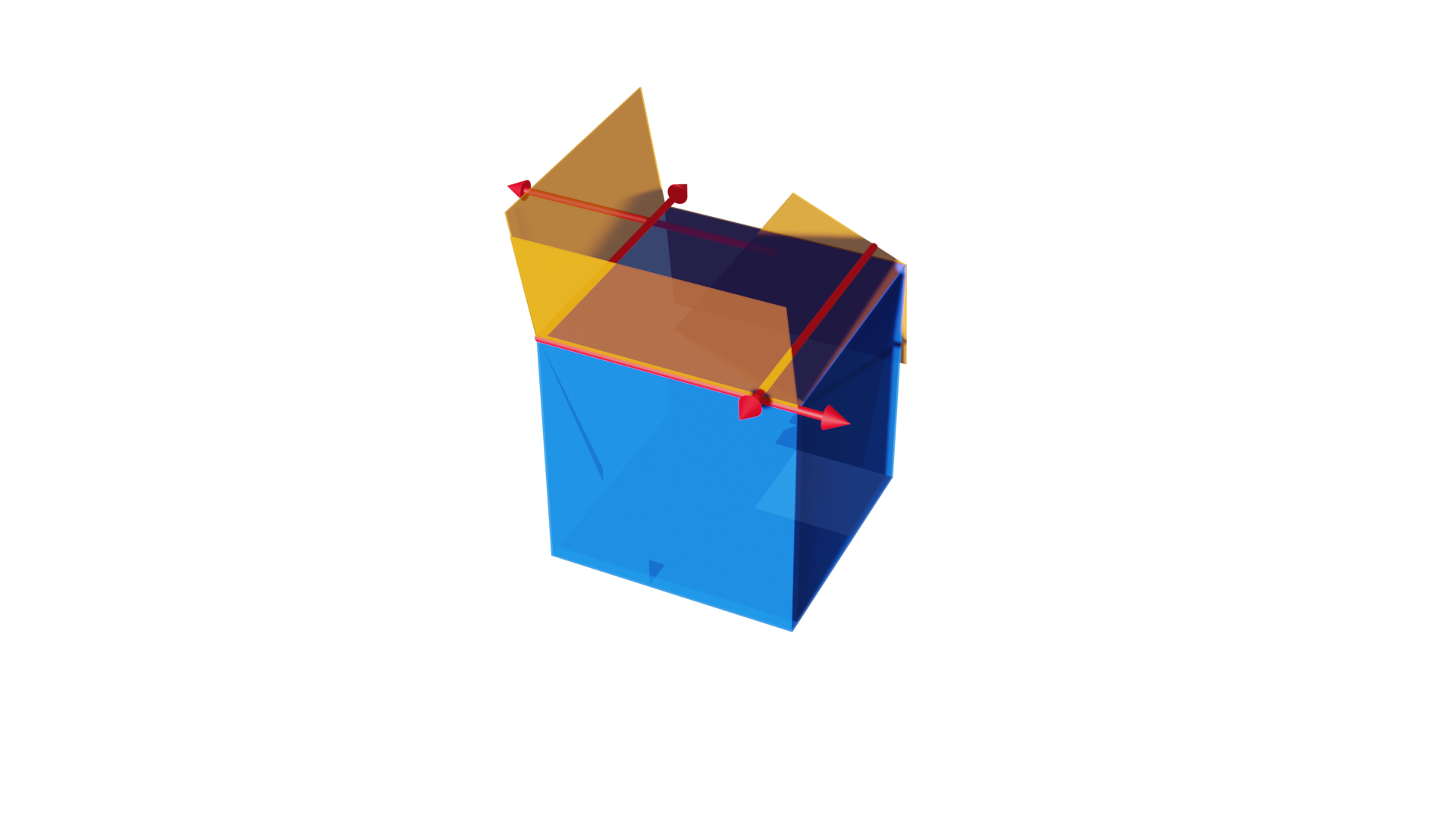}
    \includegraphics[width=0.33\linewidth]{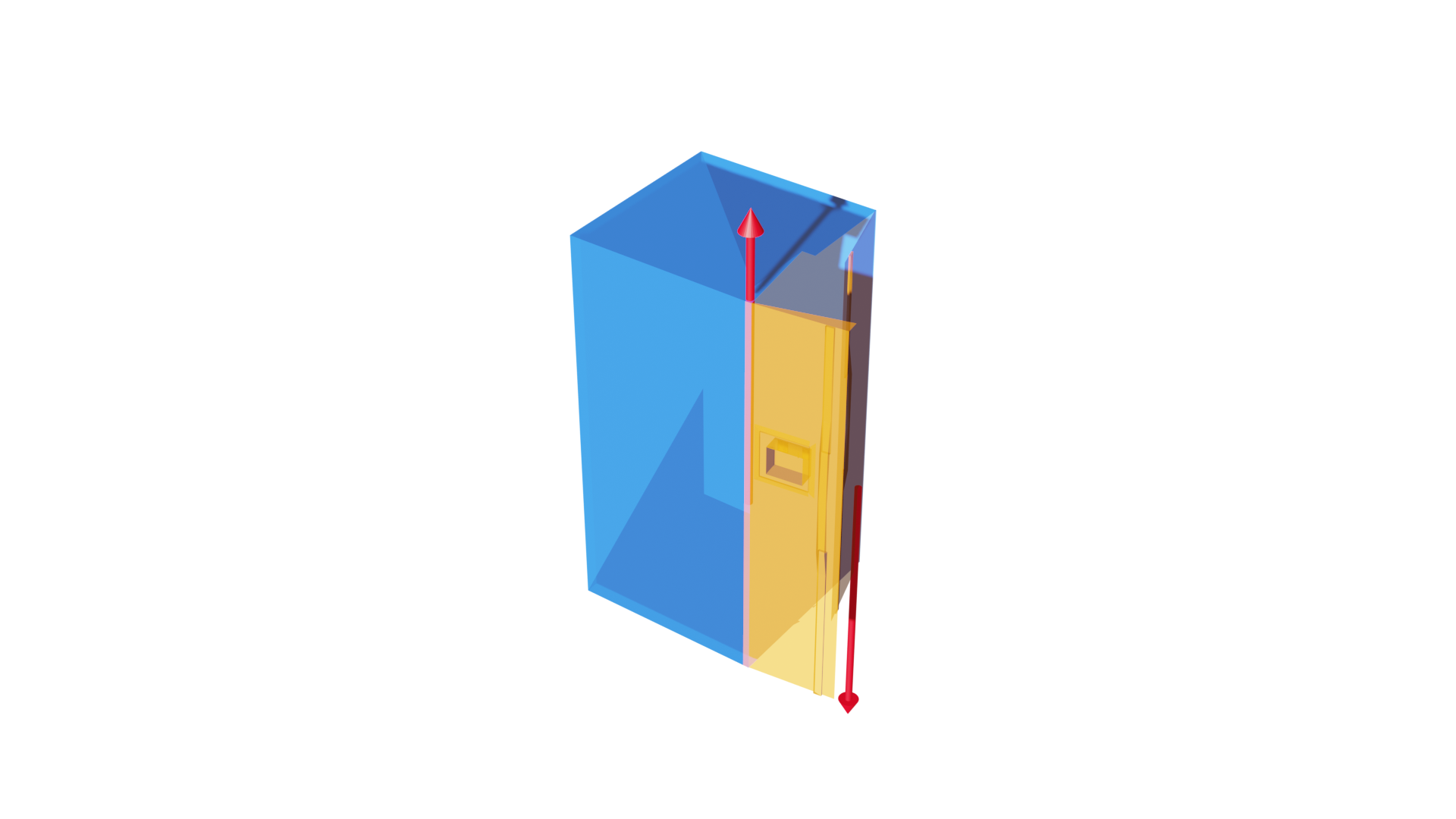}
    \\
    \includegraphics[width=0.33\linewidth]{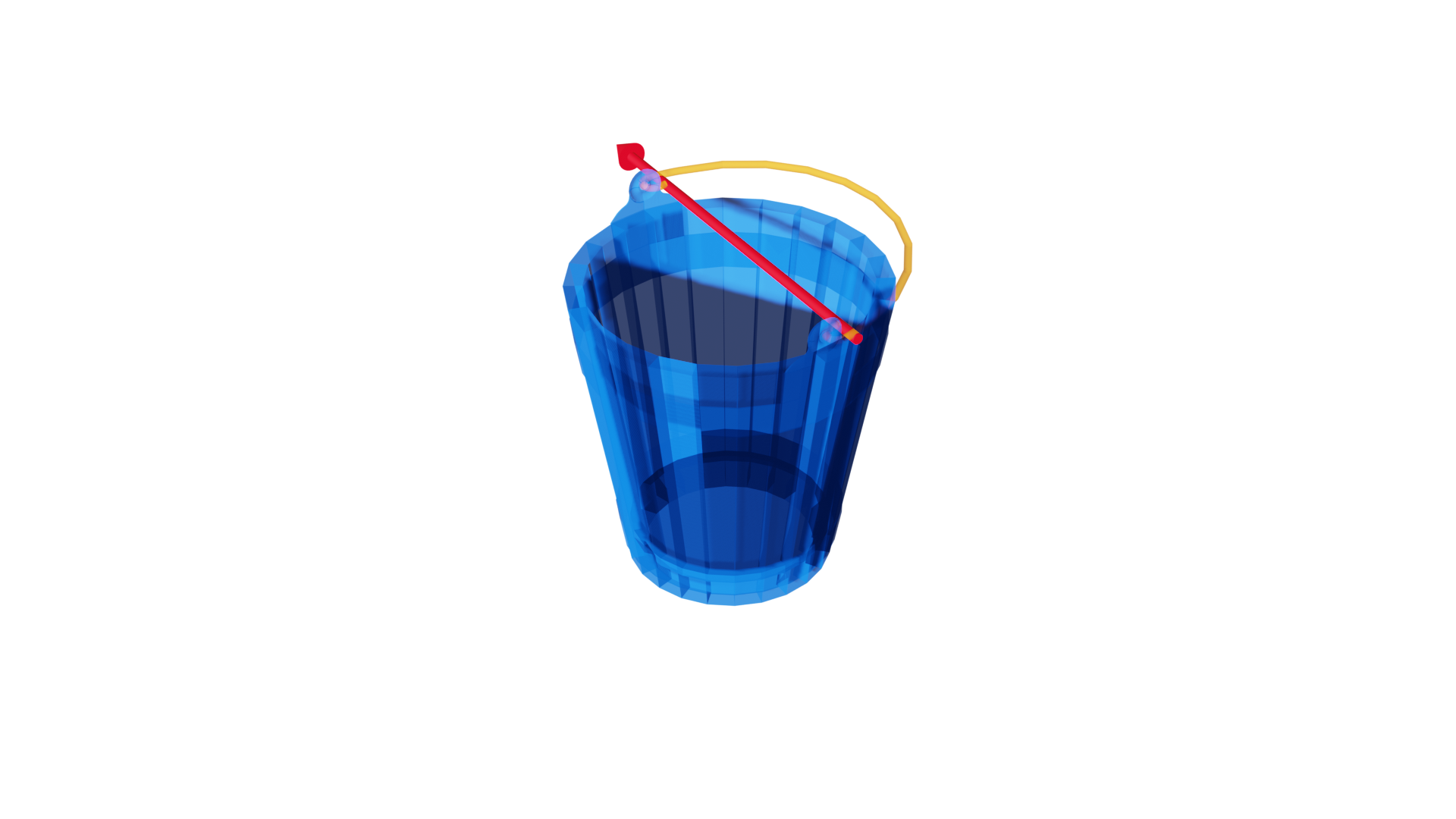}
    \includegraphics[width=0.33\linewidth]{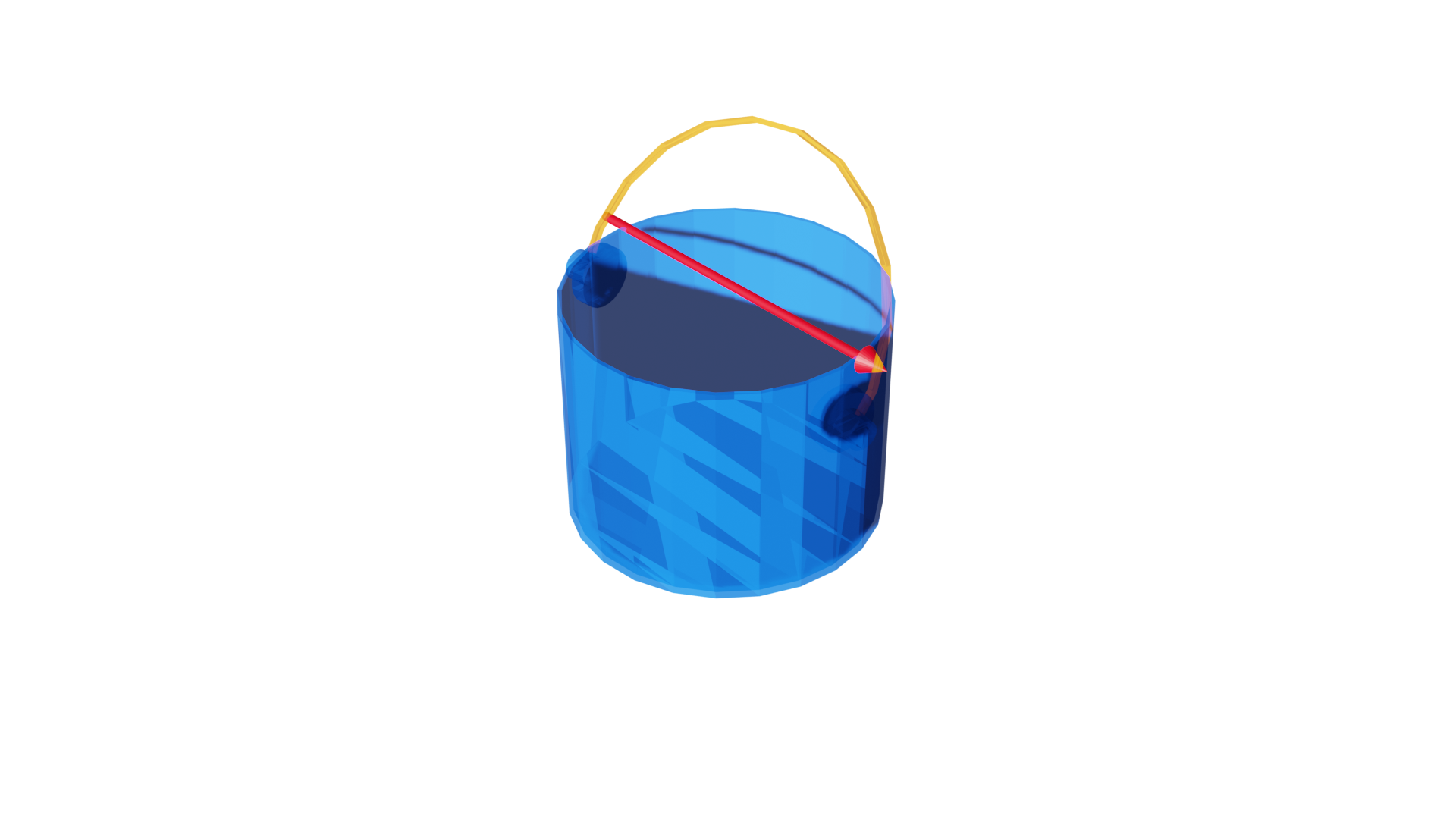}
    \includegraphics[width=0.33\linewidth]{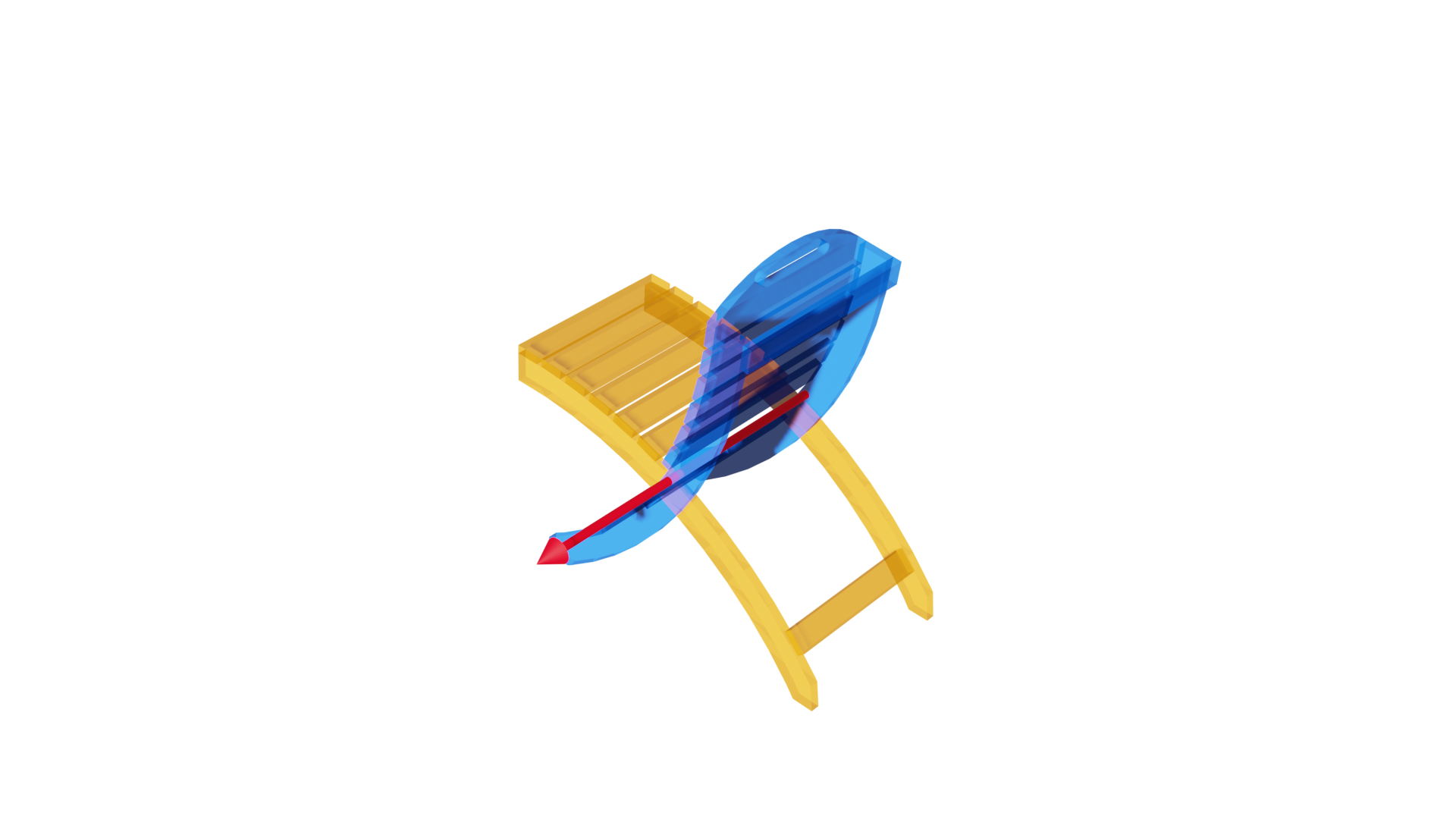}
    \\
    \includegraphics[width=0.33\linewidth]{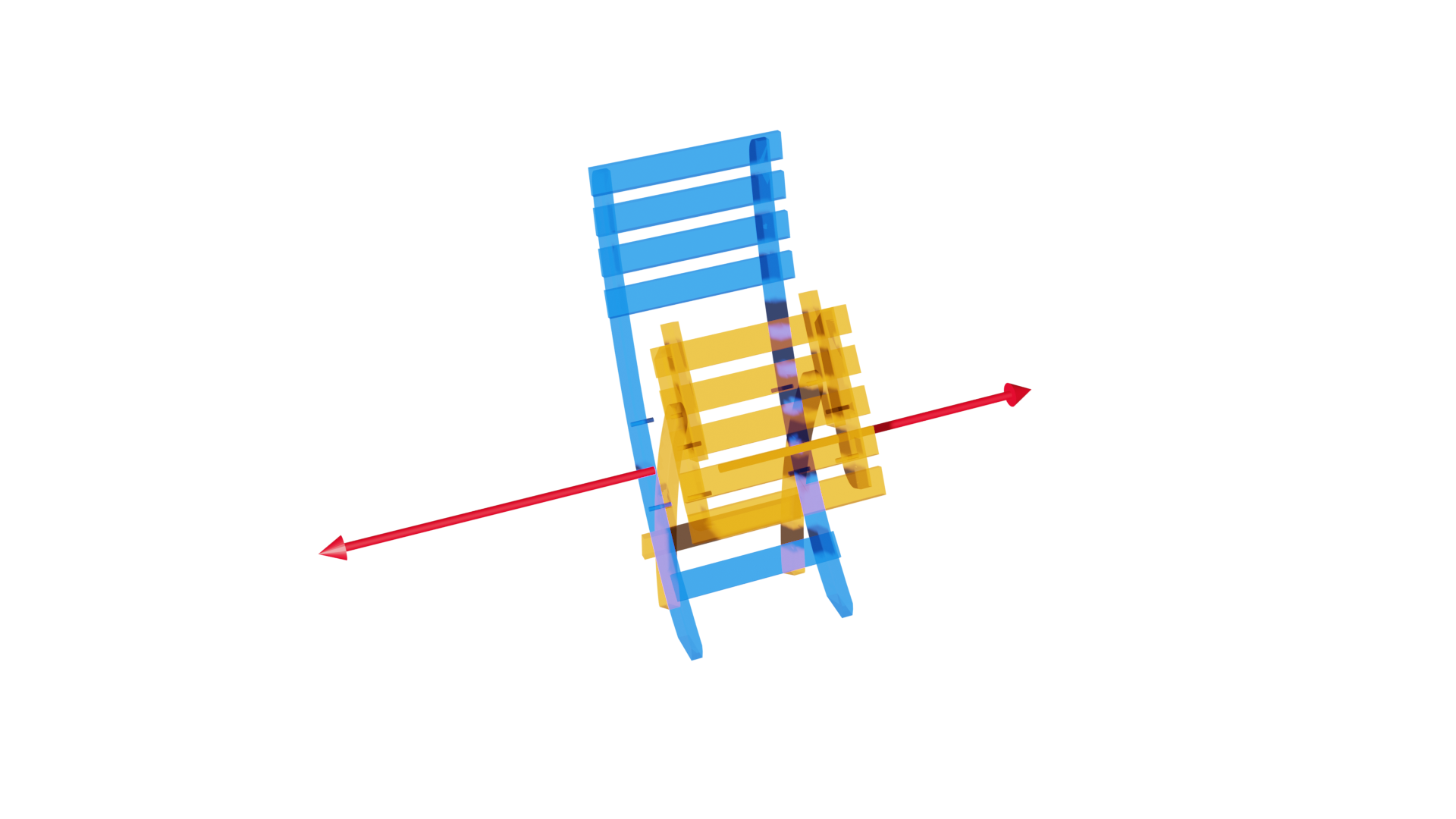}
    \includegraphics[width=0.33\linewidth]{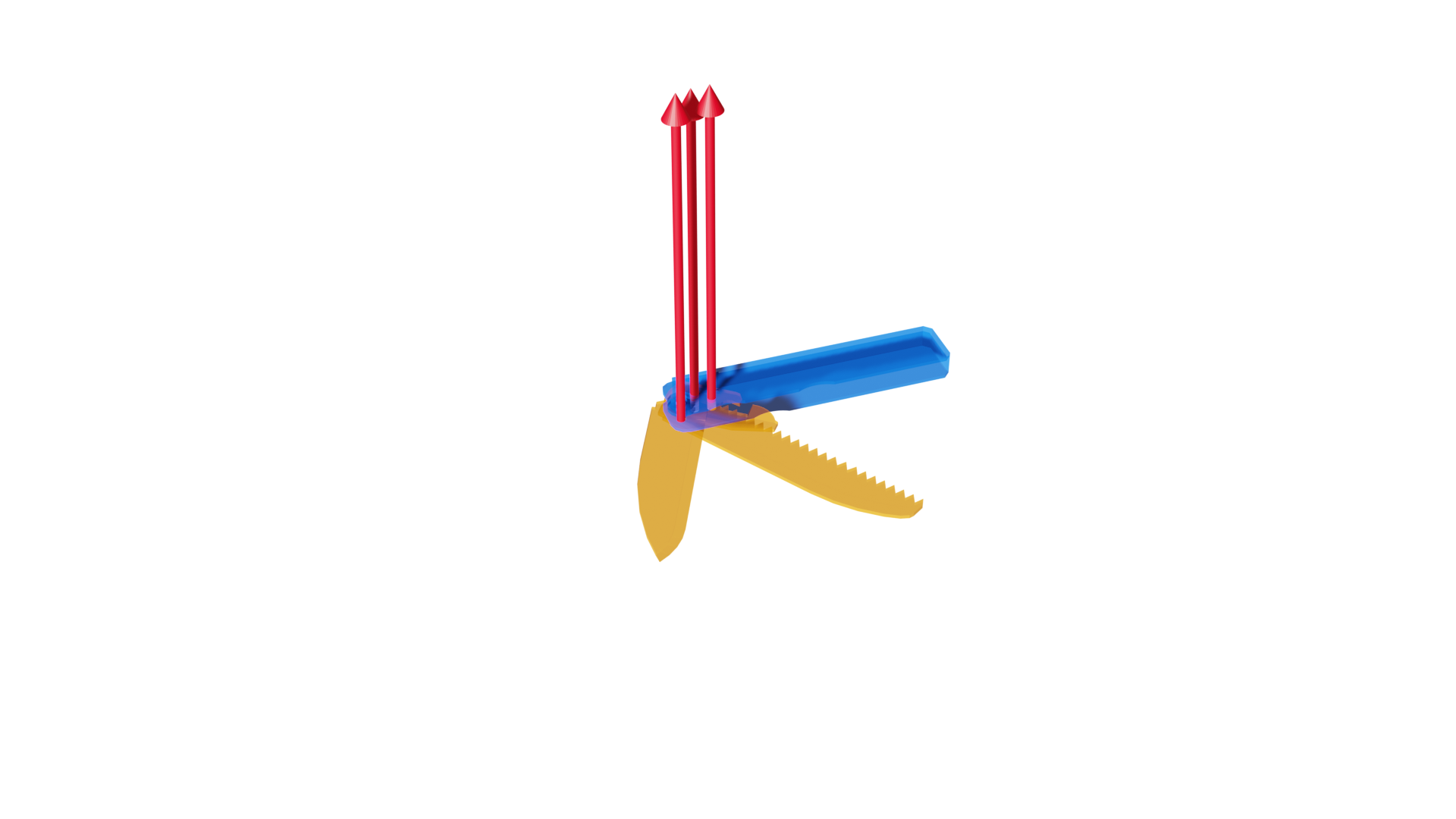}
    \includegraphics[width=0.33\linewidth]{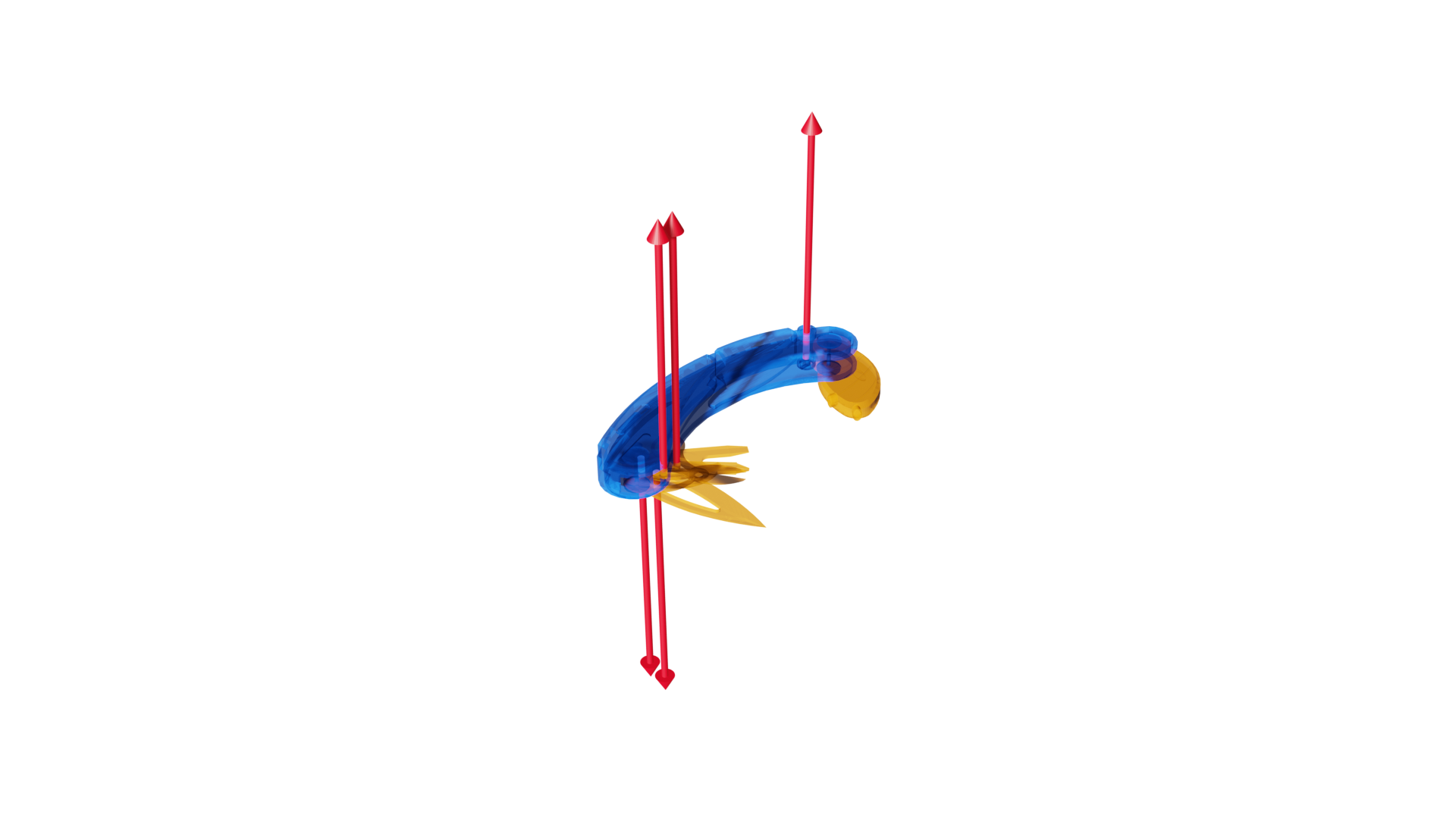}
    \\
    \includegraphics[width=0.33\linewidth]{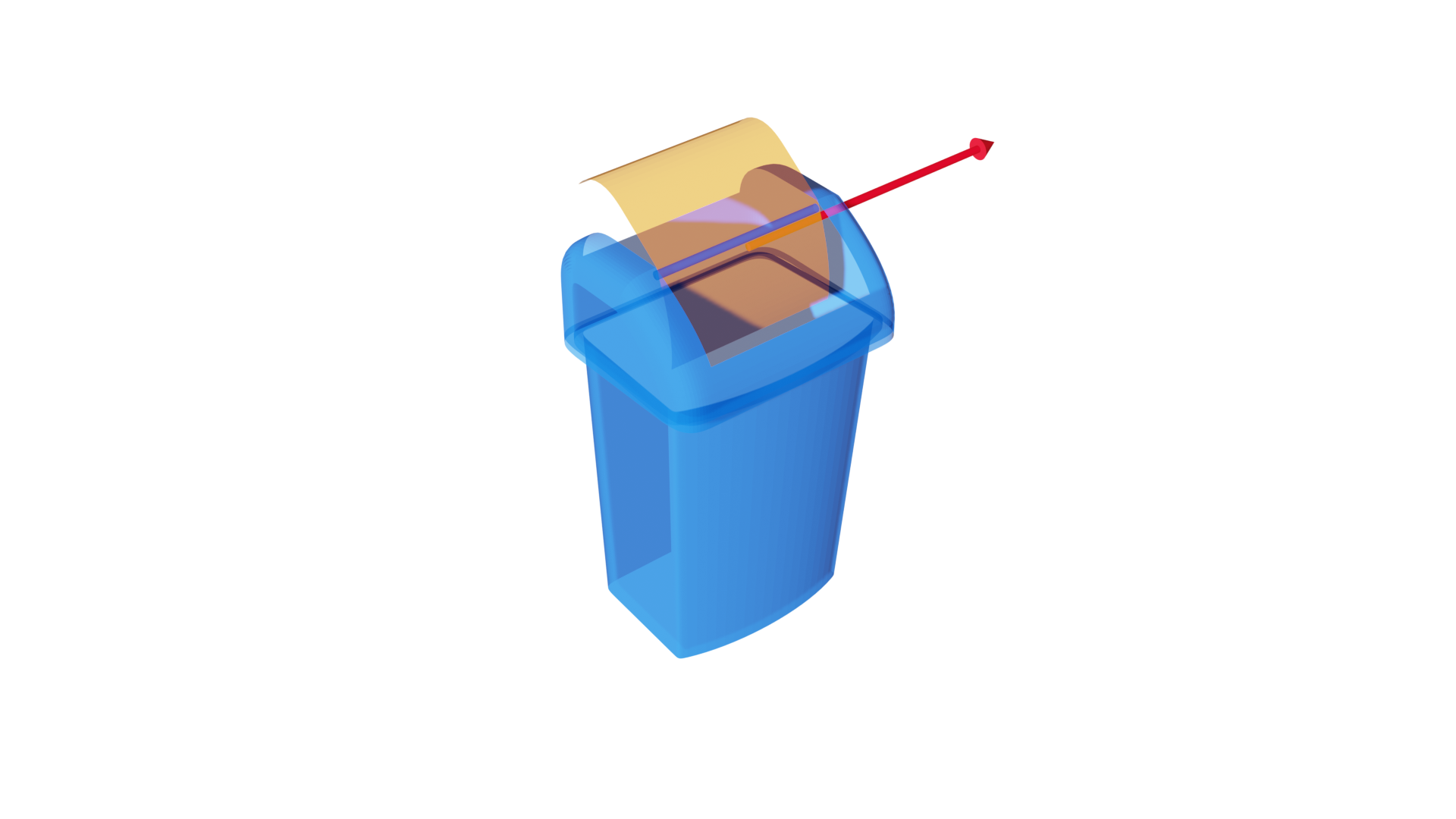}
    \includegraphics[width=0.33\linewidth]{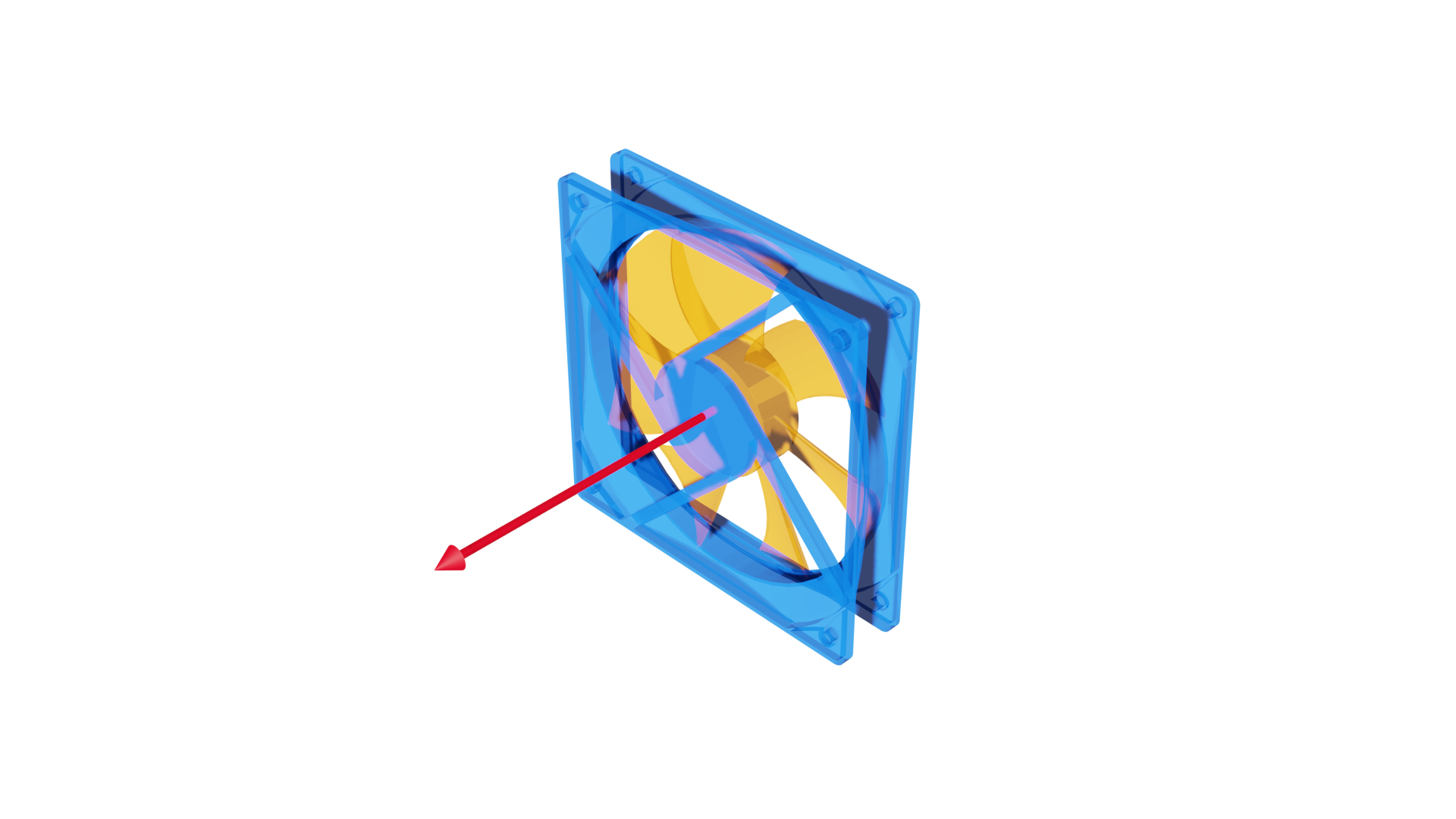}
    \includegraphics[width=0.33\linewidth]{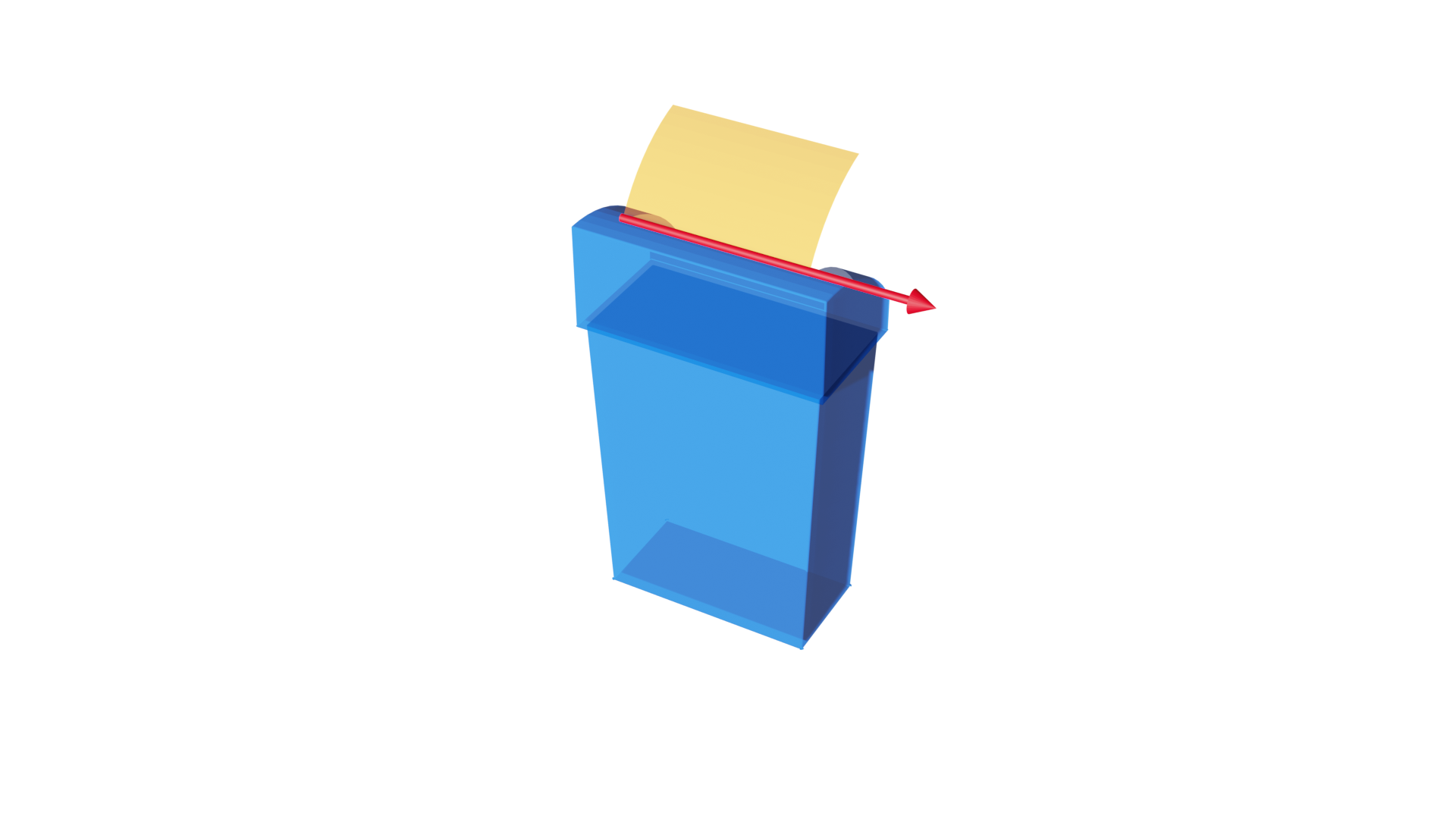}
    \\
    \includegraphics[width=0.33\linewidth]{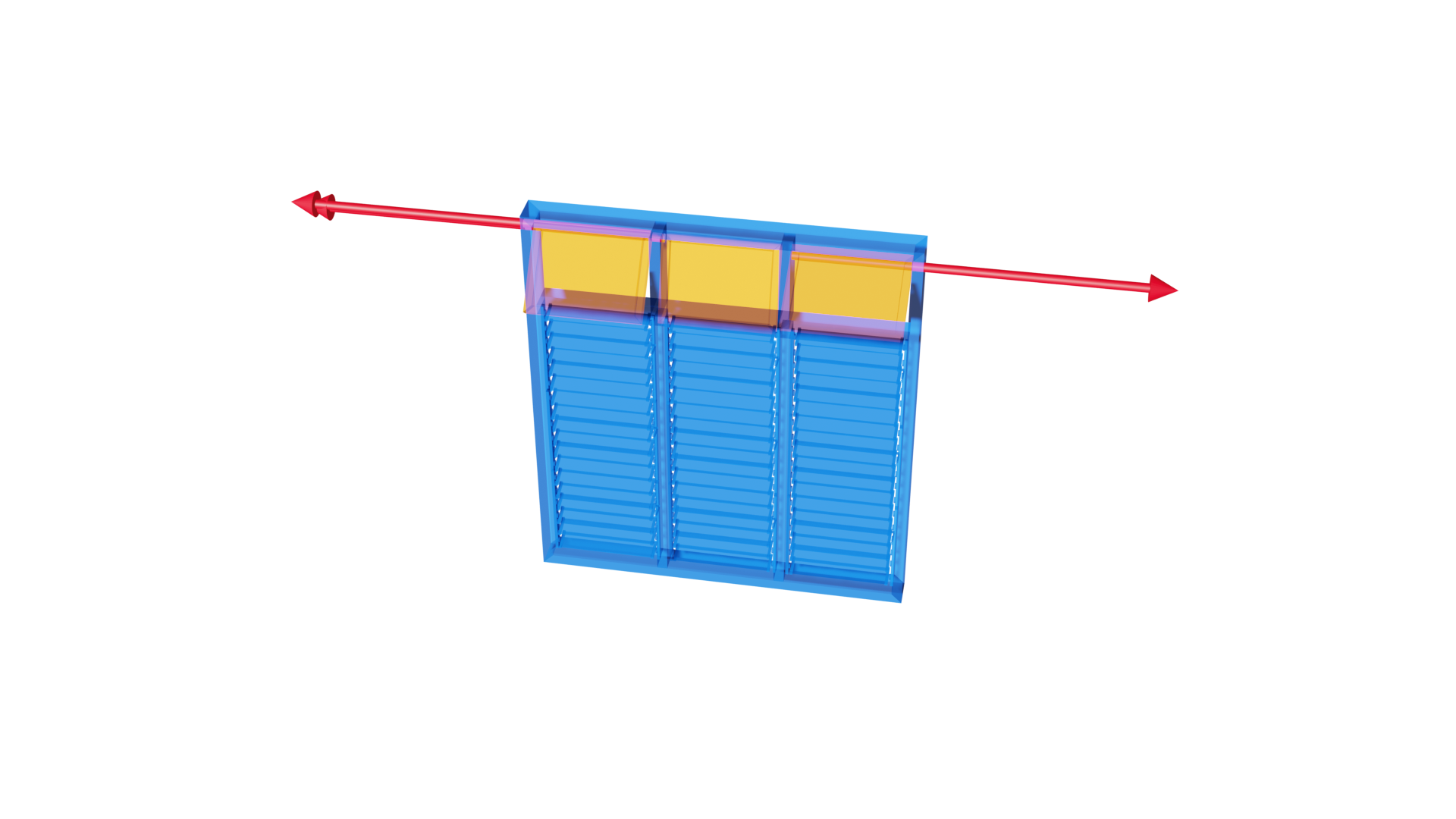}
    \includegraphics[width=0.33\linewidth]{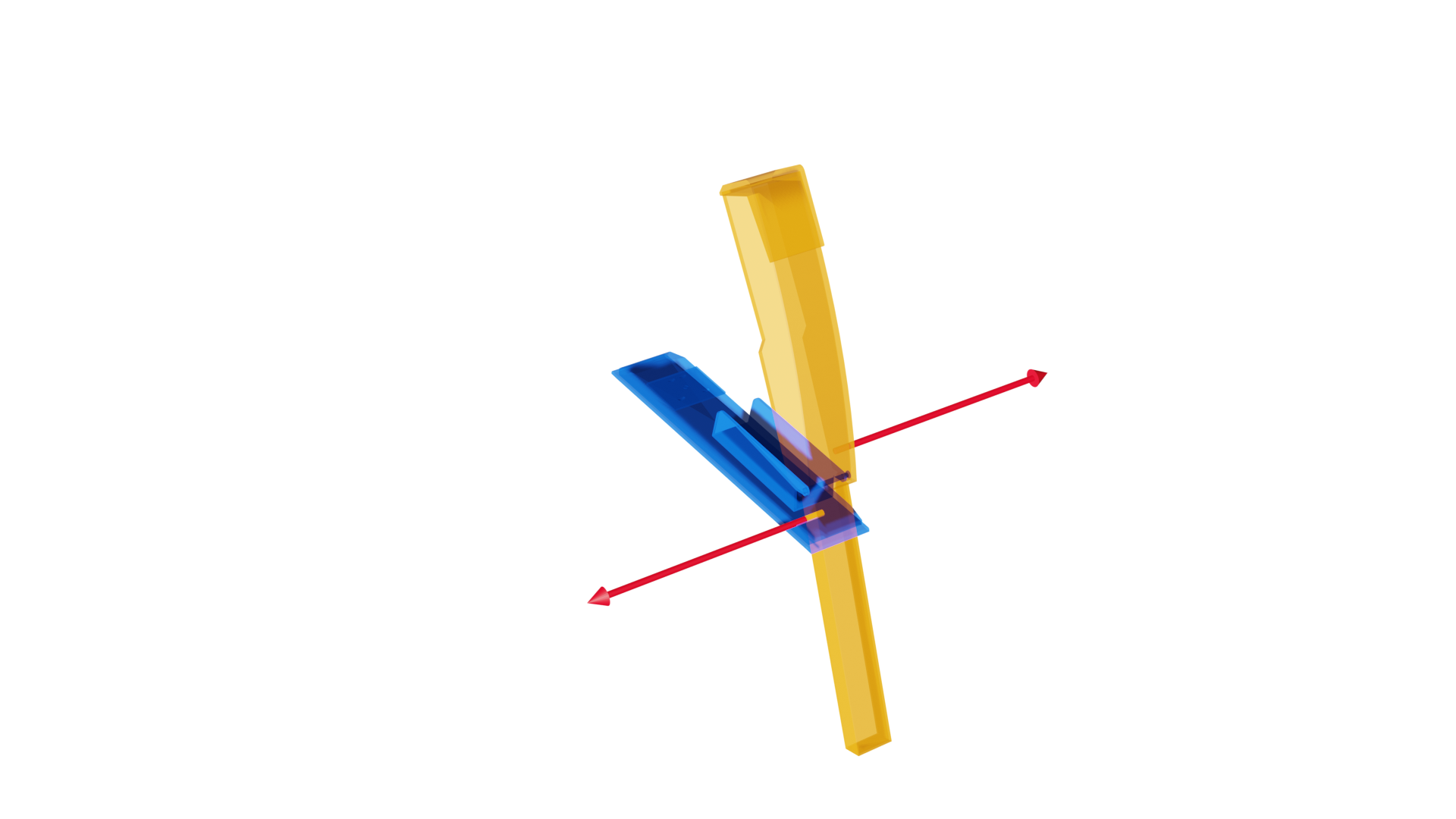}
    \includegraphics[width=0.33\linewidth]{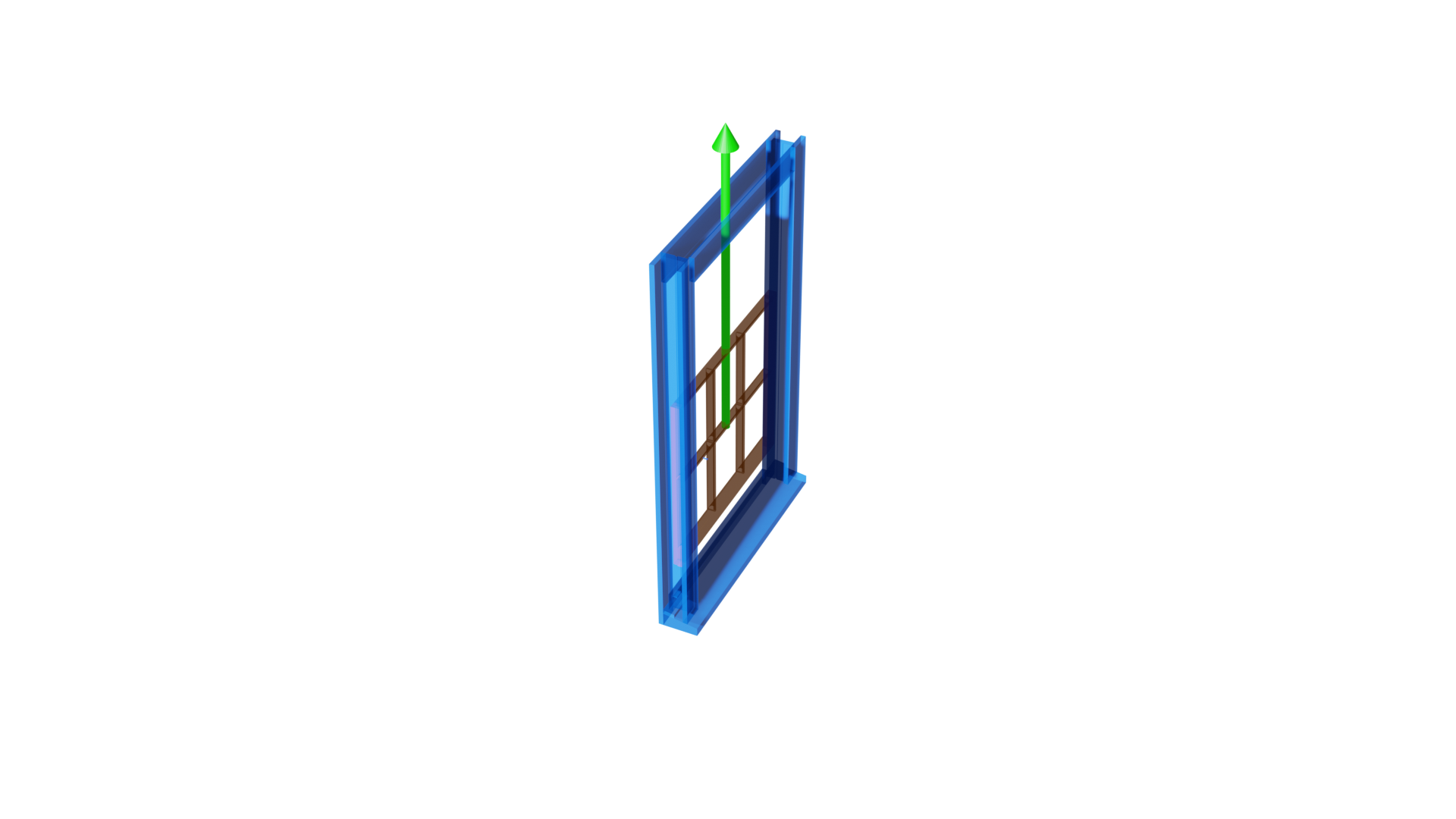}
    \\
    \end{tabular}
    \caption{
    Additional annotated shapes
    }
    \label{figure:qualitative1}
\end{figure*}

\begin{figure*}[ht!]
    \centering
    \setlength{\tabcolsep}{1pt}
    \begin{tabular}{ccc}
        
    \includegraphics[width=0.33\linewidth]{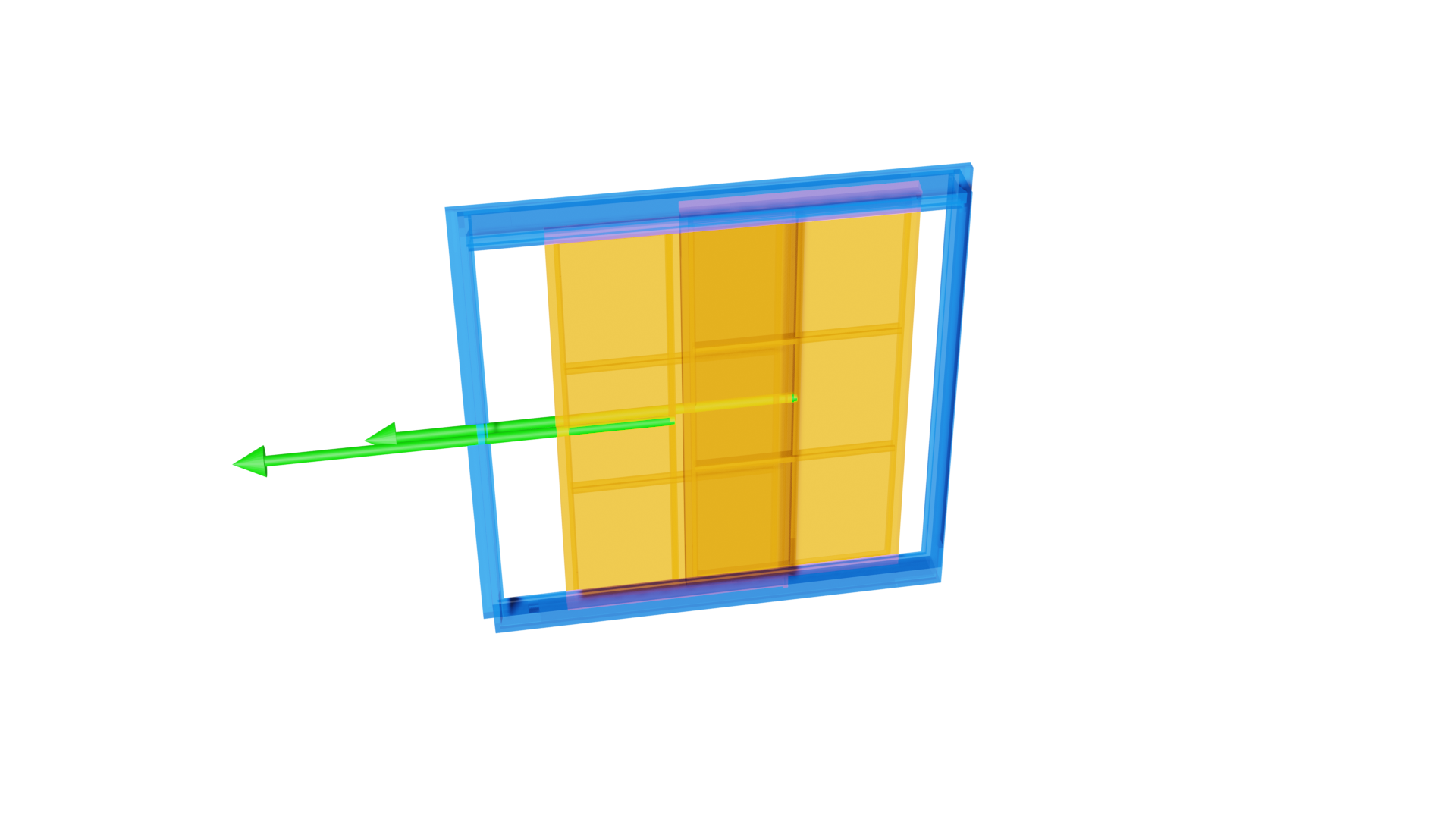}
    \includegraphics[width=0.33\linewidth]{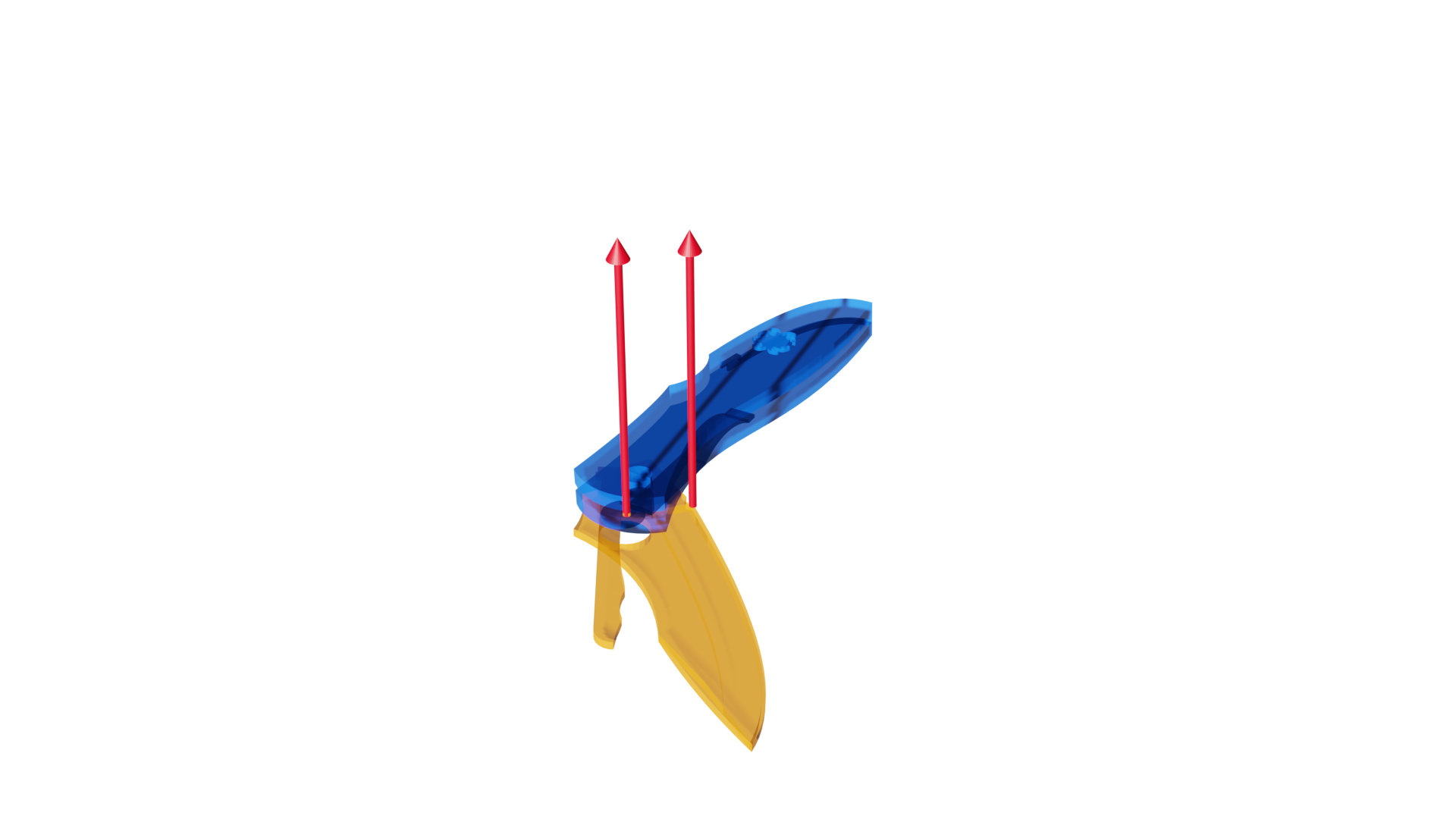}
    \includegraphics[width=0.33\linewidth]{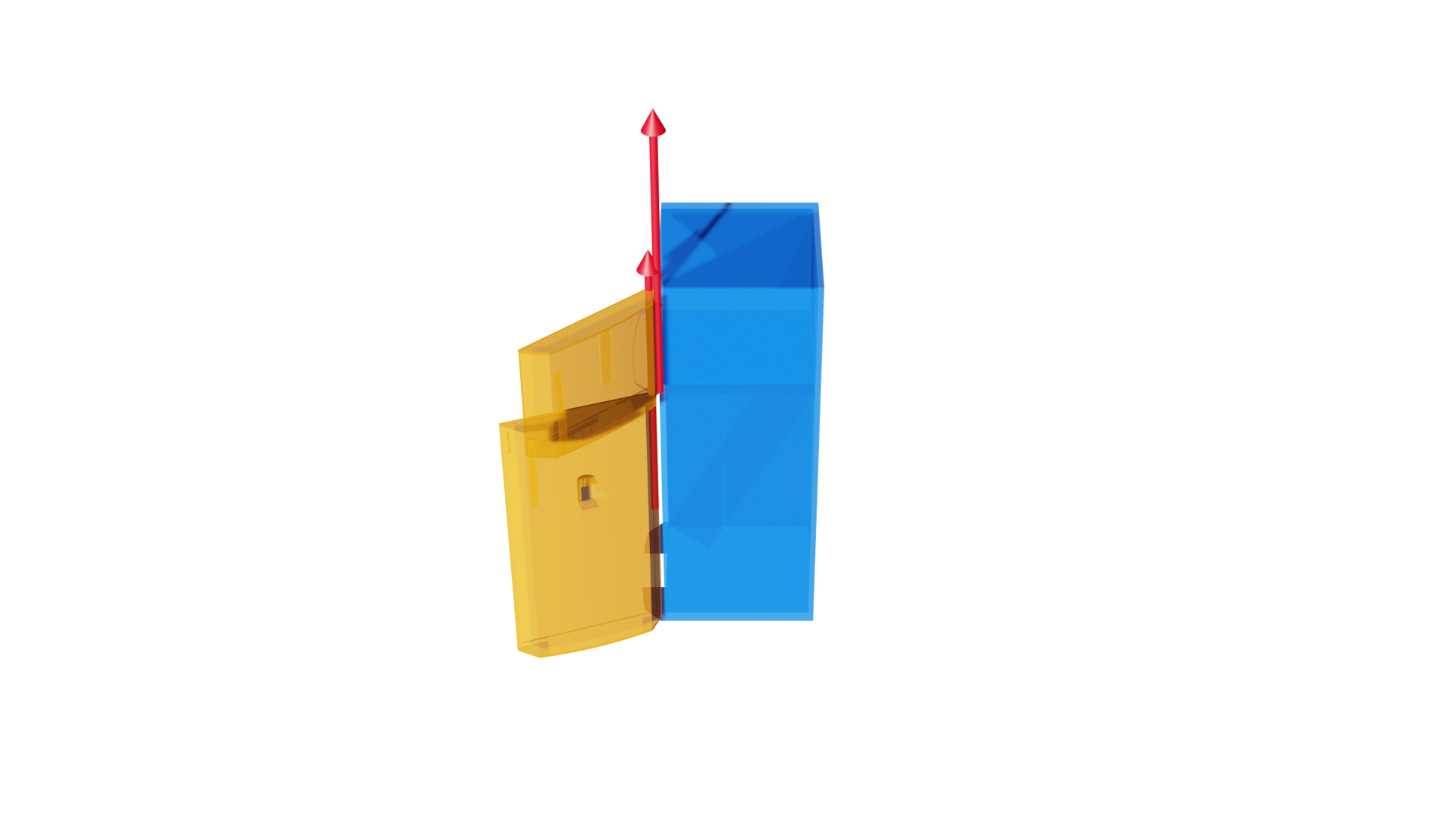}
    \\
    \includegraphics[width=0.33\linewidth]{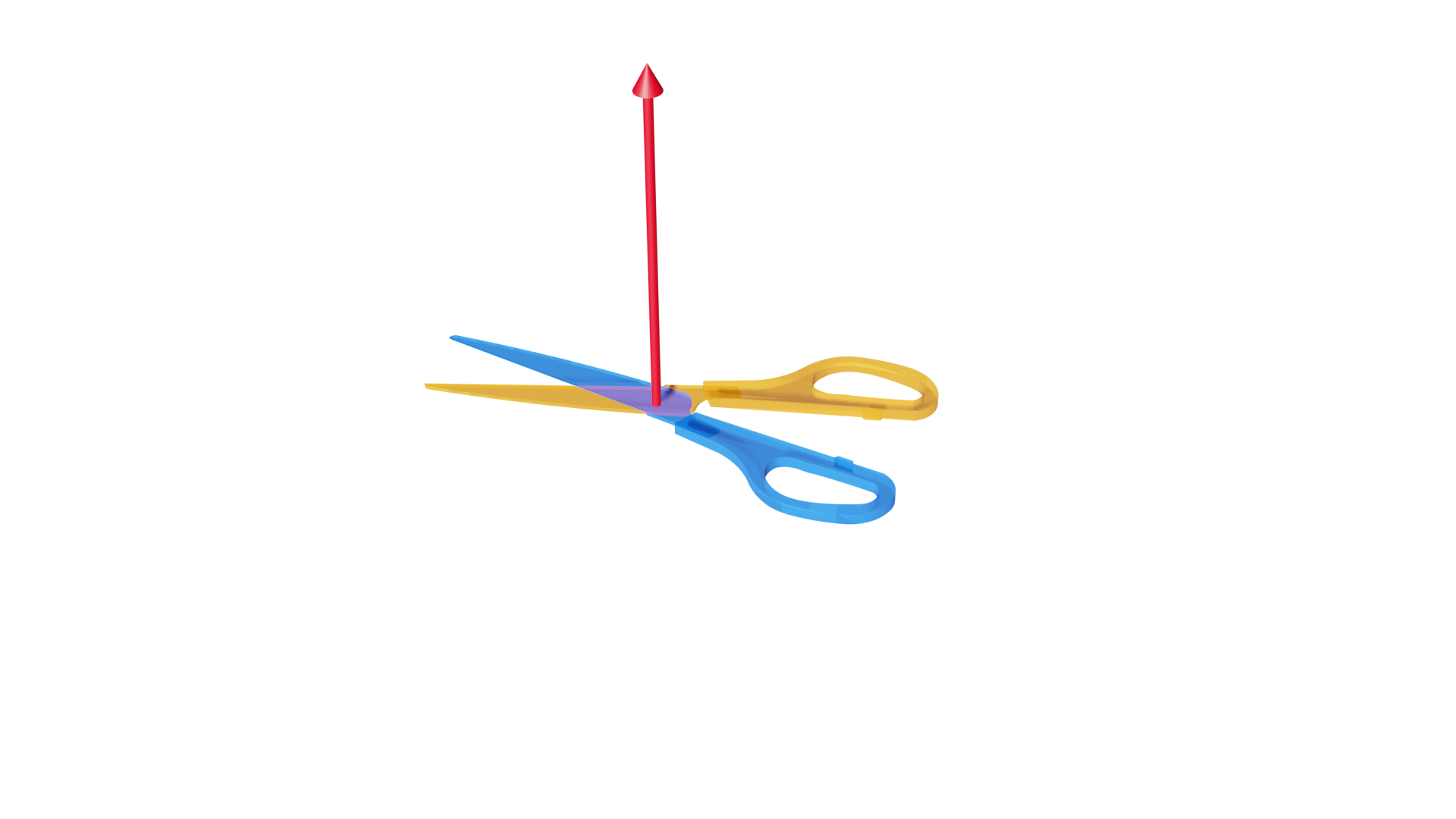}
    \includegraphics[width=0.33\linewidth]{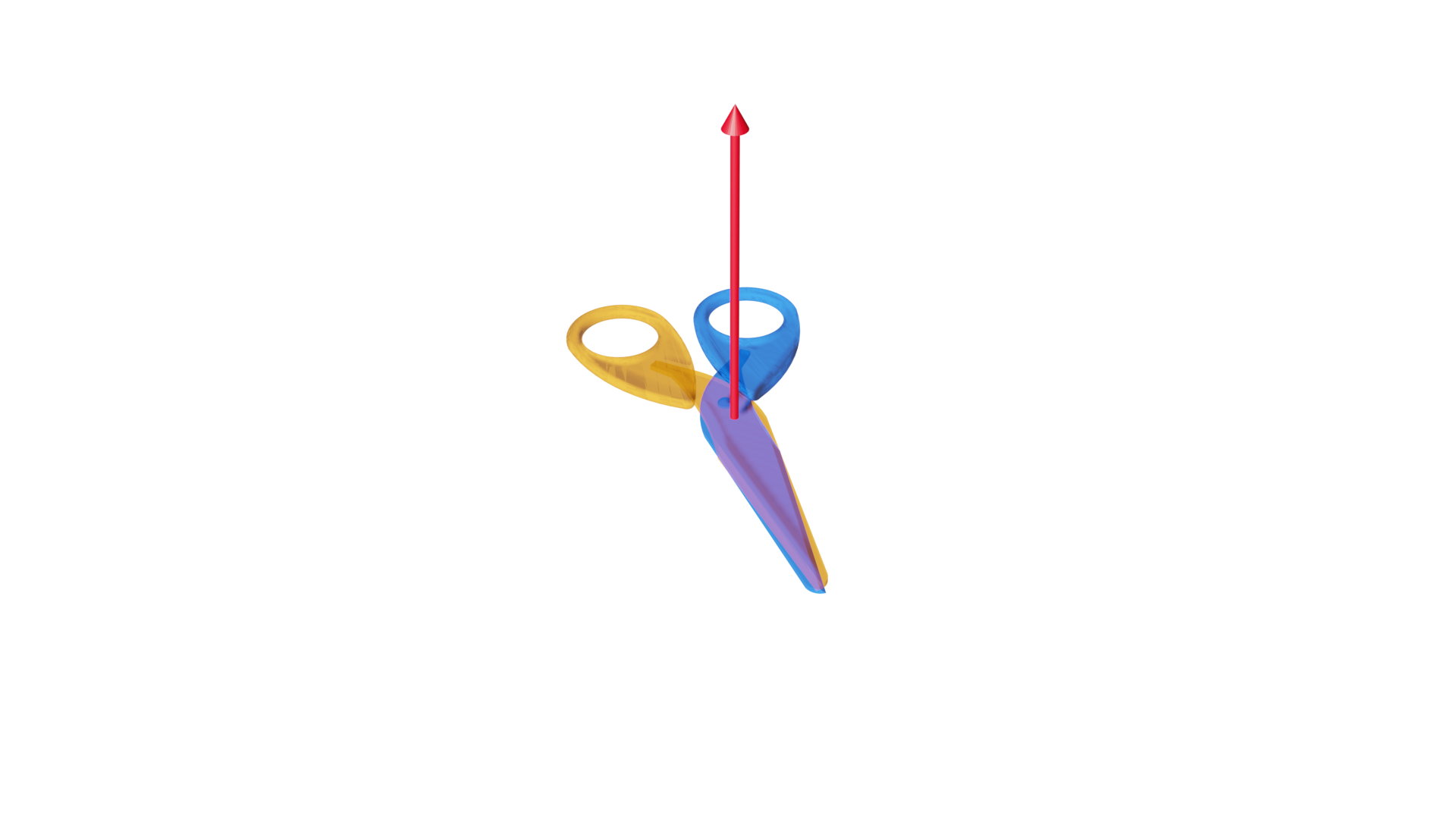}
    \includegraphics[width=0.33\linewidth]{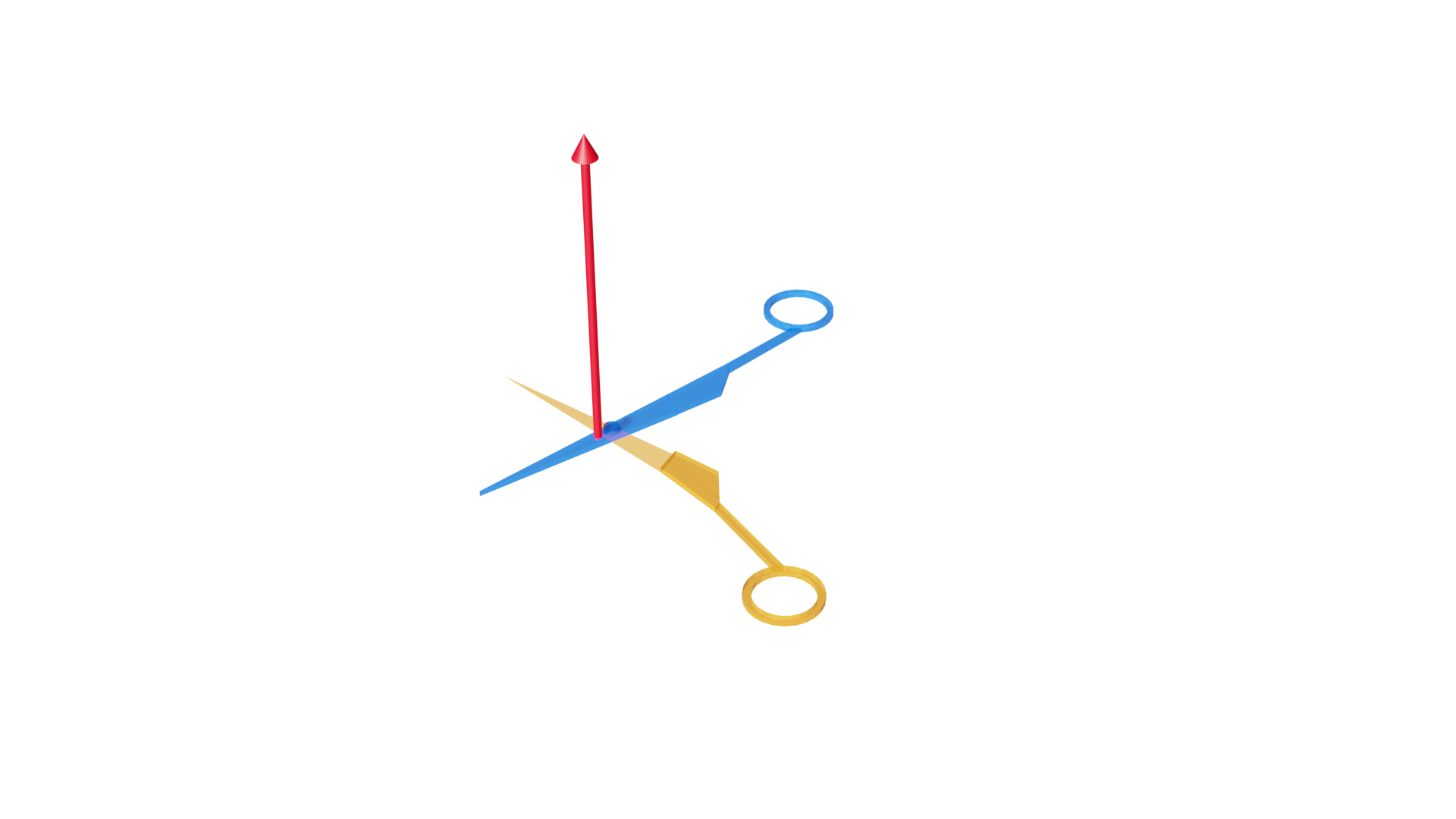}
    \\
    \includegraphics[width=0.33\linewidth]{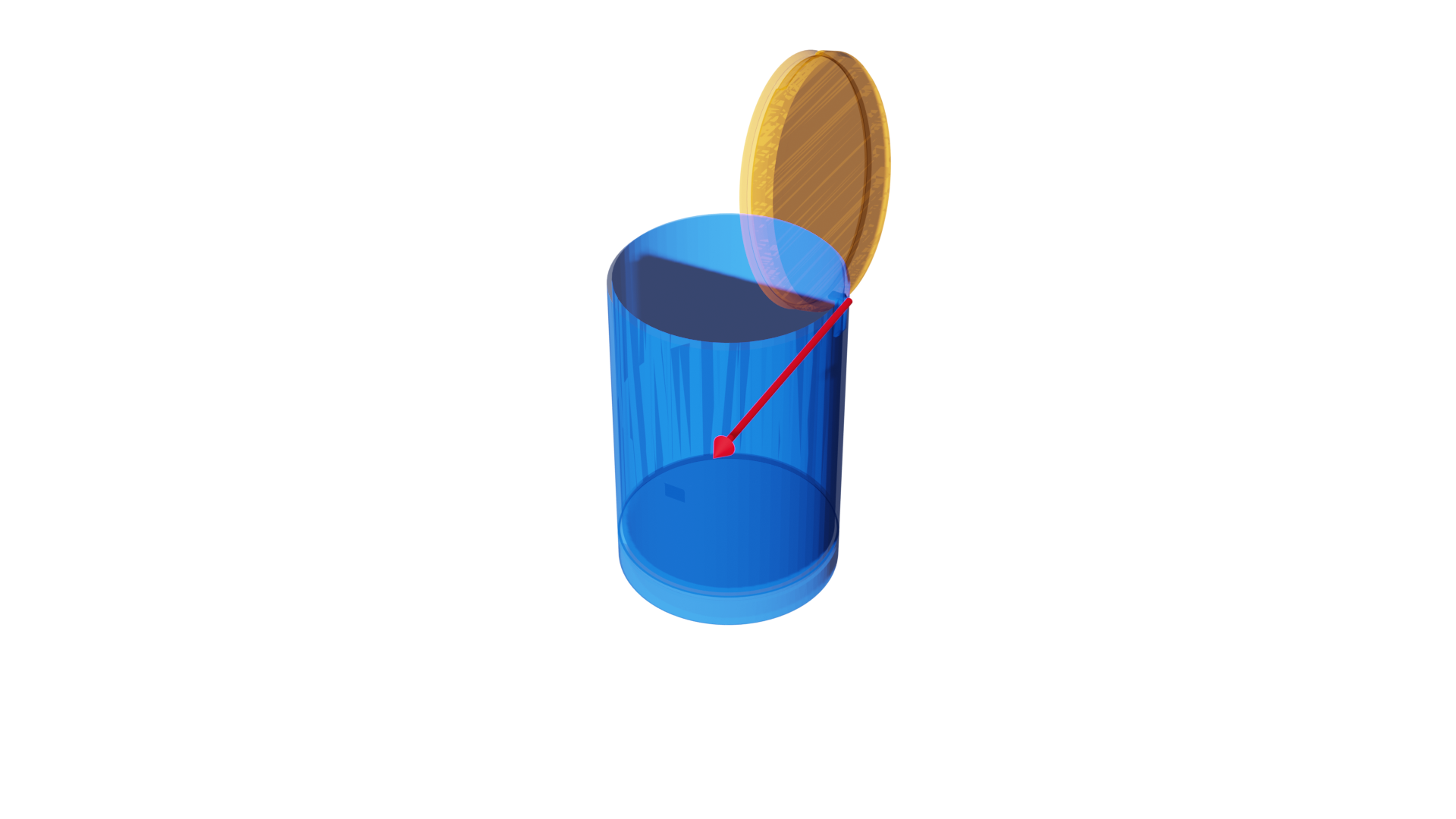}
    \includegraphics[width=0.33\linewidth]{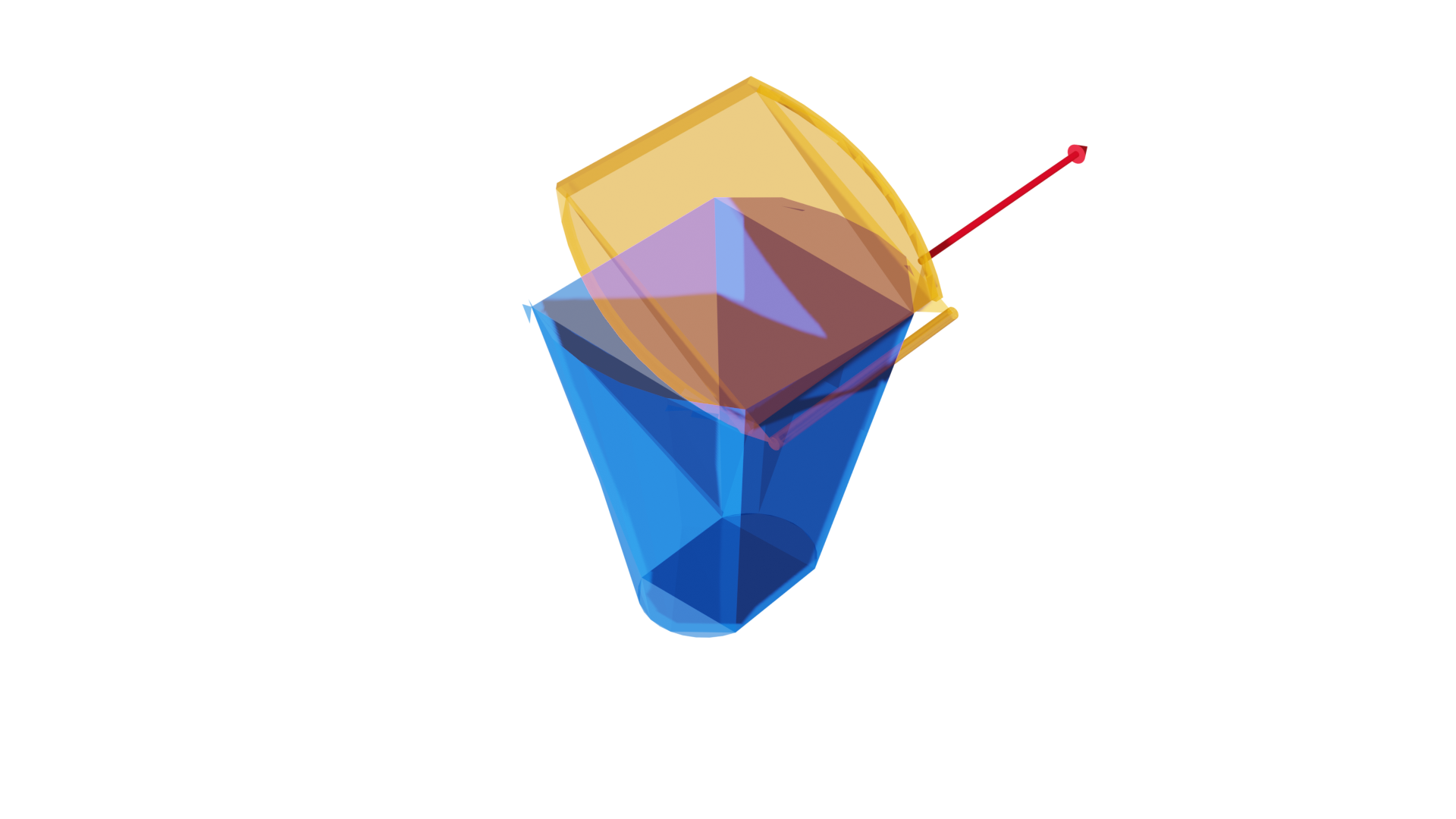}
    \\
    \includegraphics[width=0.33\linewidth]{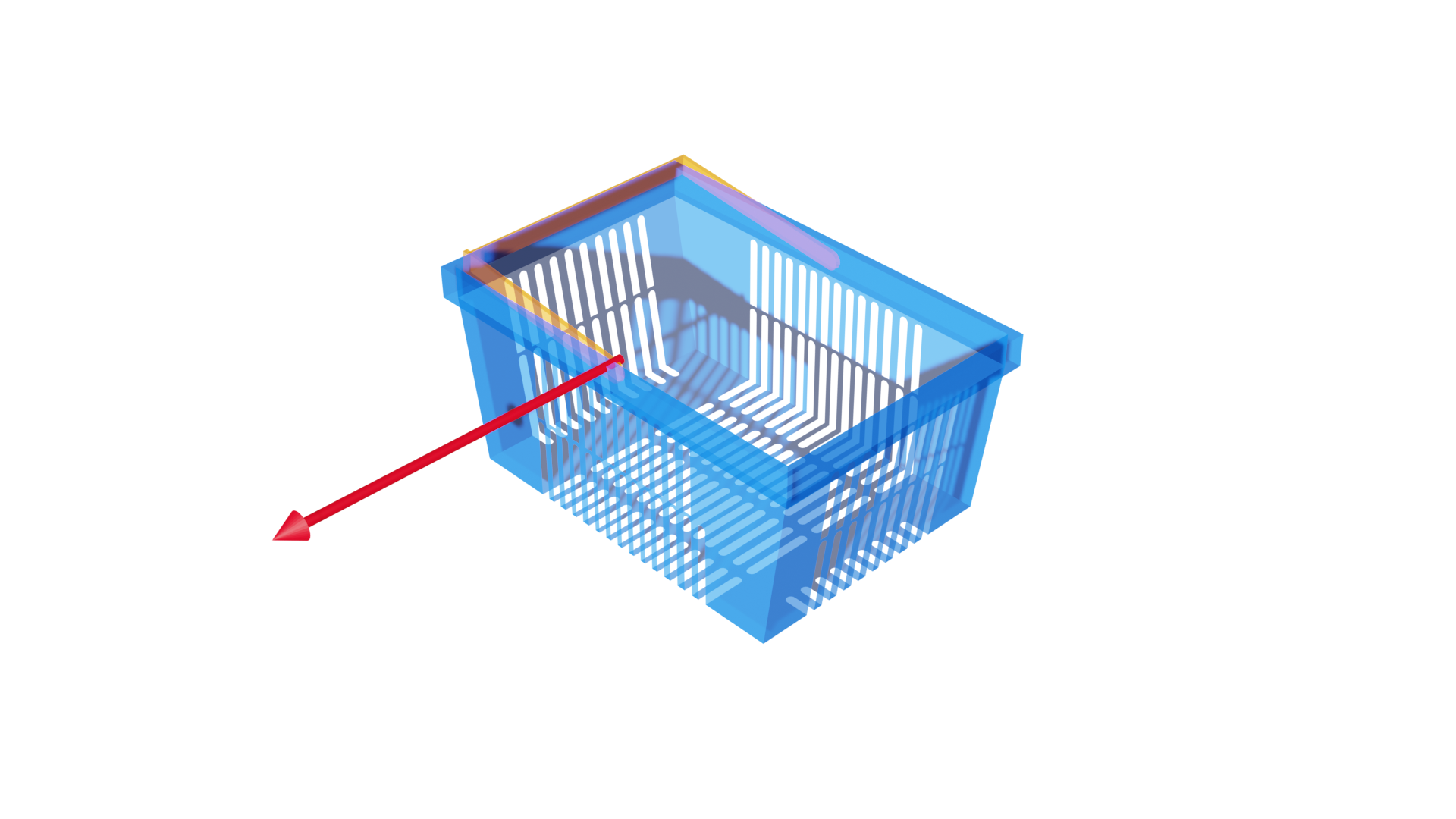}
    \includegraphics[width=0.33\linewidth]{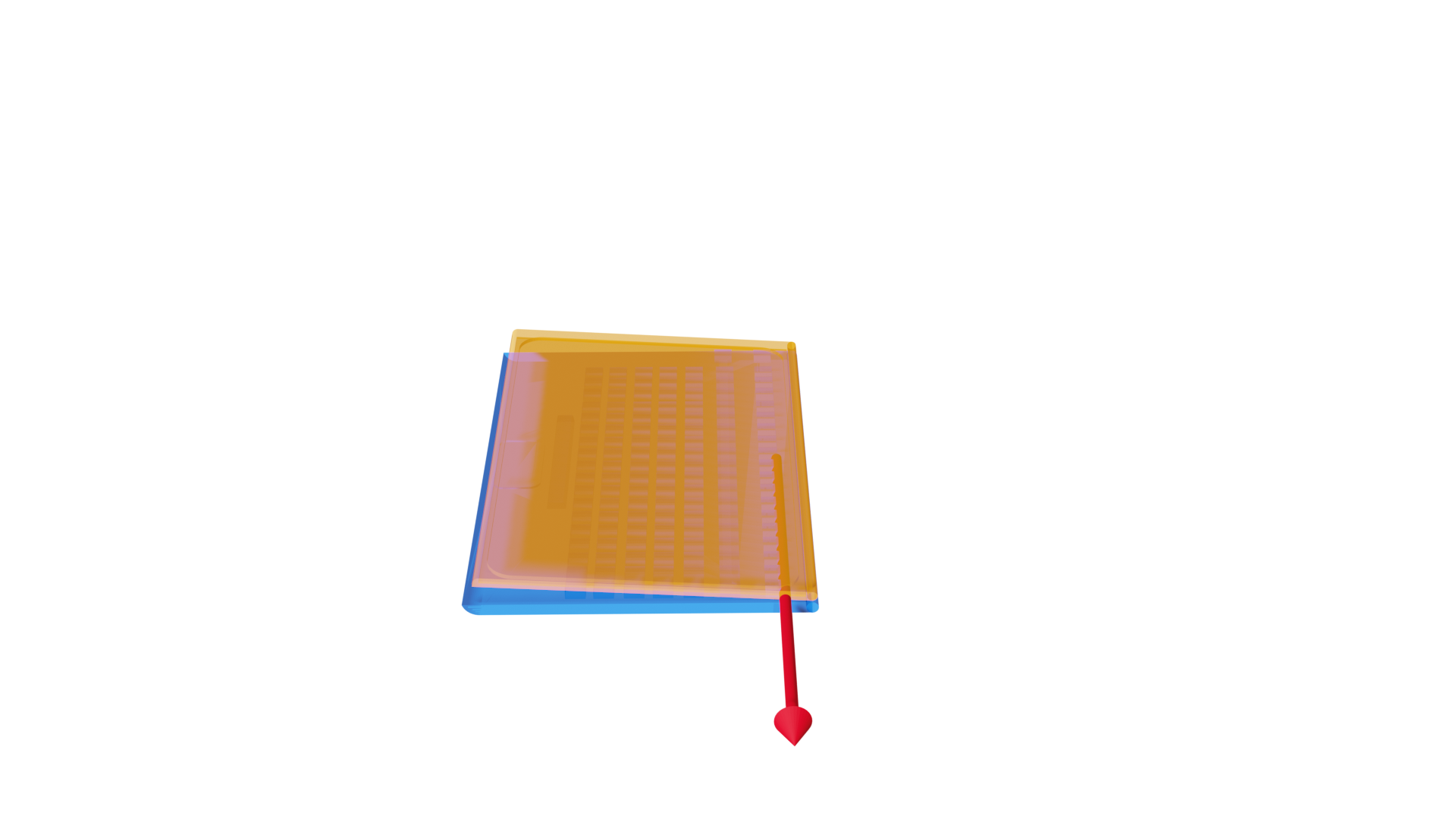}
    \includegraphics[width=0.33\linewidth]{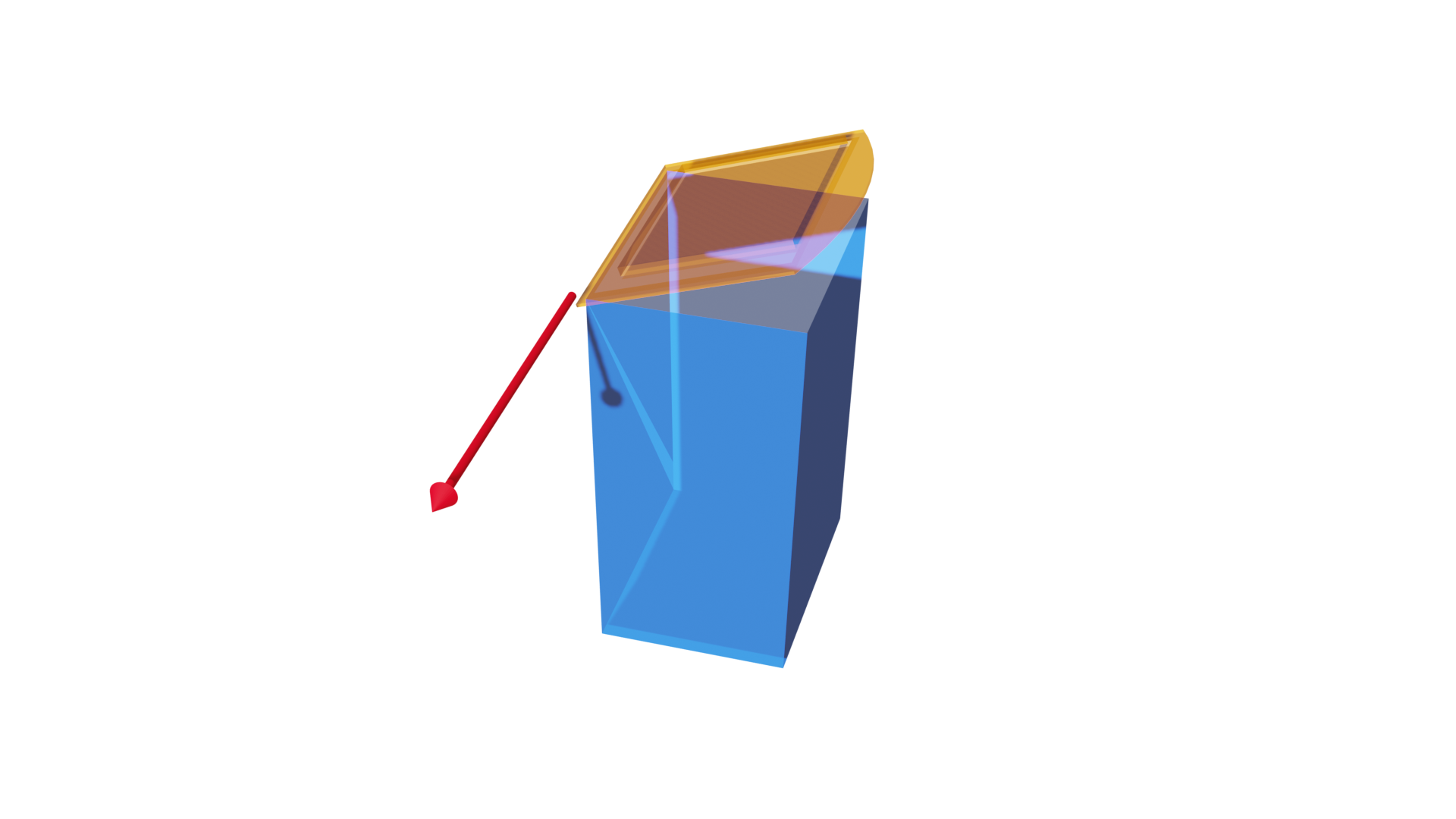}
    \\
    \includegraphics[width=0.33\linewidth]{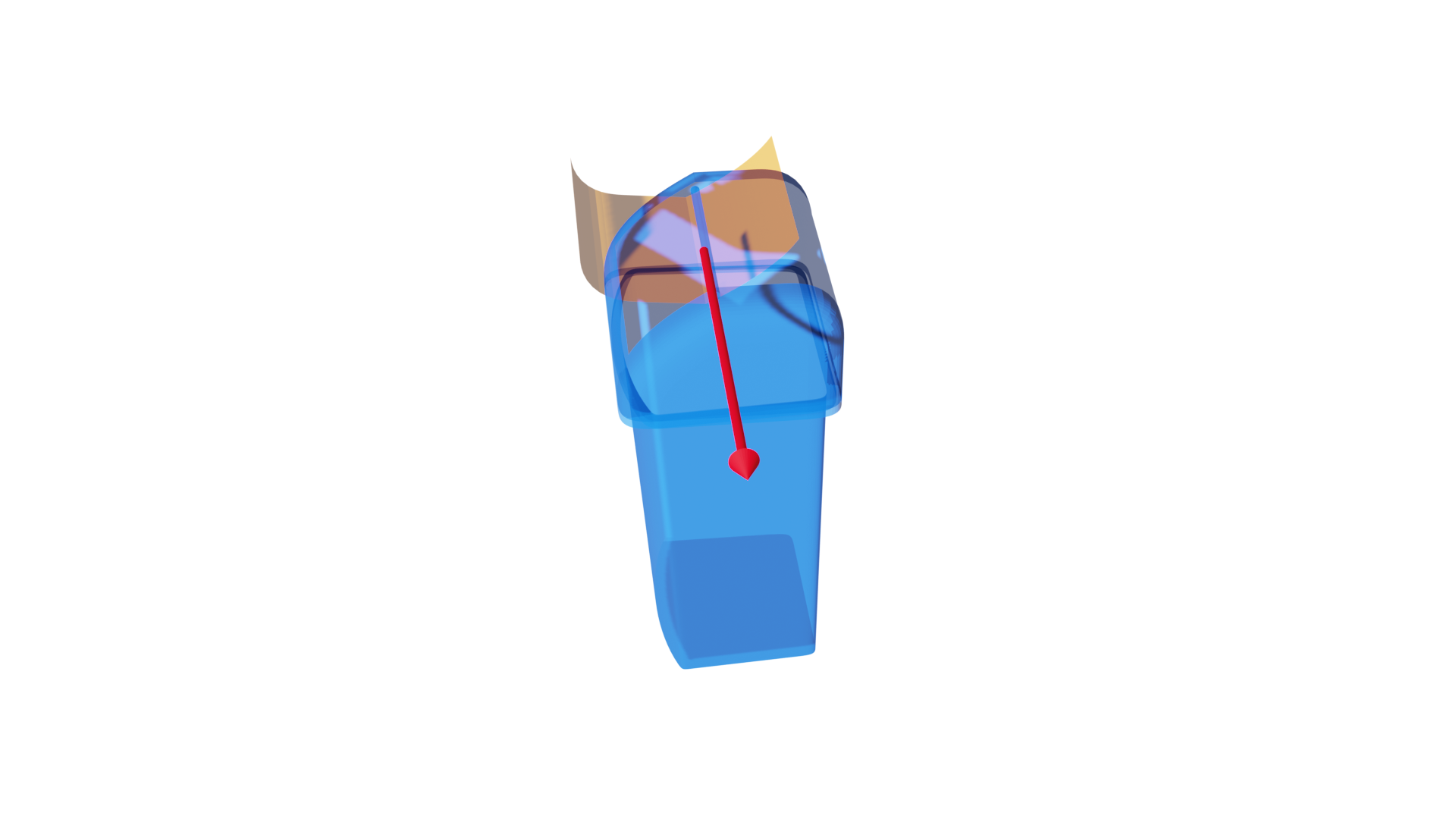}
    \includegraphics[width=0.33\linewidth]{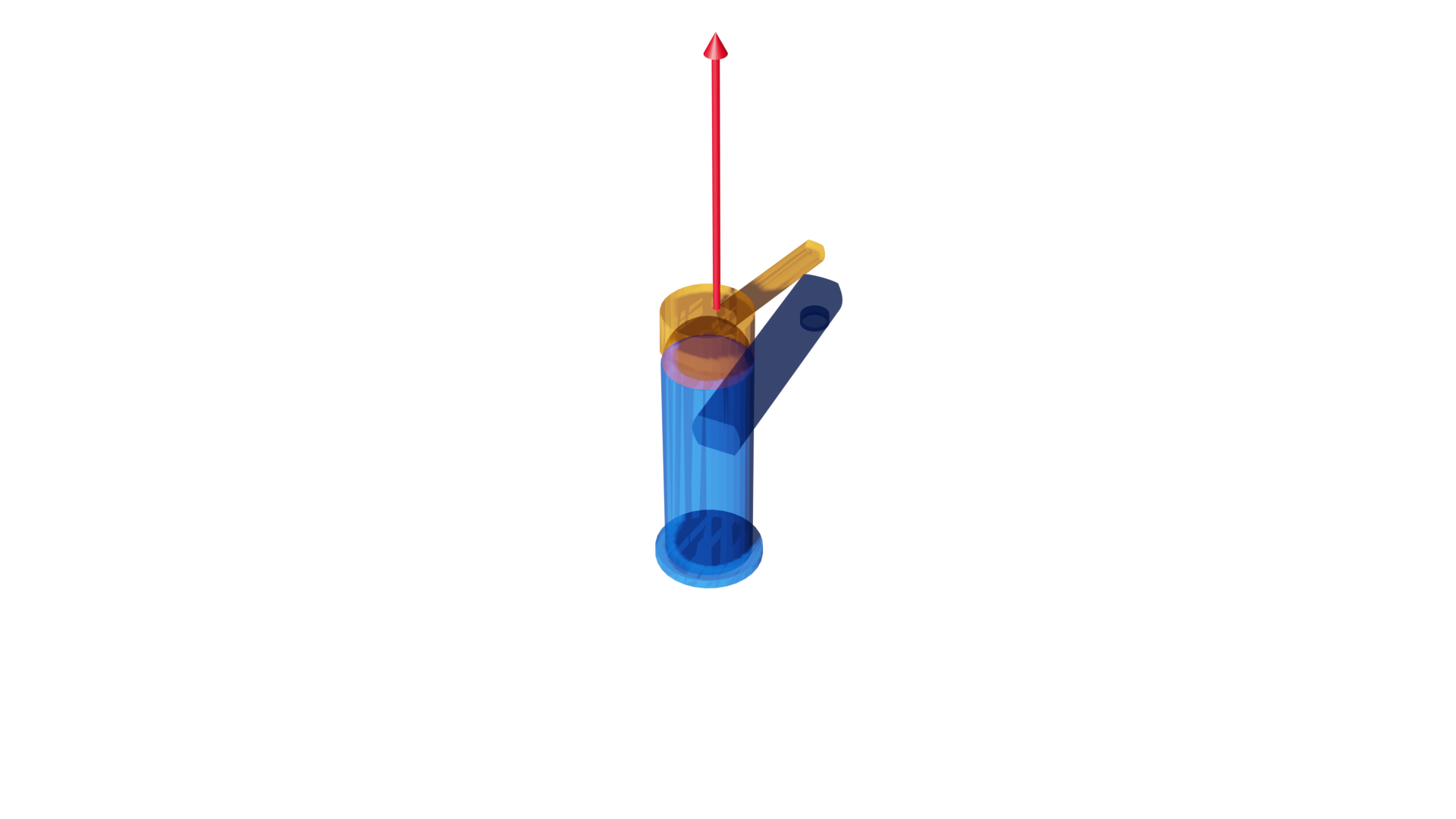}
    \includegraphics[width=0.33\linewidth]{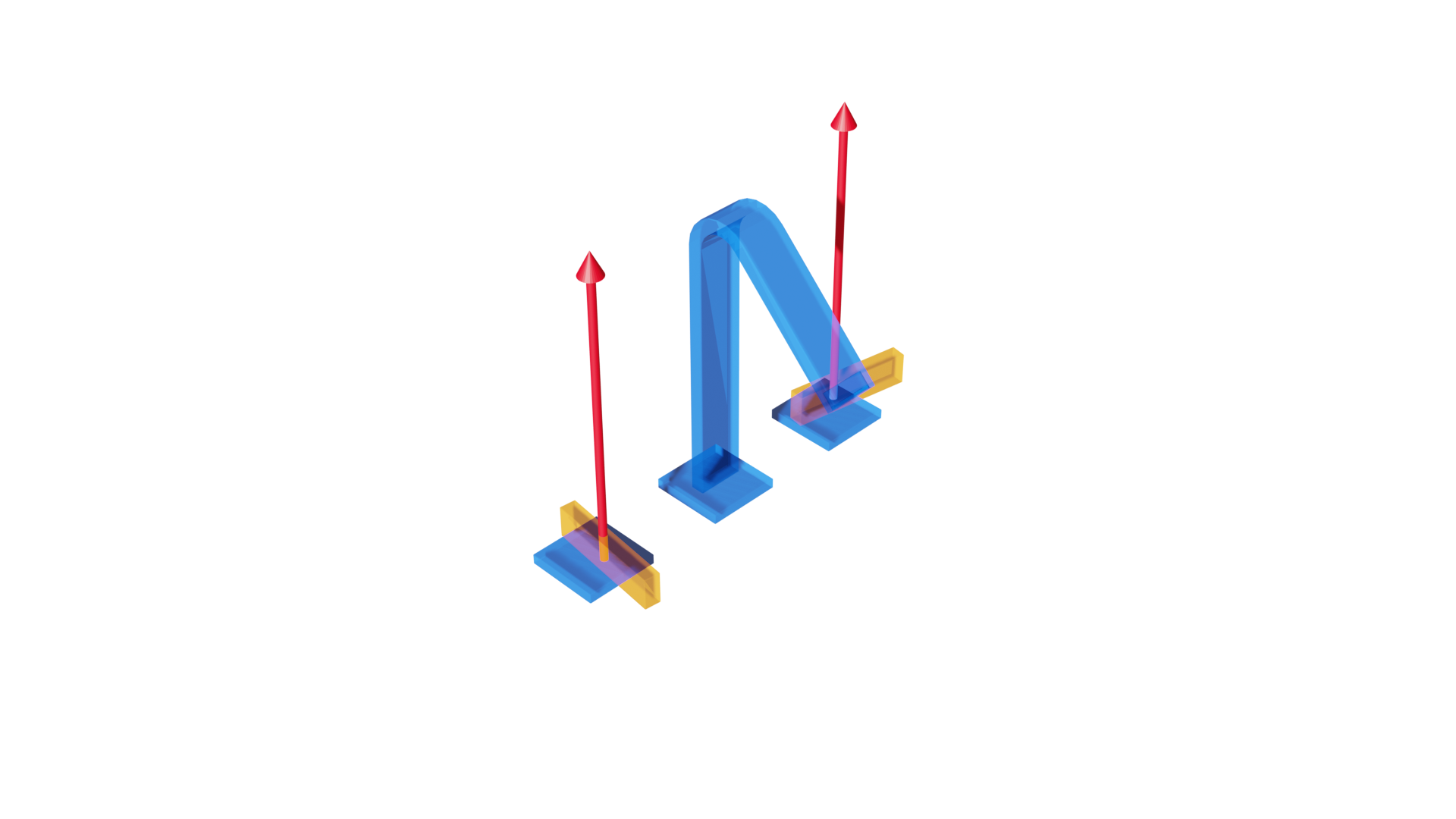}
    \\
    \end{tabular}
    \caption{
    Additional annotated shapes
    }
    \label{figure:qualitative2}
\end{figure*}

\begin{figure*}[ht!]
    \centering
    \setlength{\tabcolsep}{1pt}
    \begin{tabular}{ccc}
        
    \includegraphics[width=0.33\linewidth]{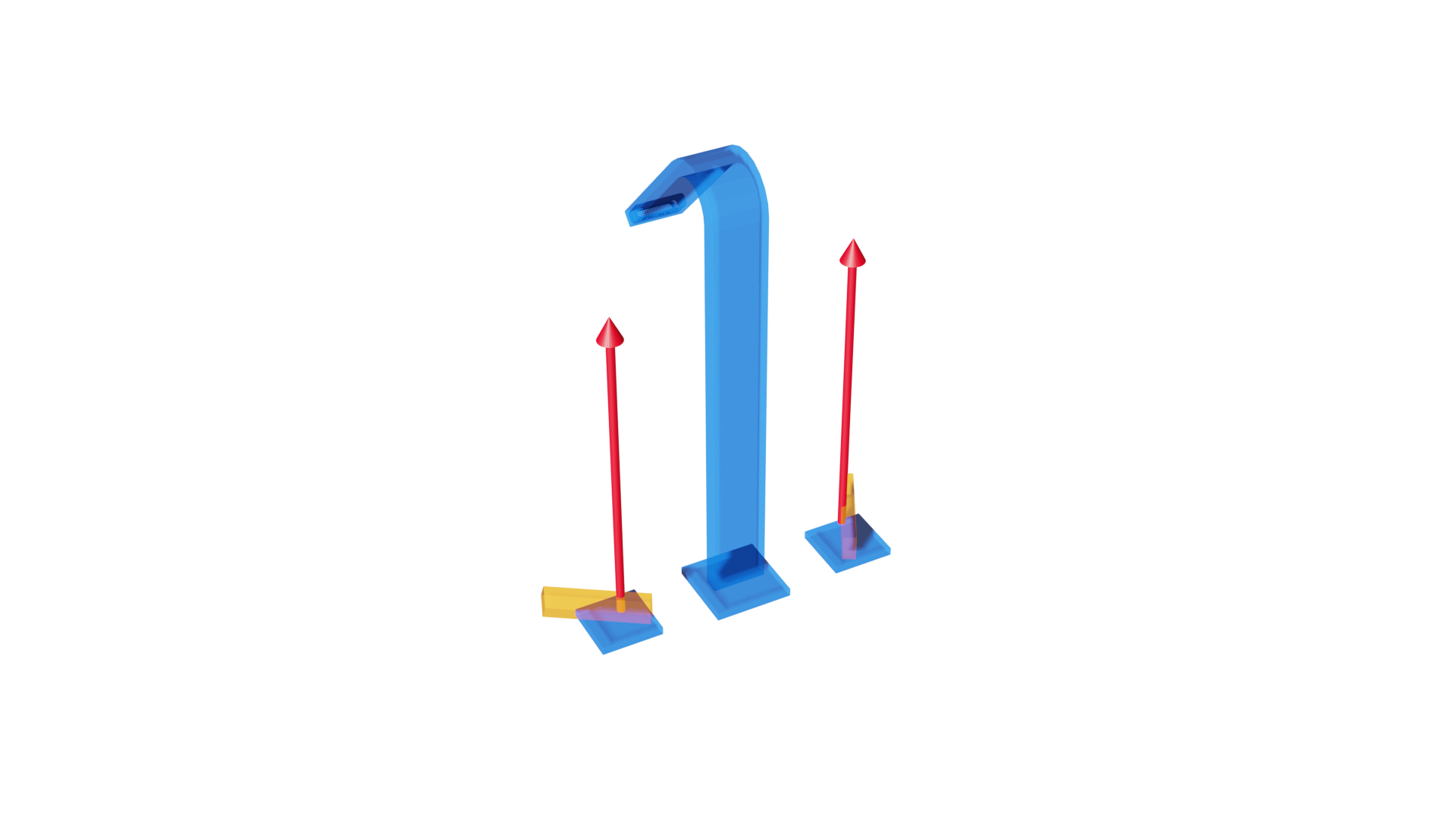}
    \includegraphics[width=0.33\linewidth]{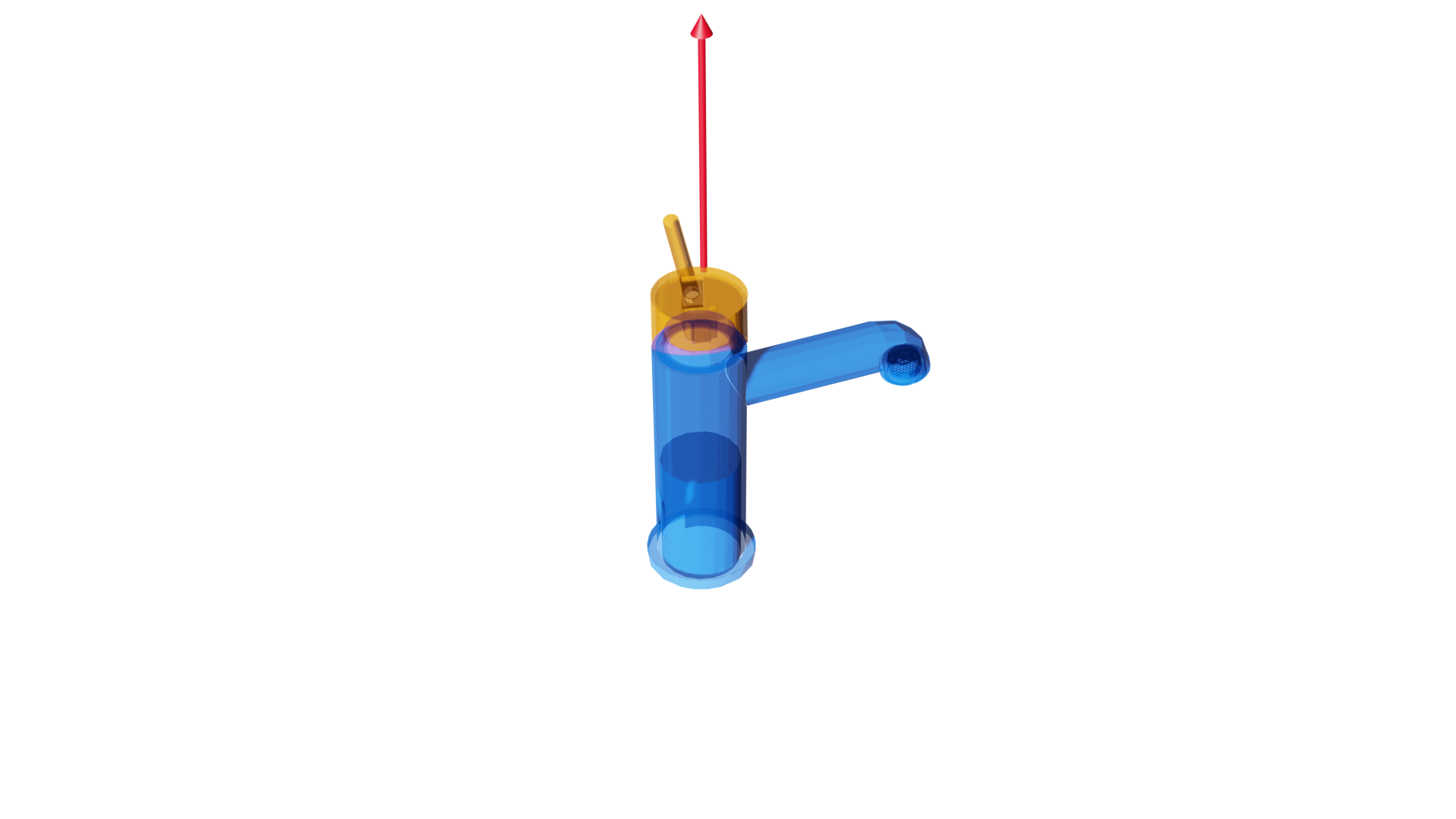}
    \includegraphics[width=0.33\linewidth]{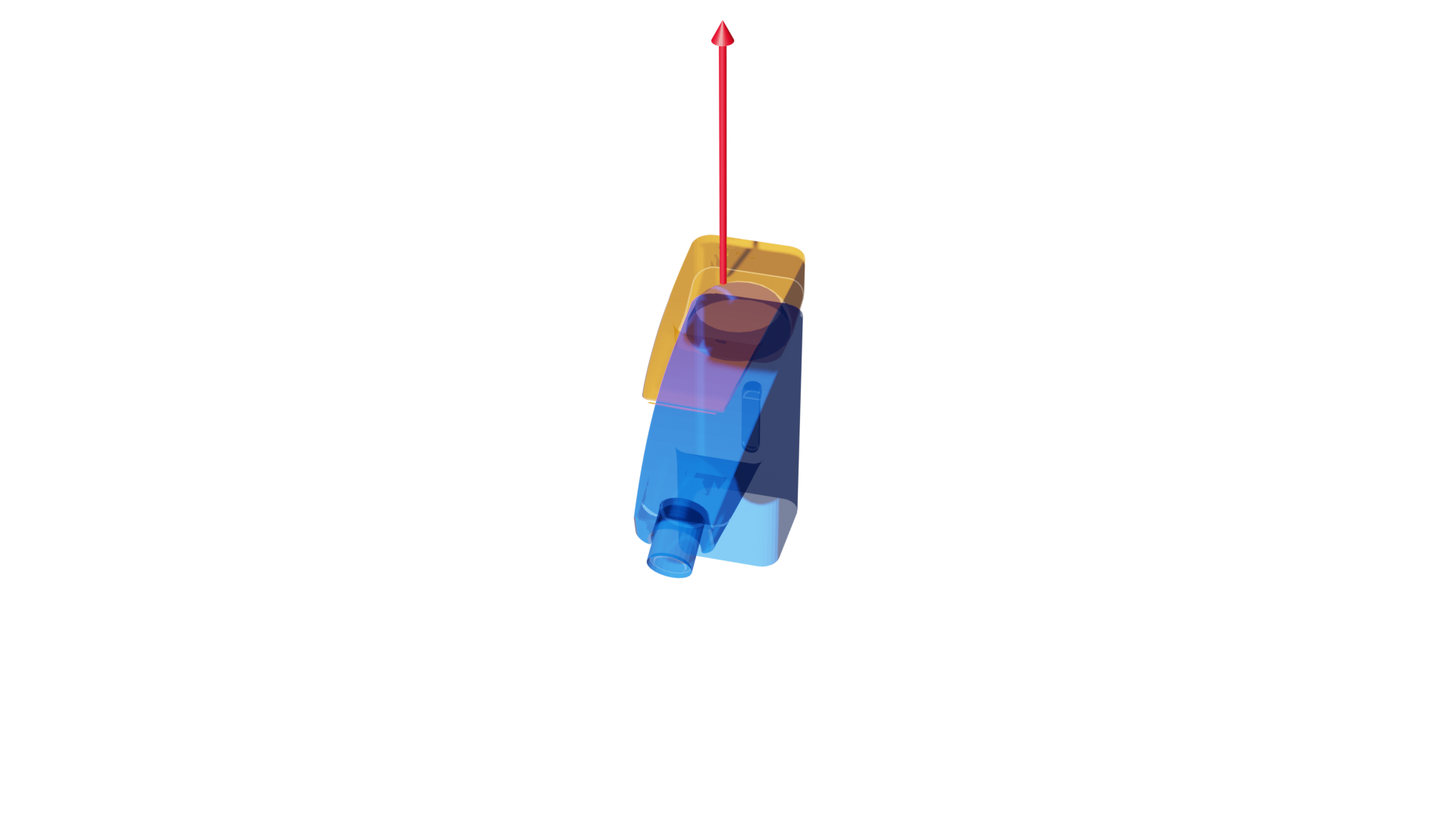}
    \\
    \includegraphics[width=0.33\linewidth]{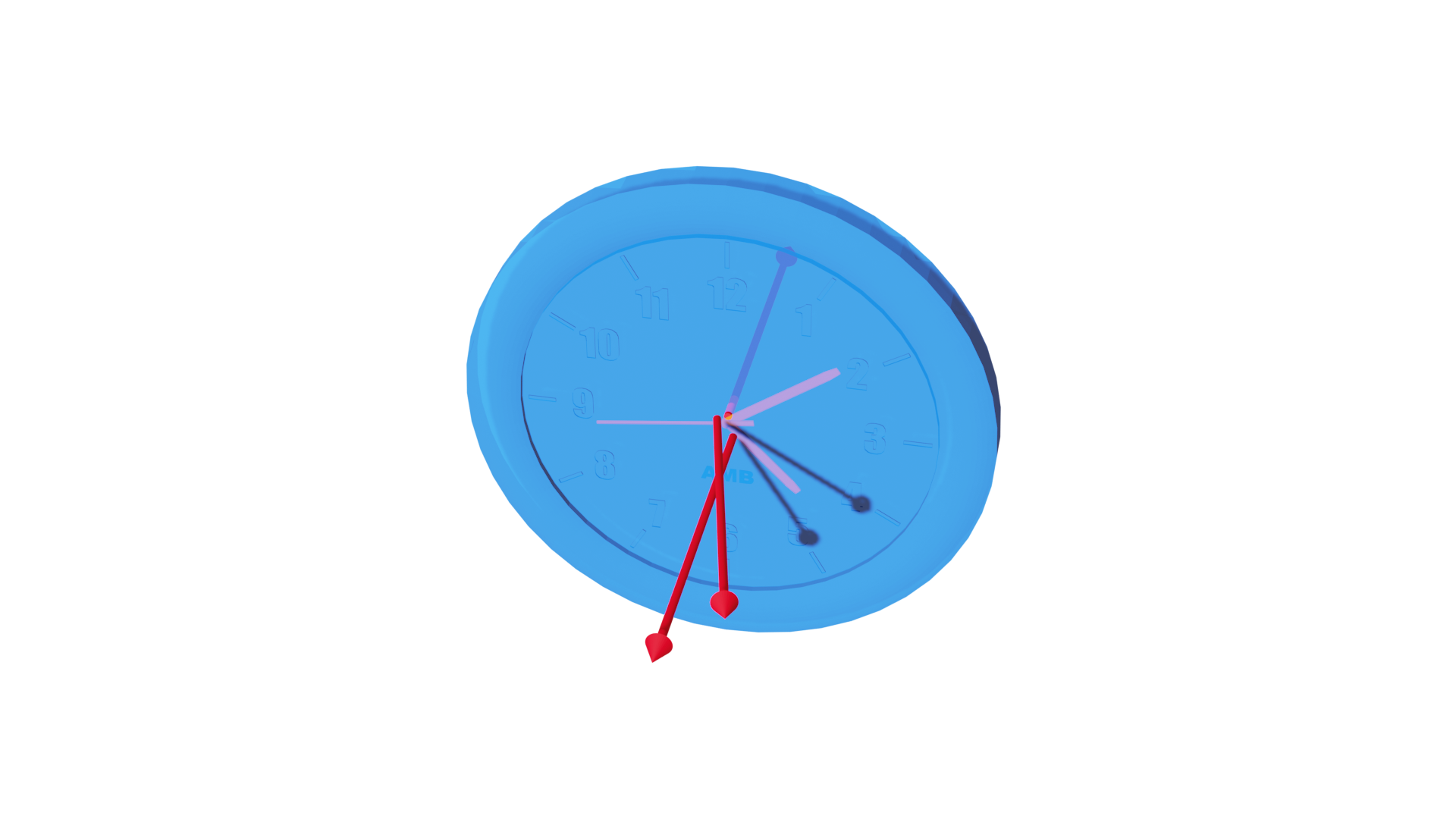}
    \includegraphics[width=0.33\linewidth]{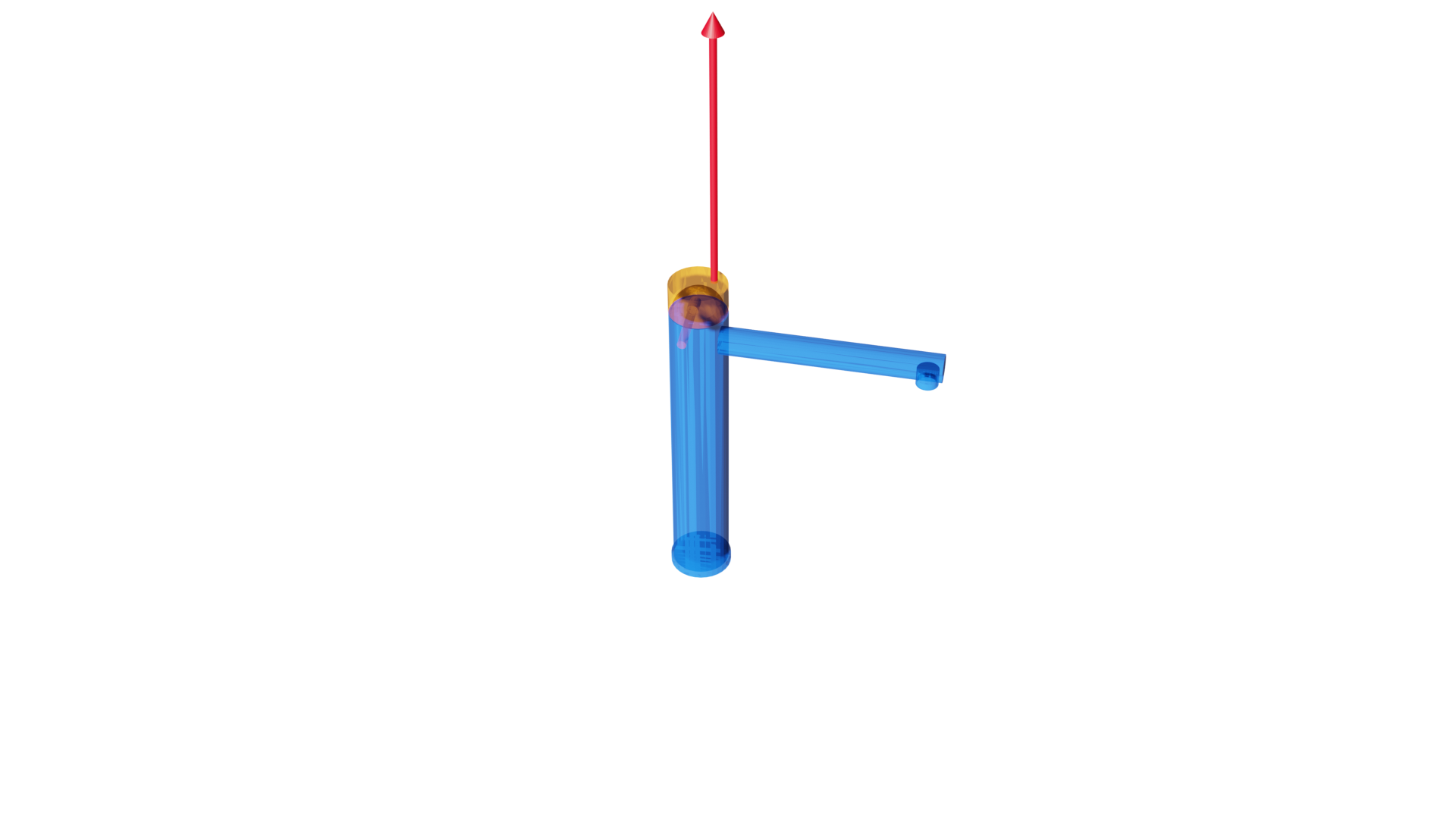}
    \includegraphics[width=0.33\linewidth]{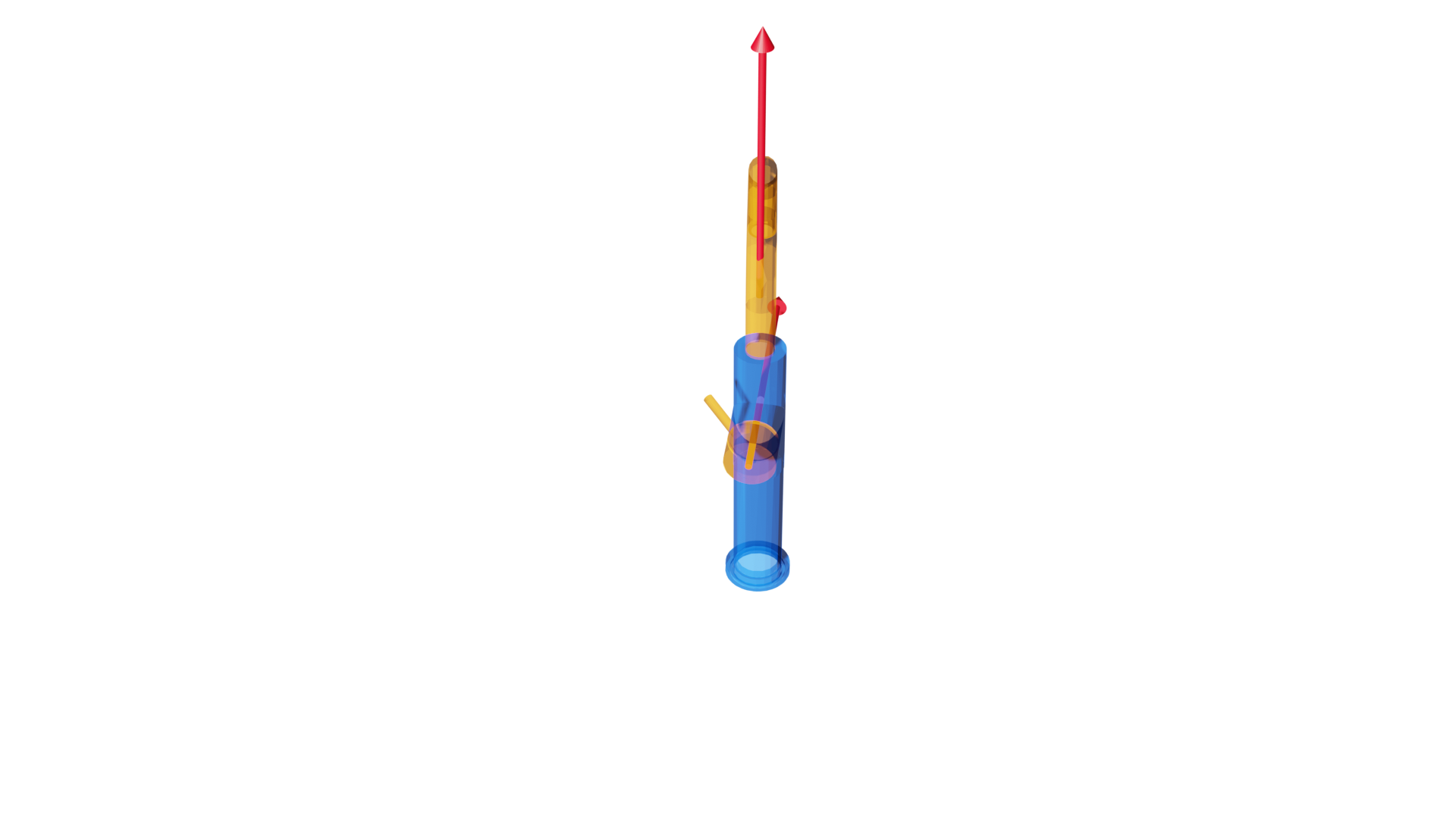}
    \\
    \includegraphics[width=0.33\linewidth]{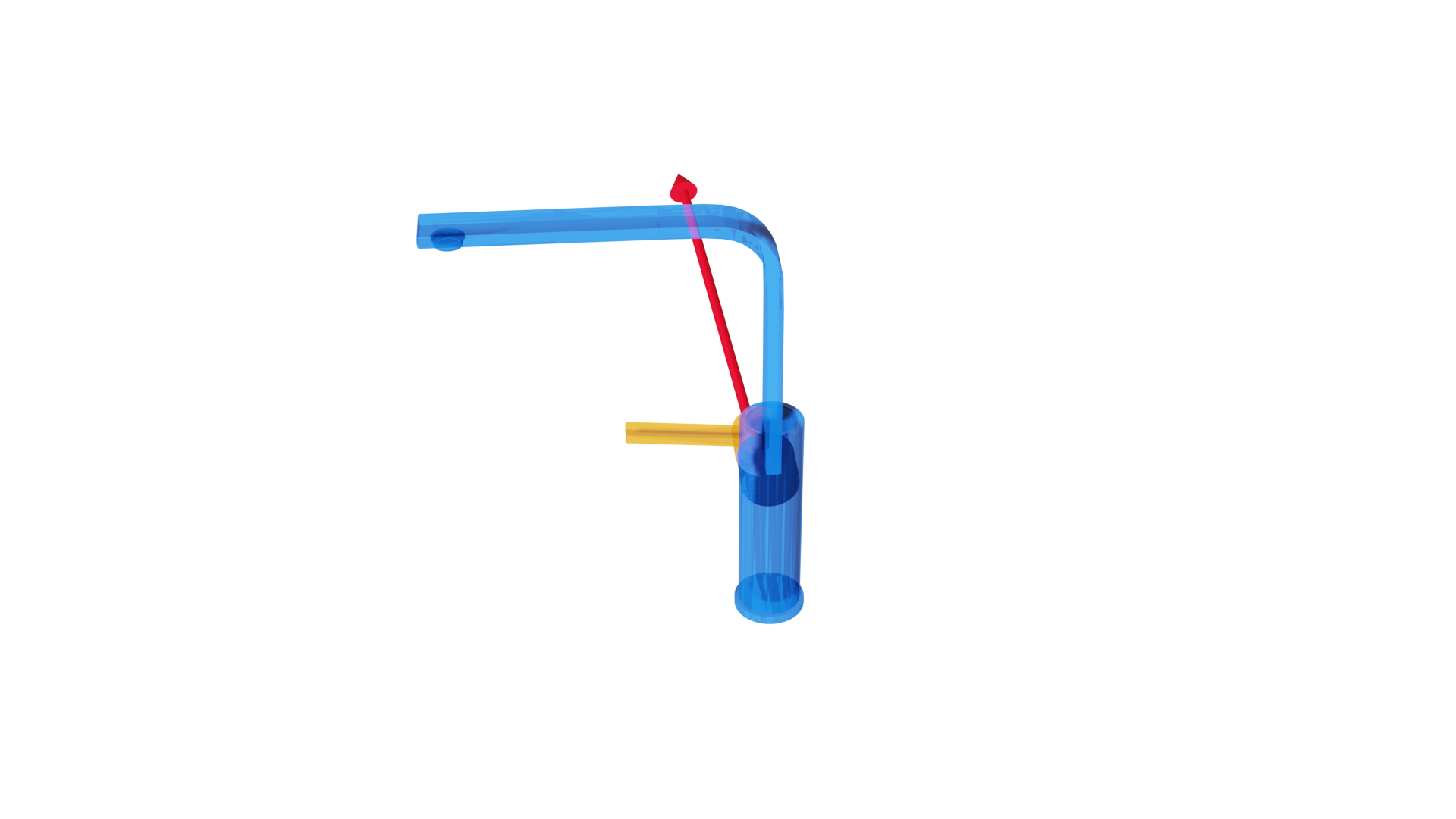}
    \includegraphics[width=0.33\linewidth]{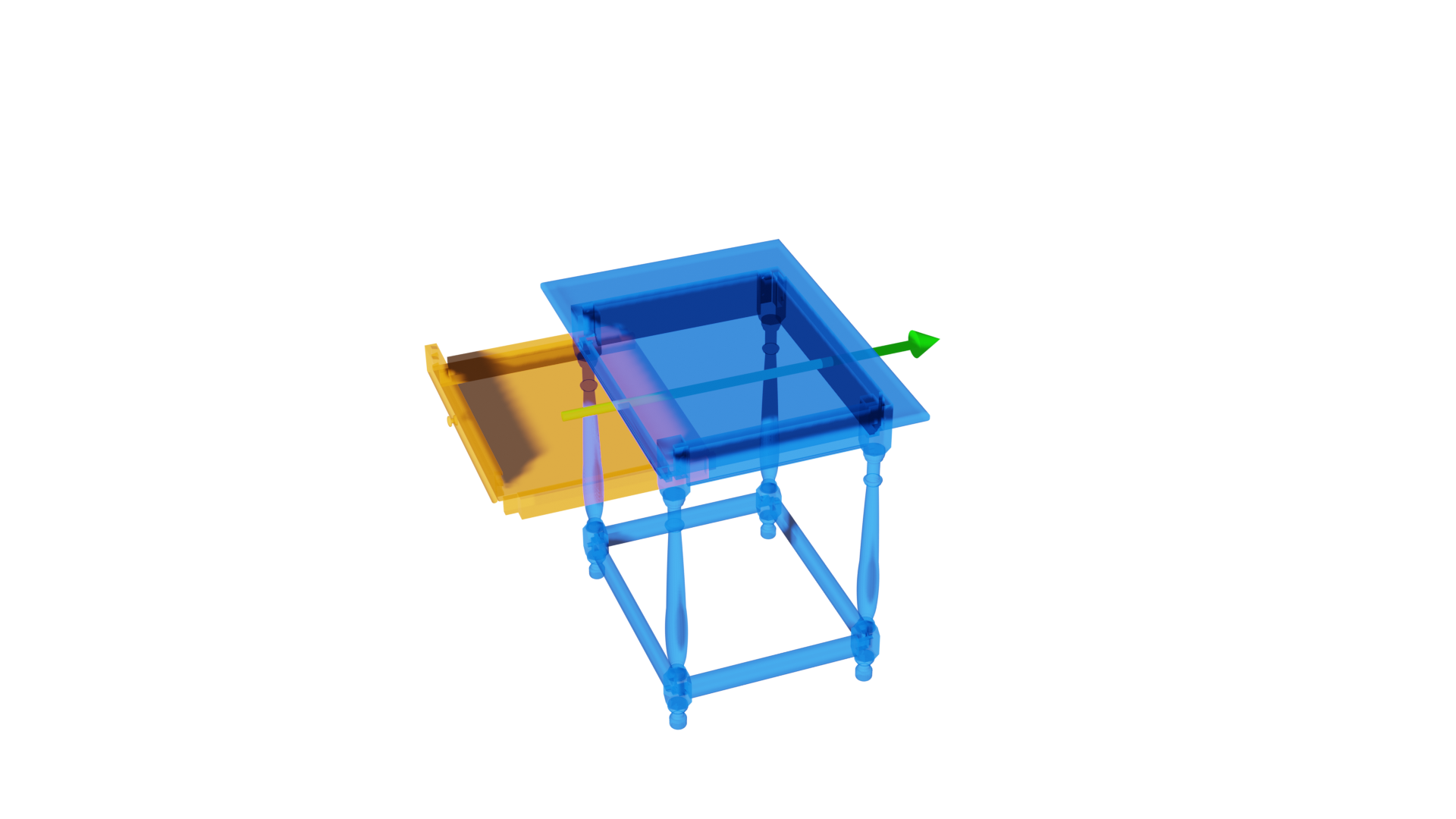}
    \includegraphics[width=0.33\linewidth]{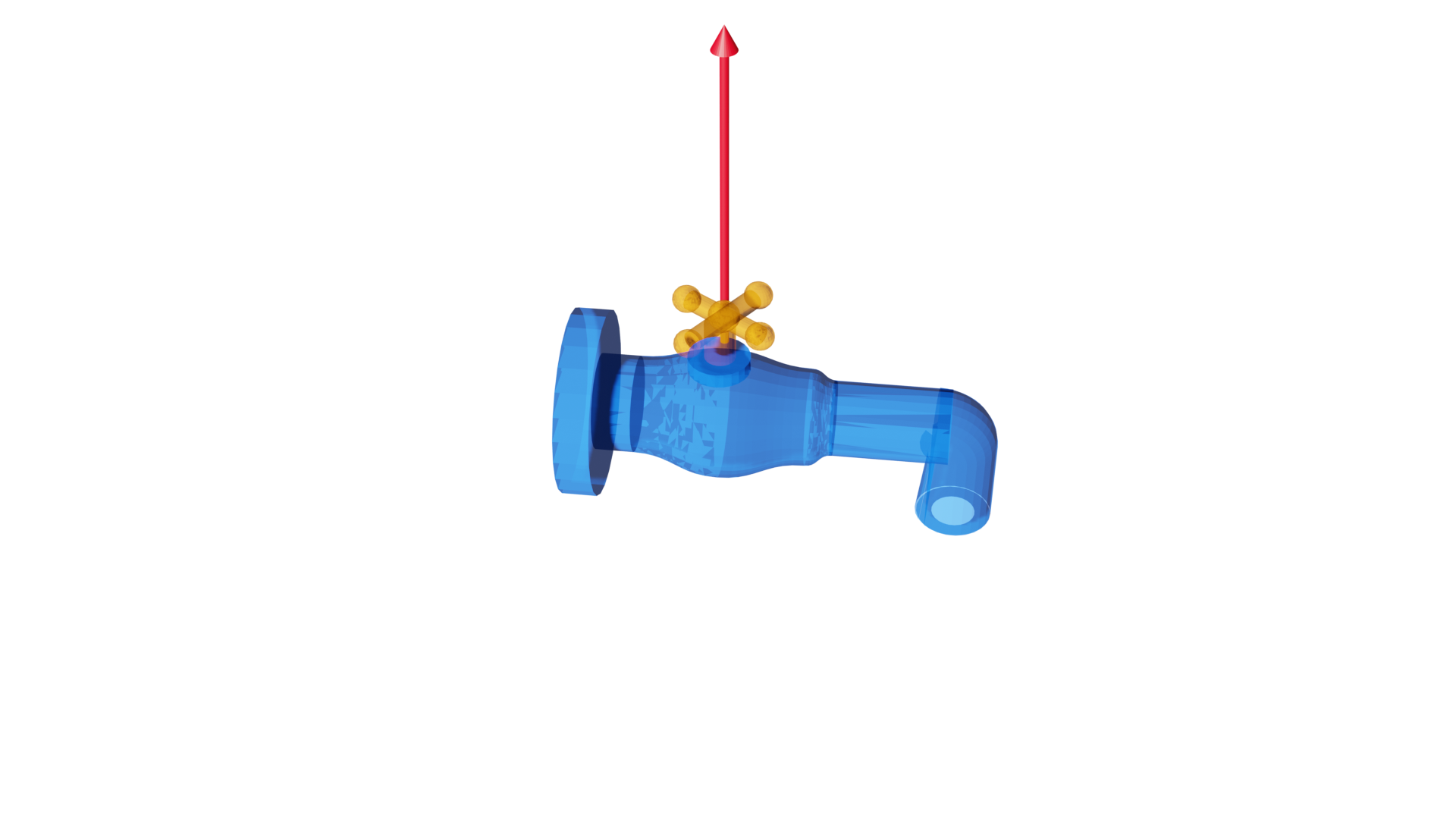}
    \\
    \includegraphics[width=0.33\linewidth]{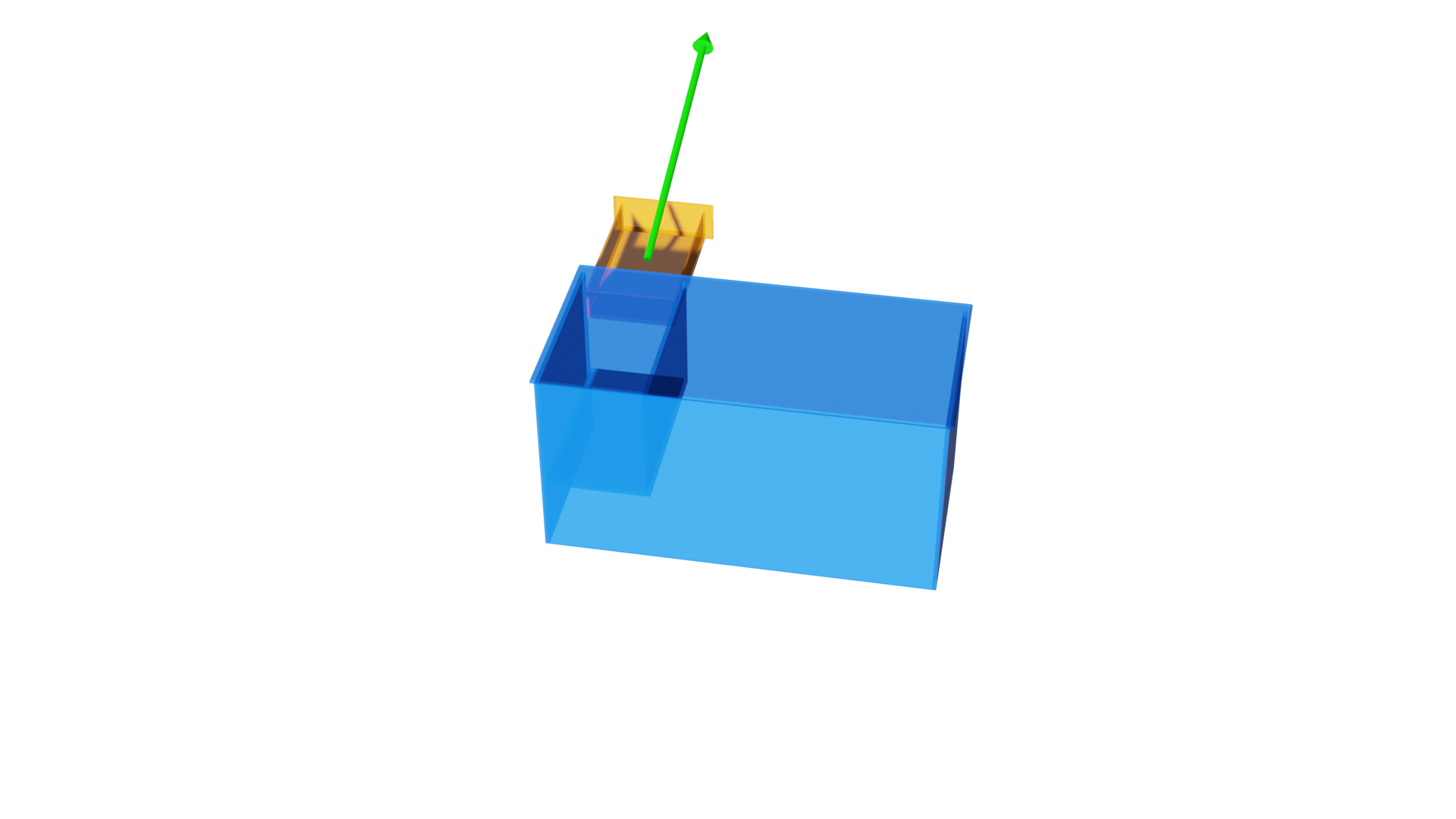}
    \includegraphics[width=0.33\linewidth]{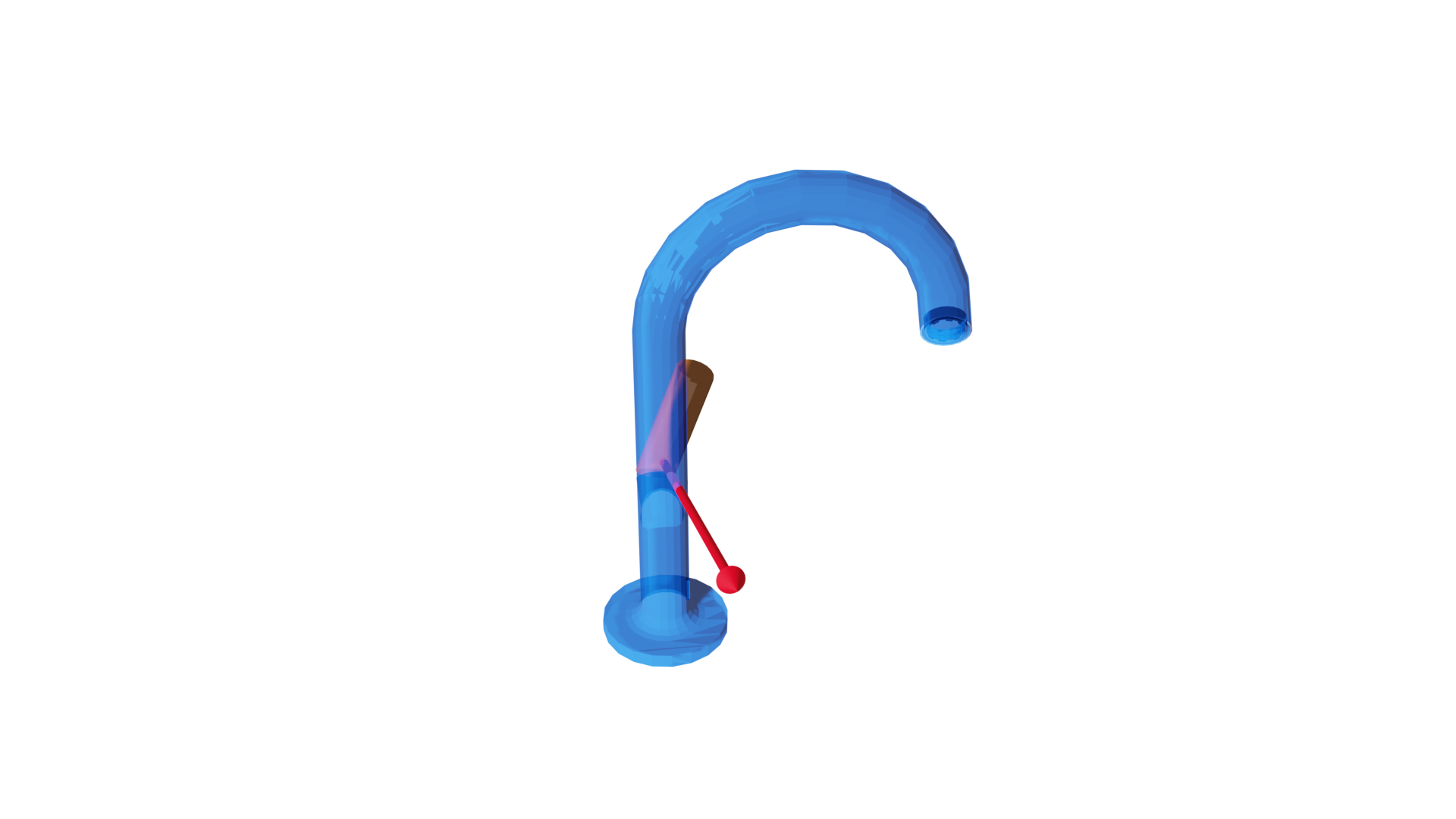}
    \\
    \includegraphics[width=0.33\linewidth]{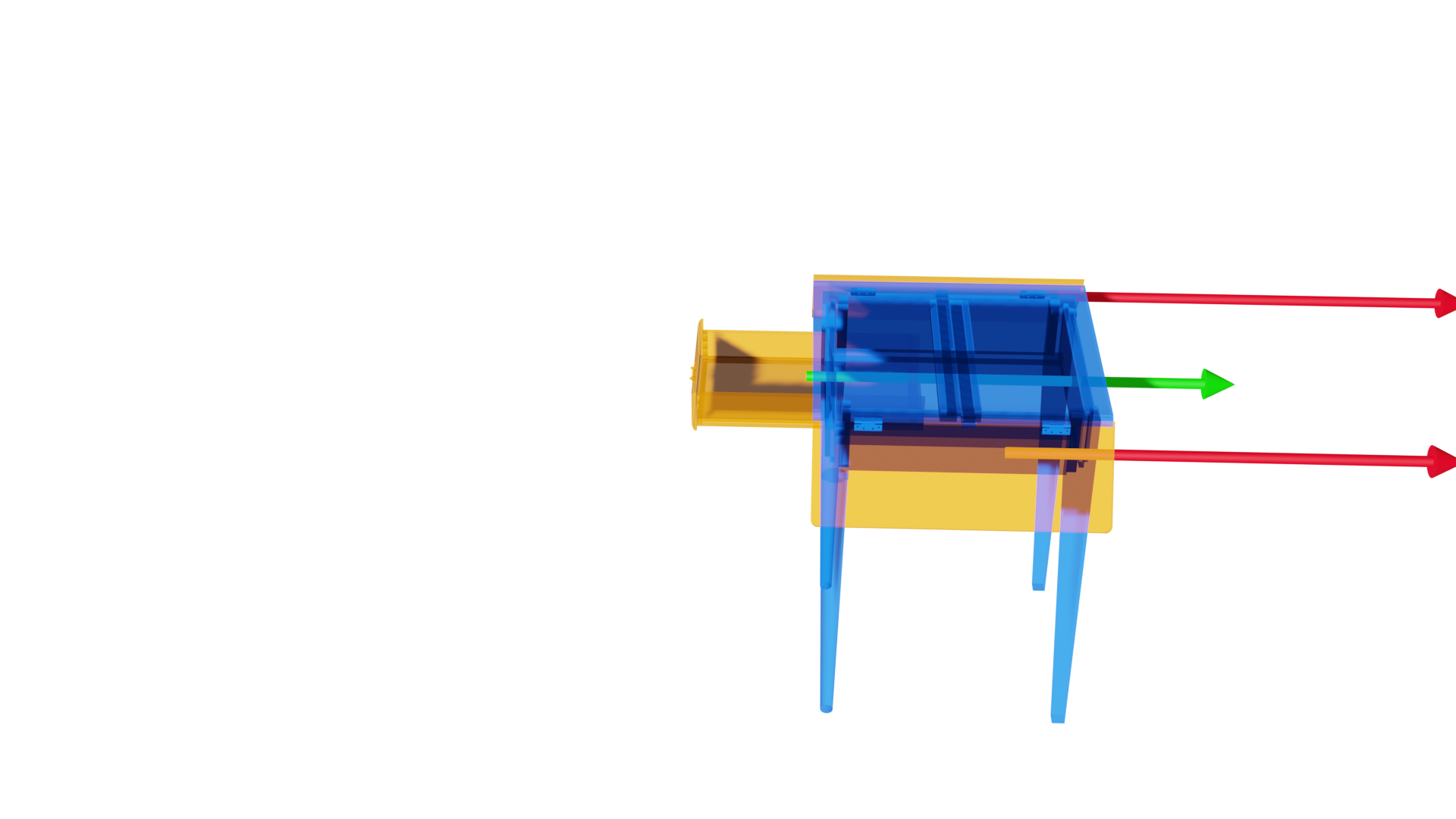}
    \includegraphics[width=0.33\linewidth]{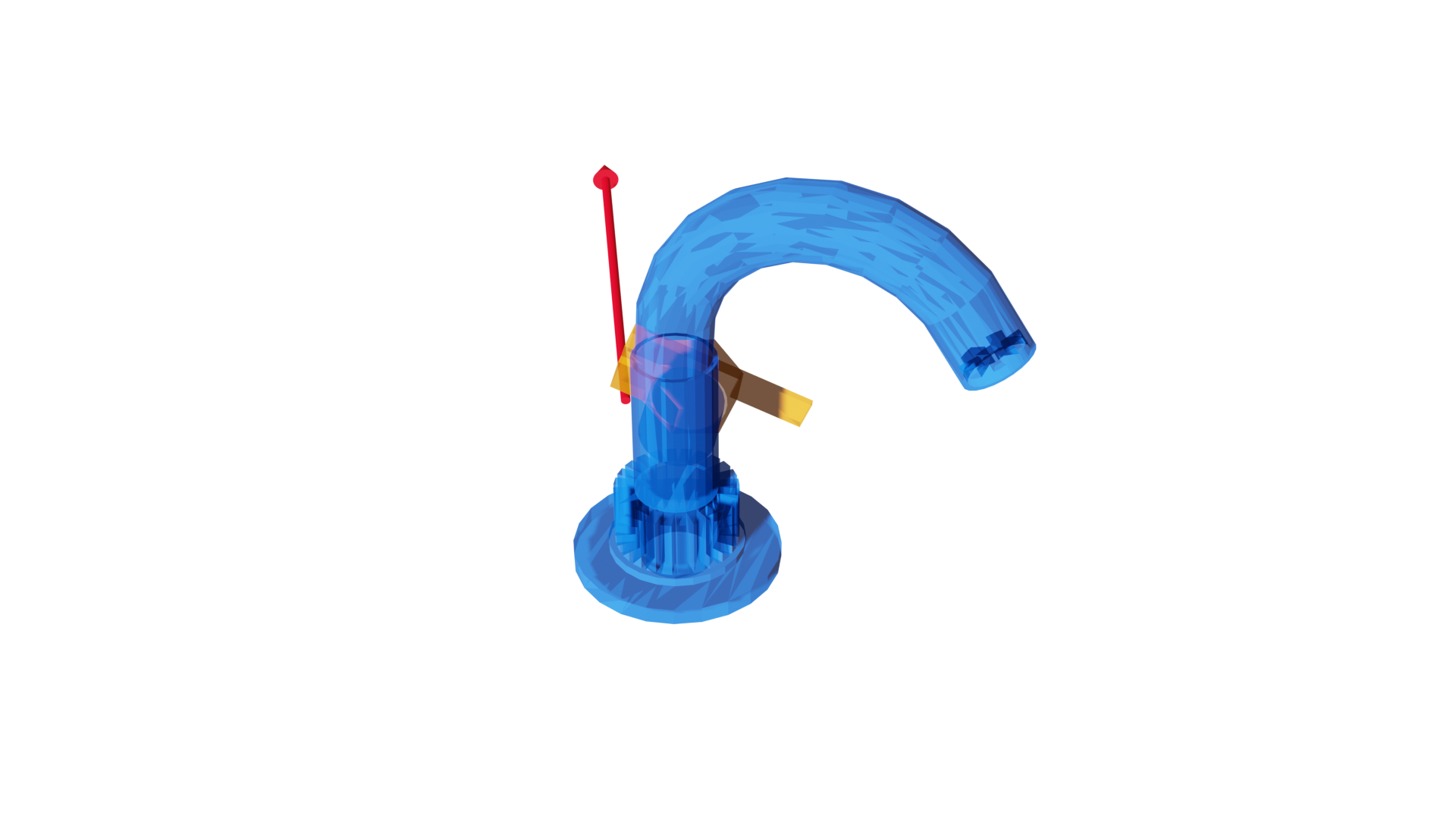}
    \includegraphics[width=0.33\linewidth]{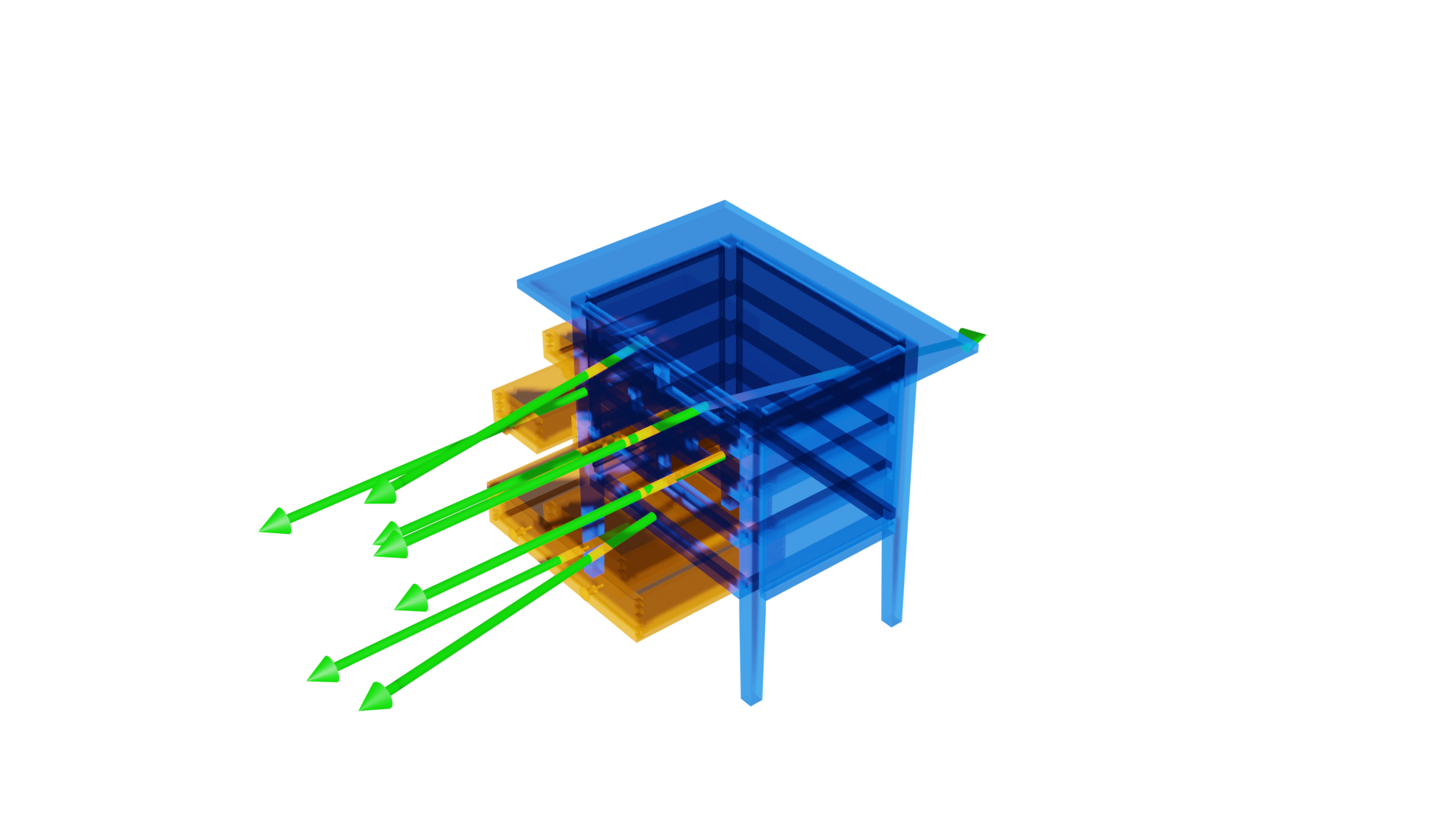}
    \\
    \end{tabular}
    \caption{
    Additional annotated shapes
    }
    \label{figure:qualitative3}
\end{figure*}

\begin{figure*}[ht!]
    \centering
    \setlength{\tabcolsep}{1pt}
    \begin{tabular}{ccc}
        
    \includegraphics[width=0.33\linewidth]{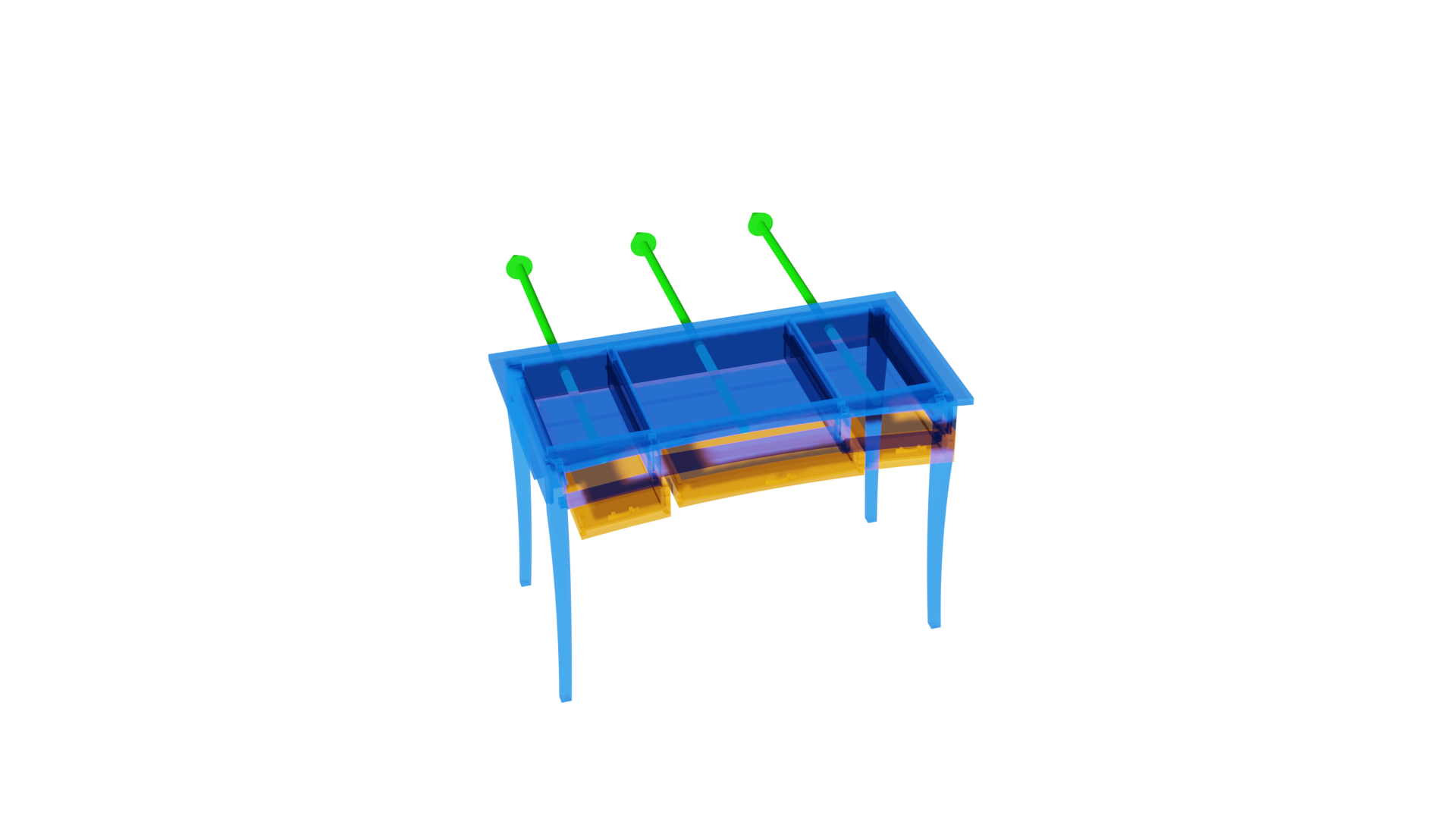}
    \includegraphics[width=0.33\linewidth]{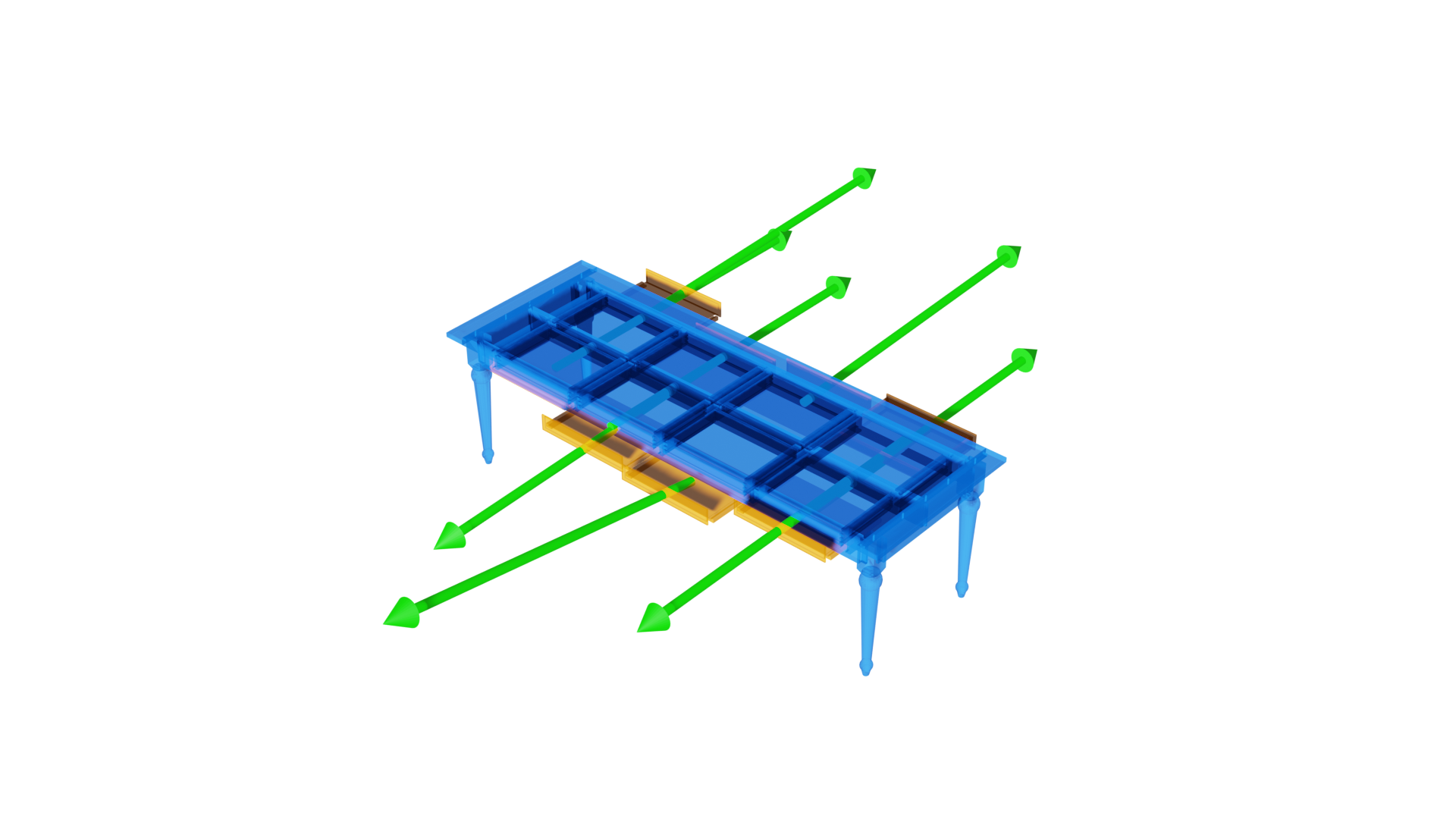}
    \includegraphics[width=0.33\linewidth]{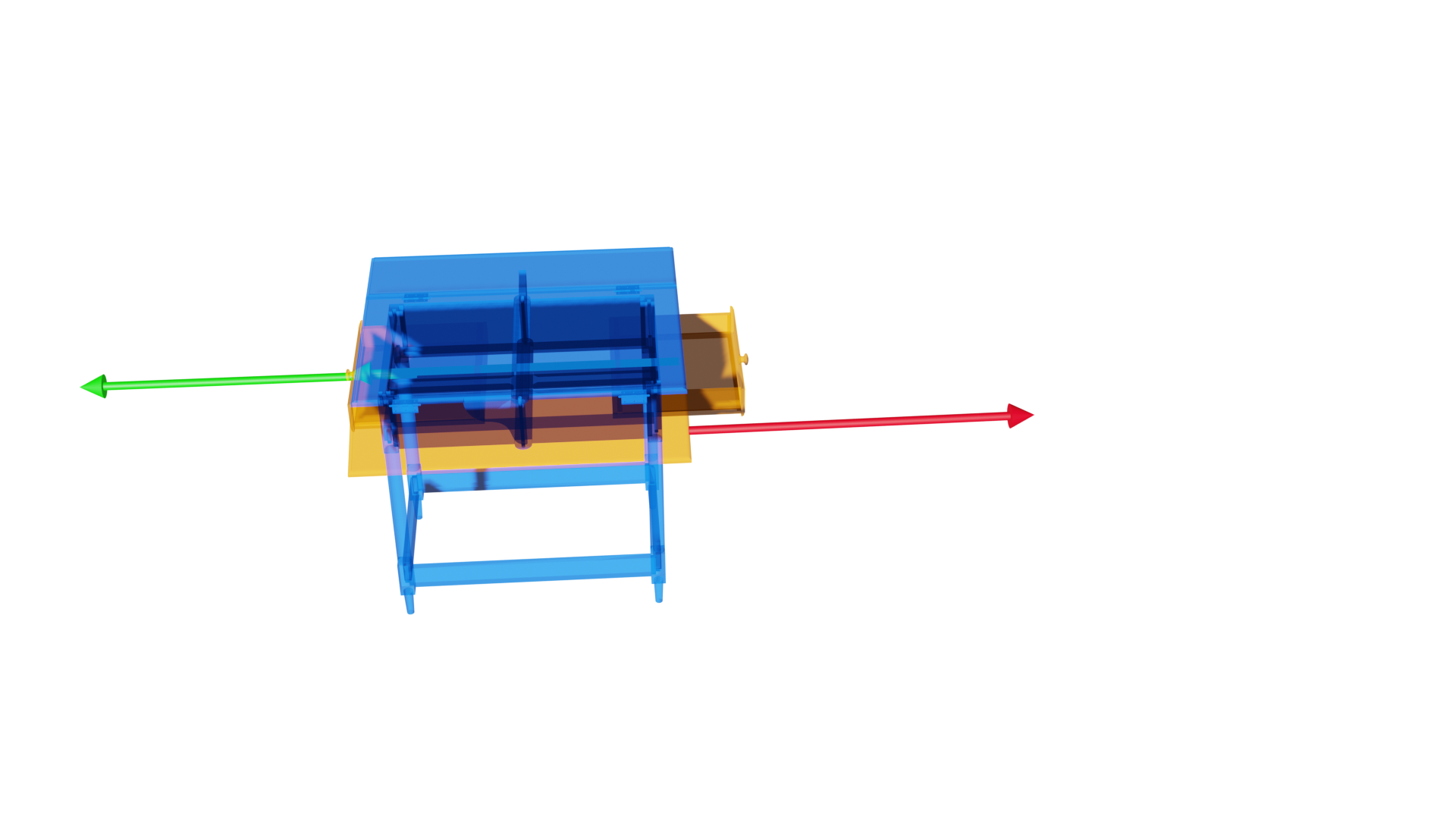}
    \\
    \includegraphics[width=0.33\linewidth]{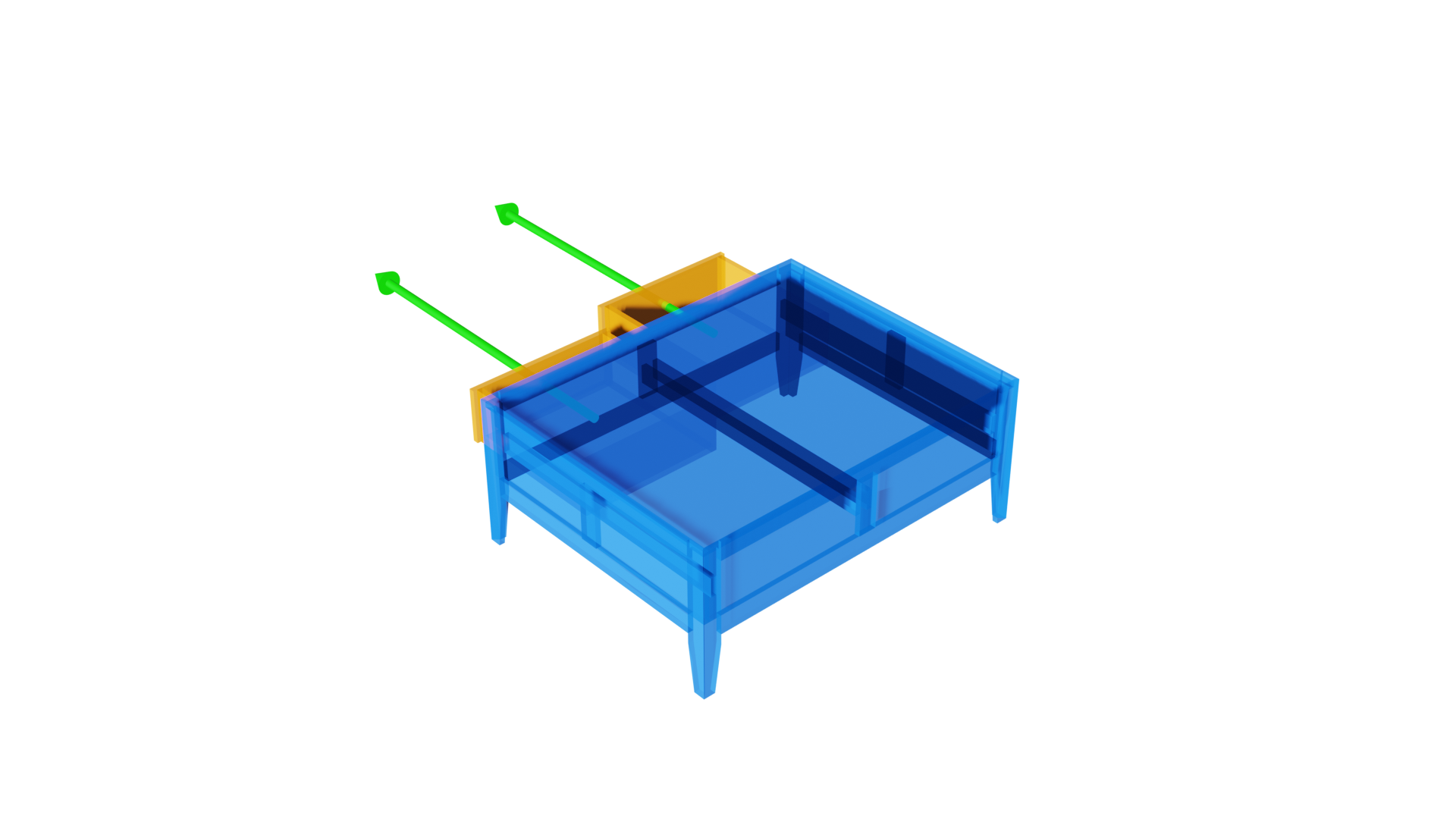}
    \includegraphics[width=0.33\linewidth]{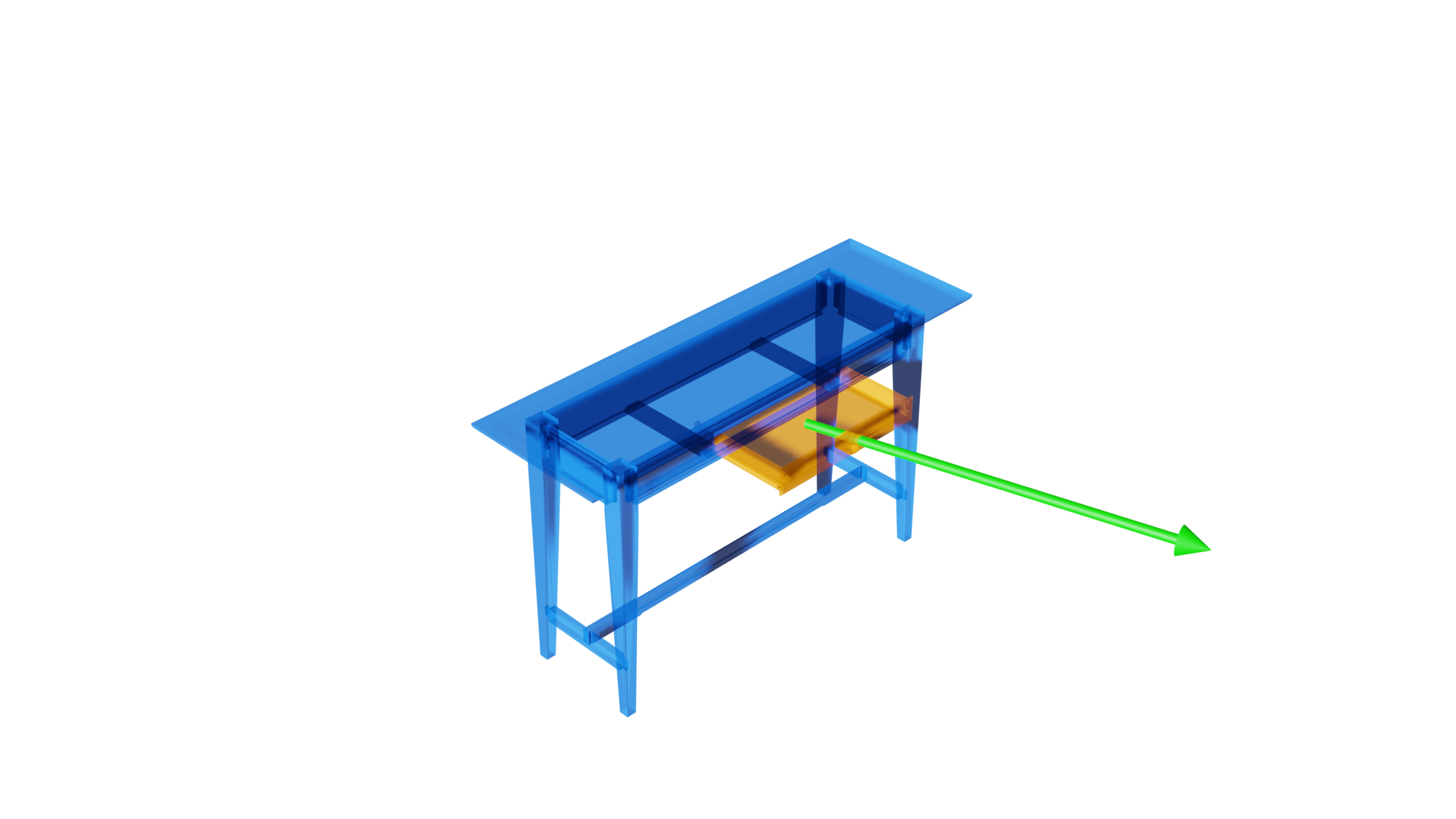}
    \includegraphics[width=0.33\linewidth]{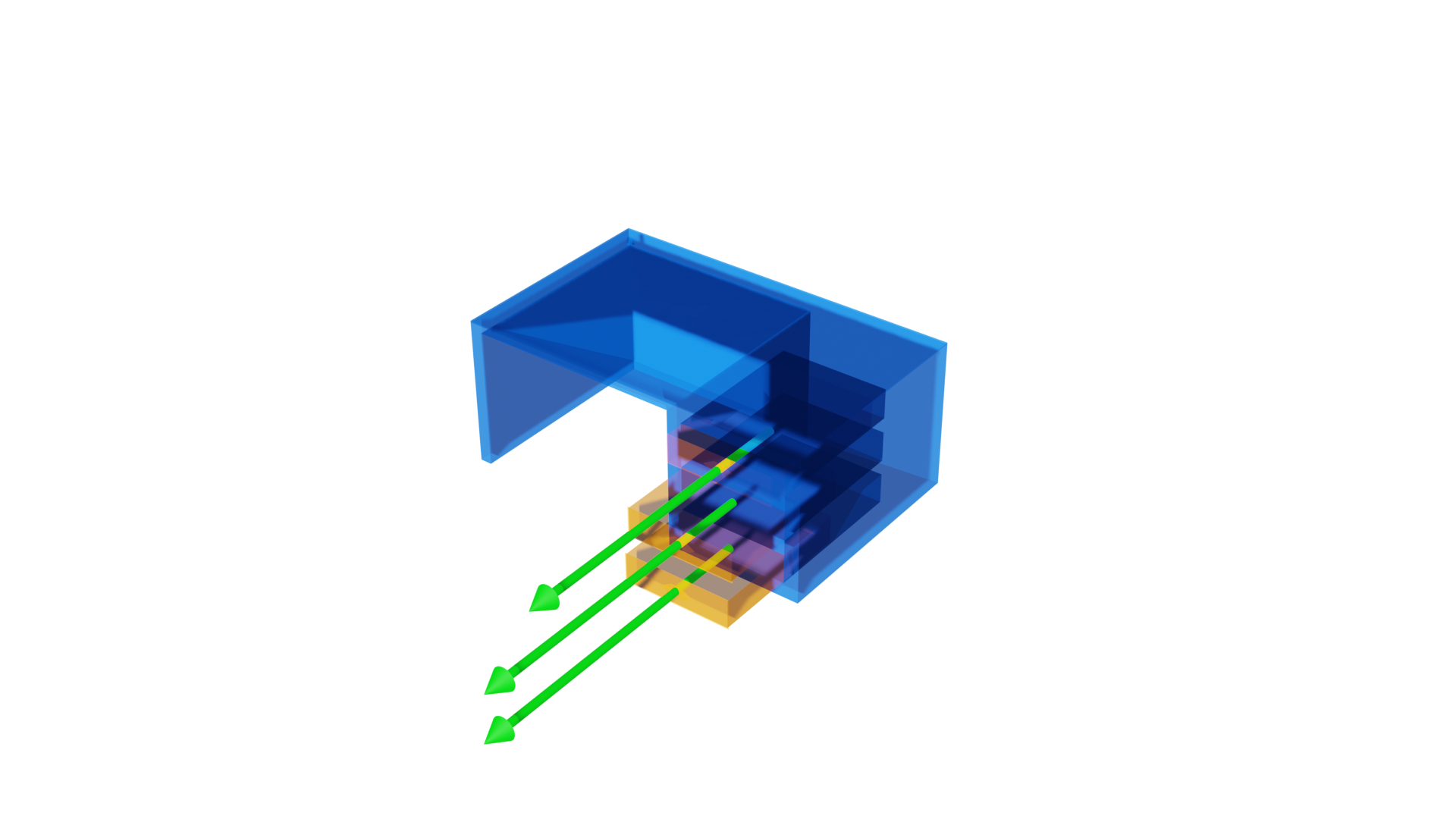}
    \\
    \includegraphics[width=0.33\linewidth]{figs/sup_shapes/final_shape_data_33116.png}
    \includegraphics[width=0.33\linewidth]{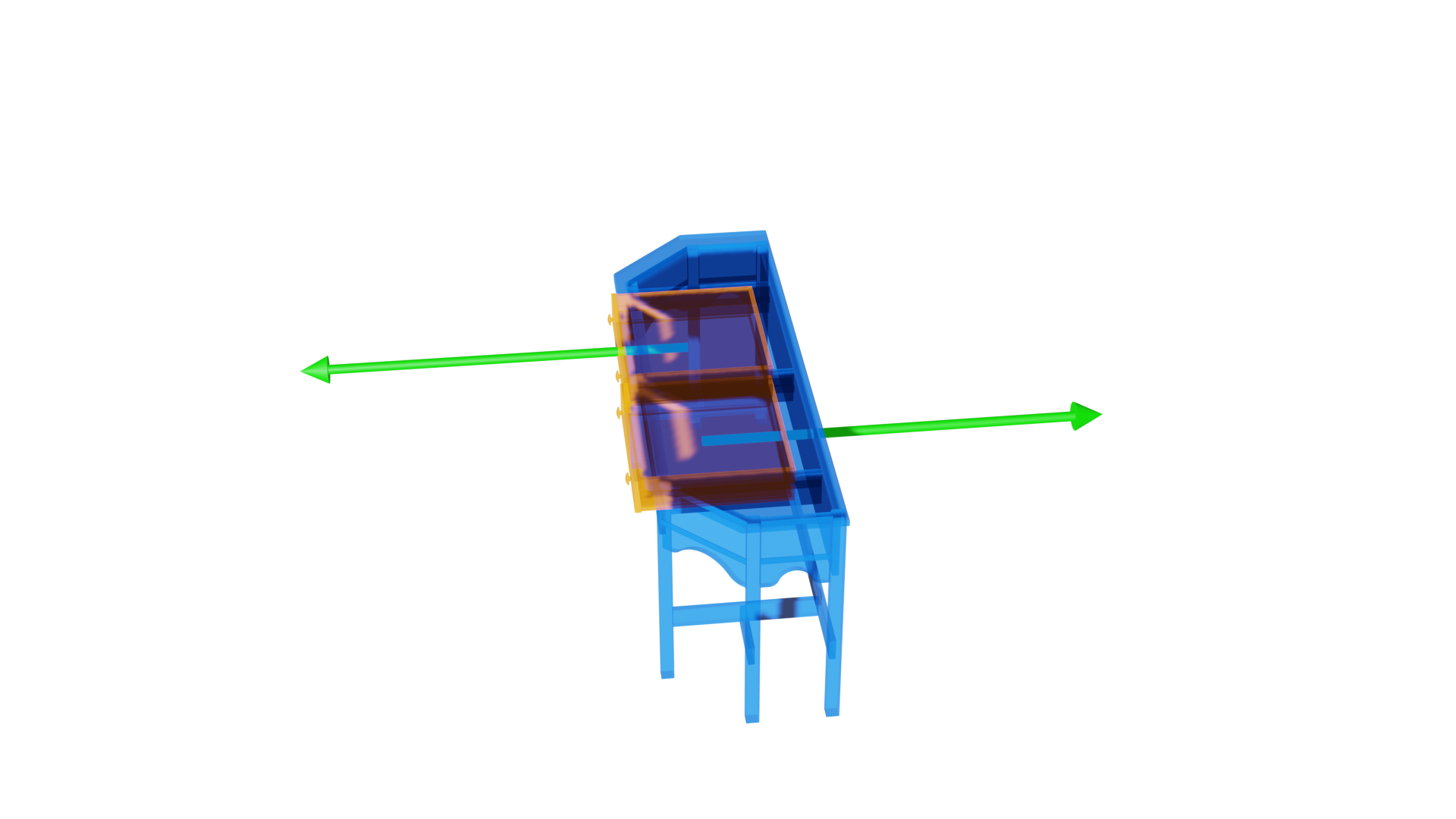}
    \includegraphics[width=0.33\linewidth]{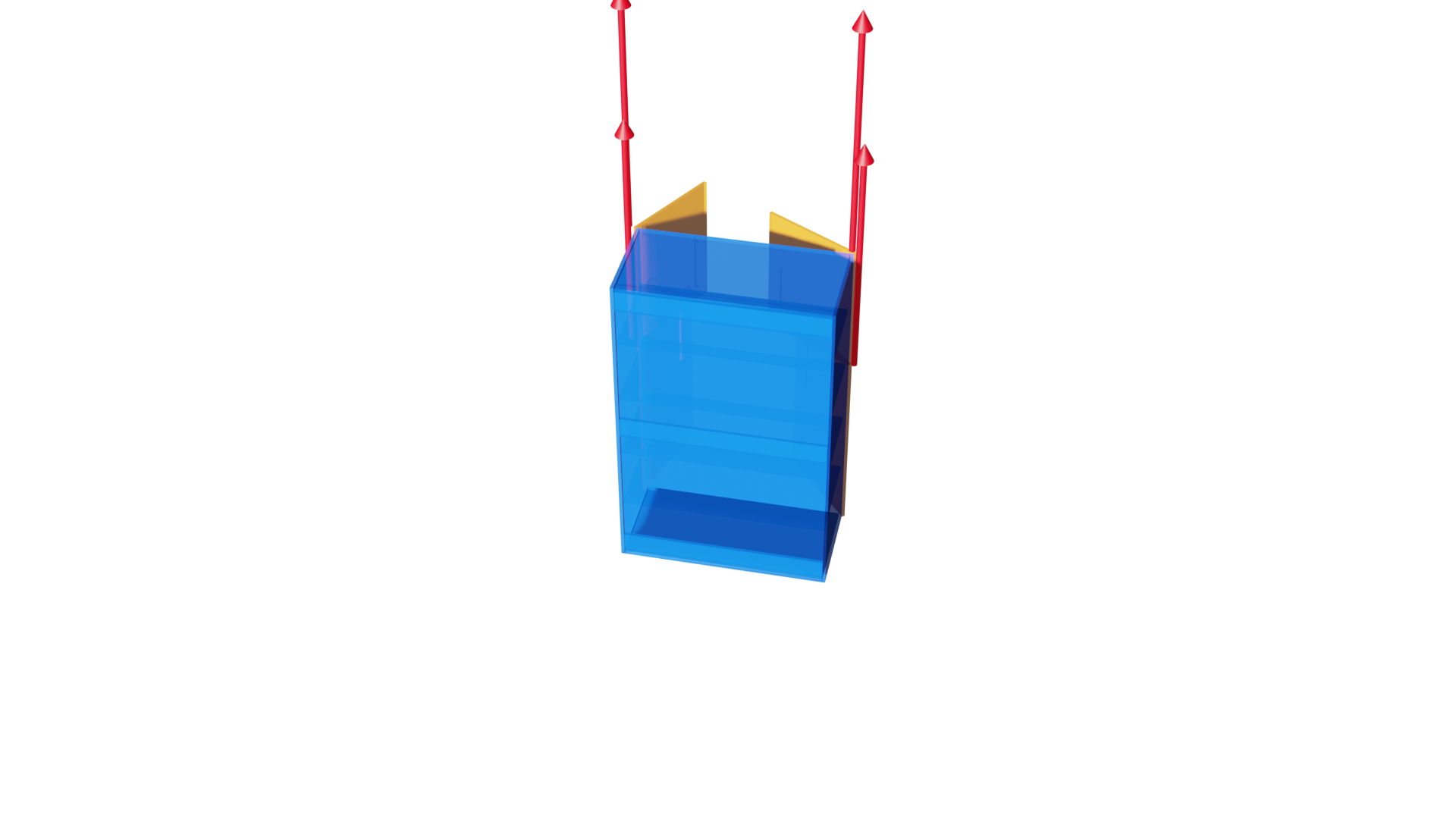}
    \\
    \includegraphics[width=0.33\linewidth]{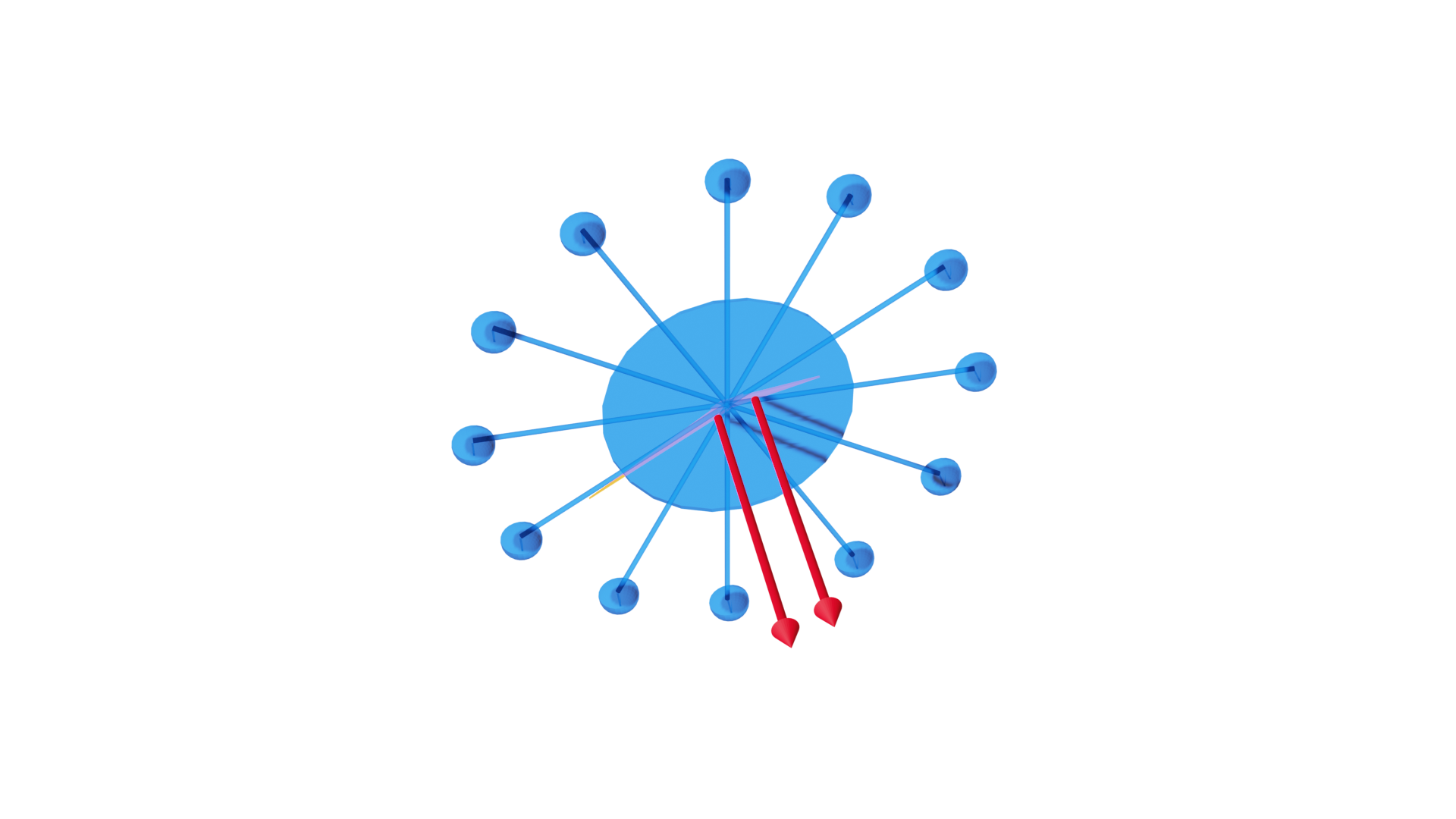}
    \includegraphics[width=0.33\linewidth]{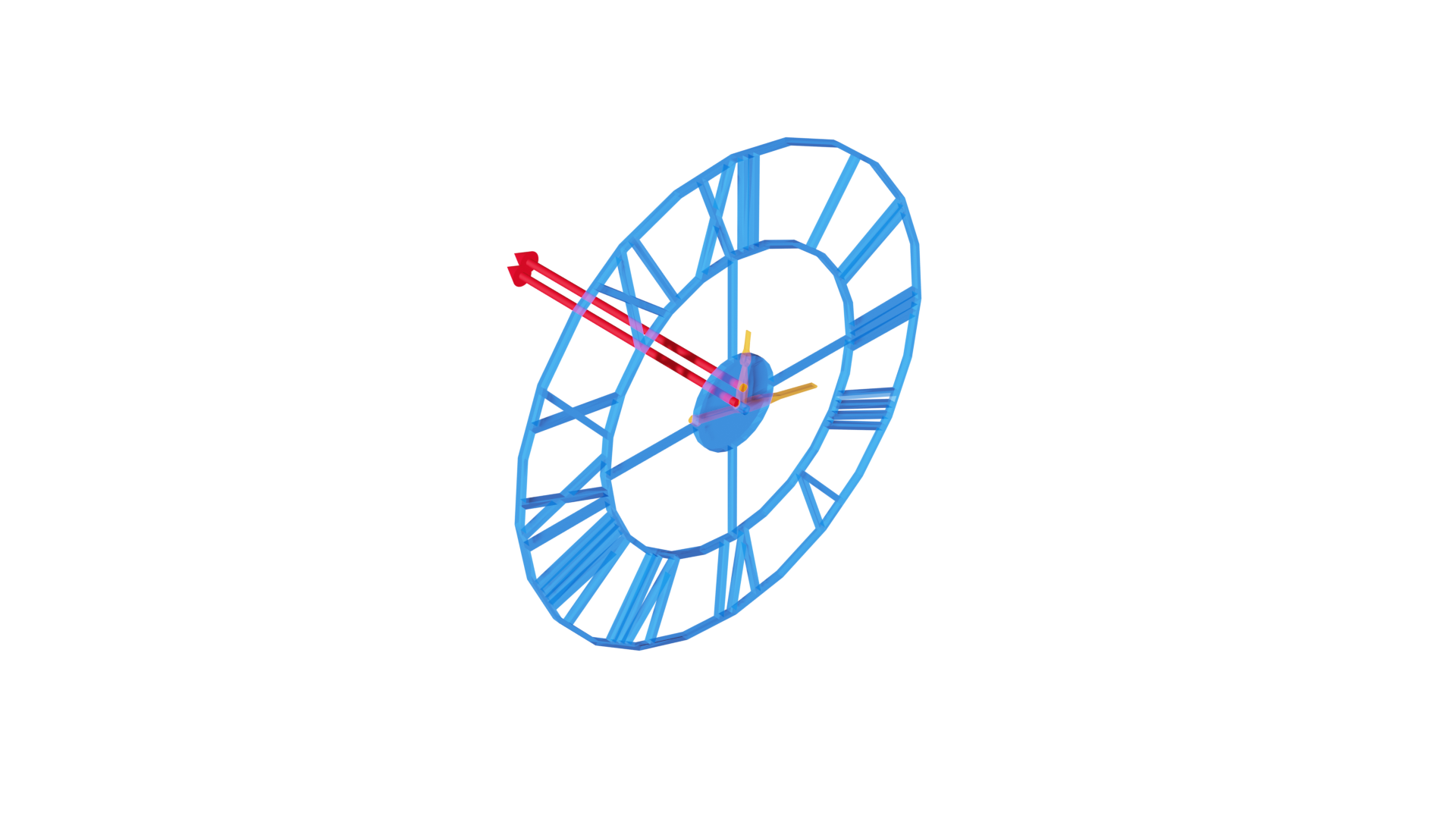}
    \includegraphics[width=0.33\linewidth]{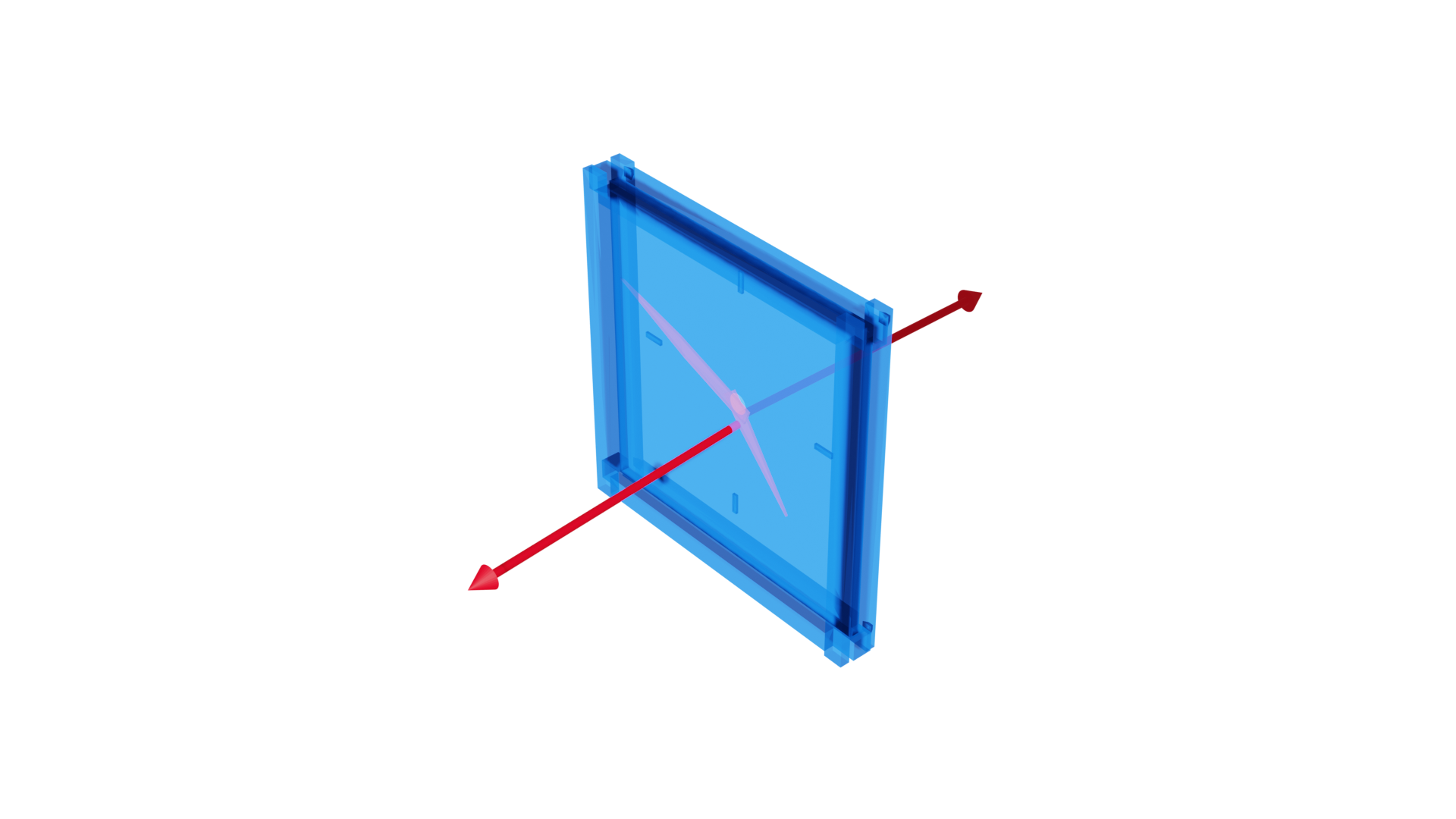}
    \\
    \includegraphics[width=0.33\linewidth]{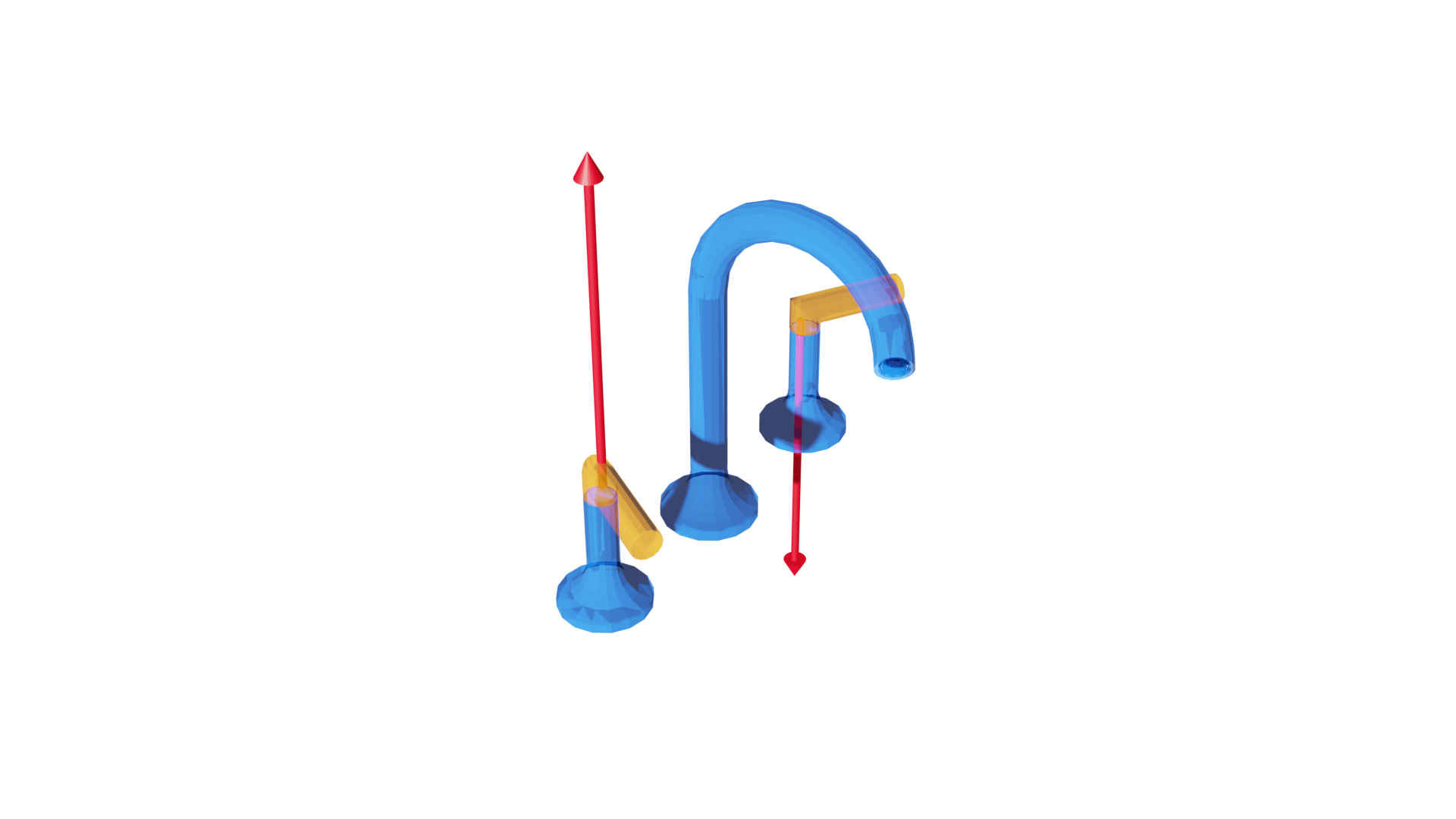}
    \includegraphics[width=0.33\linewidth]{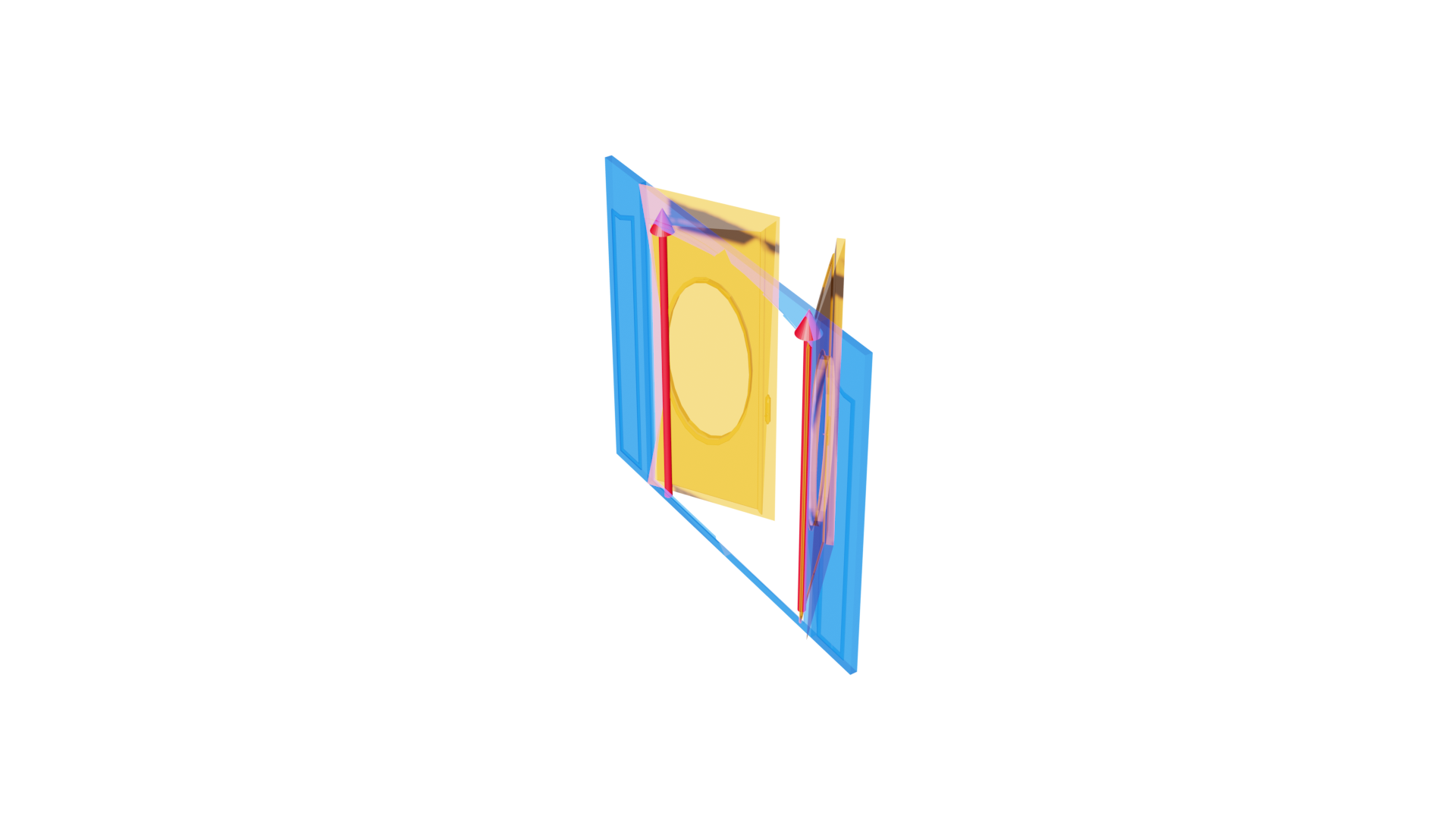}
    \includegraphics[width=0.33\linewidth]{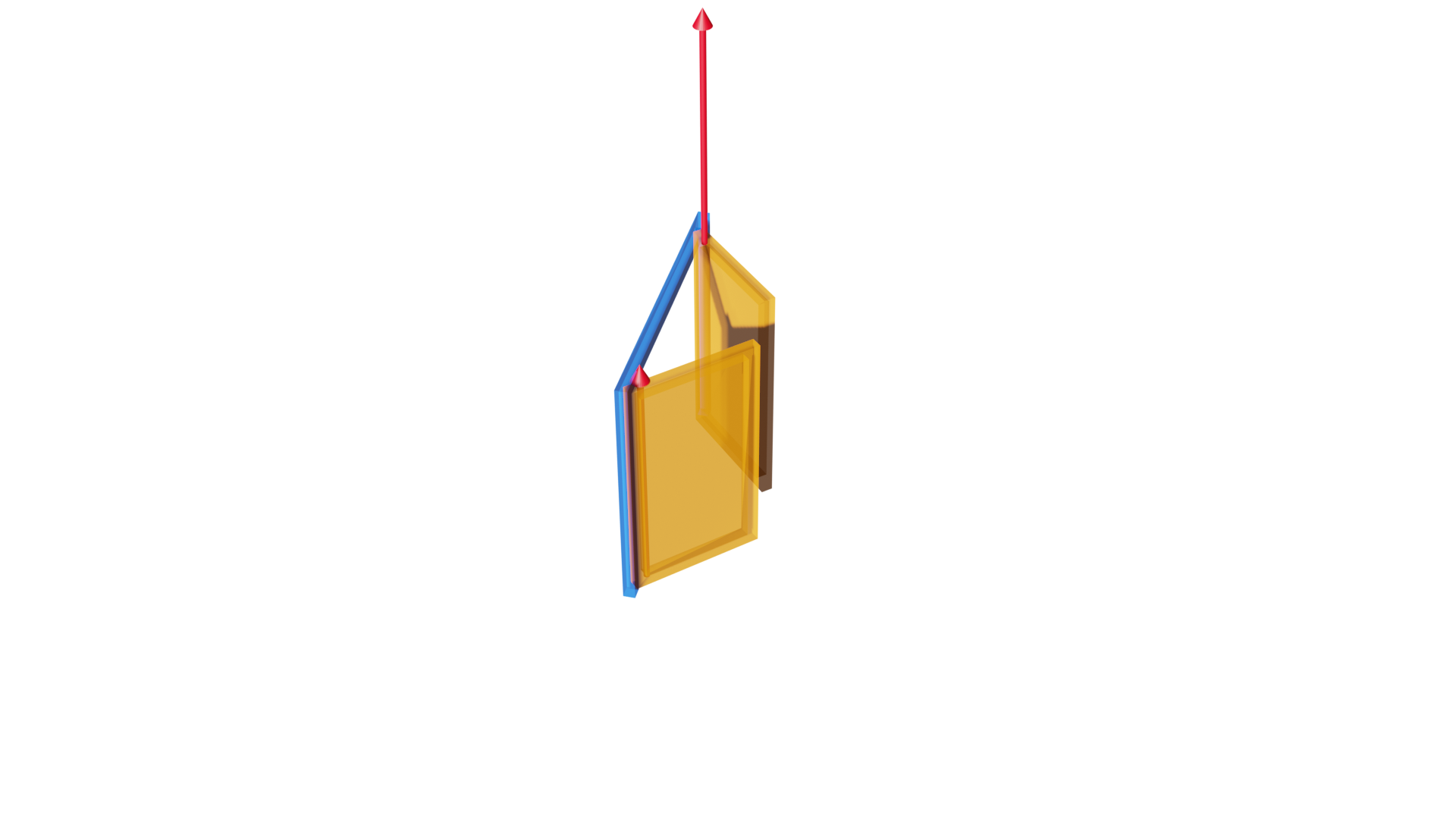}
    \\
    \end{tabular}
    \caption{
    Additional annotated shapes
    }
    \label{figure:qualitative4}
\end{figure*}

\begin{figure*}[ht!]
    \centering
    \setlength{\tabcolsep}{1pt}
    \begin{tabular}{ccc}
        
    \includegraphics[width=0.33\linewidth]{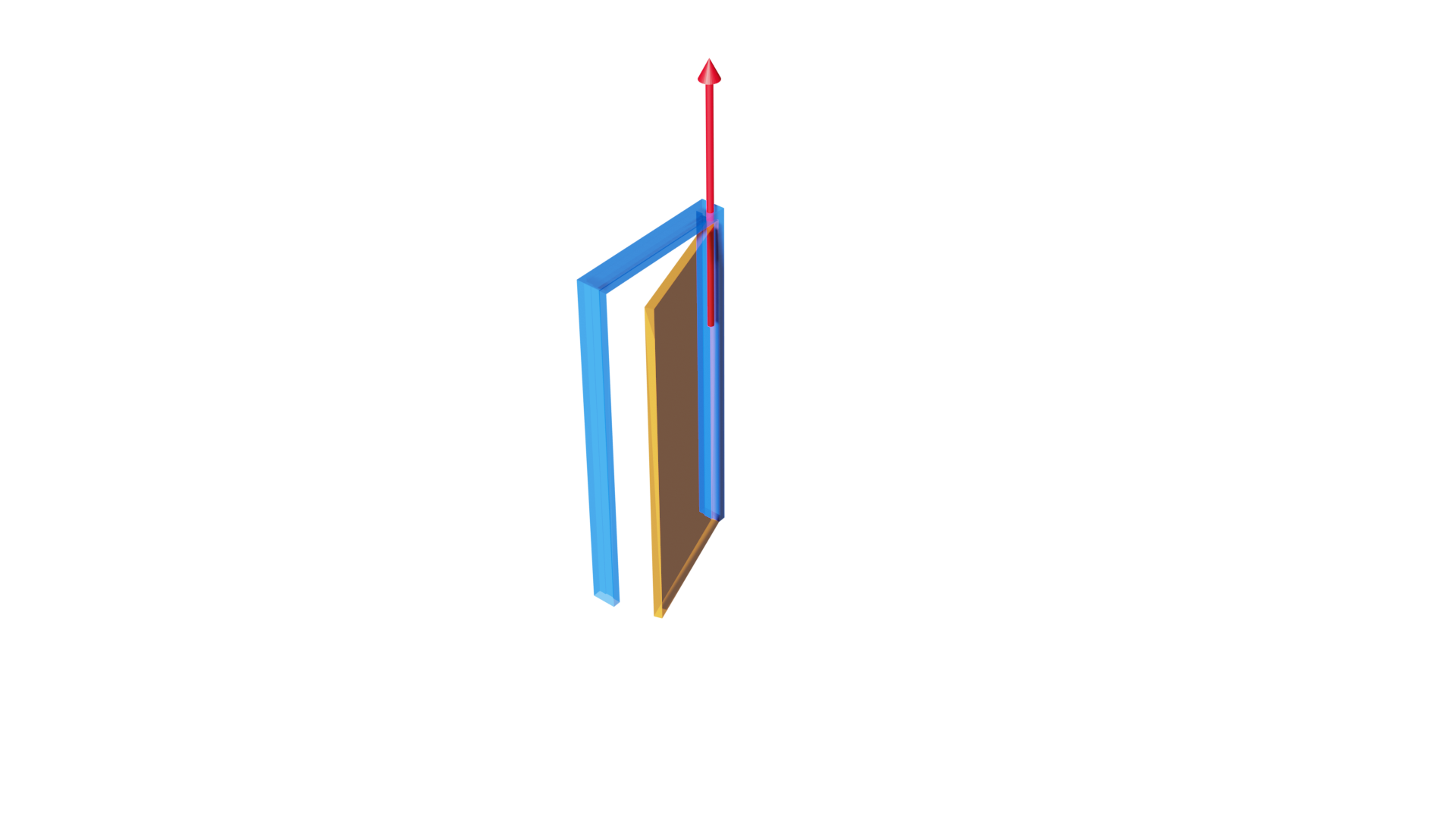}
    \includegraphics[width=0.33\linewidth]{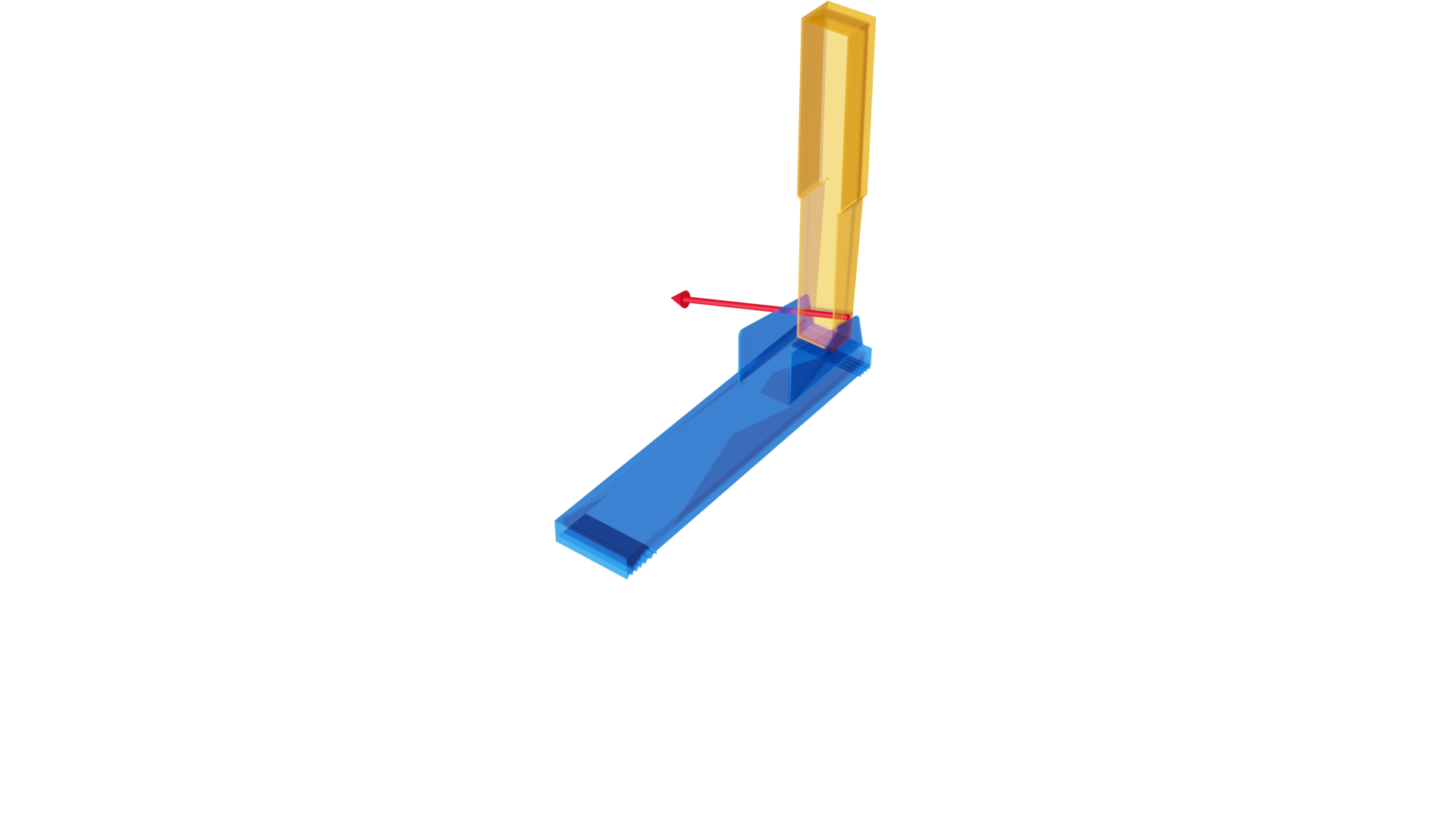}
    \includegraphics[width=0.33\linewidth]{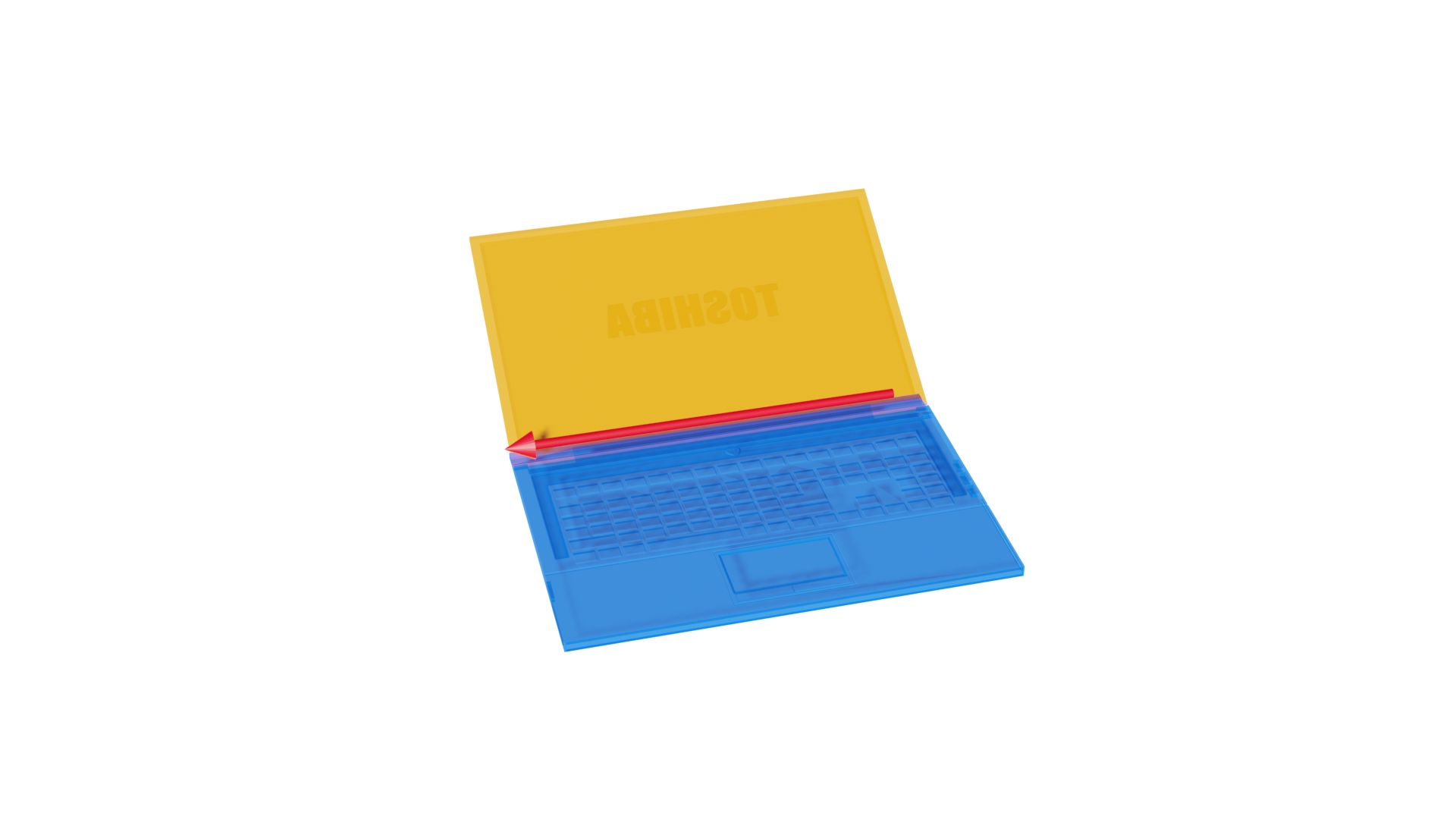}
    \\
    \includegraphics[width=0.33\linewidth]{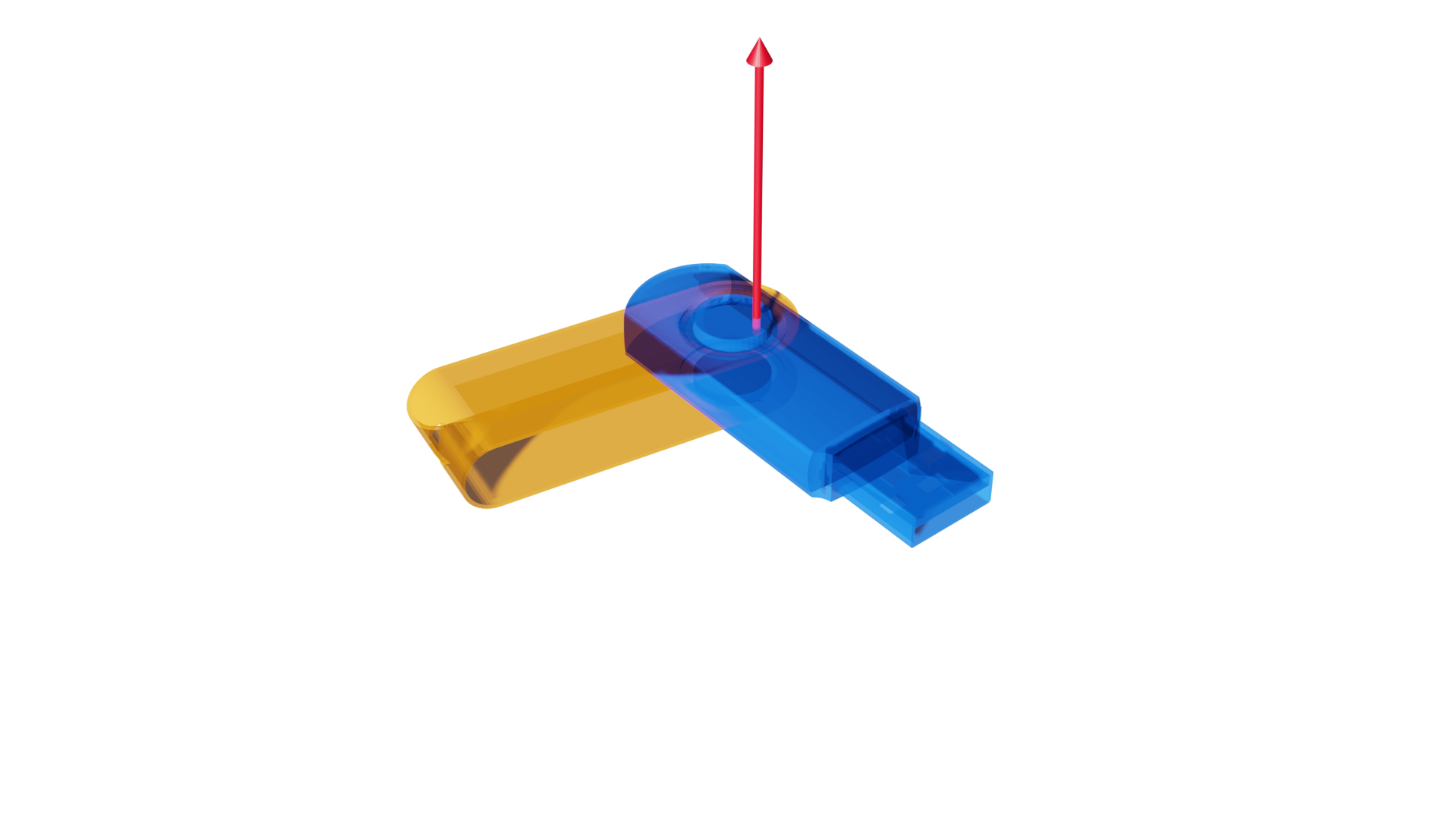}
    \includegraphics[width=0.33\linewidth]{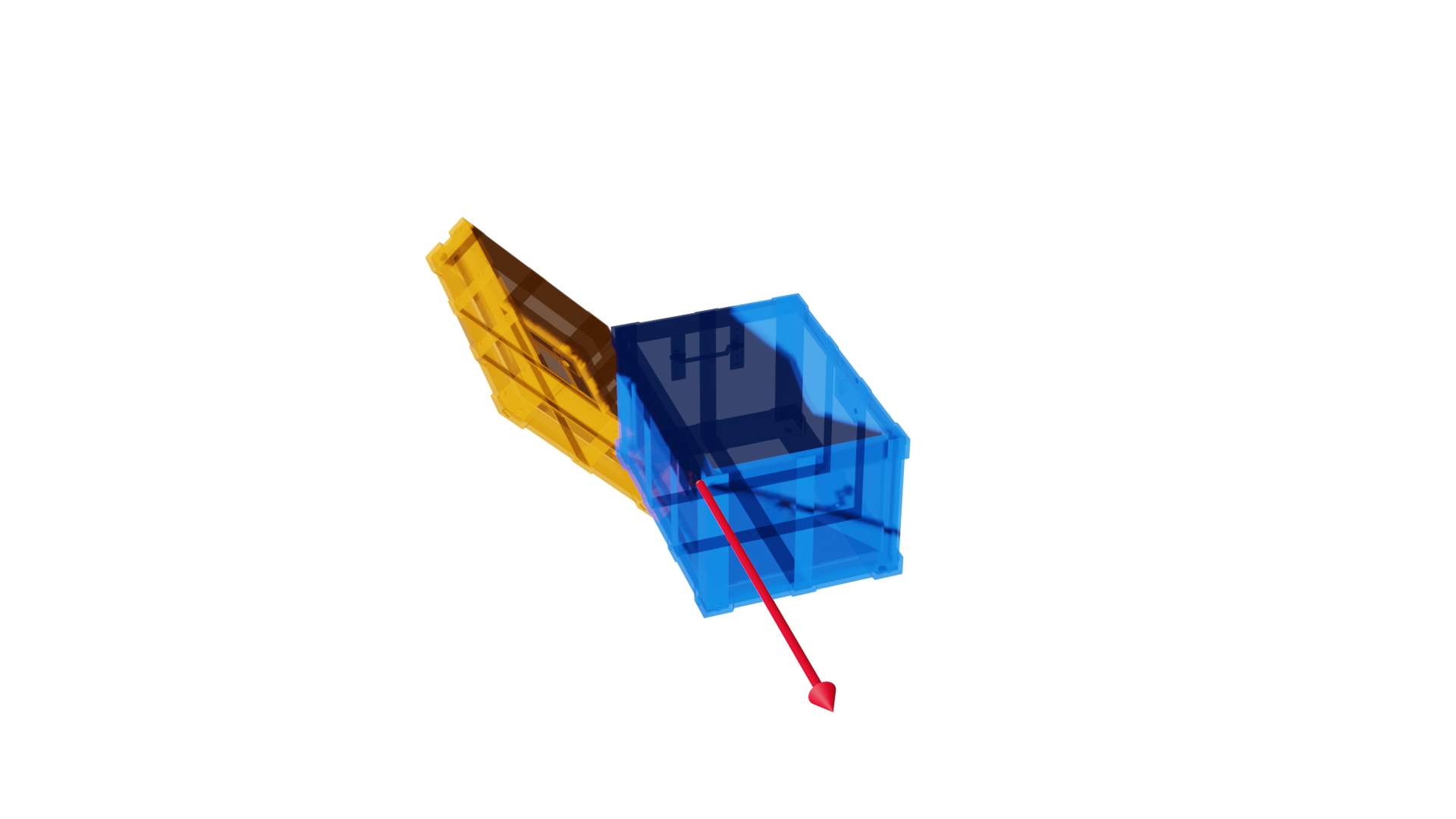}
    \includegraphics[width=0.33\linewidth]{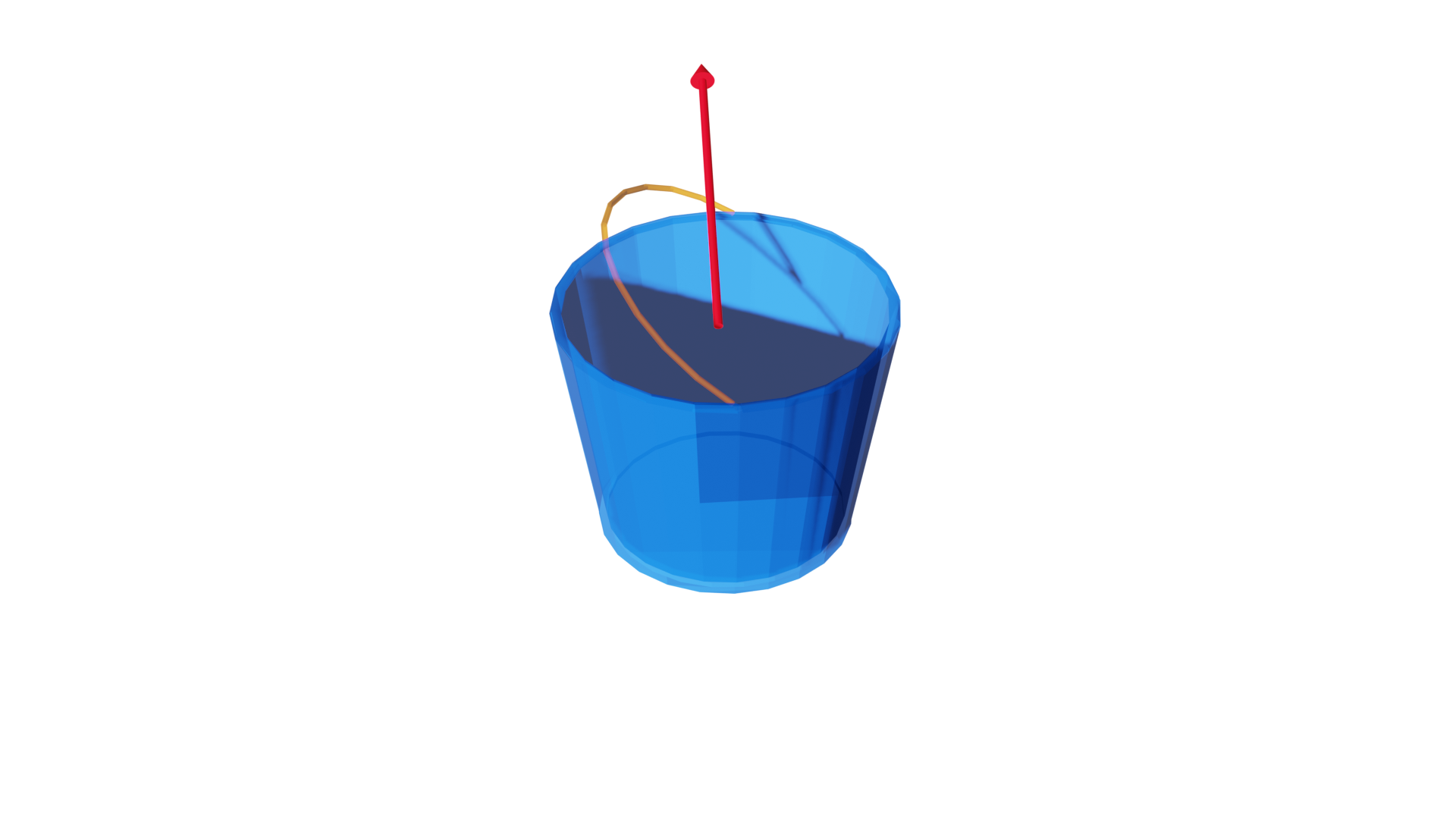}
    \\
    \includegraphics[width=0.33\linewidth]{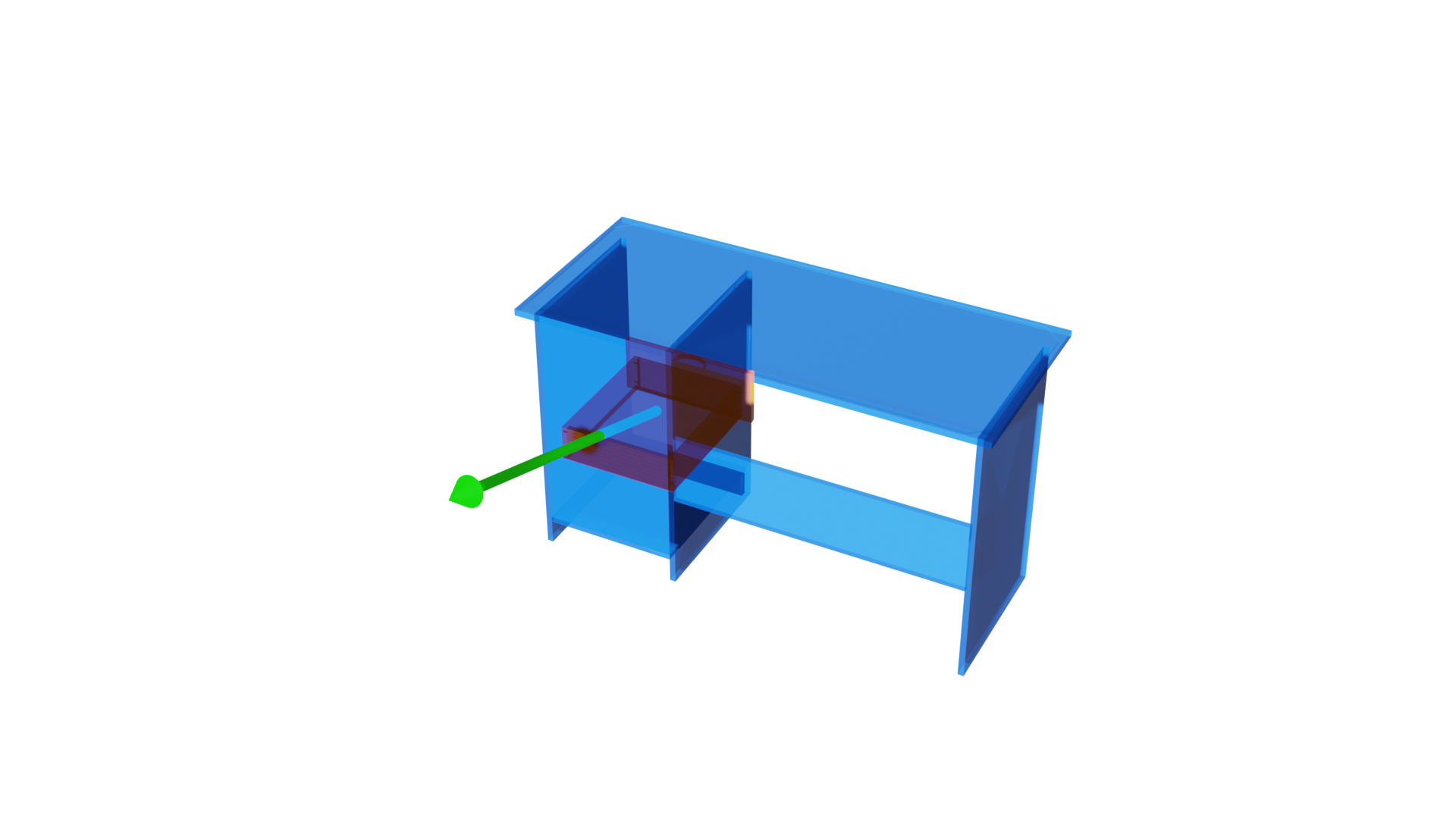}
    \includegraphics[width=0.33\linewidth]{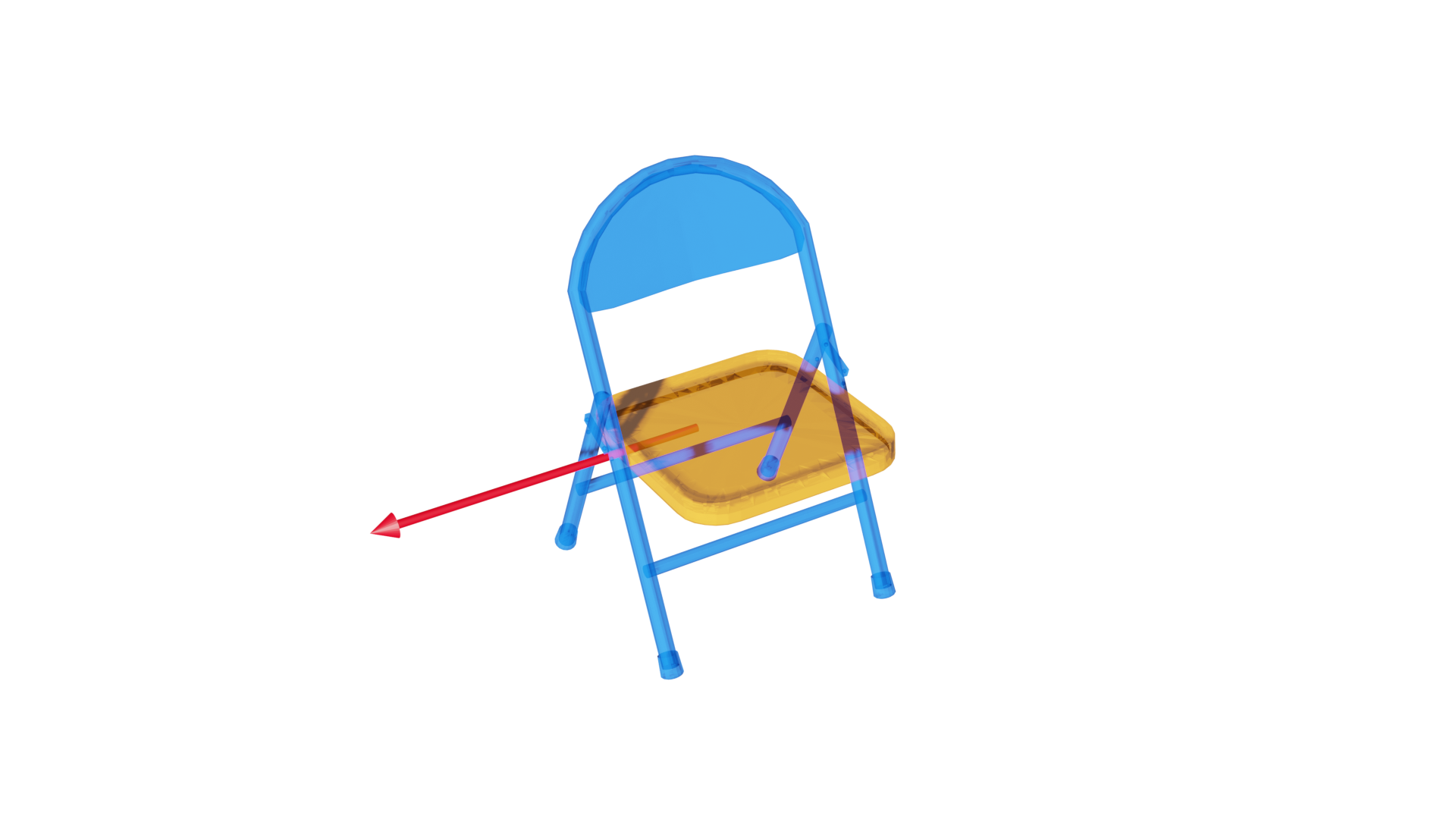}
    \includegraphics[width=0.33\linewidth]{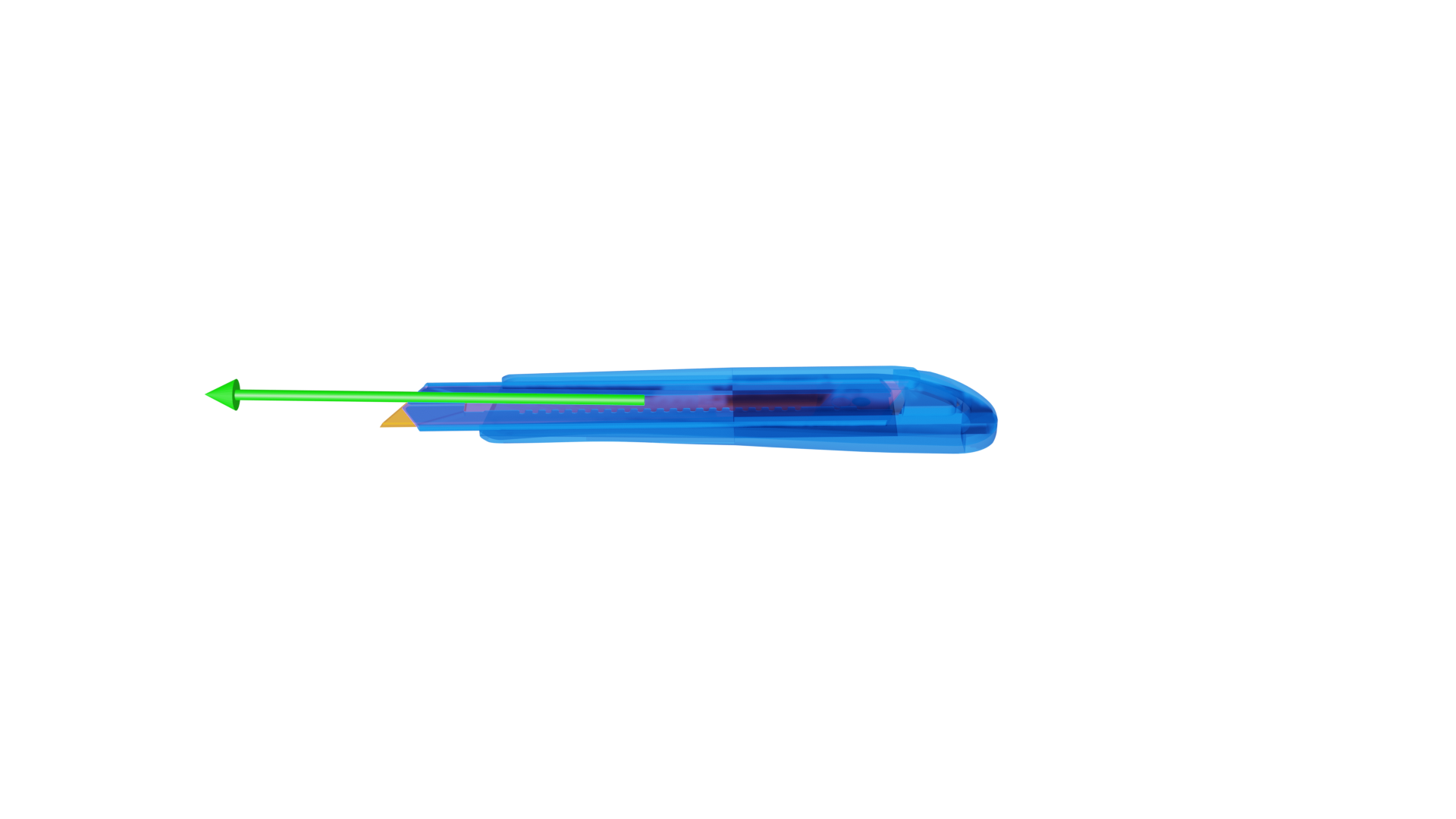}
    \\
    \includegraphics[width=0.33\linewidth]{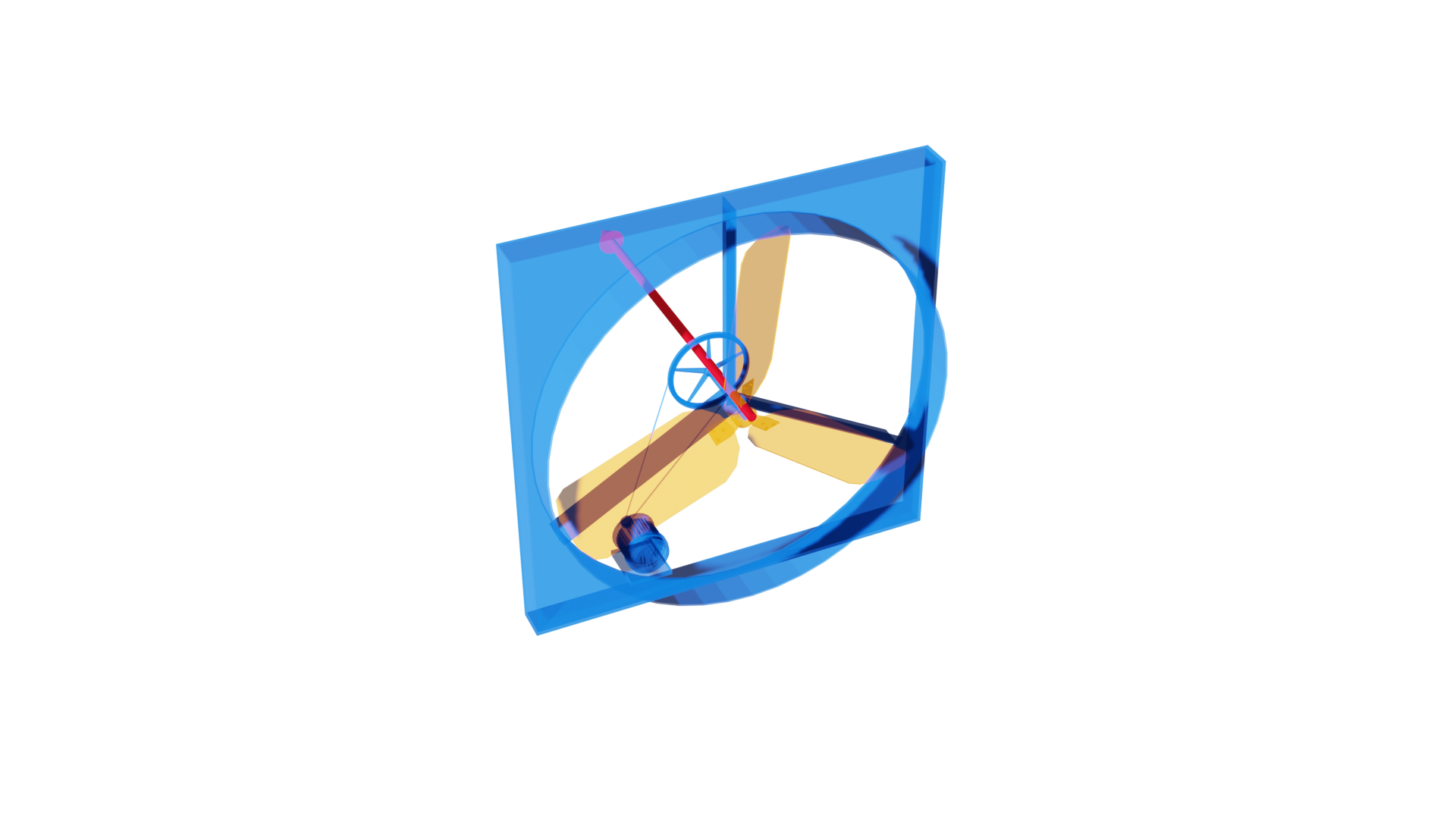}
    \\
    \includegraphics[width=0.33\linewidth]{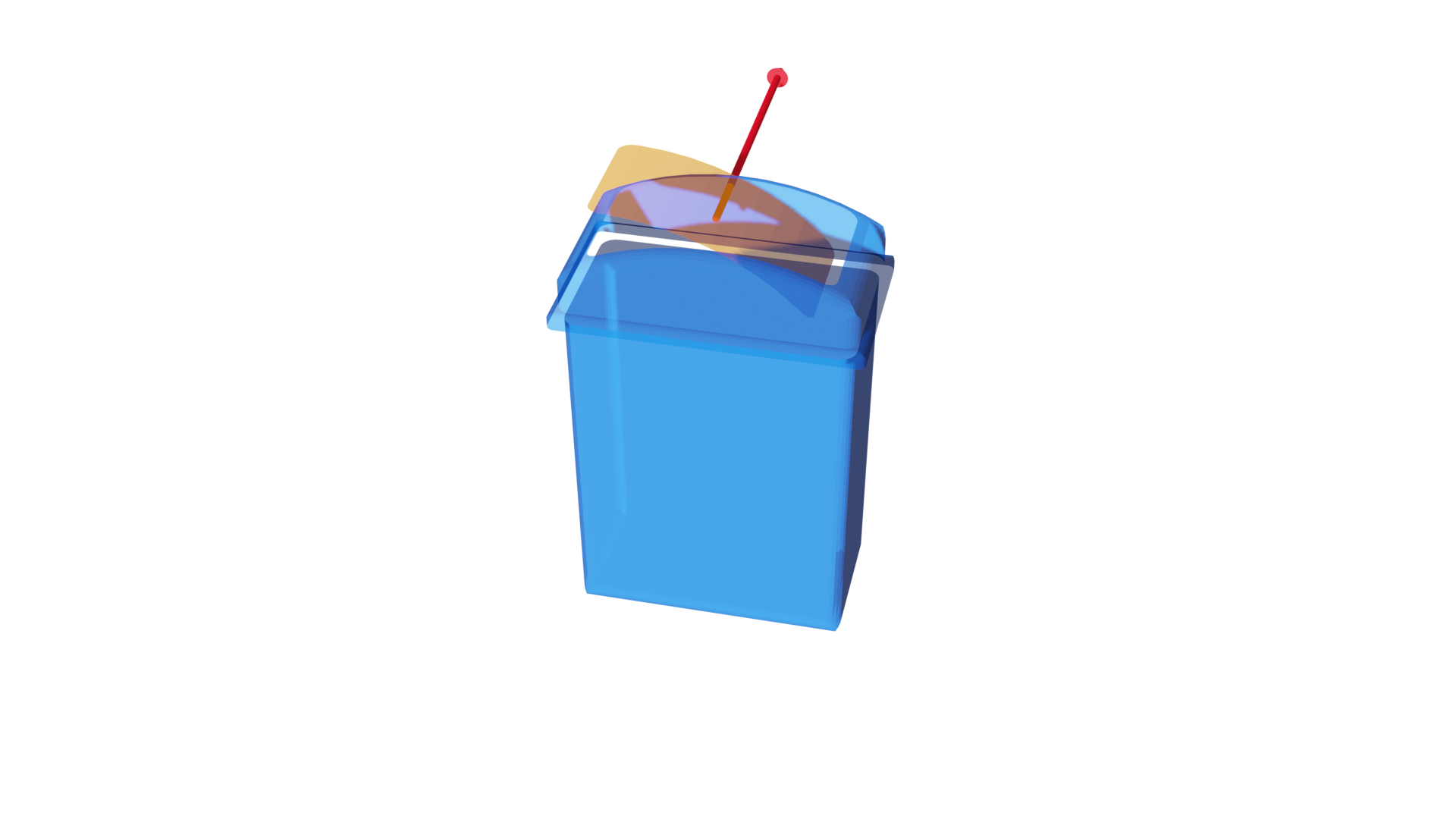}
    \includegraphics[width=0.33\linewidth]{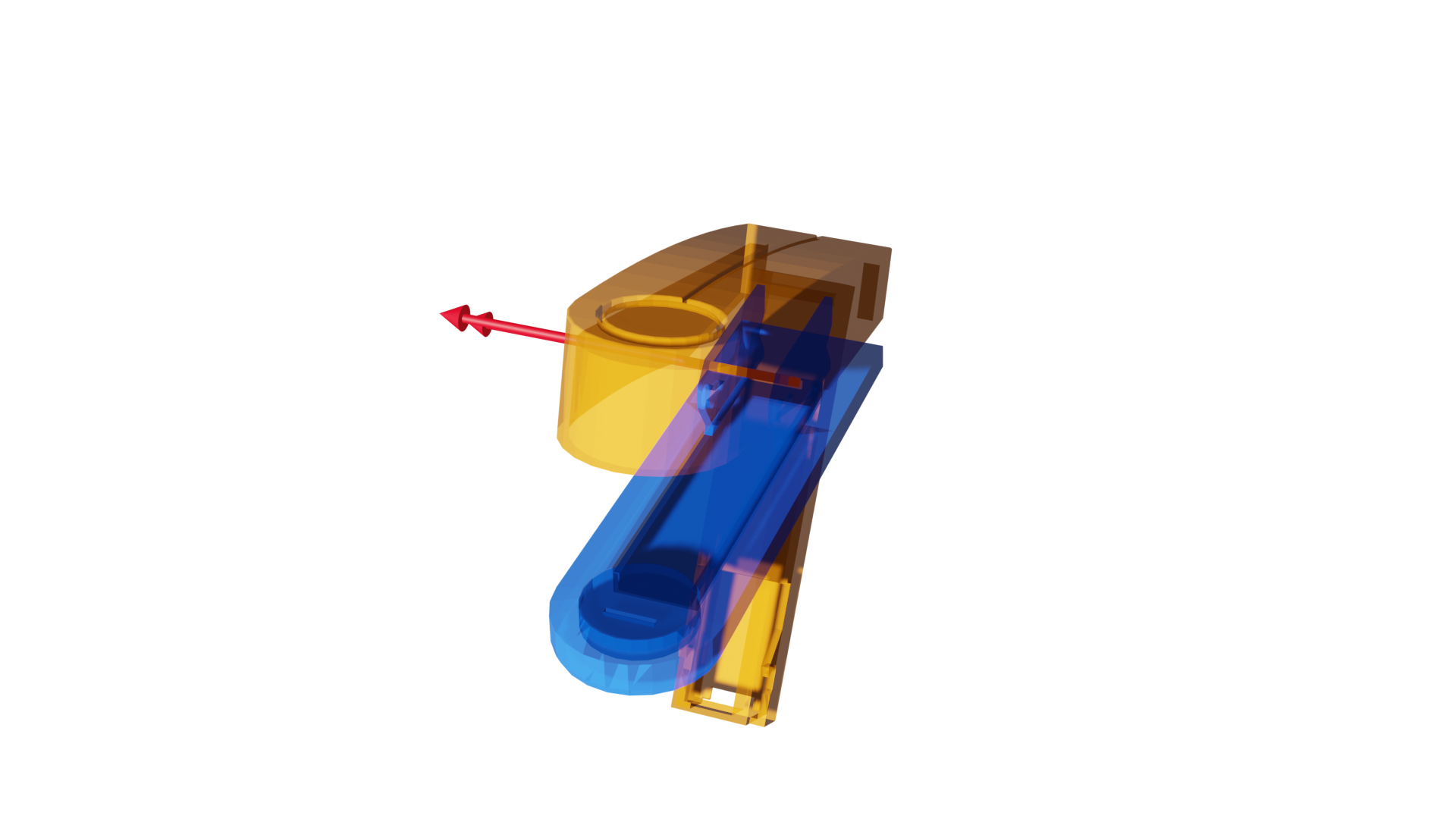}
    \includegraphics[width=0.33\linewidth]{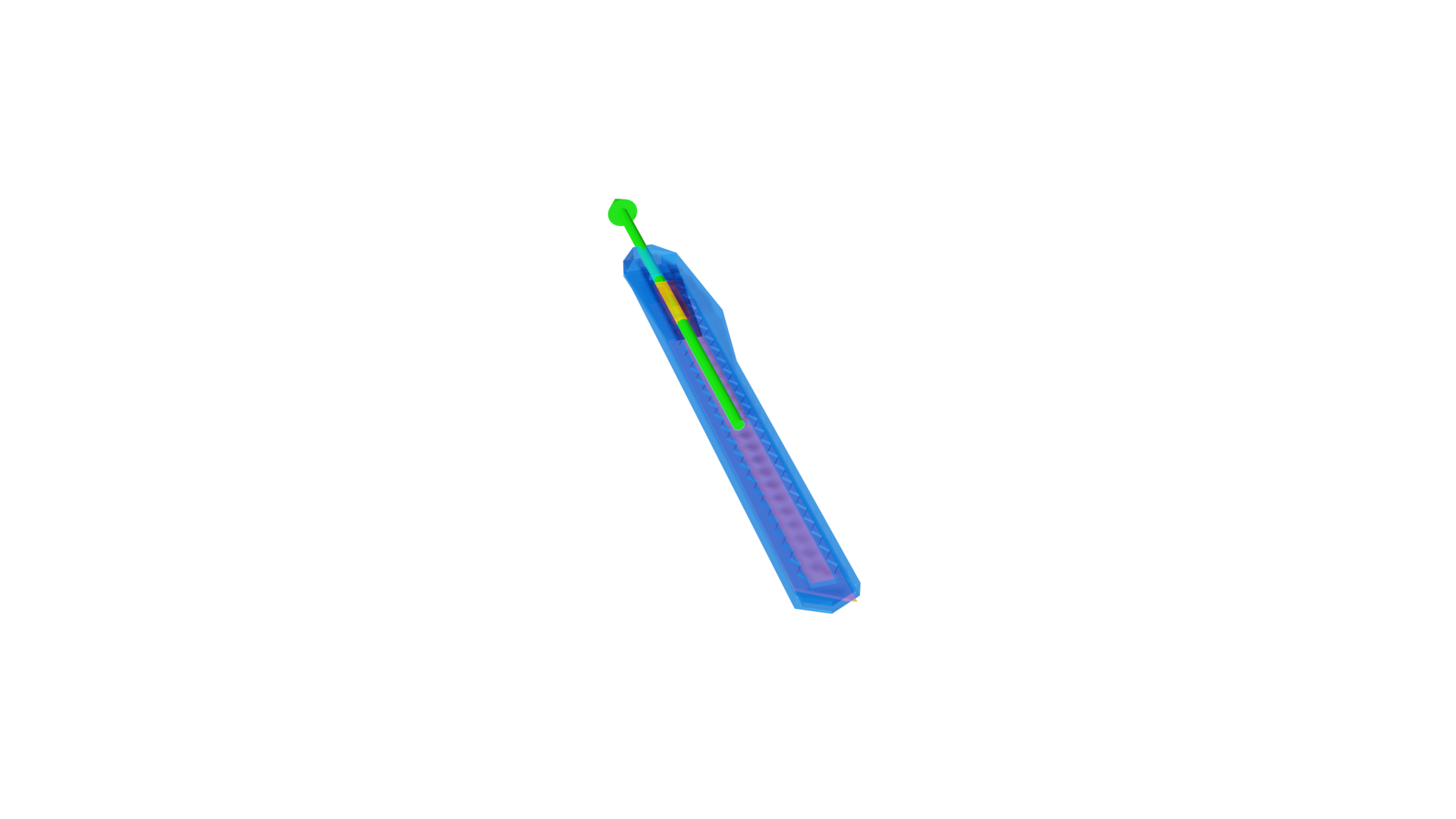}
    \\
    \end{tabular}
    \caption{
    Additional annotated shapes
    }
    \label{figure:qualitative5}
\end{figure*}

Please see Figure~\ref{figure:qualitative1},~\ref{figure:qualitative2},~\ref{figure:qualitative3},~\ref{figure:qualitative4},~\ref{figure:qualitative5}, for additional annotated shapes by our system.

\begin{figure*}[ht!]
    \centering
    \setlength{\tabcolsep}{1pt}
    \begin{tabular}{ccc}
    \textbf{BaseNet} & \textbf{Ours} & \textbf{GT}
        \\
    \includegraphics[width=0.33\linewidth]{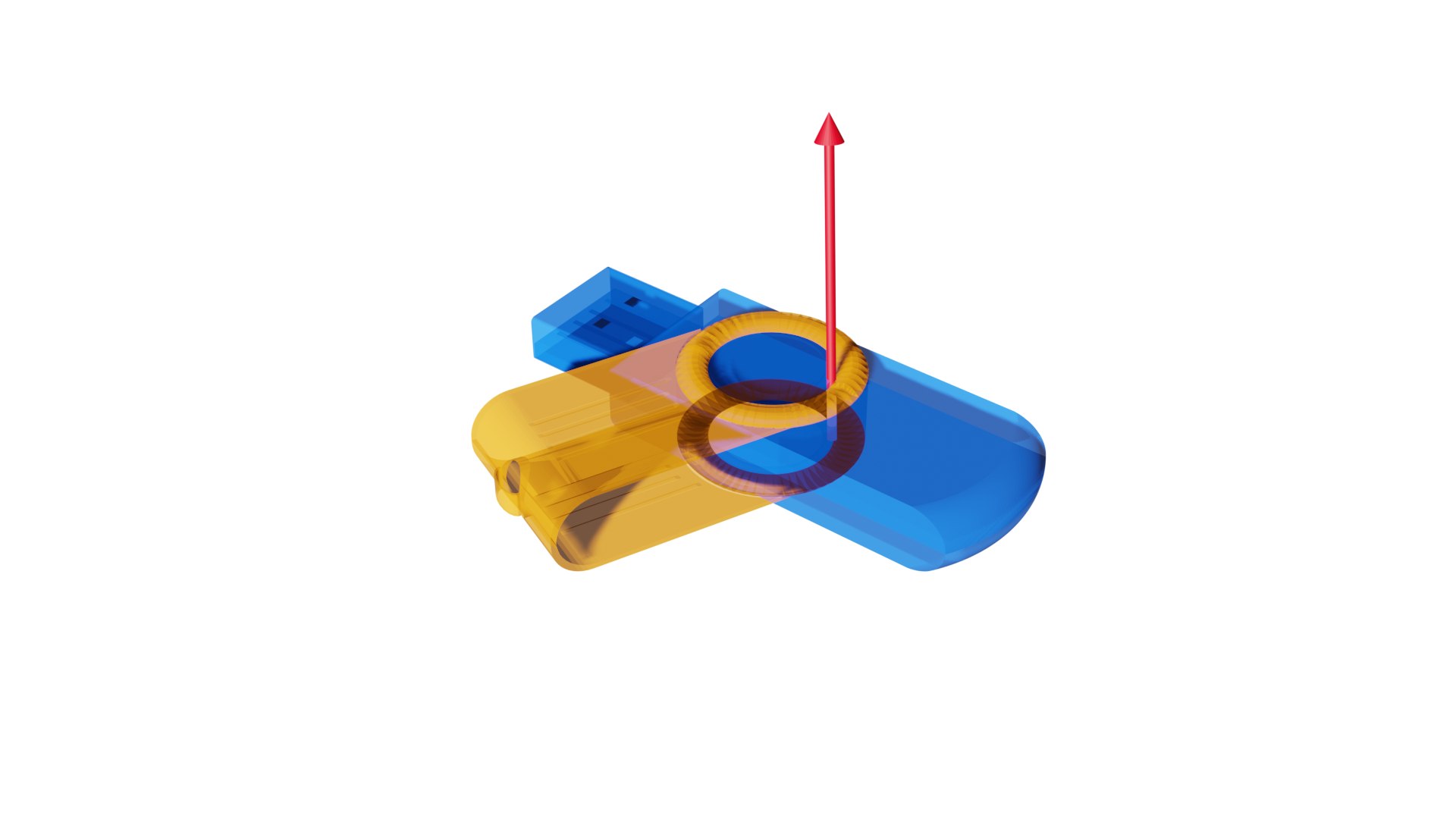} &
    \includegraphics[width=0.33\linewidth]{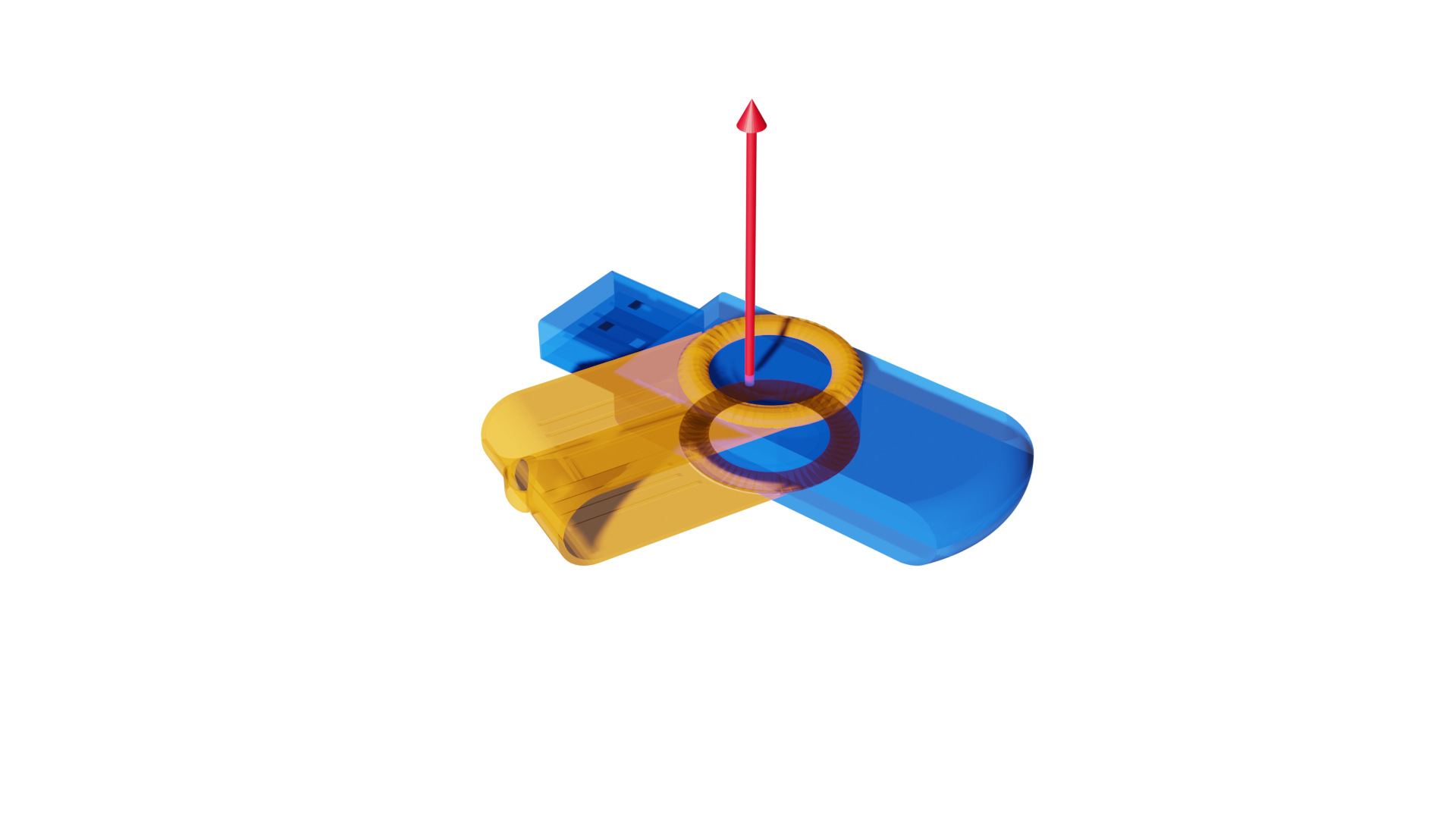} &
    \includegraphics[width=0.33\linewidth]{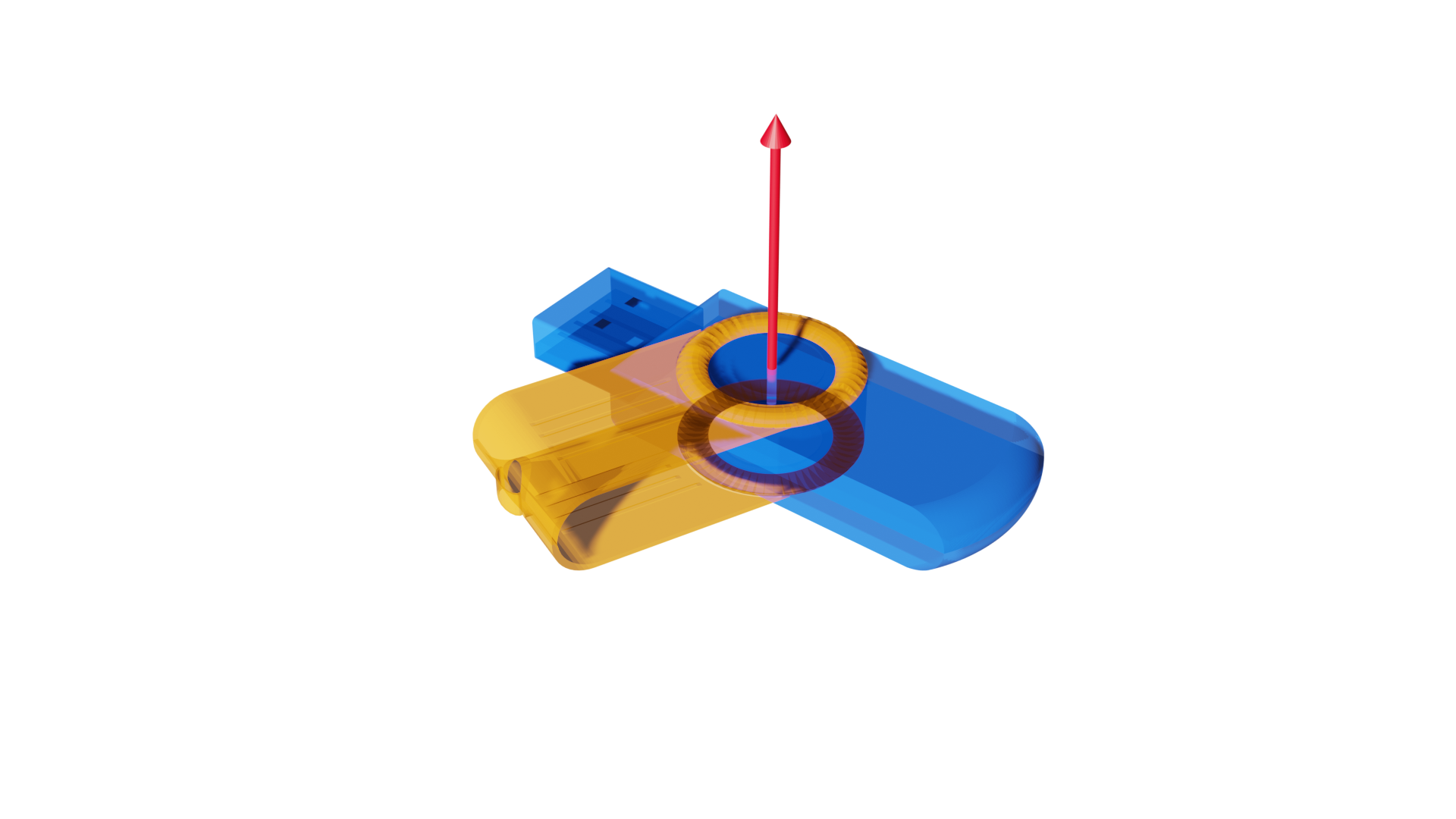}
    \\
    \includegraphics[width=0.33\linewidth]{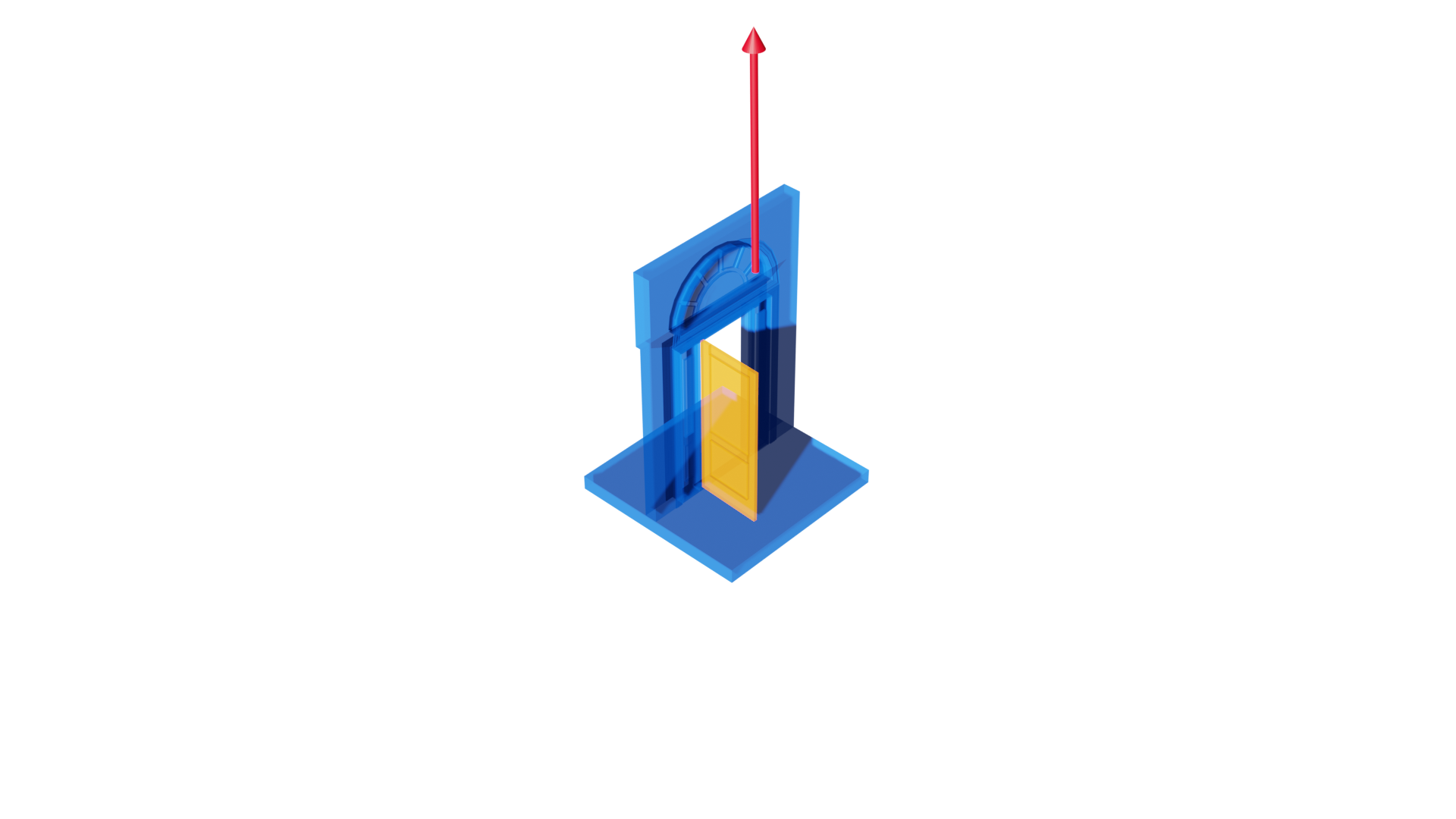} &
    \includegraphics[width=0.33\linewidth]{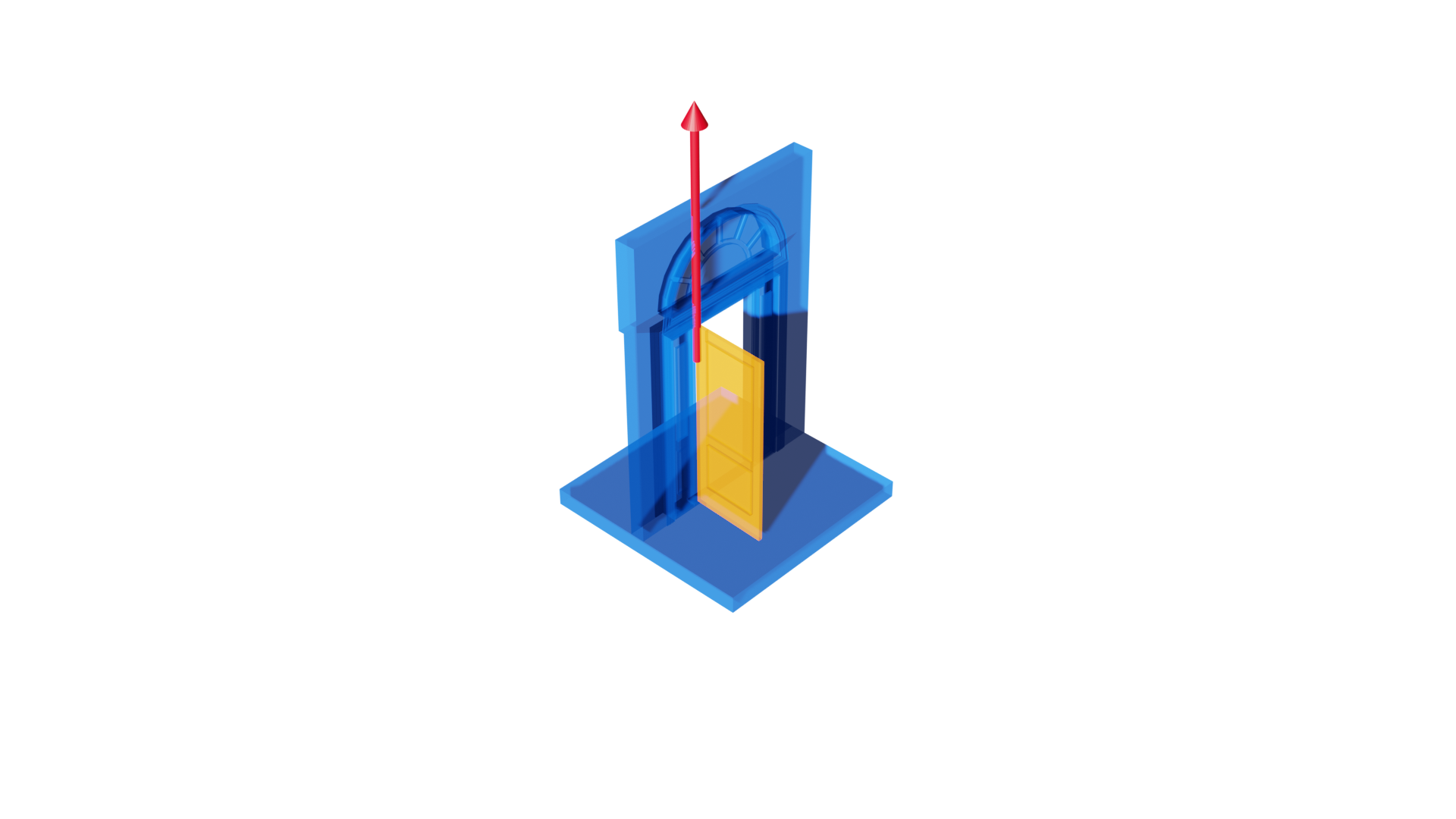} &
    \includegraphics[width=0.33\linewidth]{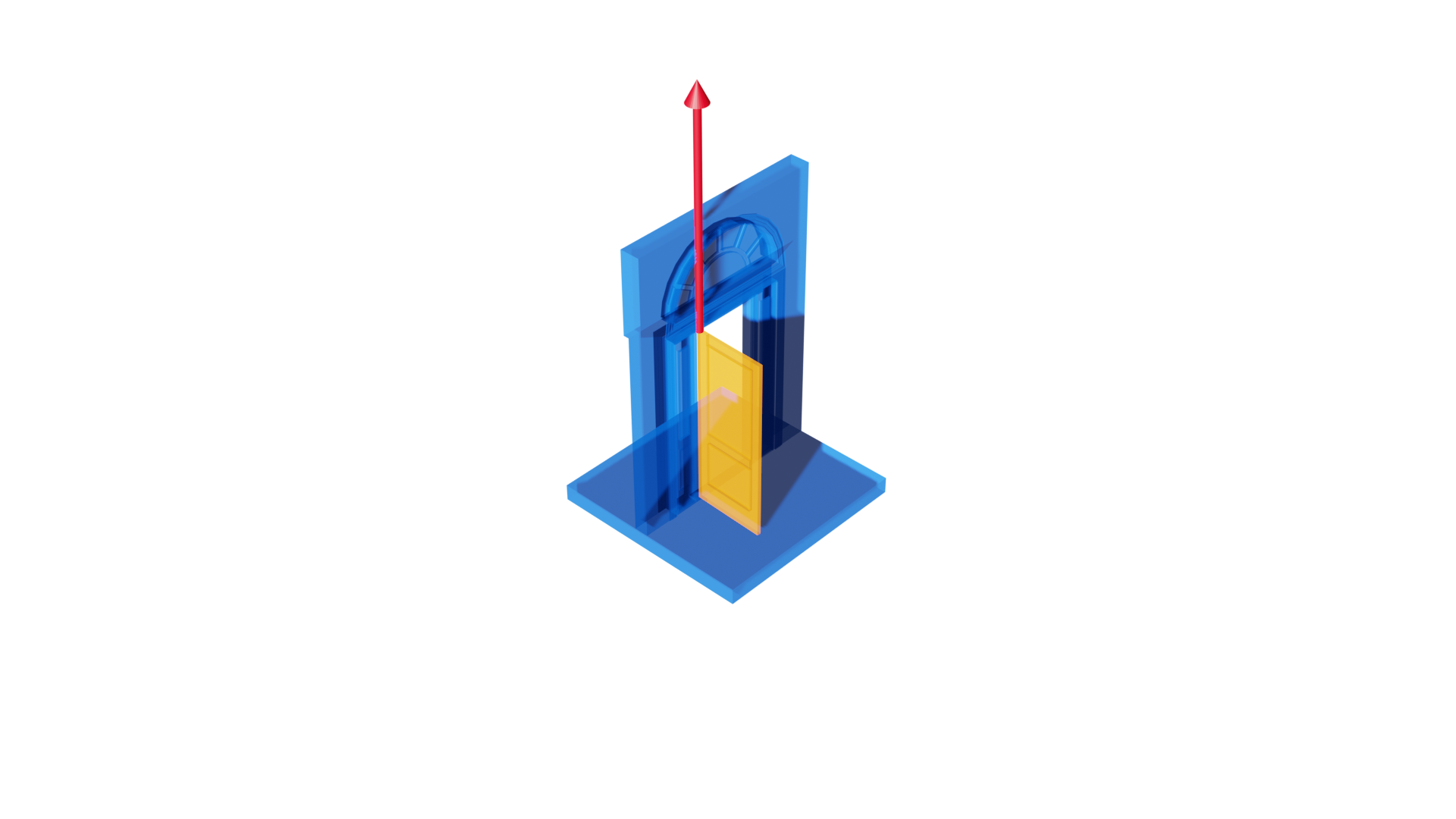}
    \\
    \includegraphics[width=0.33\linewidth]{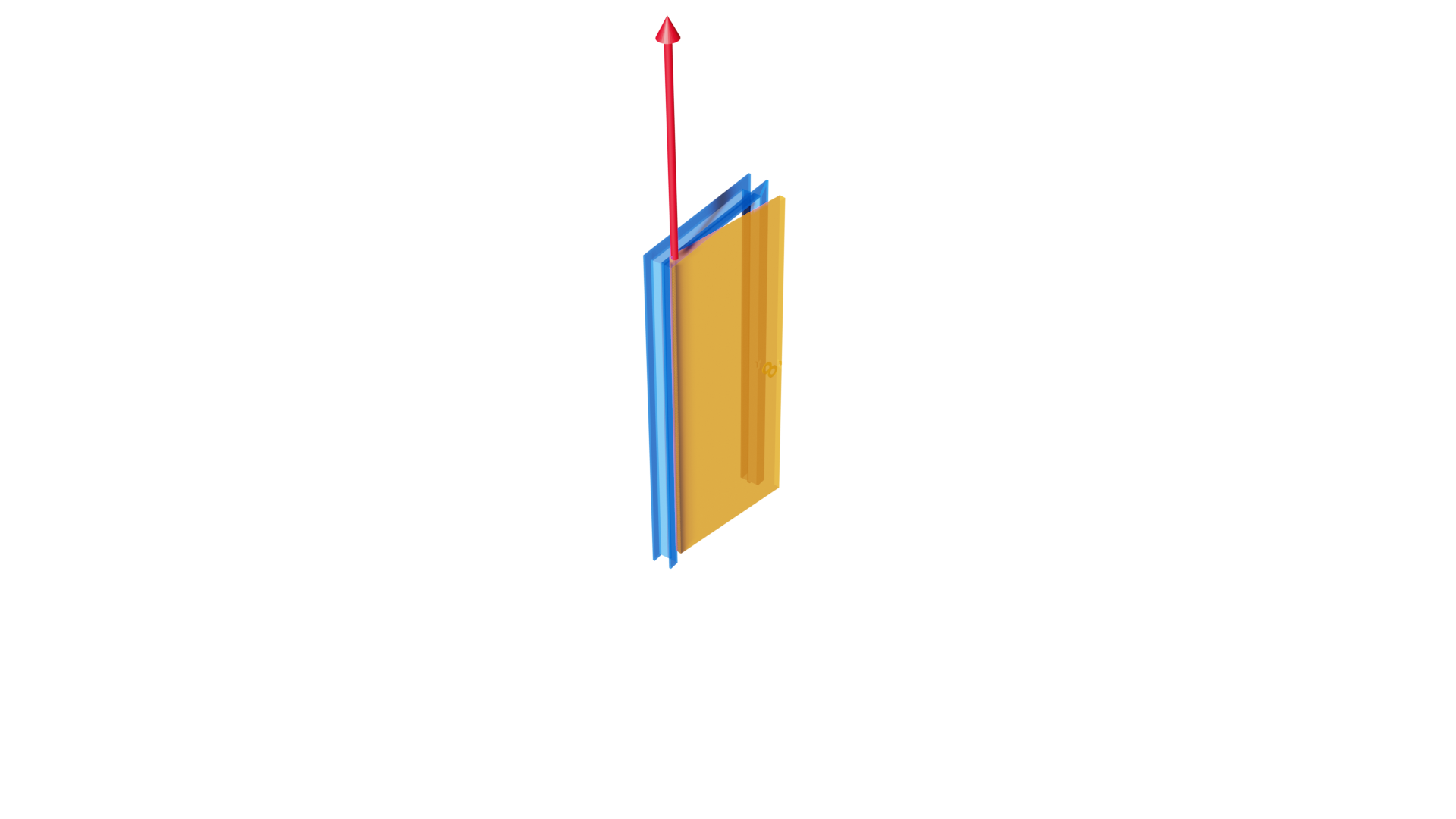} &
    \includegraphics[width=0.33\linewidth]{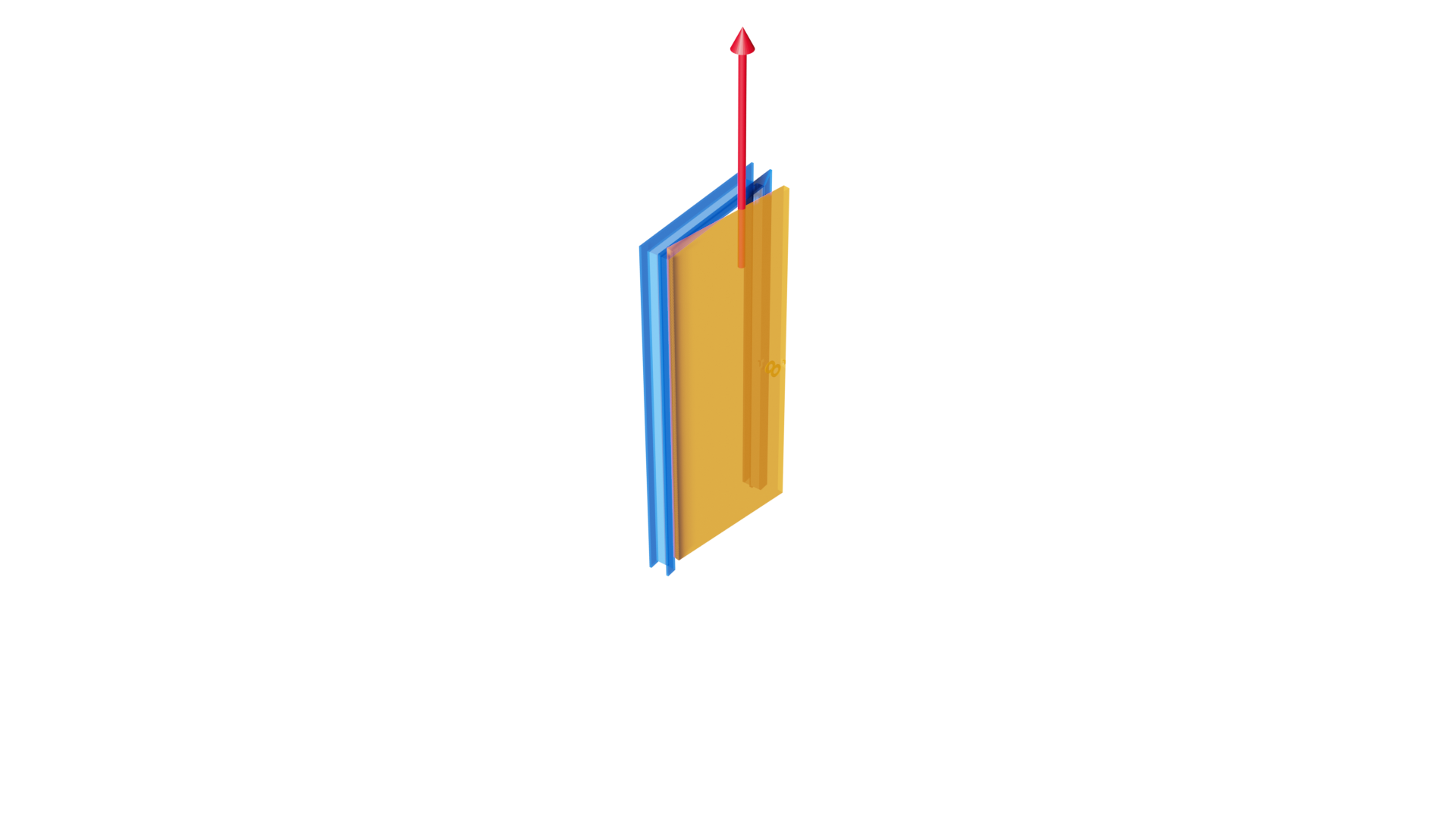} &
    \includegraphics[width=0.33\linewidth]{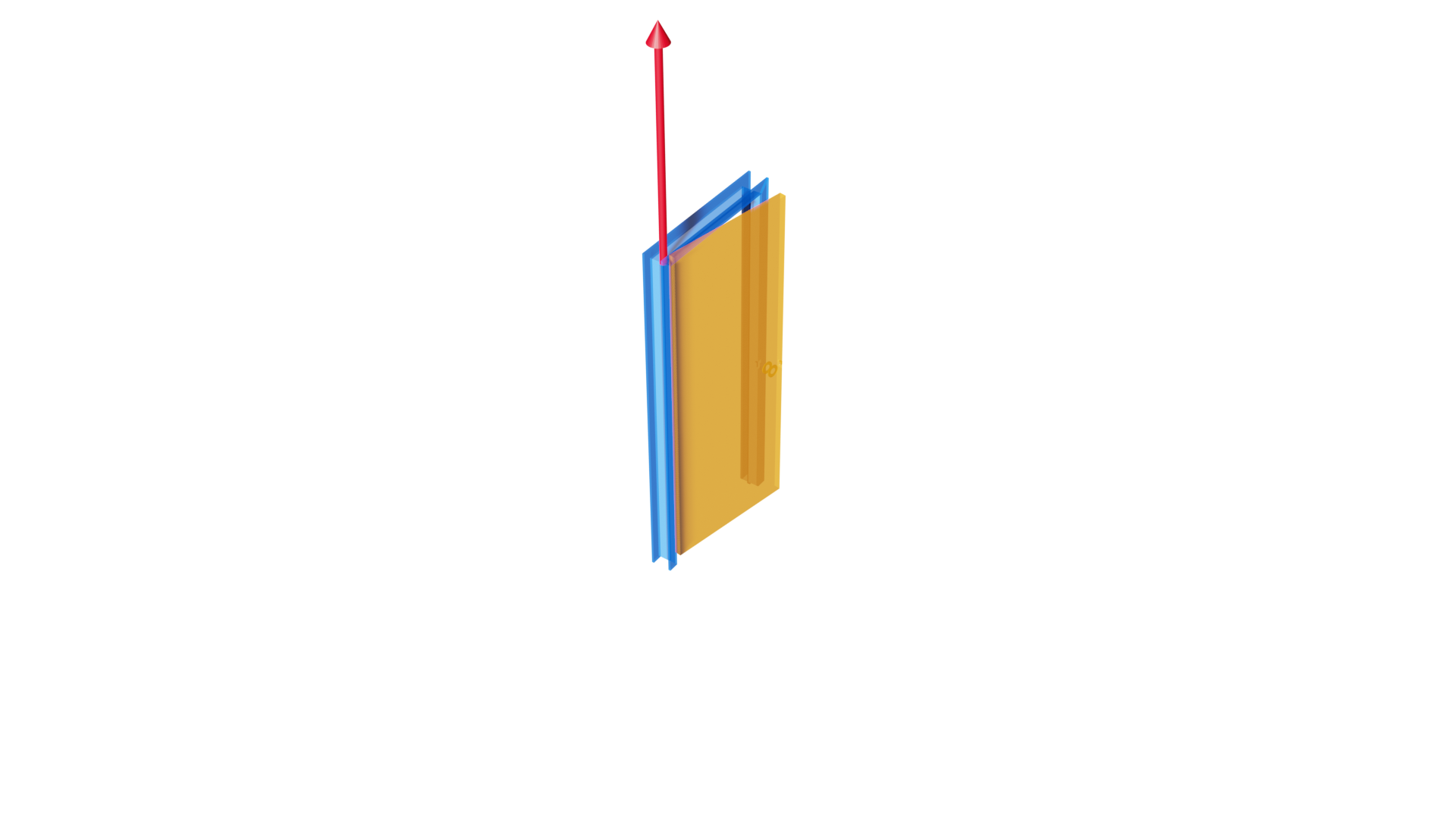}
    \\
    \includegraphics[width=0.33\linewidth]{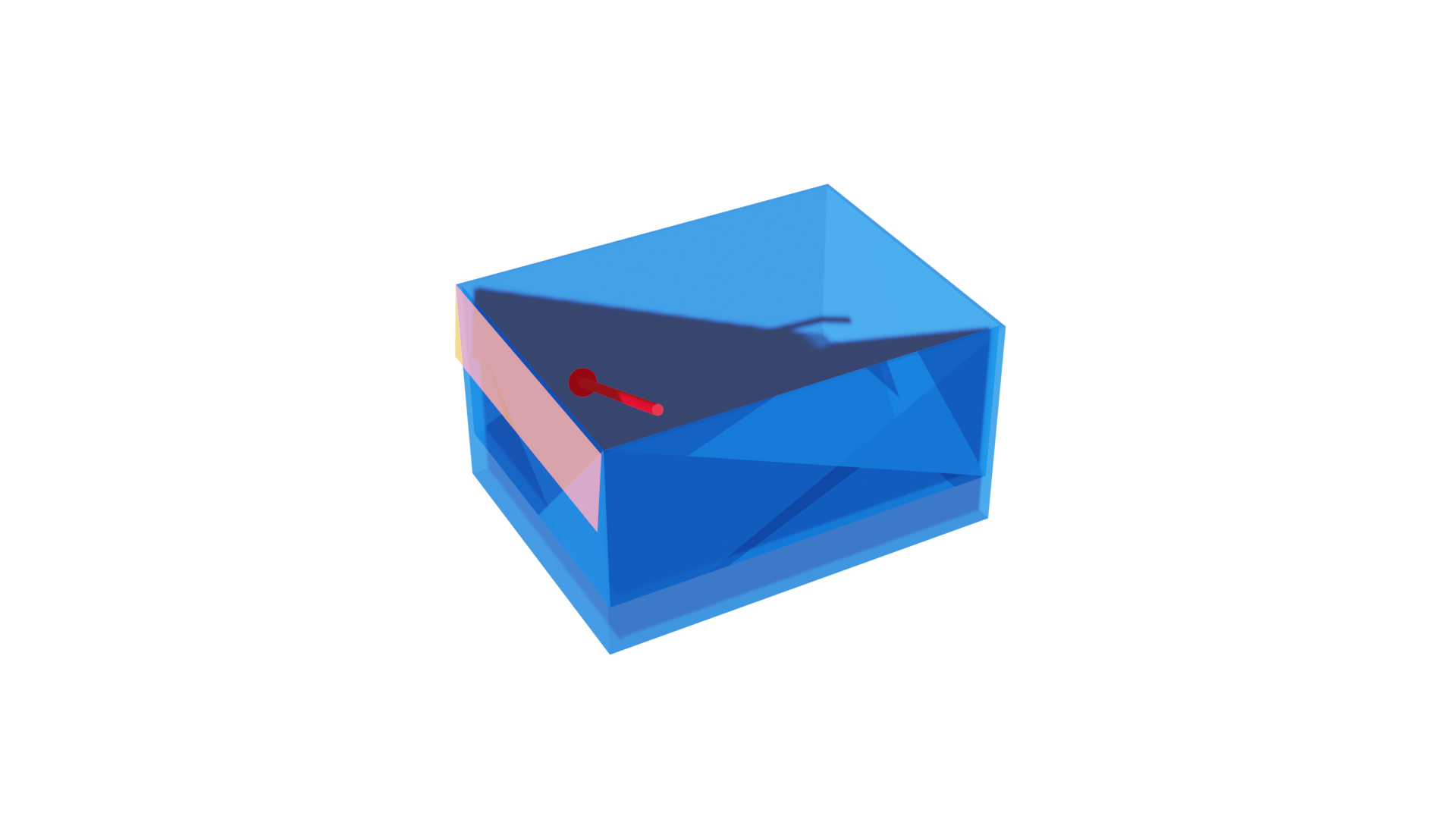} &
    \includegraphics[width=0.33\linewidth]{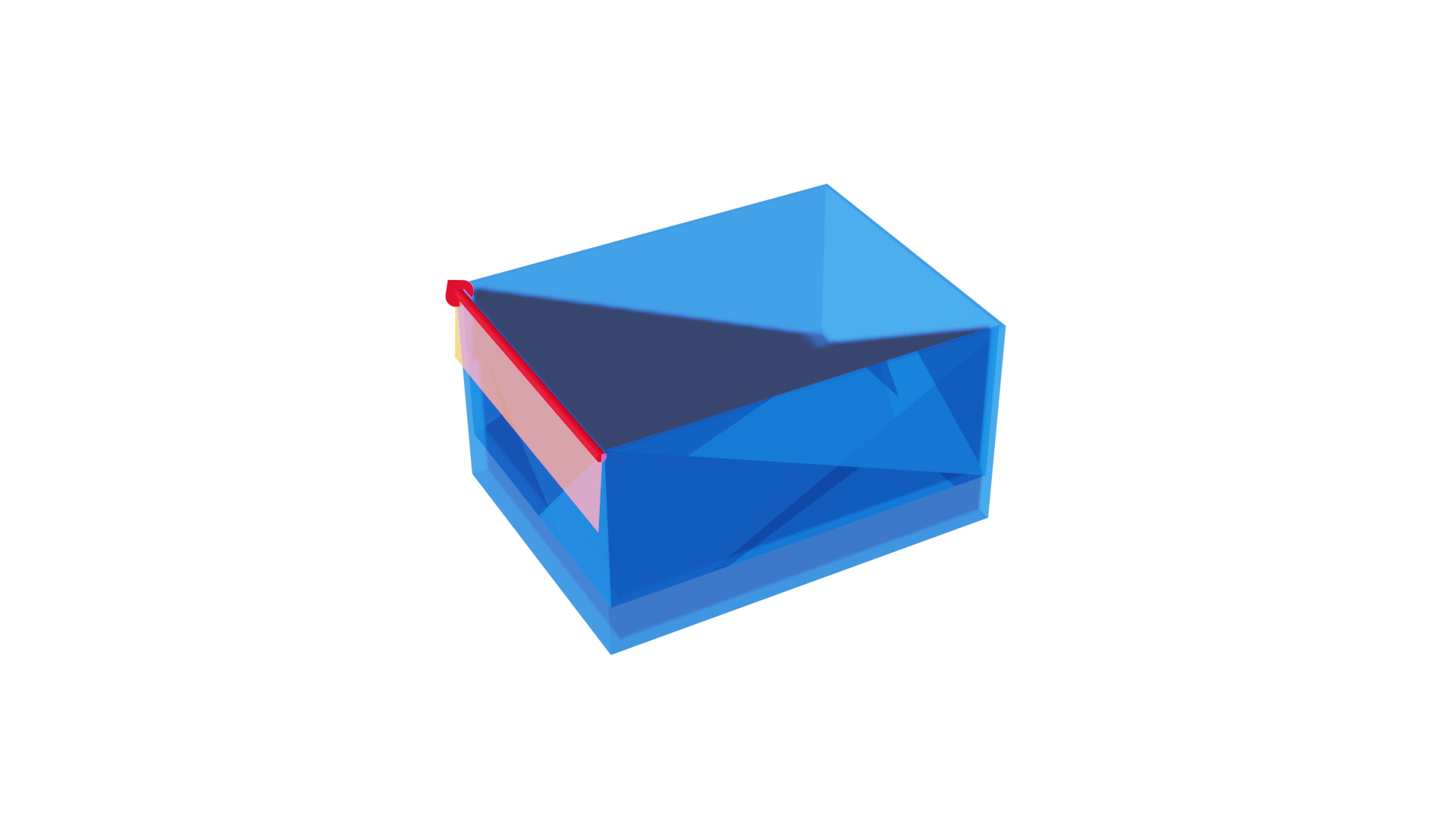} &
    \includegraphics[width=0.33\linewidth]{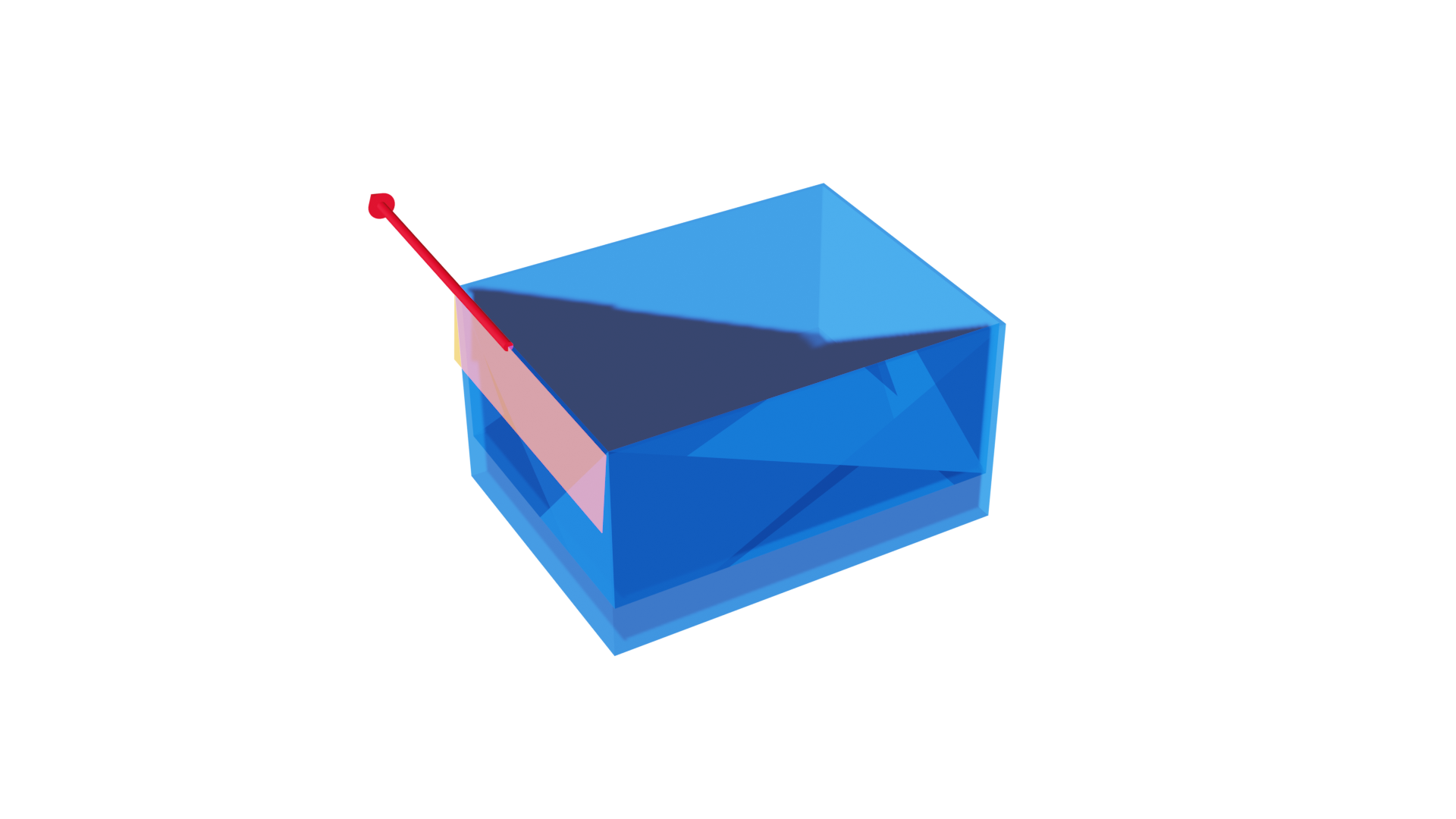}
    \\
    \includegraphics[width=0.33\linewidth]{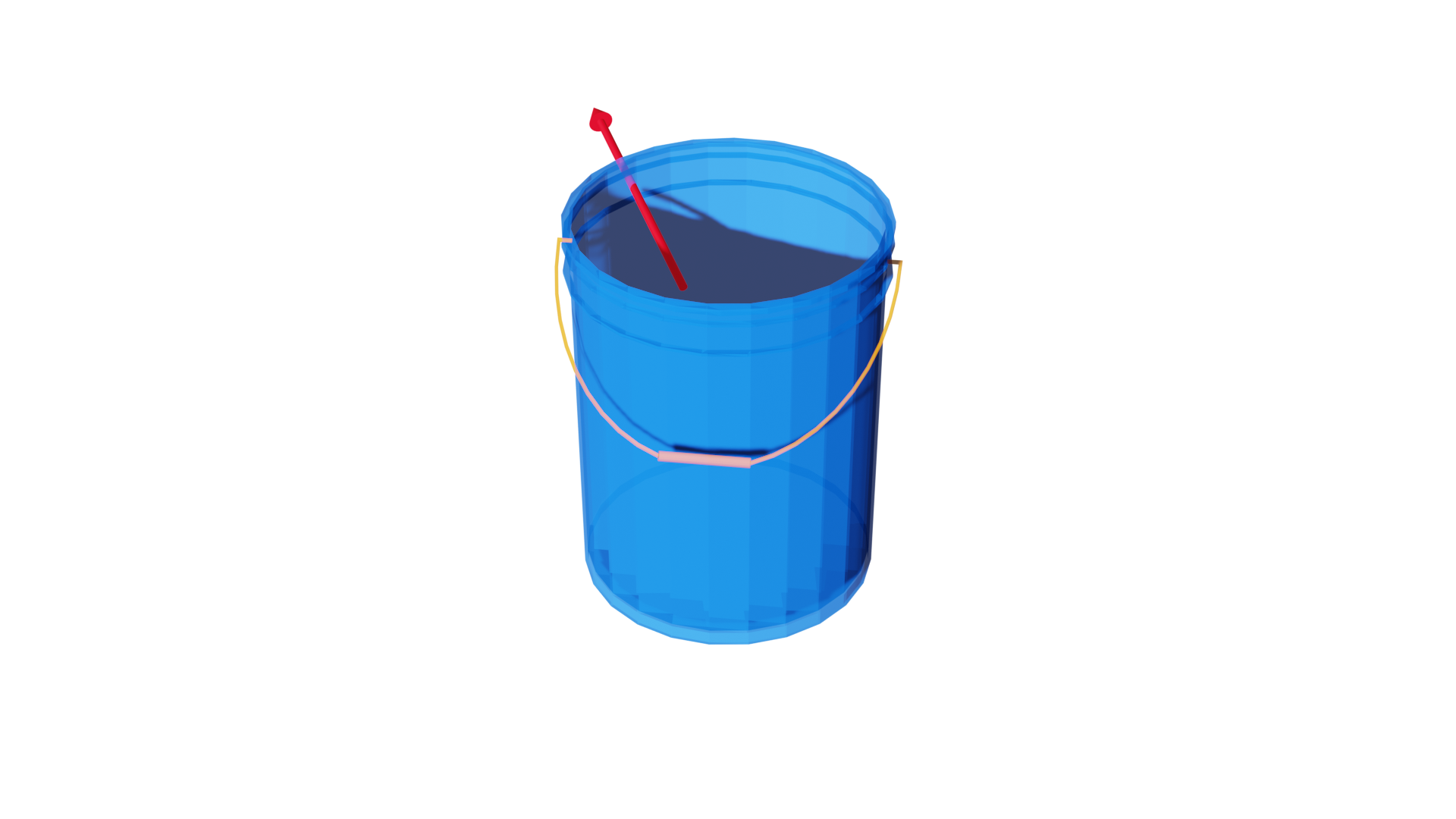} &
    \includegraphics[width=0.33\linewidth]{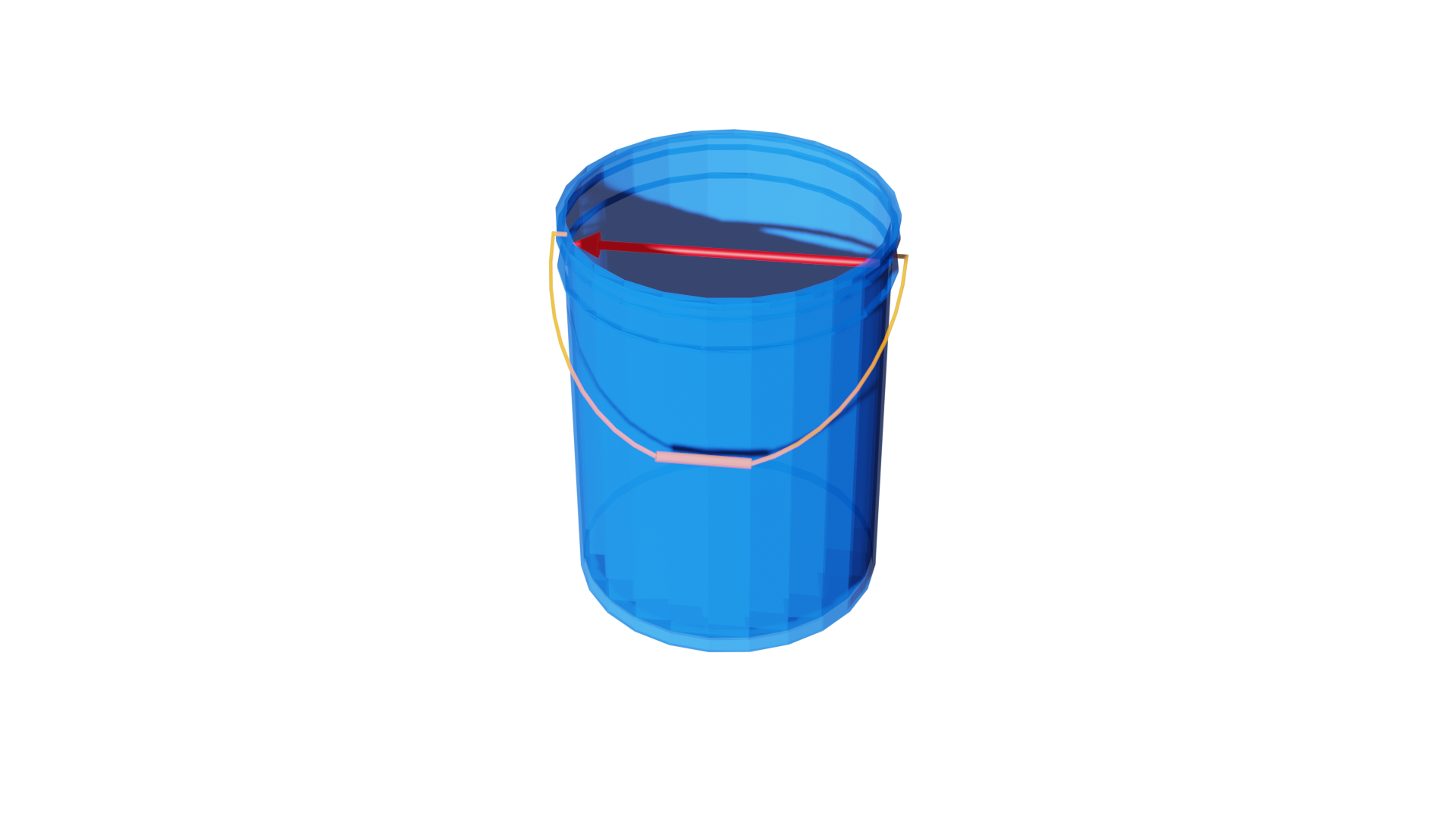} &
    \includegraphics[width=0.33\linewidth]{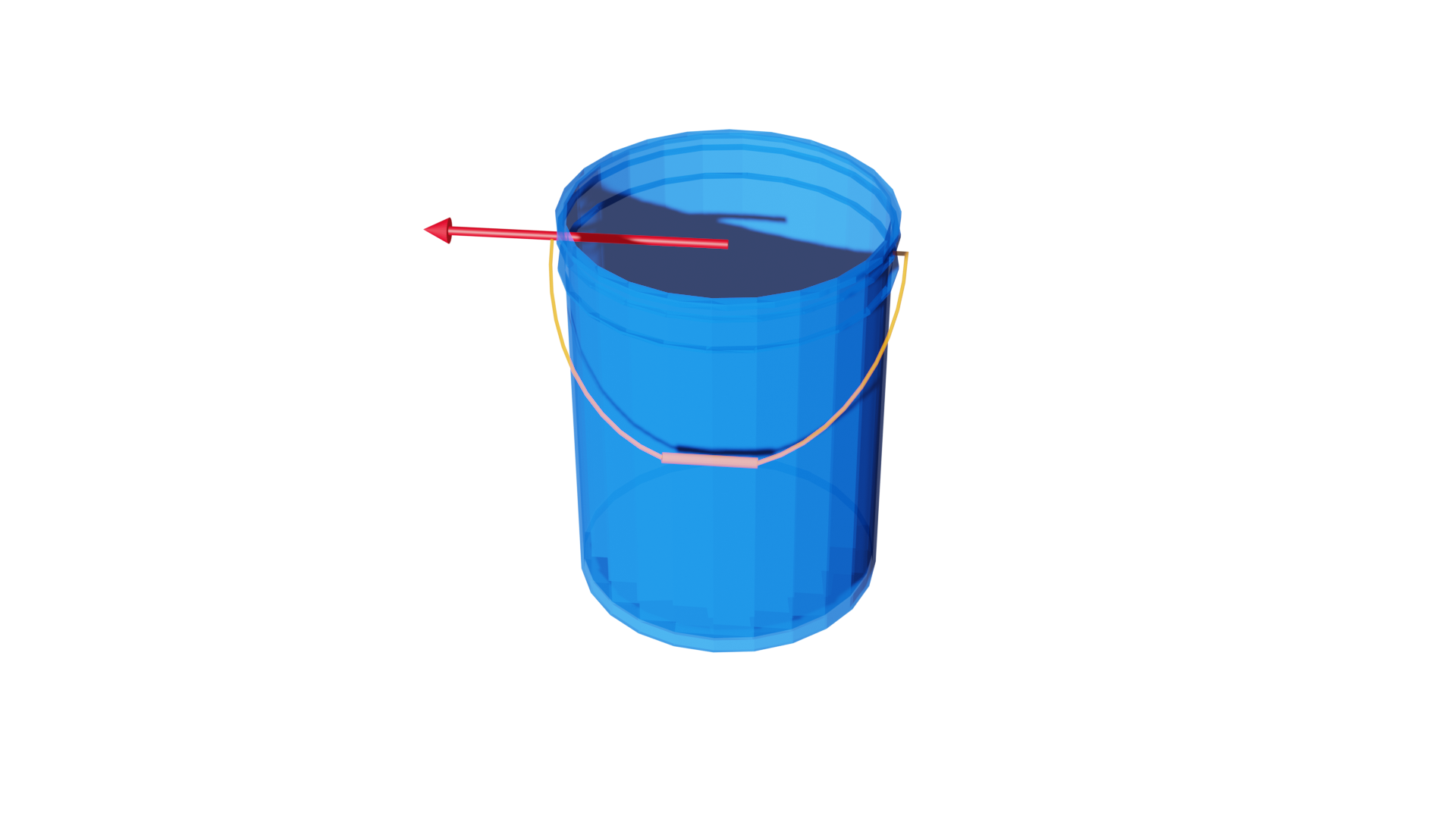}
    \\
    \end{tabular}
    \caption{
    Additional qualitative comparison of our method with the supervised BaseNet baseline
    }
    \label{figure:qualitative_comparison1}
\end{figure*}

\begin{figure*}[ht!]
    \centering
    \setlength{\tabcolsep}{1pt}
    \begin{tabular}{ccc}
    \textbf{BaseNet} & \textbf{Ours} & \textbf{GT}
        \\
    \includegraphics[width=0.33\linewidth]{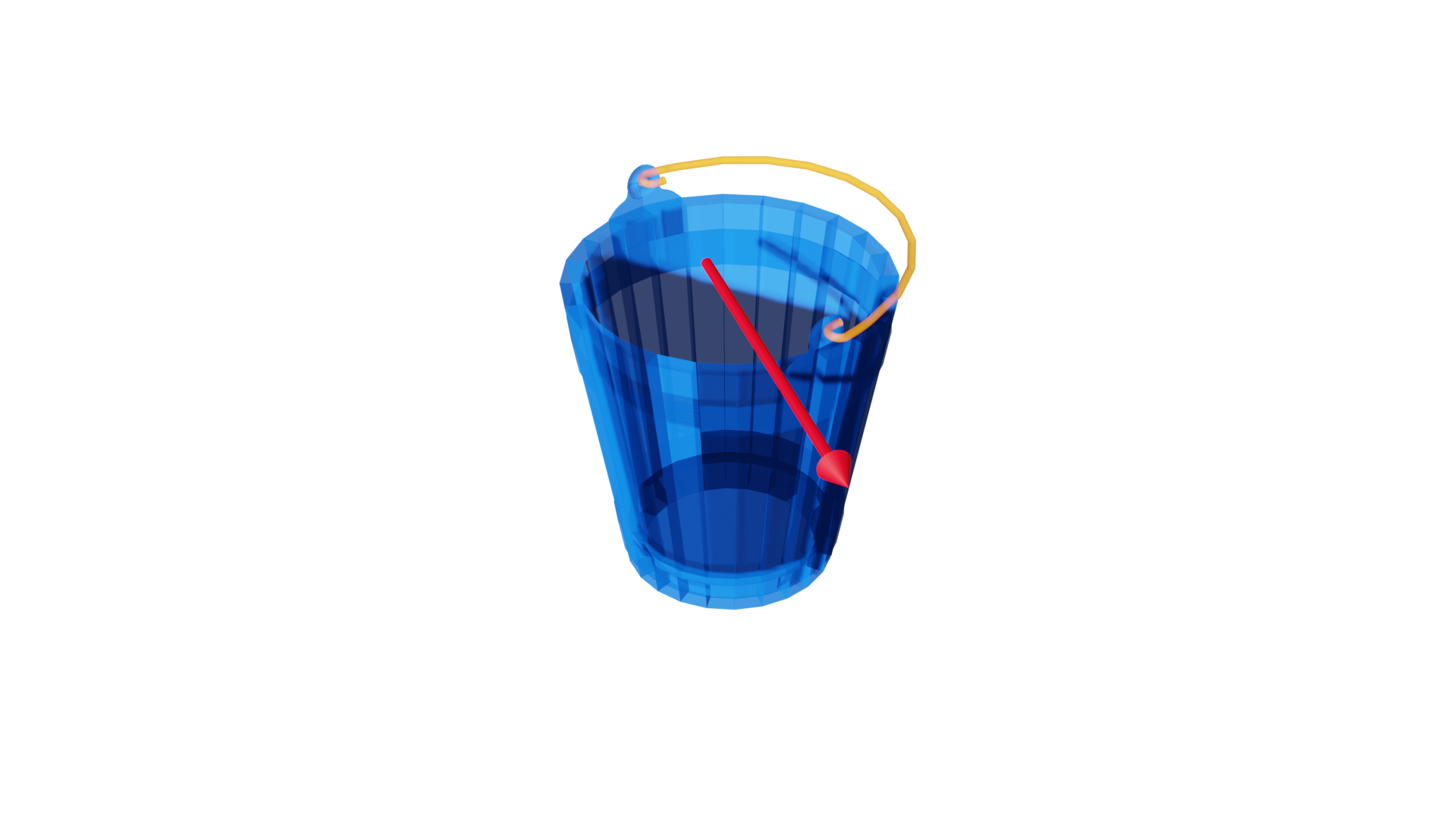} &
    \includegraphics[width=0.33\linewidth]{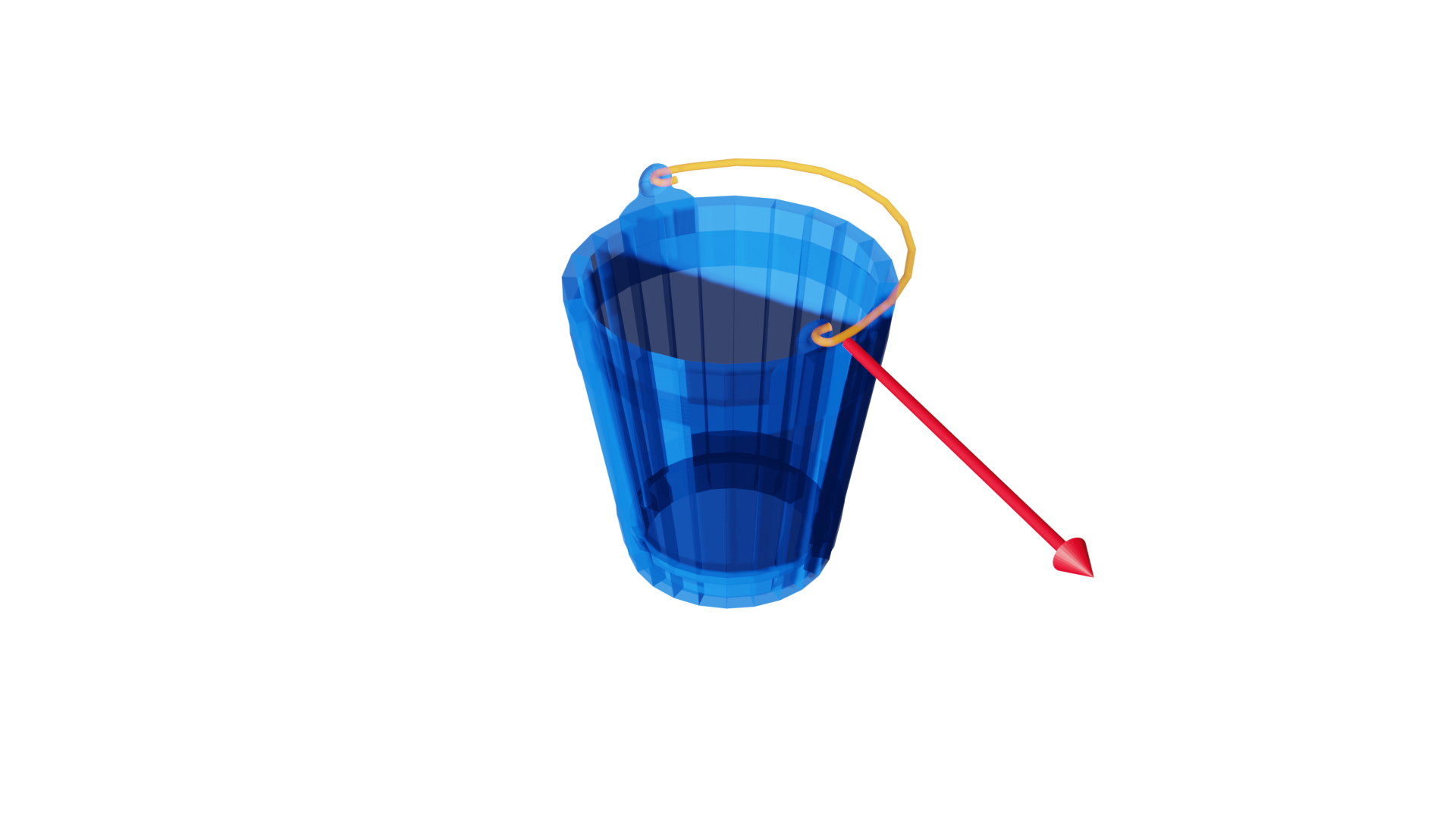} &
    \includegraphics[width=0.33\linewidth]{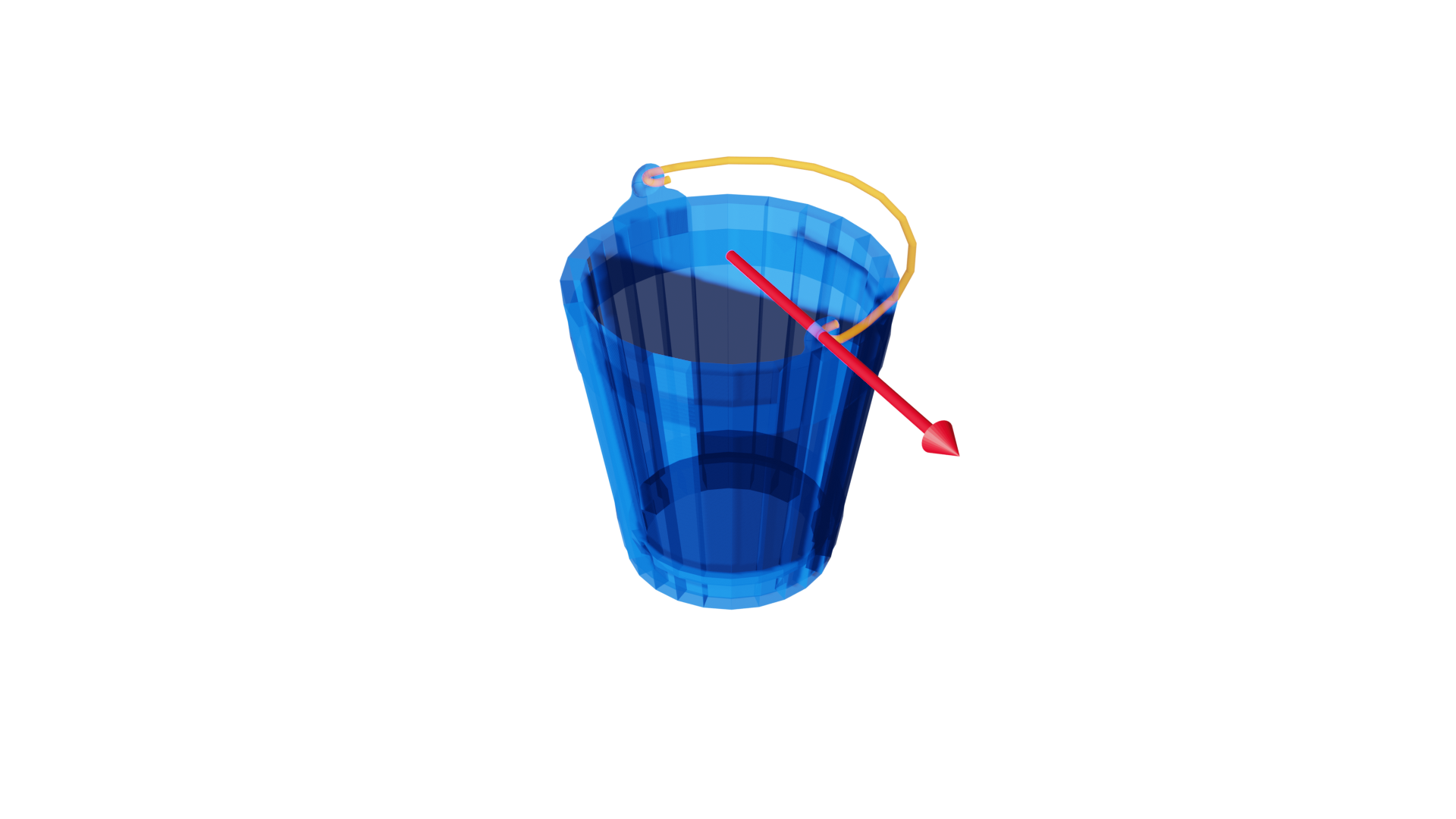}
    \\
    \includegraphics[width=0.33\linewidth]{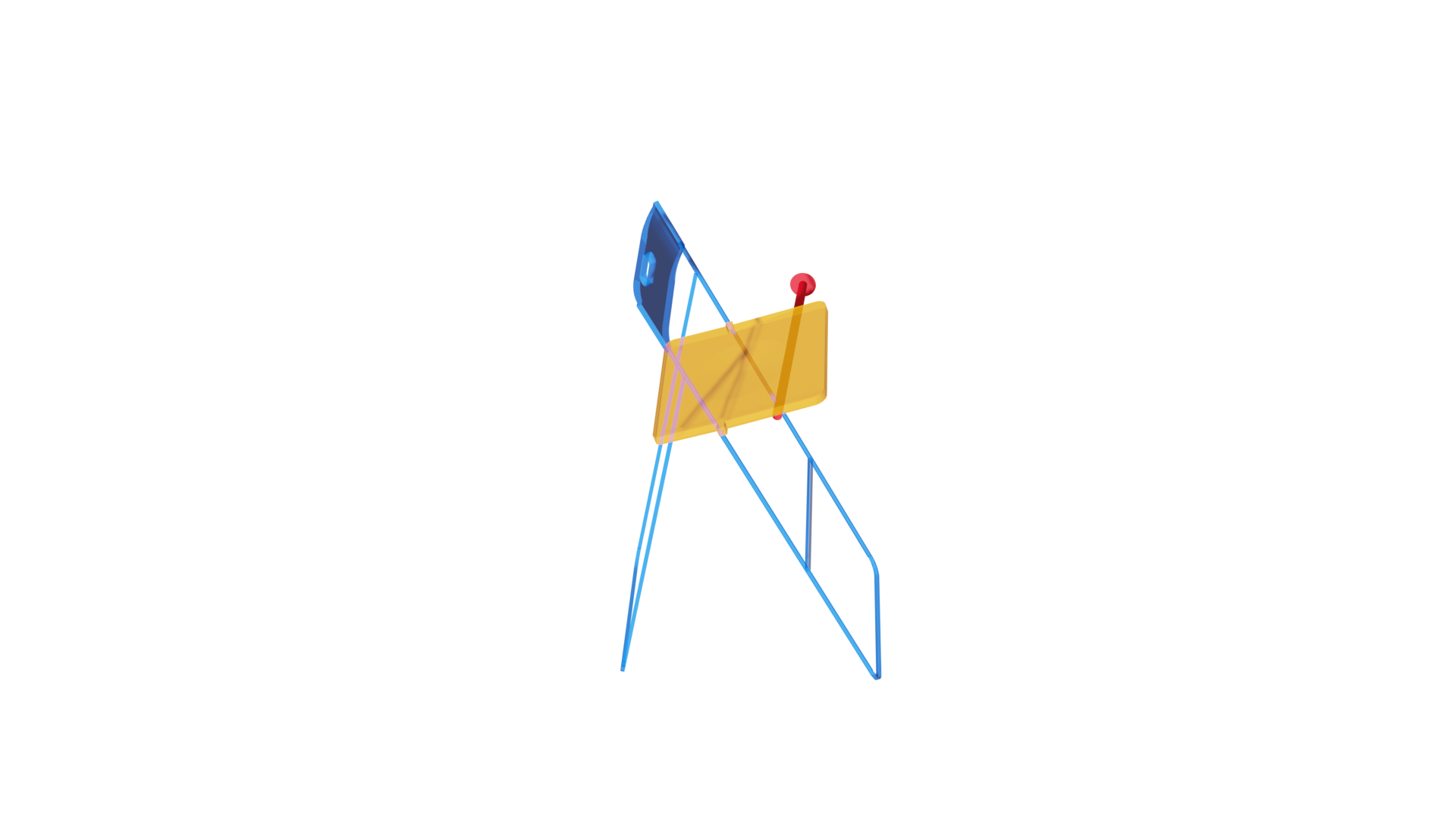} &
    \includegraphics[width=0.33\linewidth]{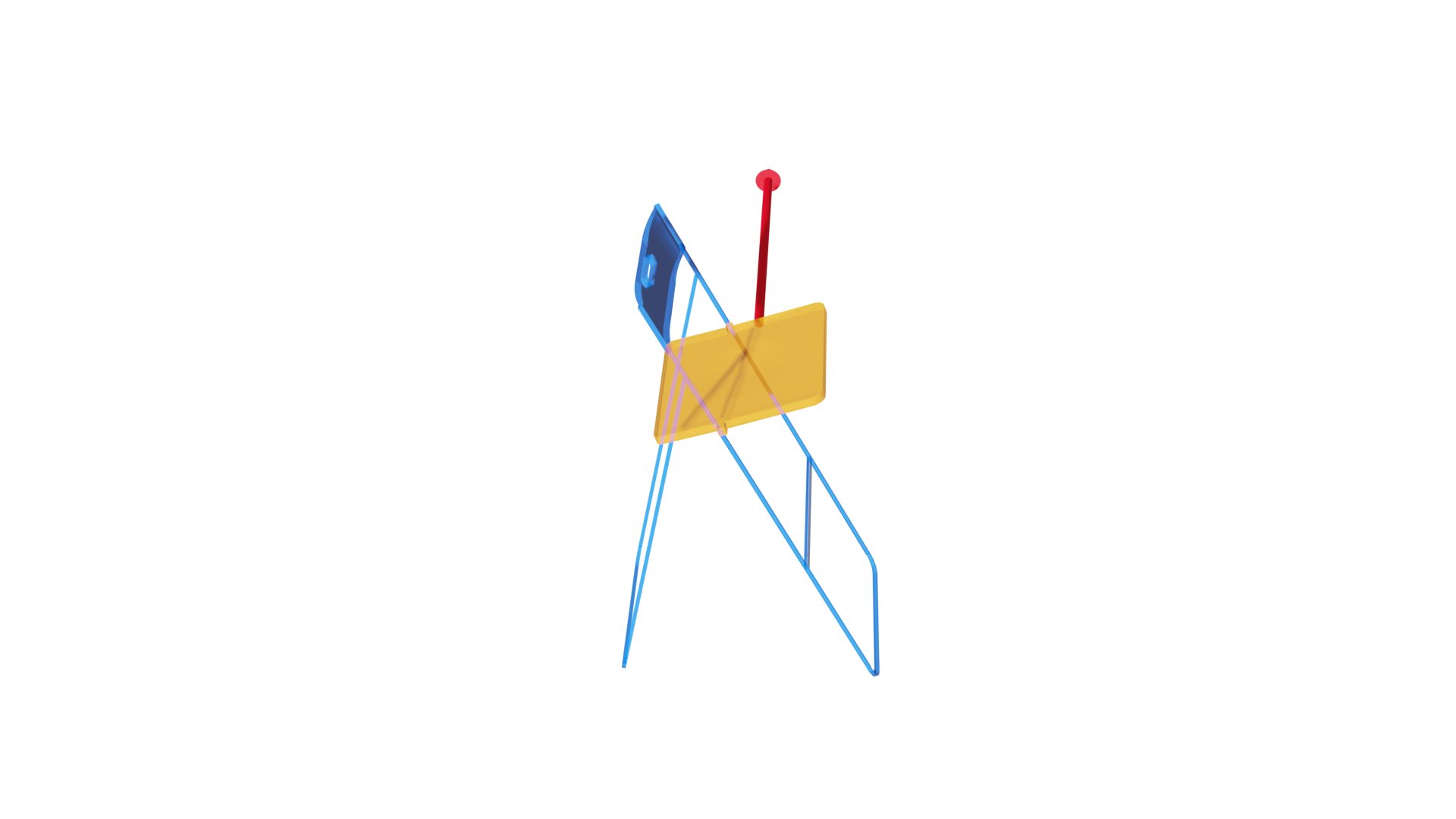} &
    \includegraphics[width=0.33\linewidth]{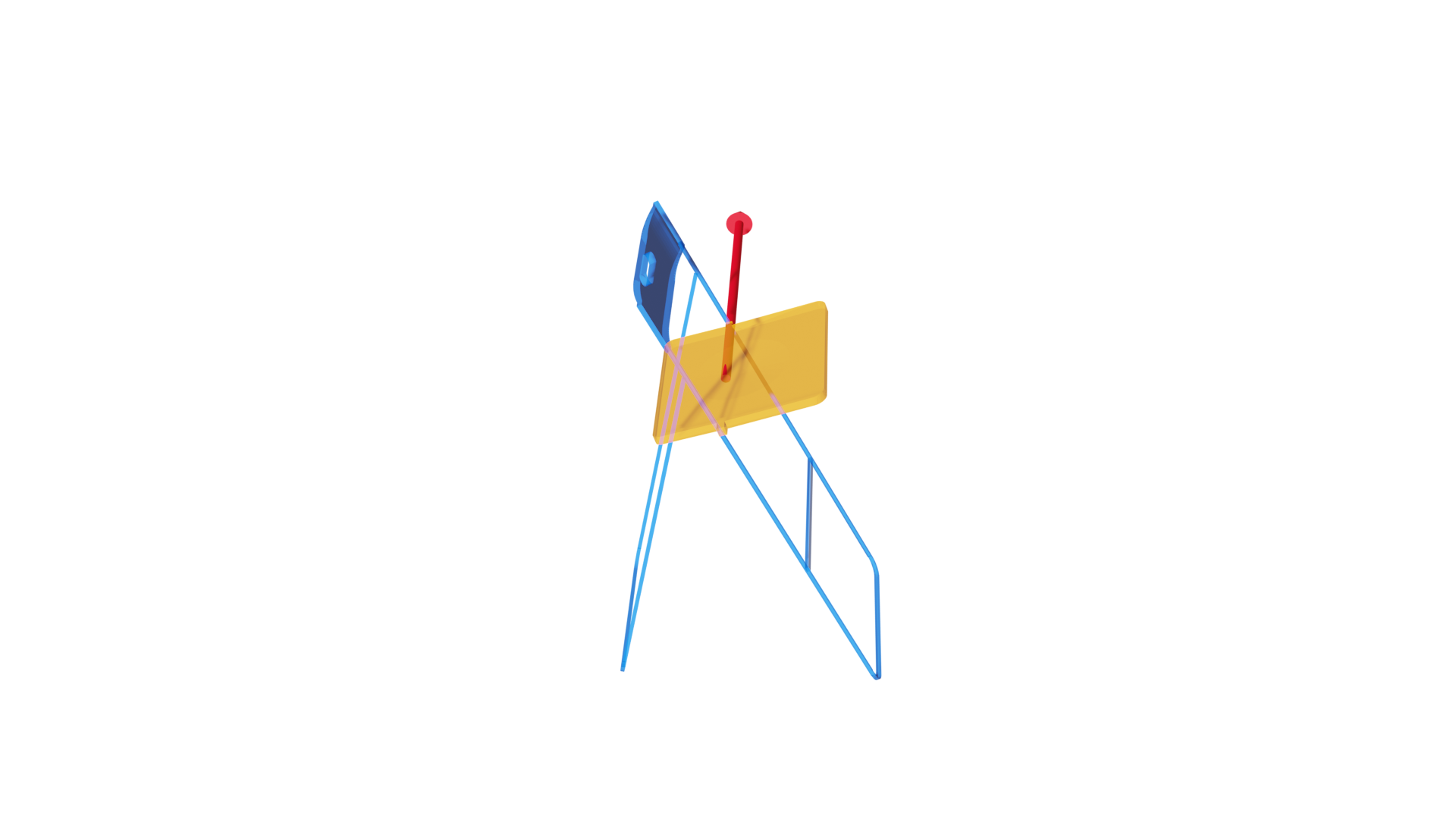}
    \\
    \includegraphics[width=0.33\linewidth]{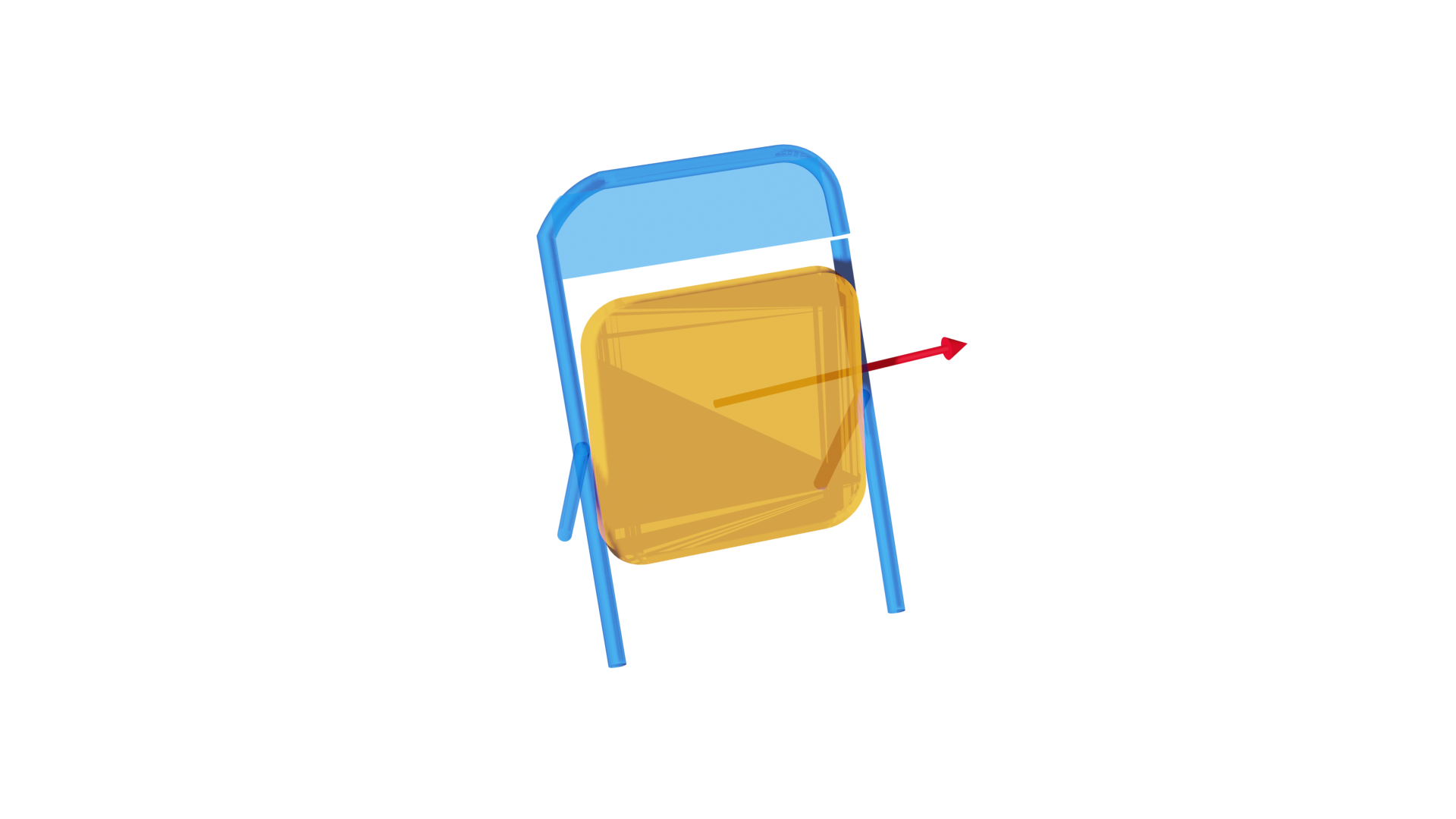} &
    \includegraphics[width=0.33\linewidth]{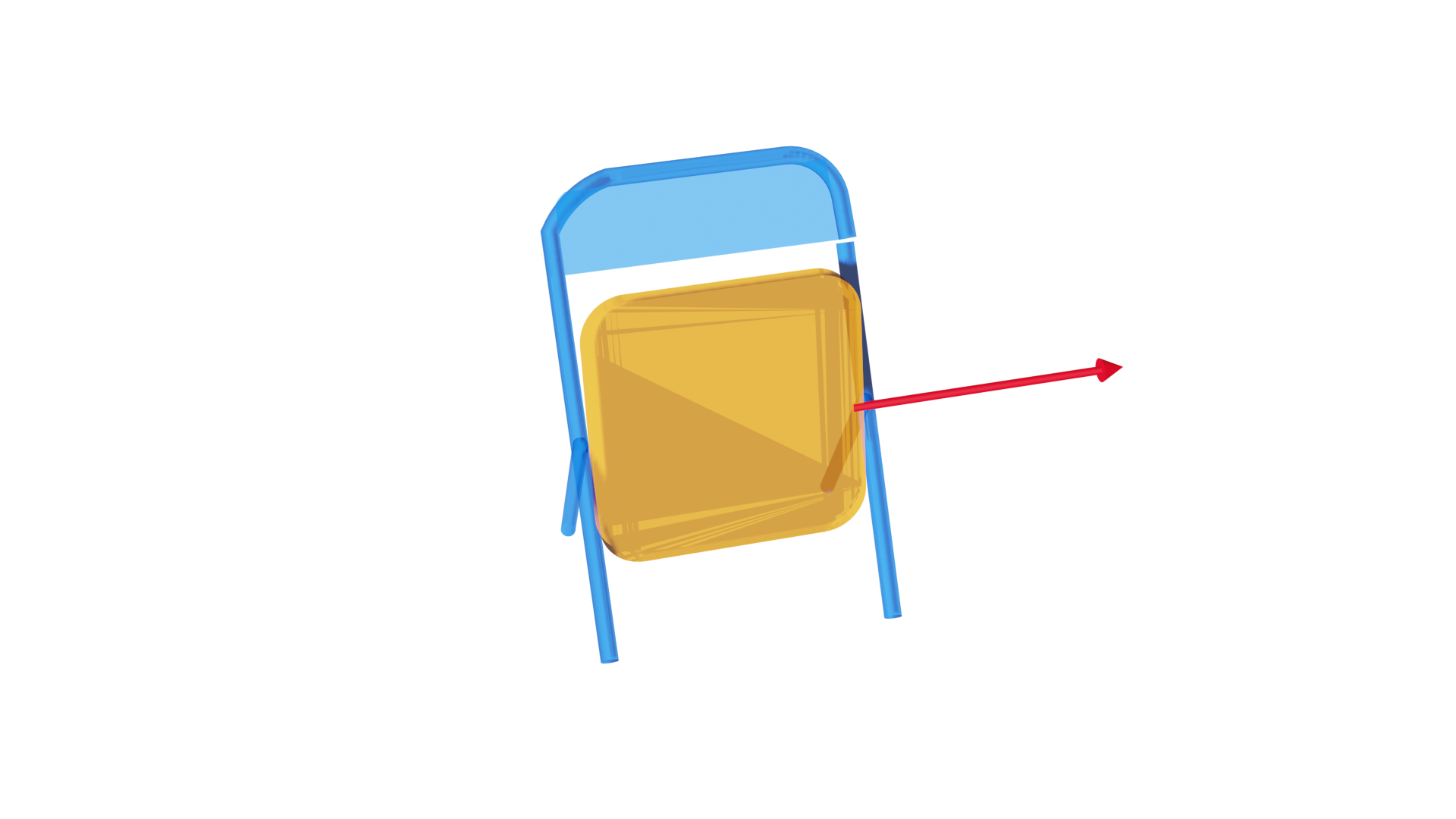} &
    \includegraphics[width=0.33\linewidth]{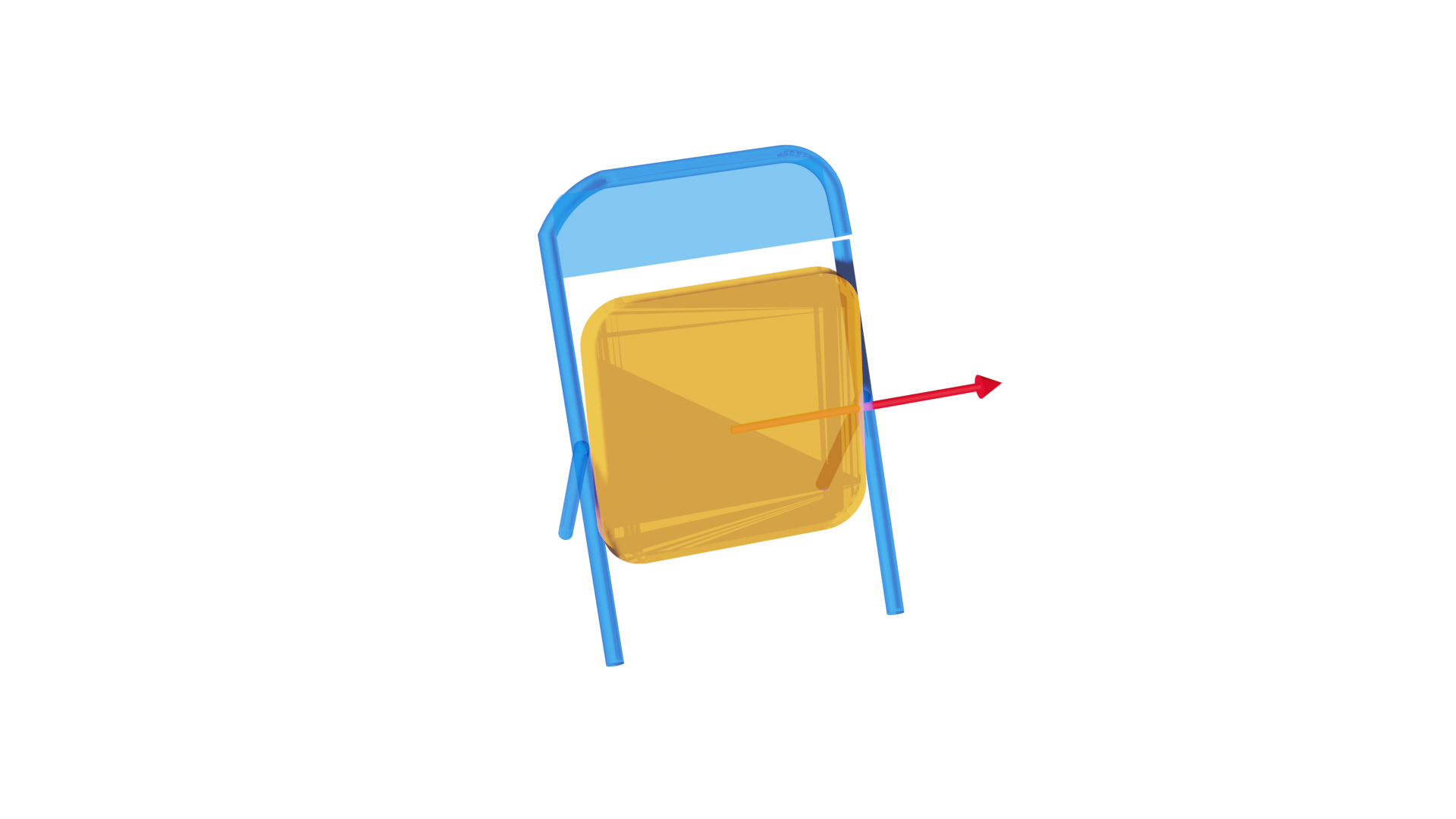}
    \\
    \includegraphics[width=0.33\linewidth]{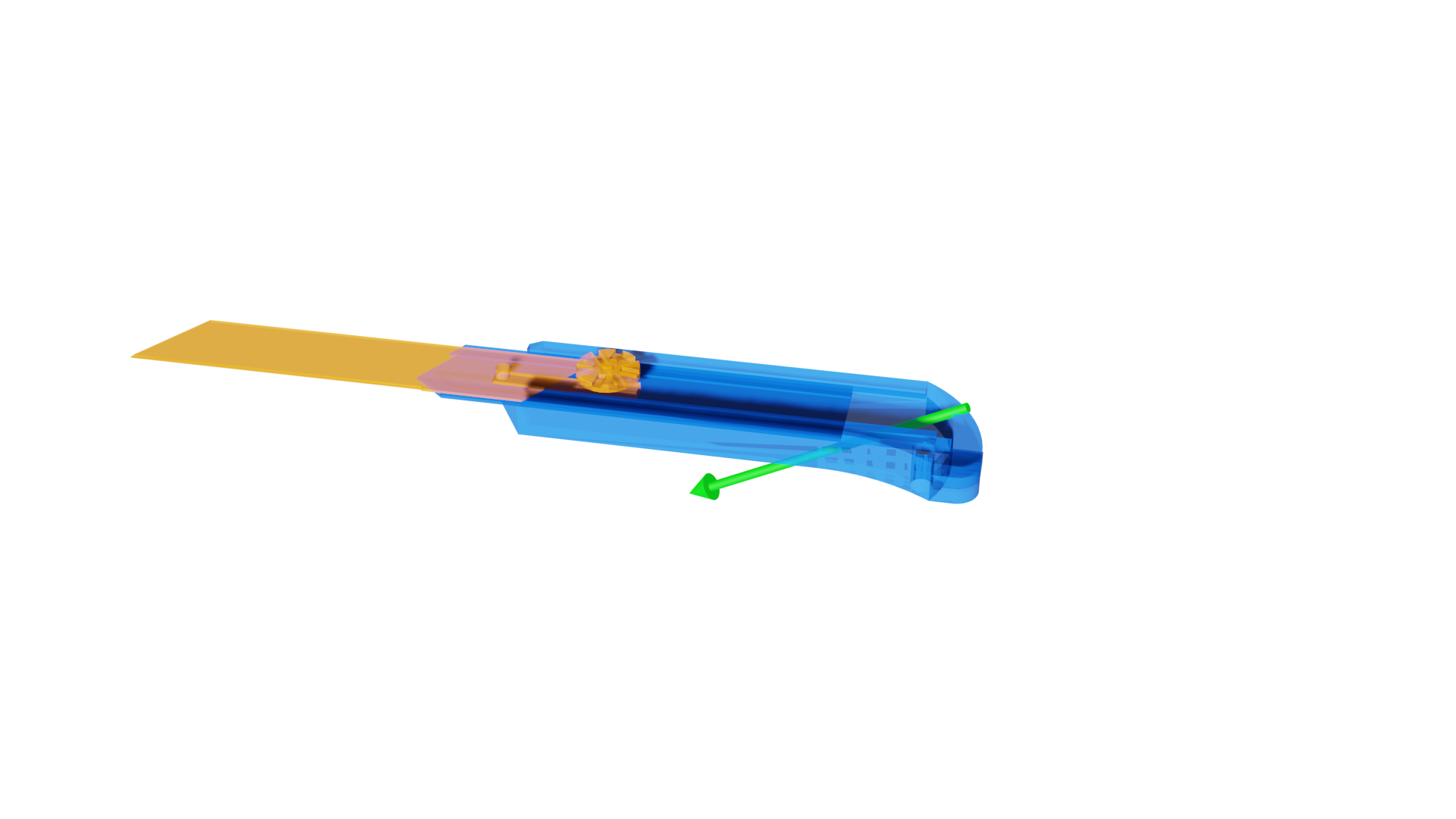} &
    \includegraphics[width=0.33\linewidth]{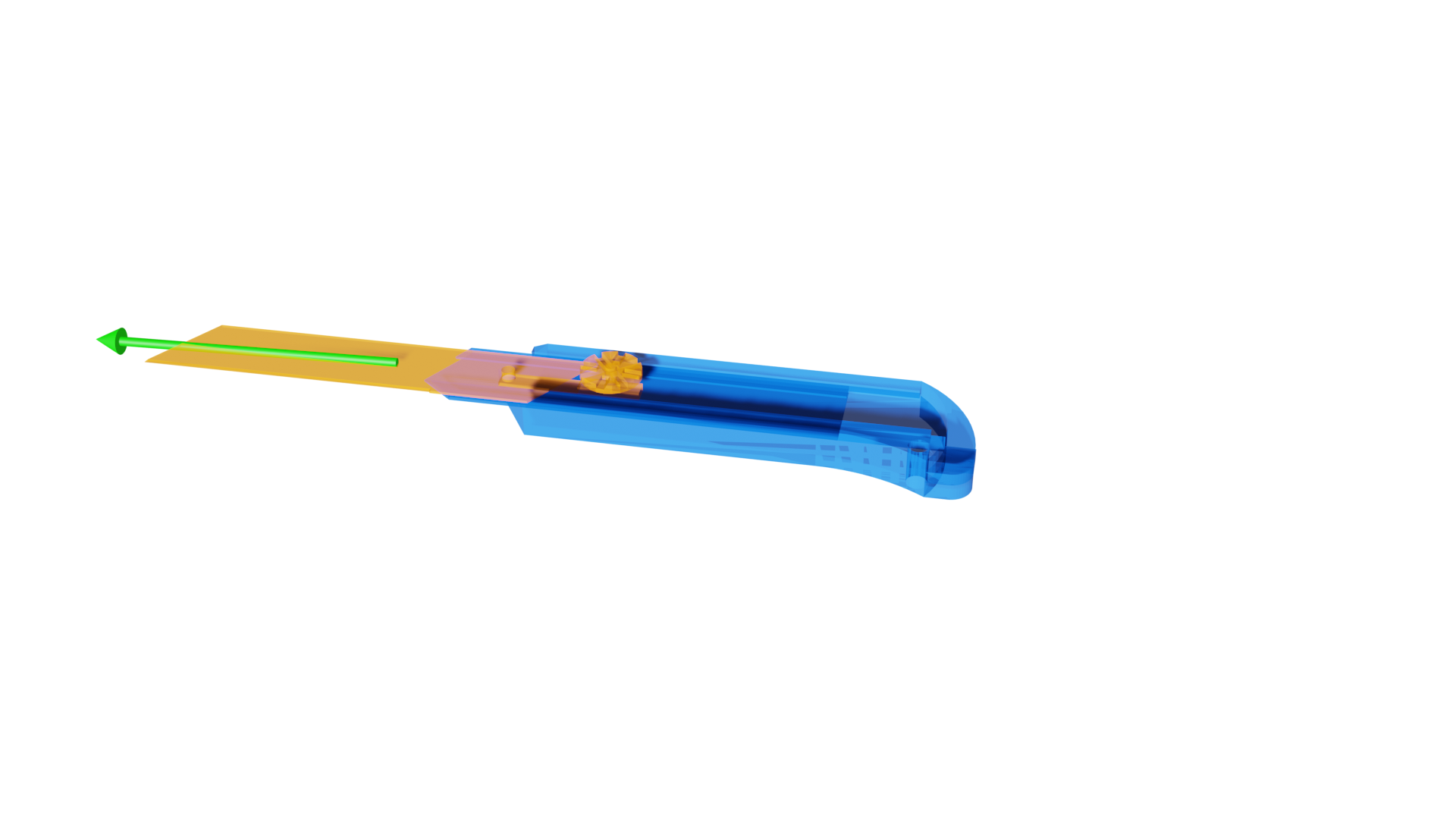} &
    \includegraphics[width=0.33\linewidth]{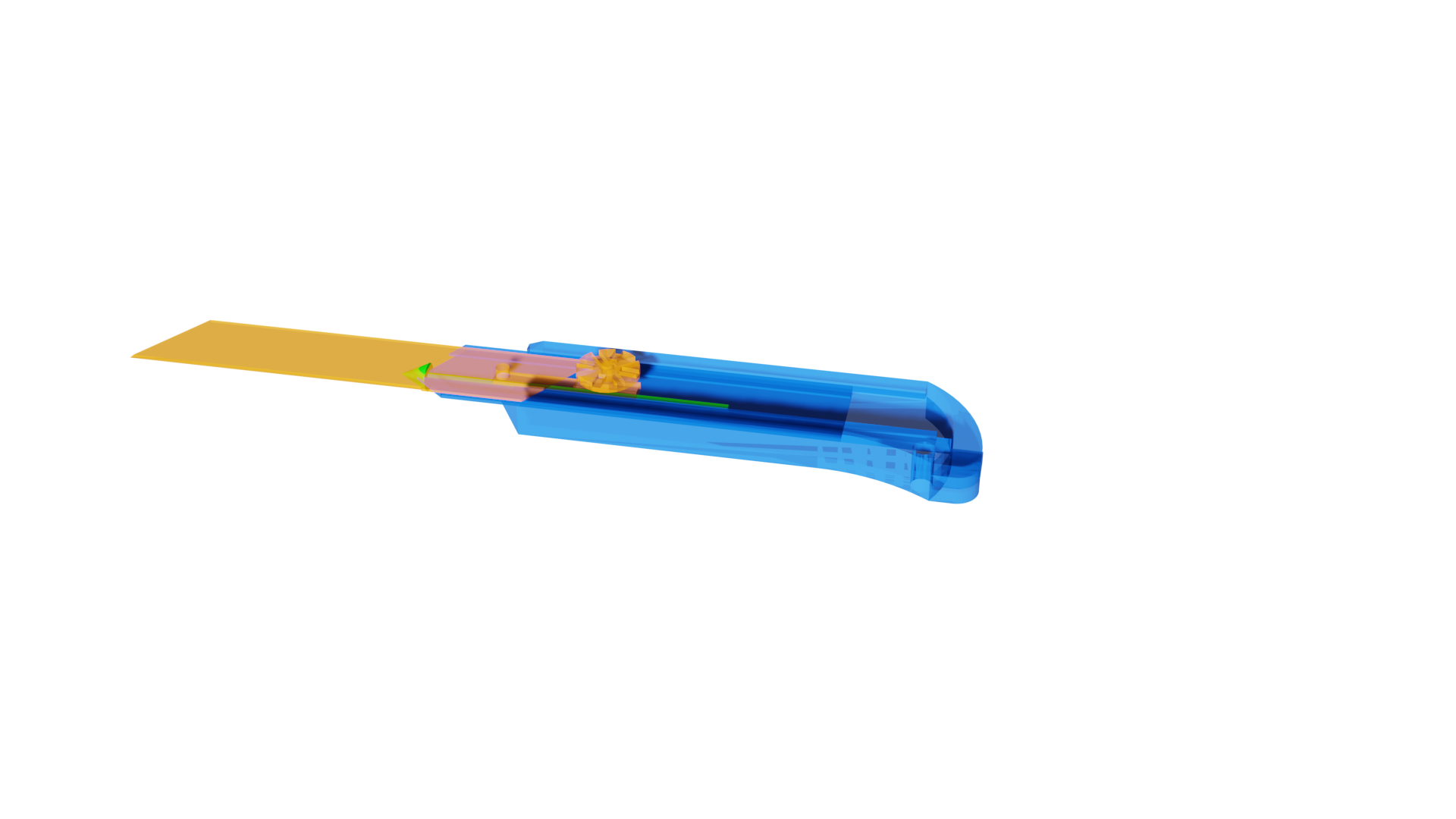}
    \\
    \includegraphics[width=0.33\linewidth]{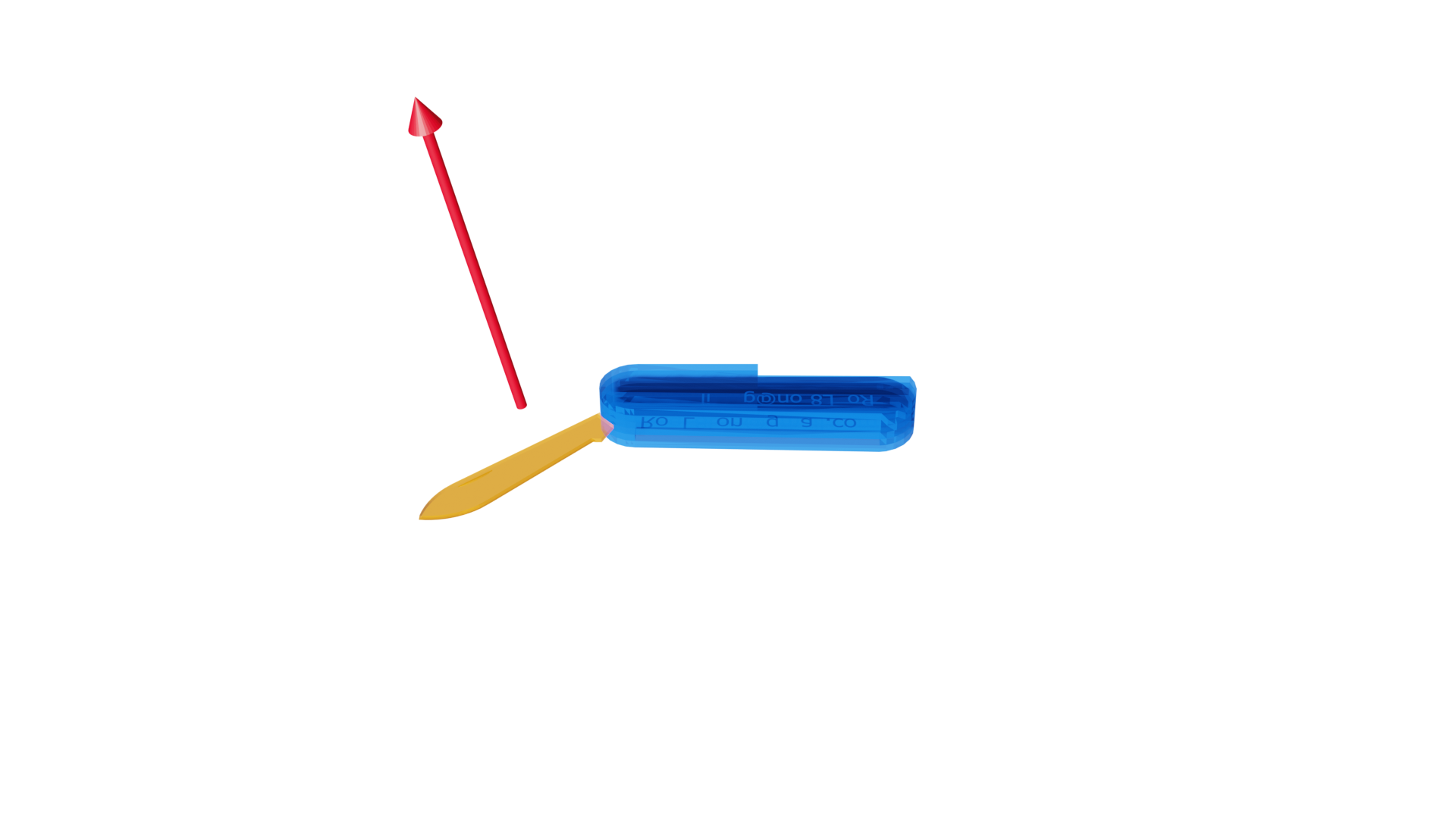} &
    \includegraphics[width=0.33\linewidth]{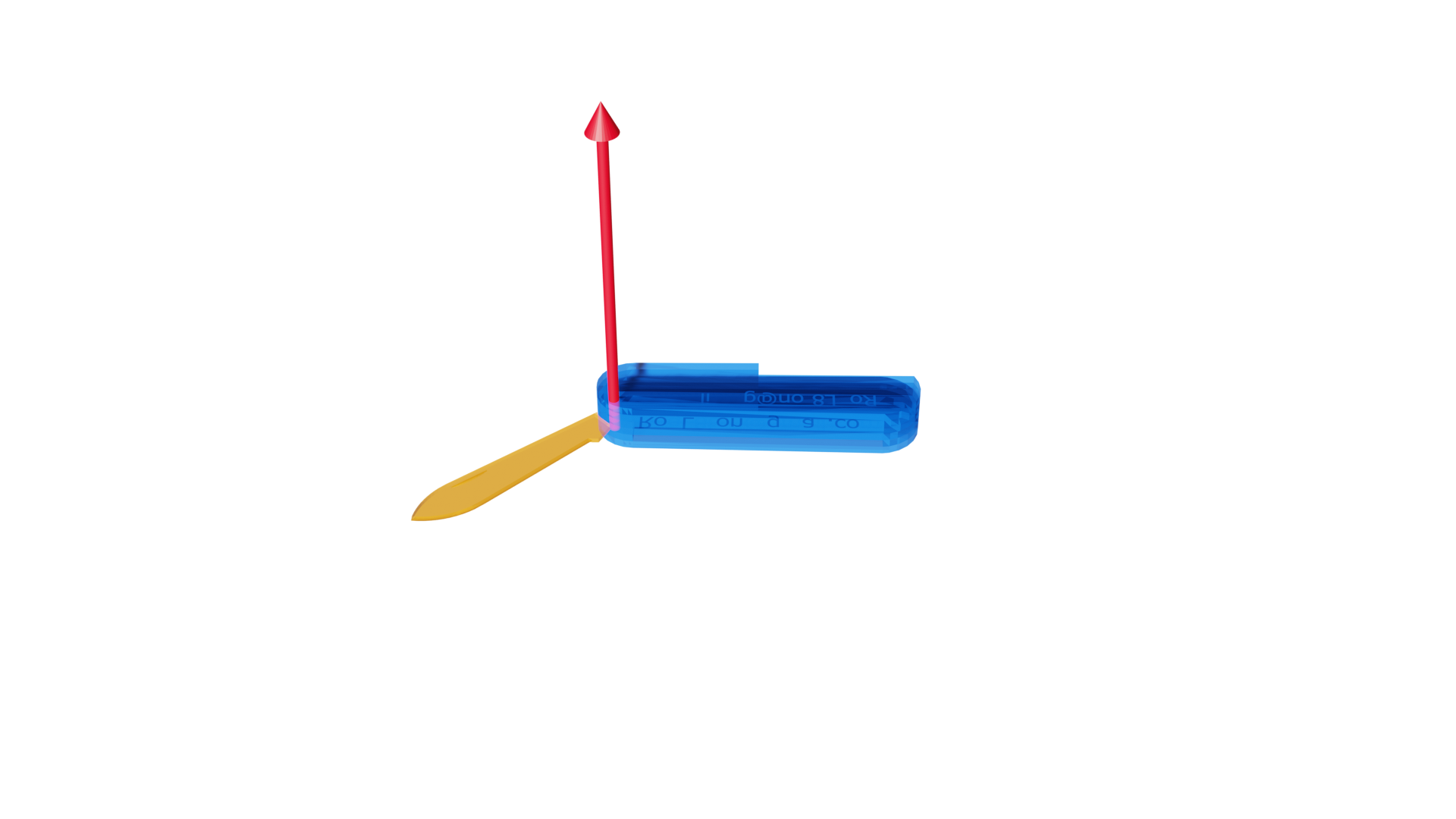} &
    \includegraphics[width=0.33\linewidth]{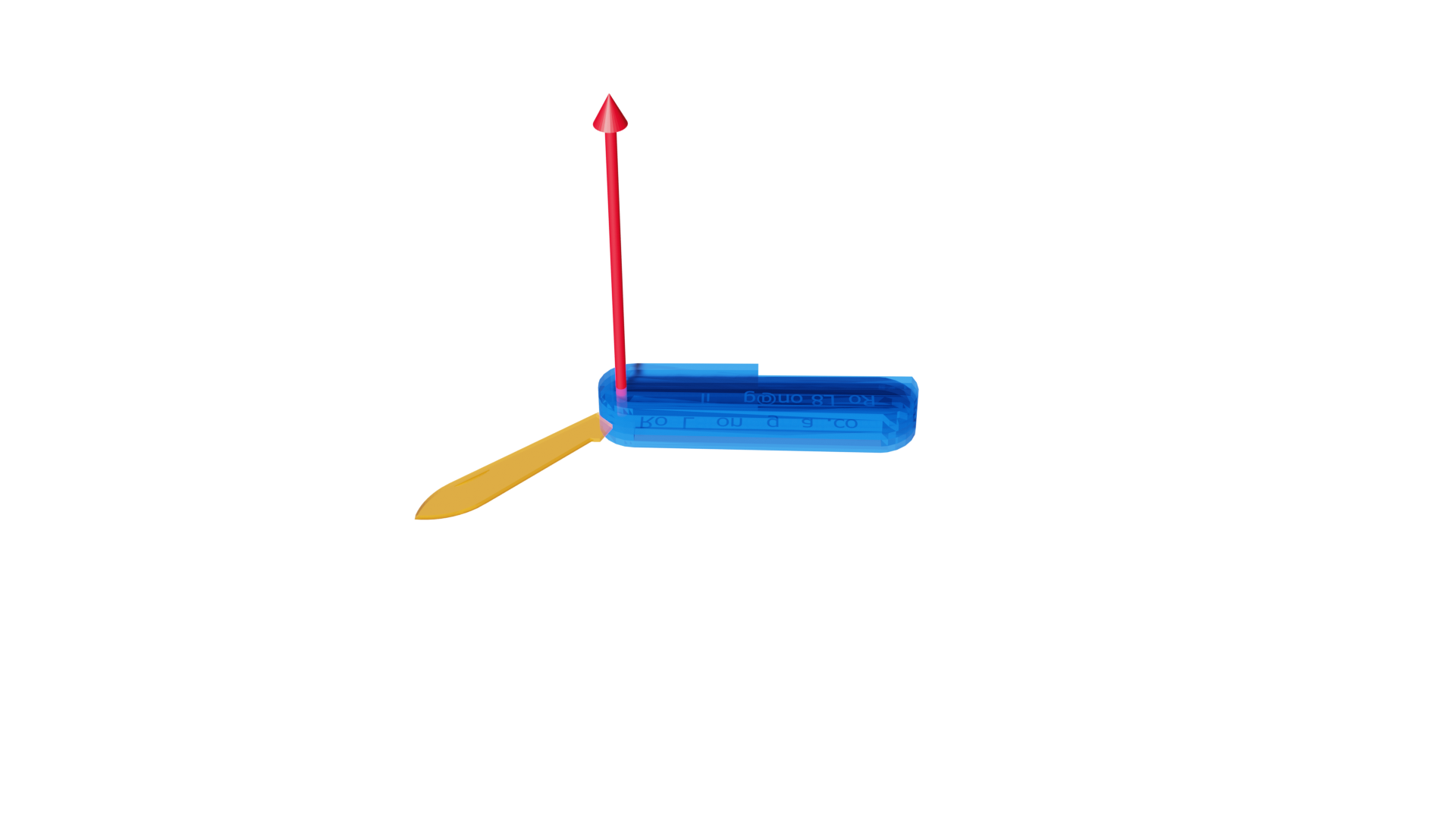}
    \\
    \end{tabular}
    \caption{
    Additional qualitative comparison of our method with the supervised BaseNet baseline
    }
    \label{figure:qualitative_comparison2}
\end{figure*}

\begin{figure*}[ht!]
    \centering
    \setlength{\tabcolsep}{1pt}
    \begin{tabular}{ccc}
    \textbf{BaseNet} & \textbf{Ours} & \textbf{GT}
        \\
    \includegraphics[width=0.33\linewidth]{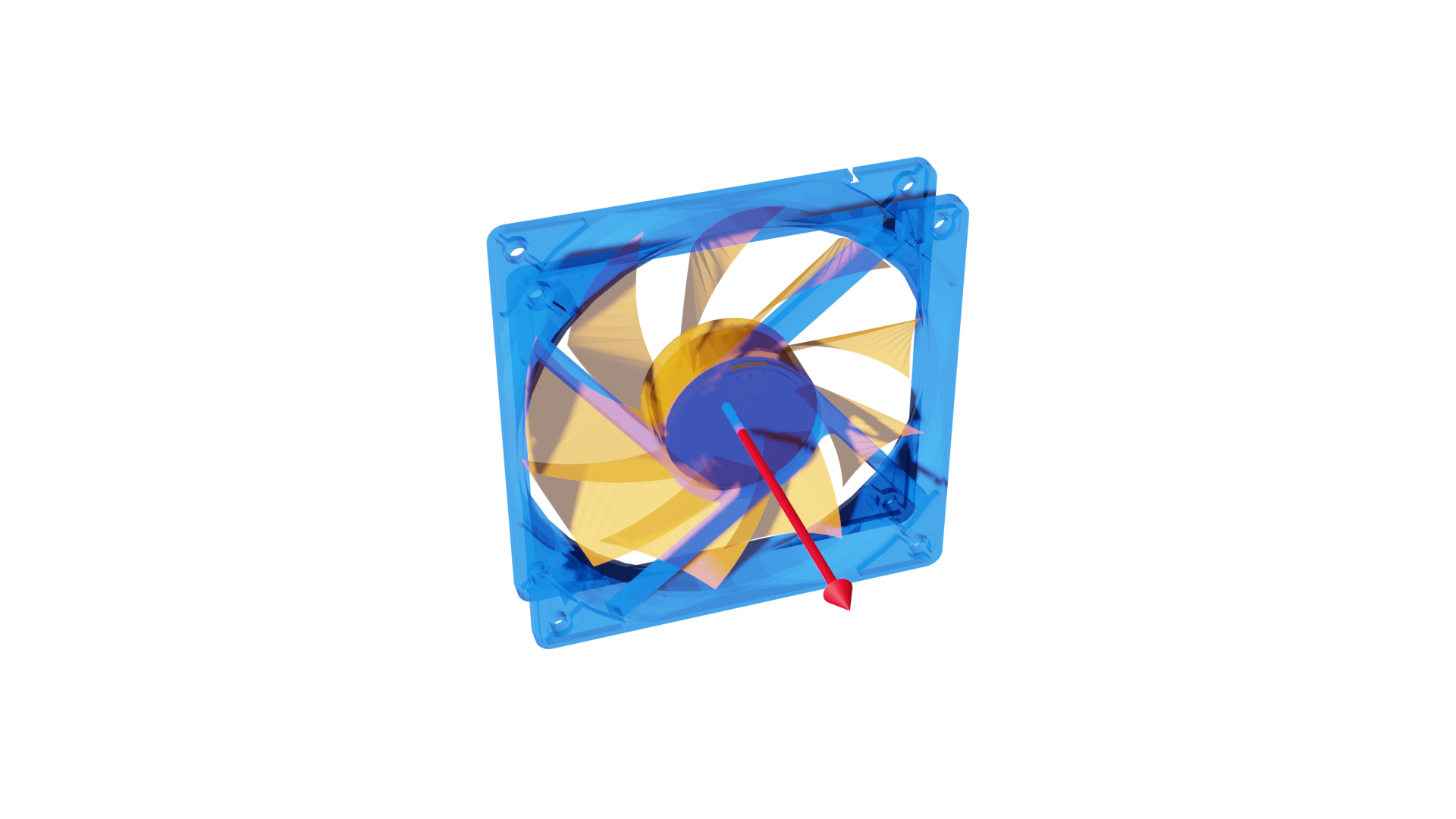} &
    \includegraphics[width=0.33\linewidth]{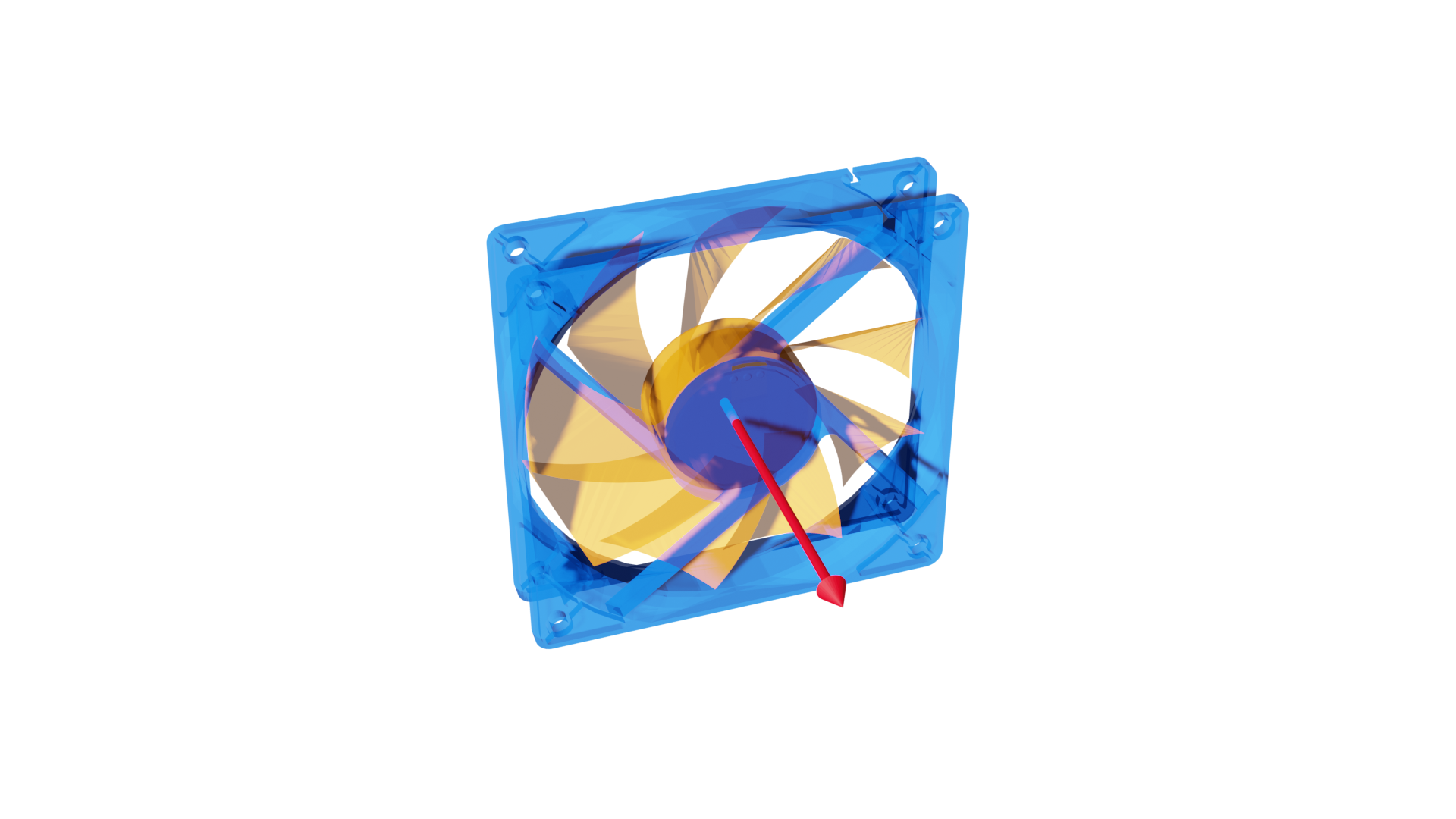} &
    \includegraphics[width=0.33\linewidth]{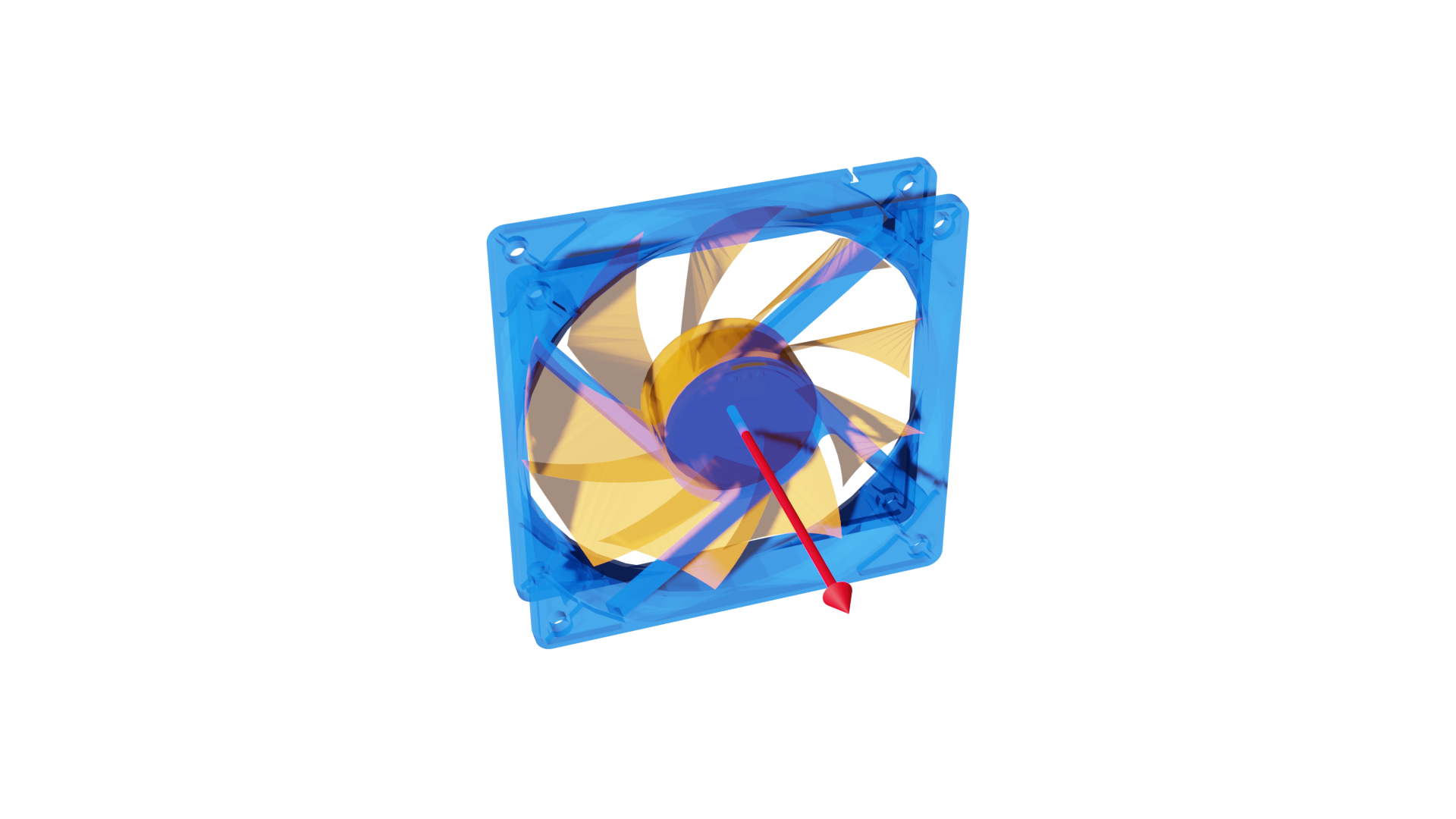}
    \\
    \includegraphics[width=0.33\linewidth]{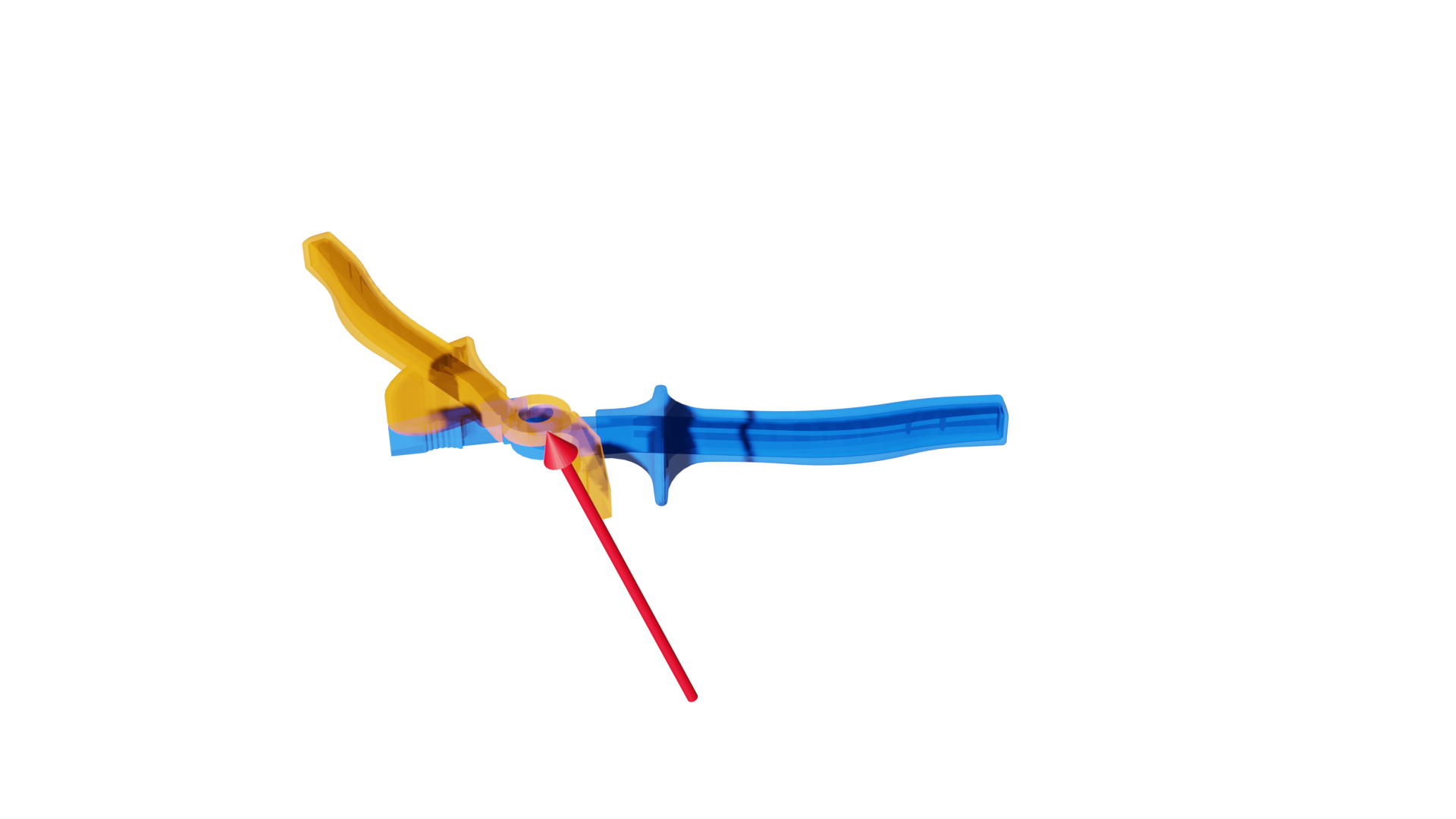} &
    \includegraphics[width=0.33\linewidth]{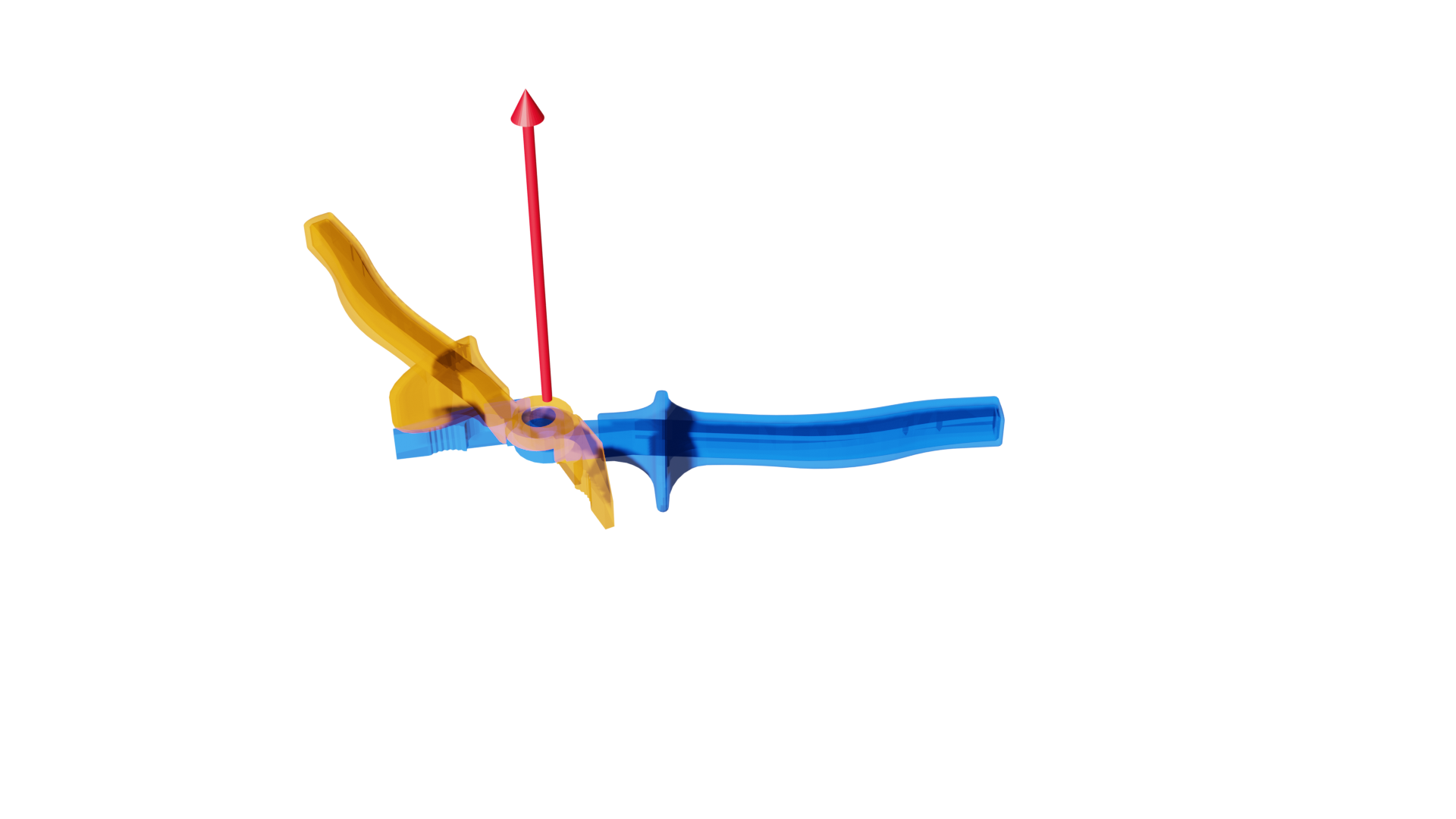} &
    \includegraphics[width=0.33\linewidth]{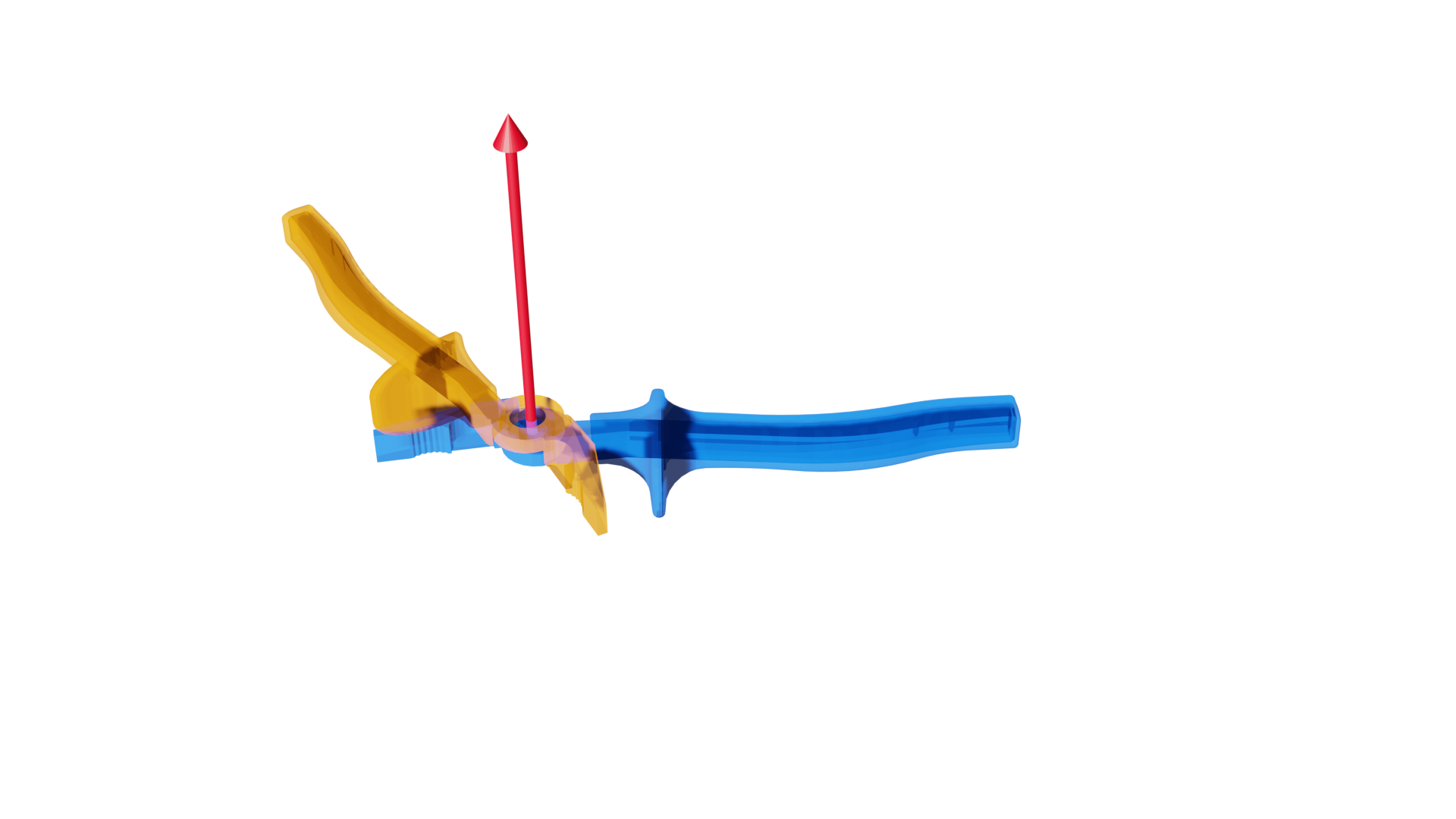}
    \\
    \includegraphics[width=0.33\linewidth]{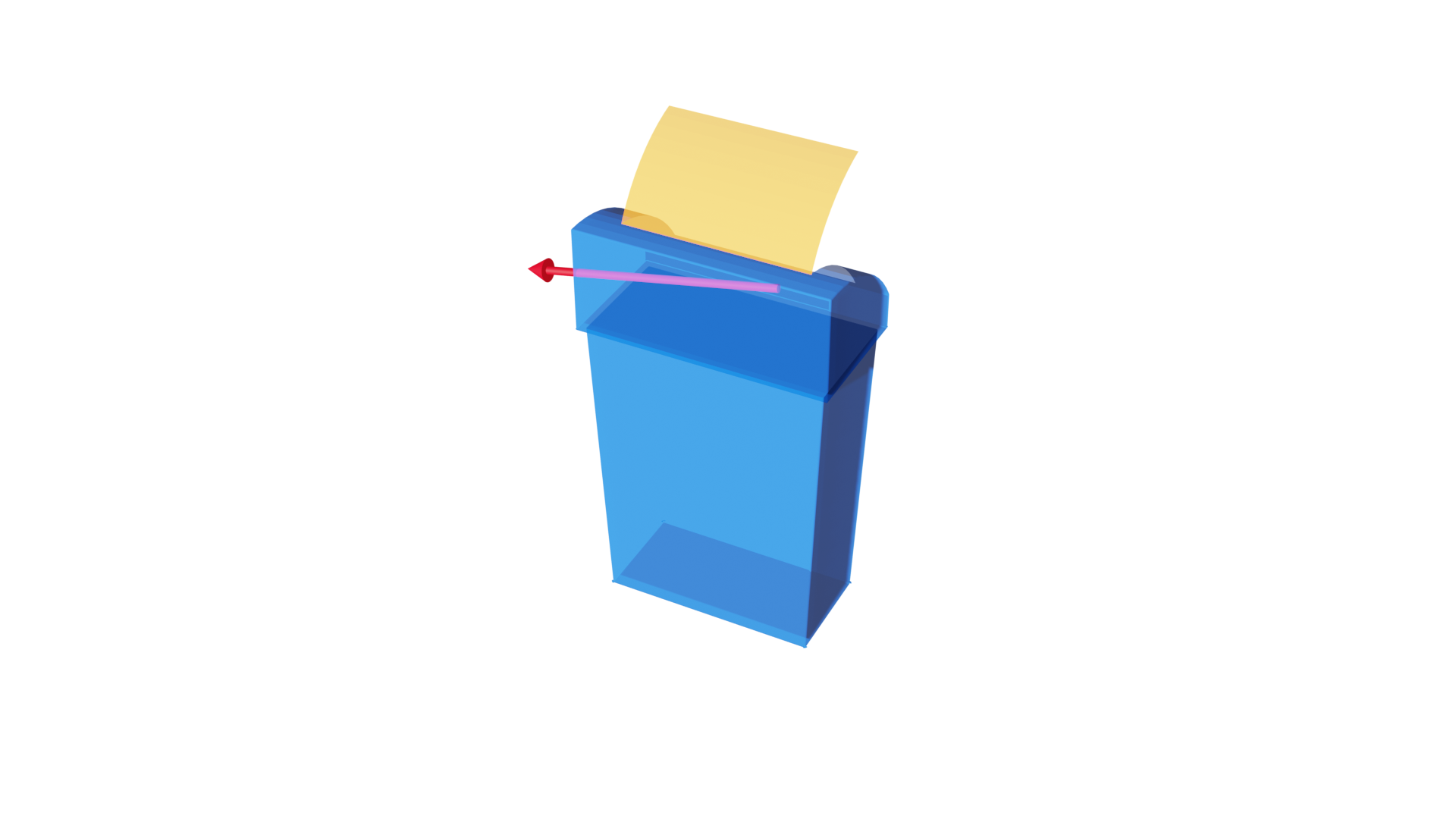} &
    \includegraphics[width=0.33\linewidth]{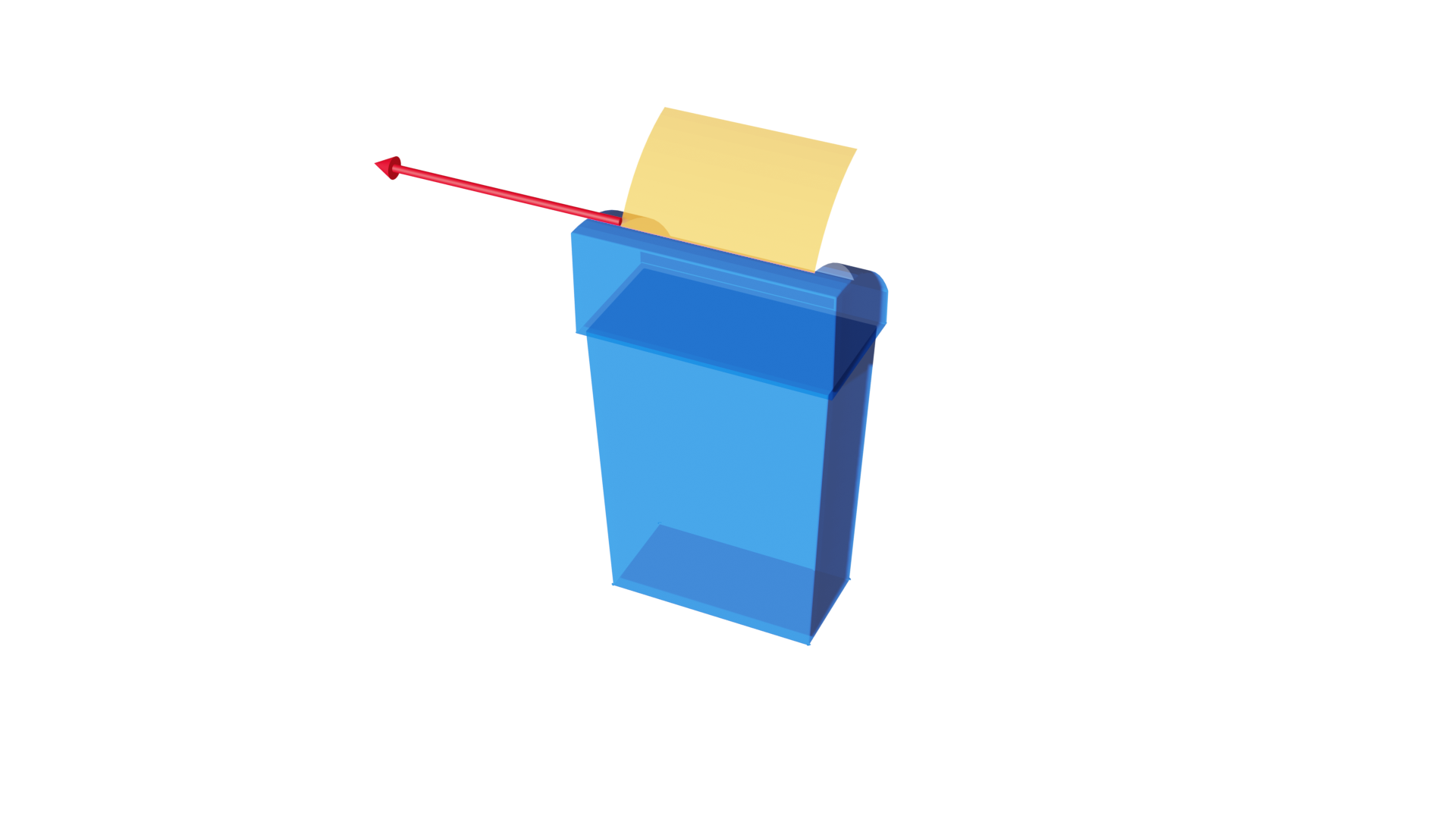} &
    \includegraphics[width=0.33\linewidth]{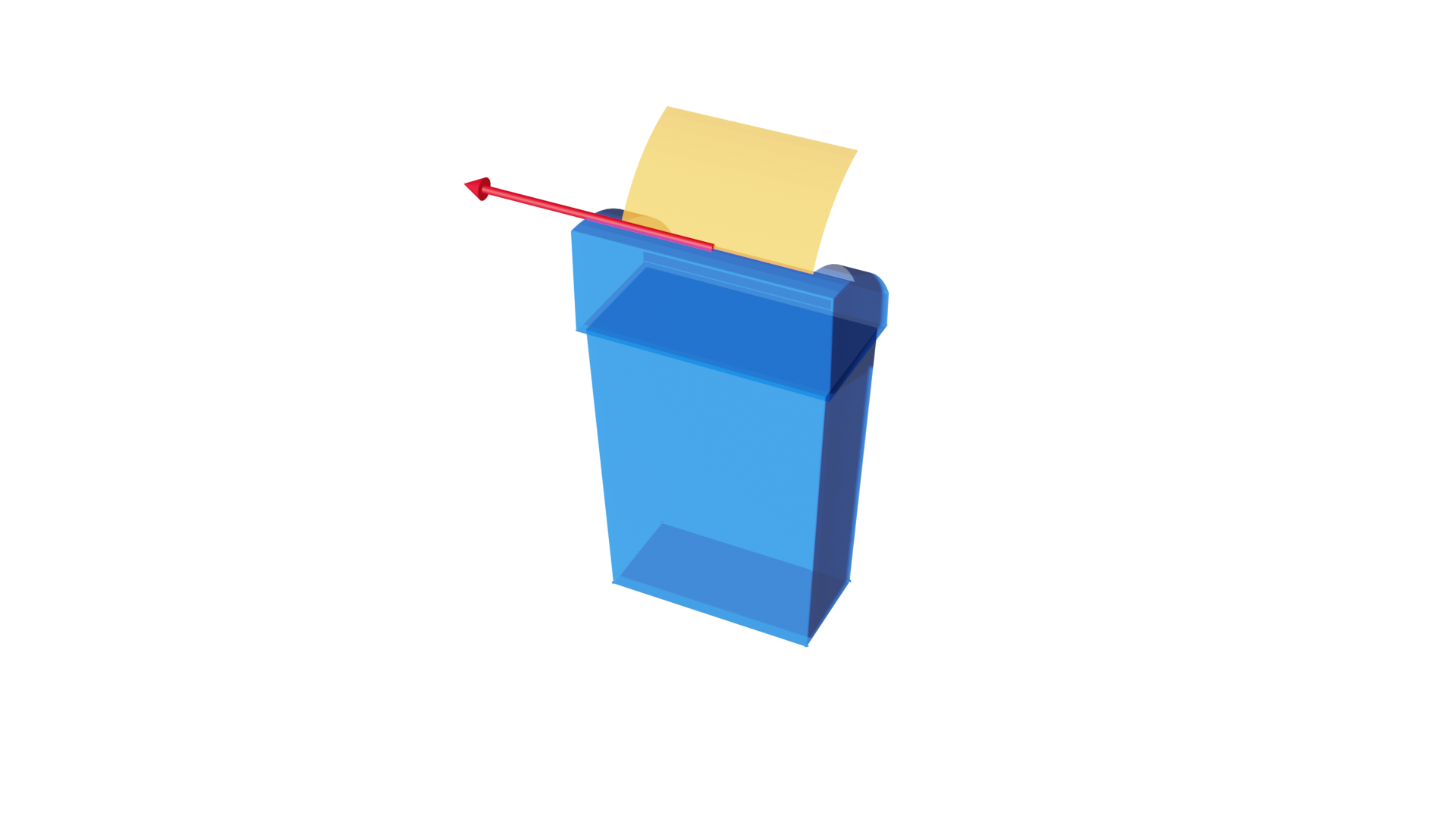}
    \\
    \includegraphics[width=0.33\linewidth]{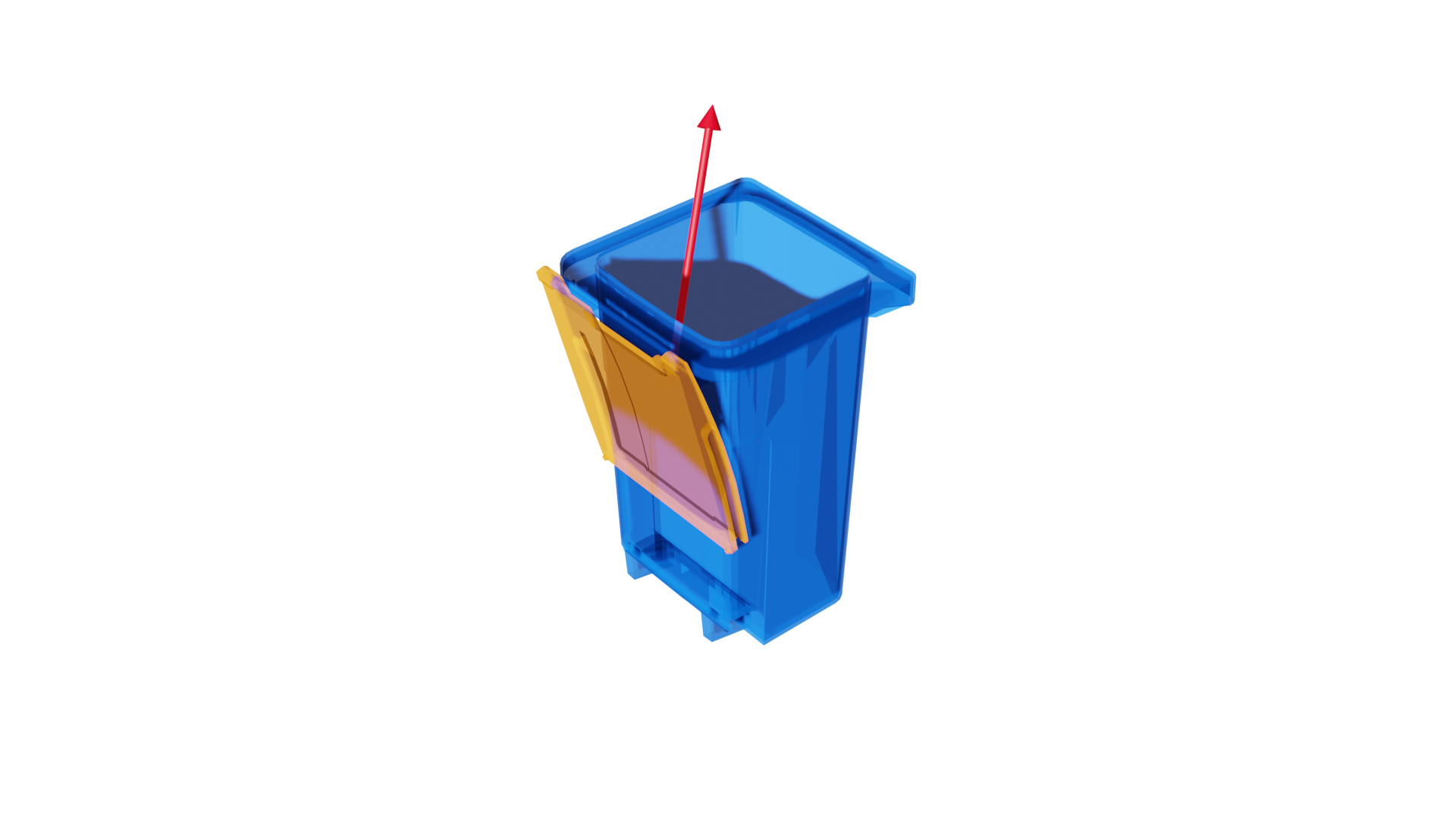} &
    \includegraphics[width=0.33\linewidth]{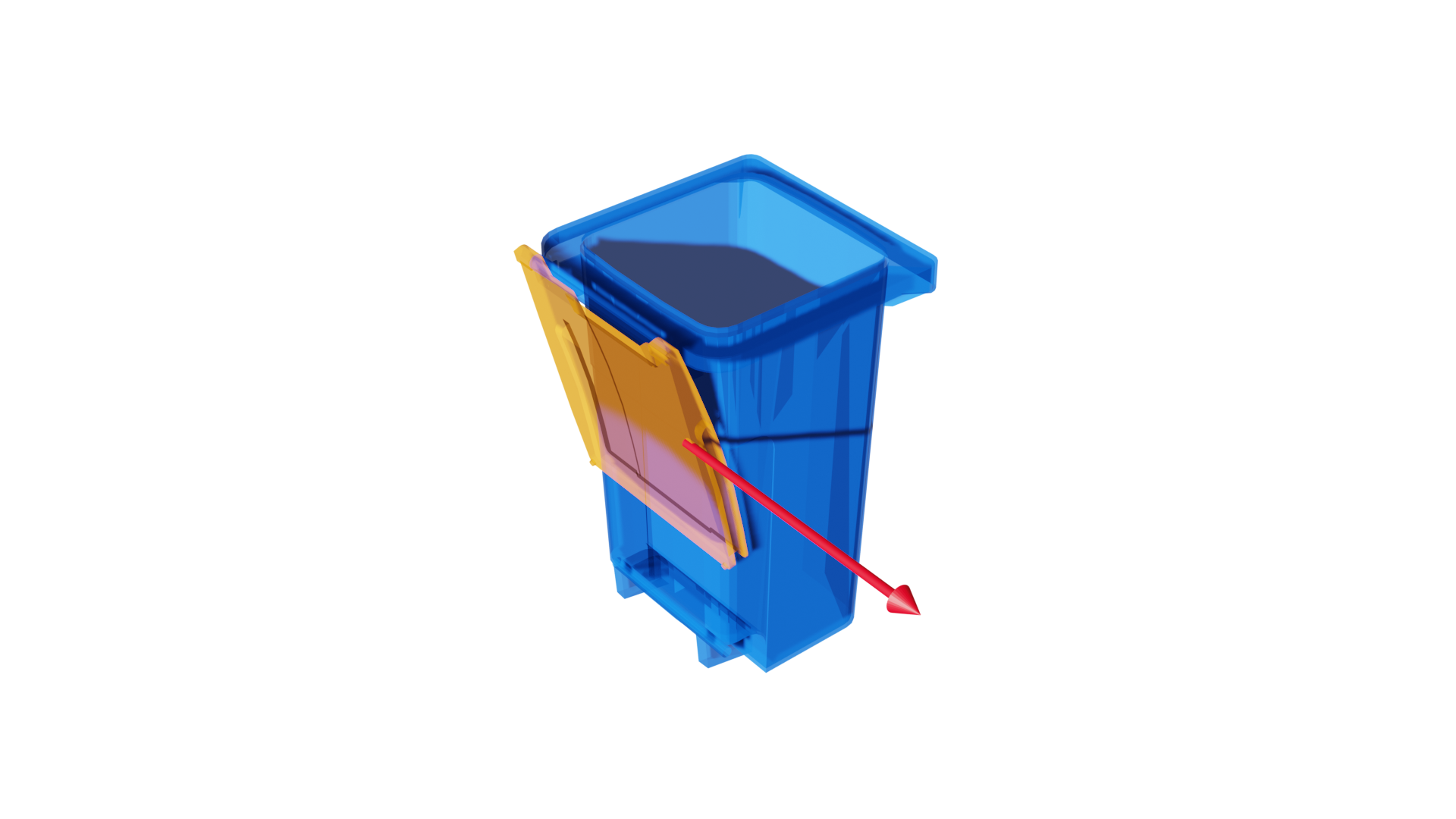} &
    \includegraphics[width=0.33\linewidth]{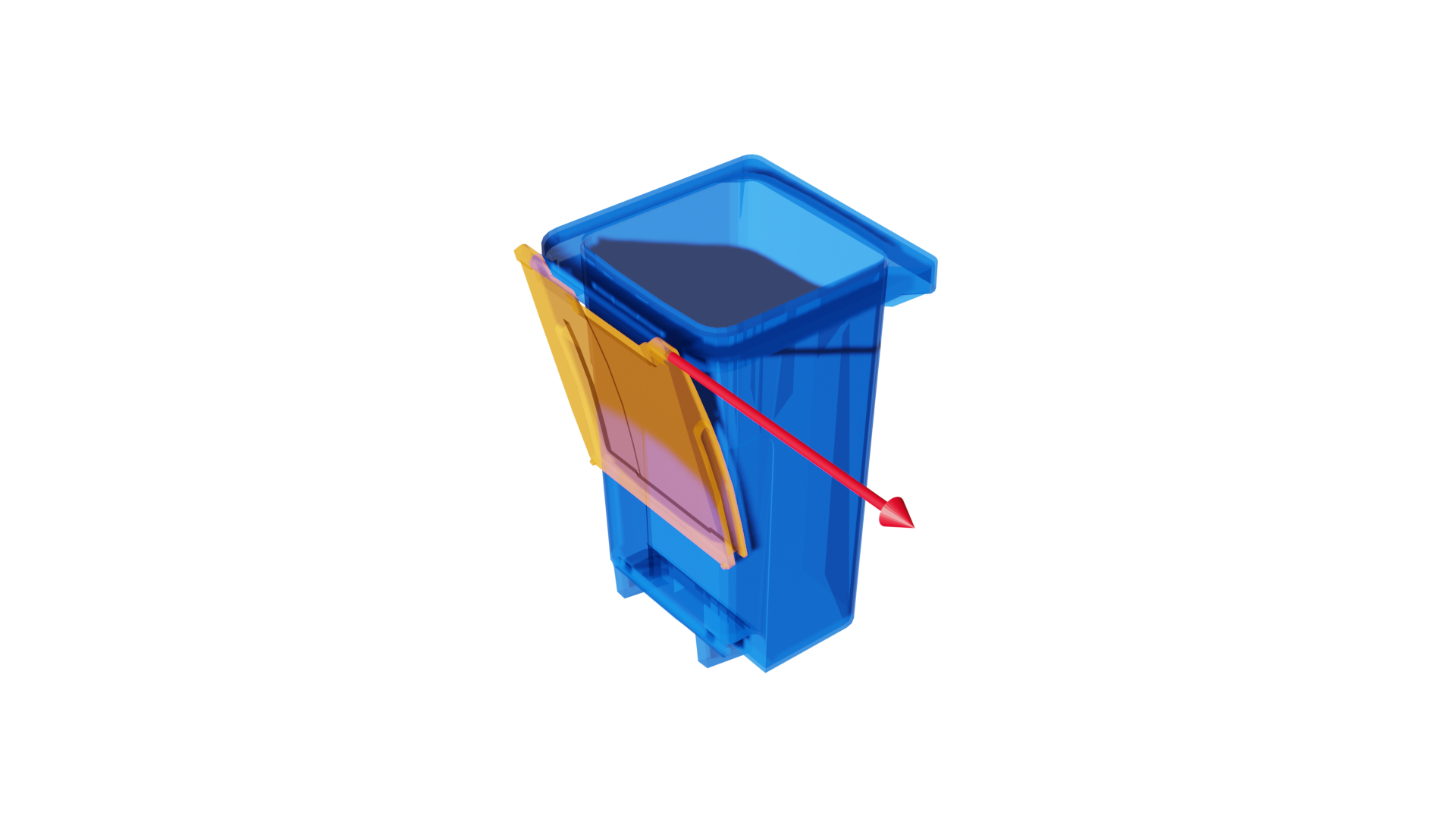}
    \\
    \includegraphics[width=0.33\linewidth]{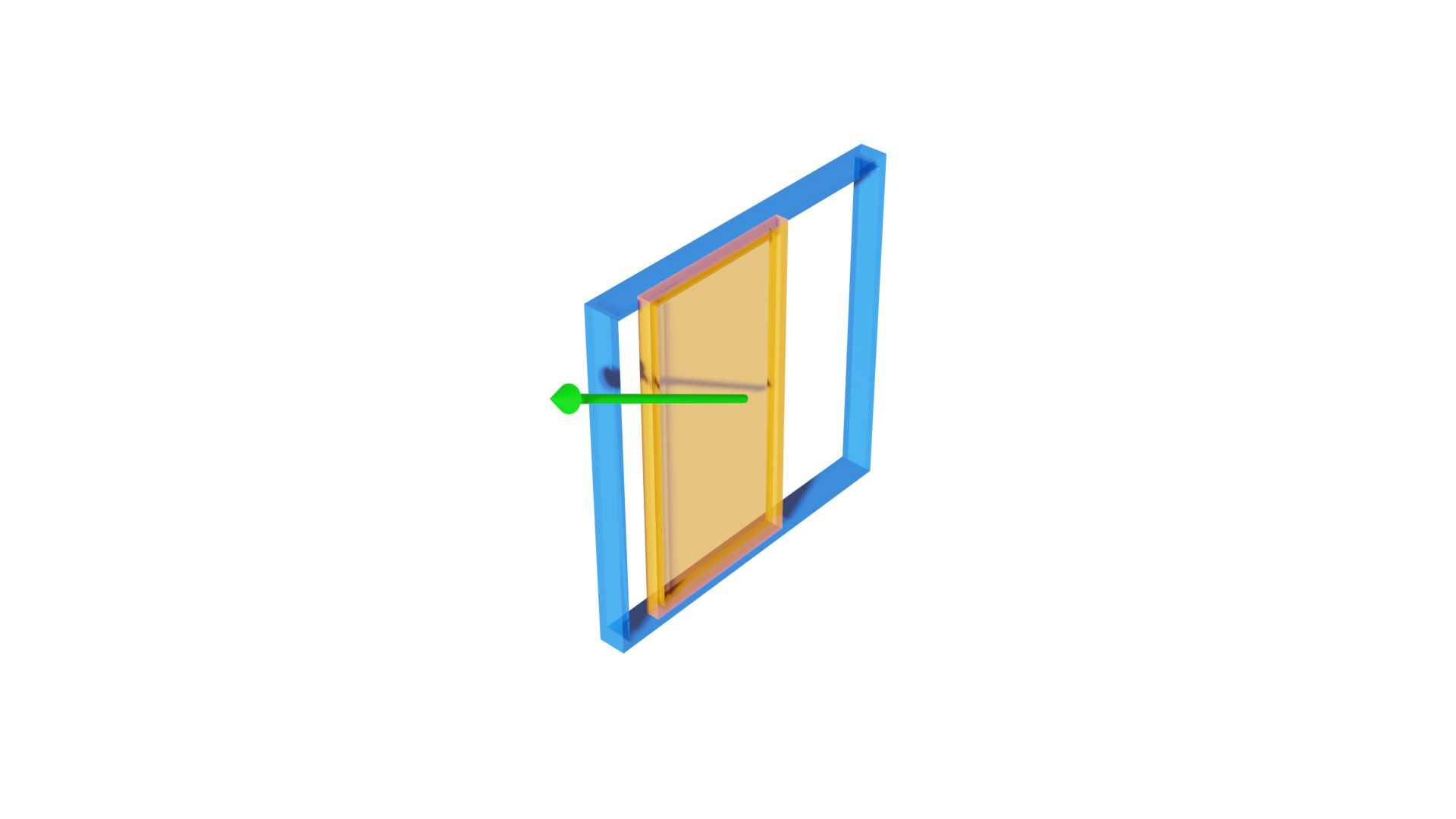} &
    \includegraphics[width=0.33\linewidth]{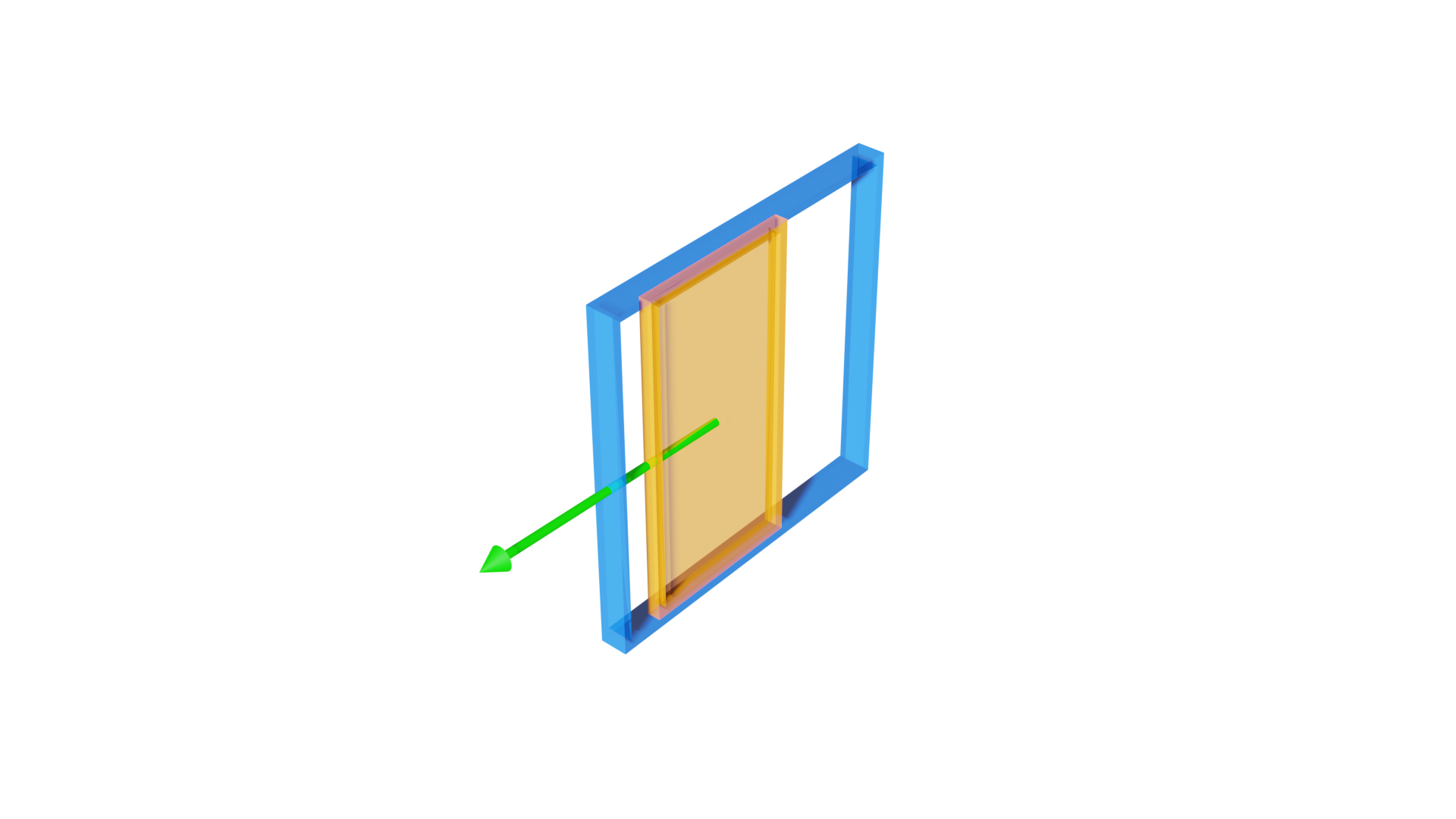} &
    \includegraphics[width=0.33\linewidth]{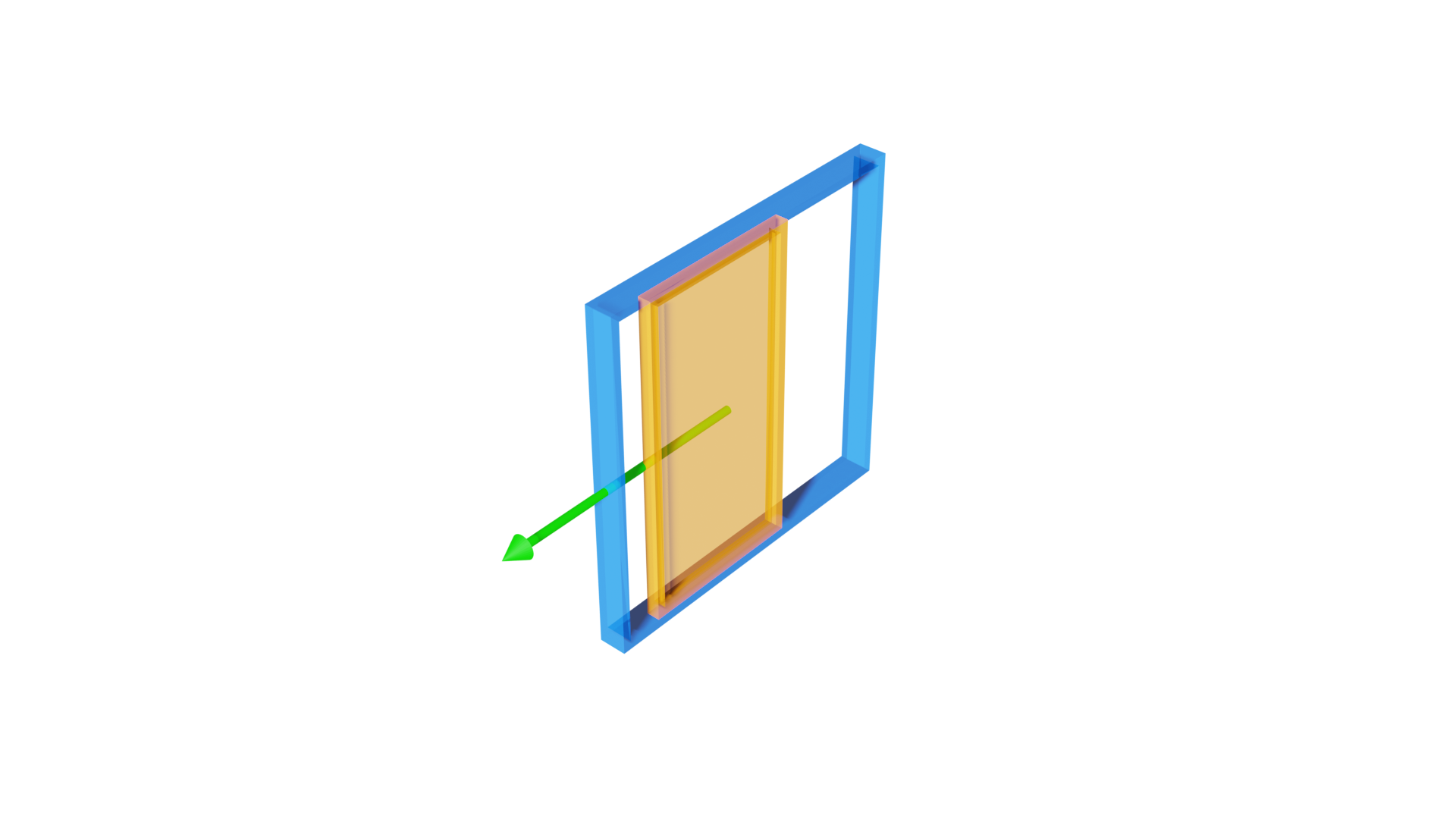}
    \\
    \end{tabular}
    \caption{
    Additional qualitative comparison of our method with the supervised BaseNet baseline
    }
    \label{figure:qualitative_comparison3}
\end{figure*}

\begin{figure*}[ht!]
    \centering
    \setlength{\tabcolsep}{1pt}
    \begin{tabular}{ccc}
    \textbf{BaseNet} & \textbf{Ours} & \textbf{GT}
        \\
    \includegraphics[width=0.33\linewidth]{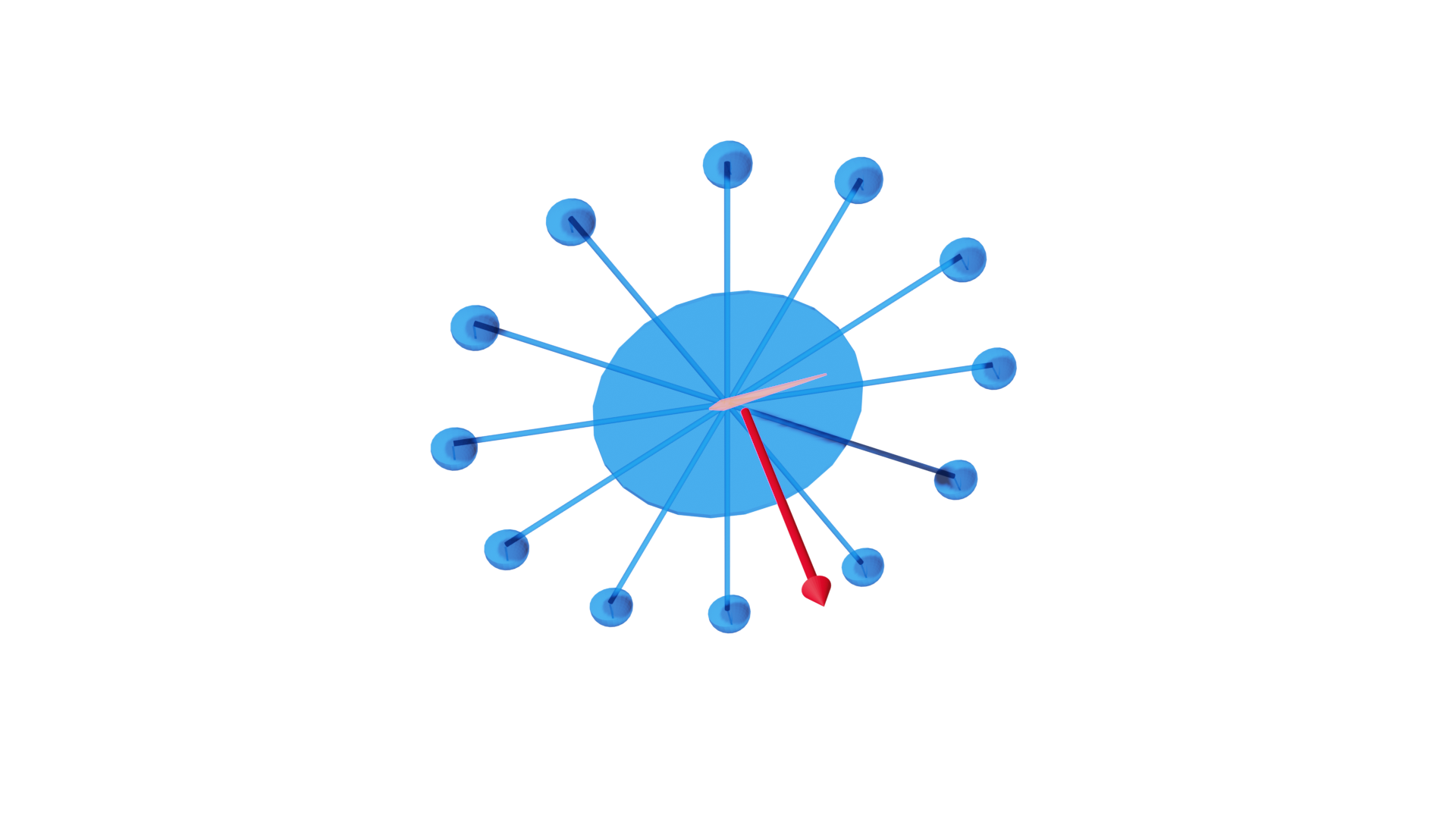} &
    \includegraphics[width=0.33\linewidth]{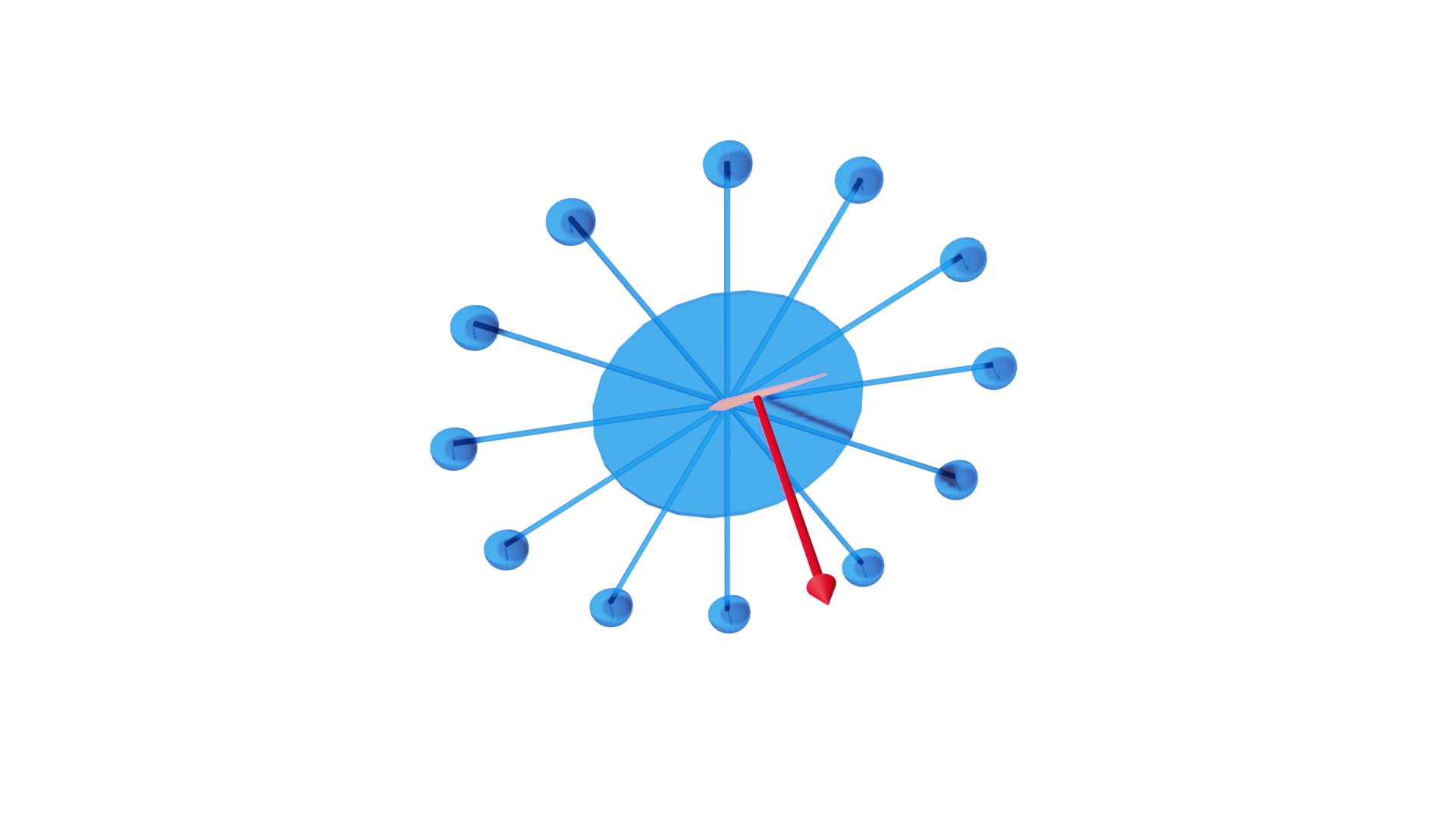} &
    \includegraphics[width=0.33\linewidth]{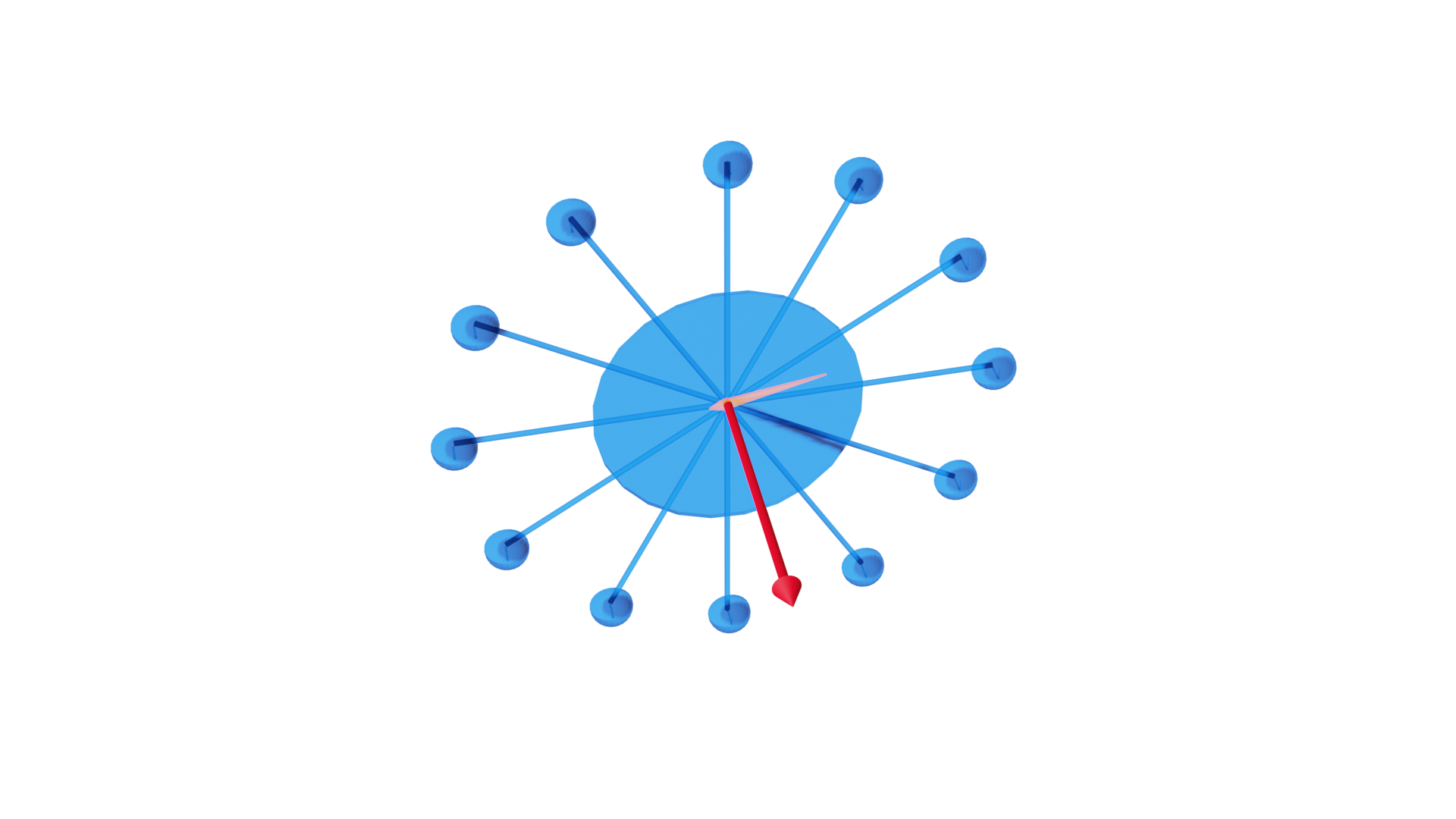}
    \\
    \includegraphics[width=0.33\linewidth]{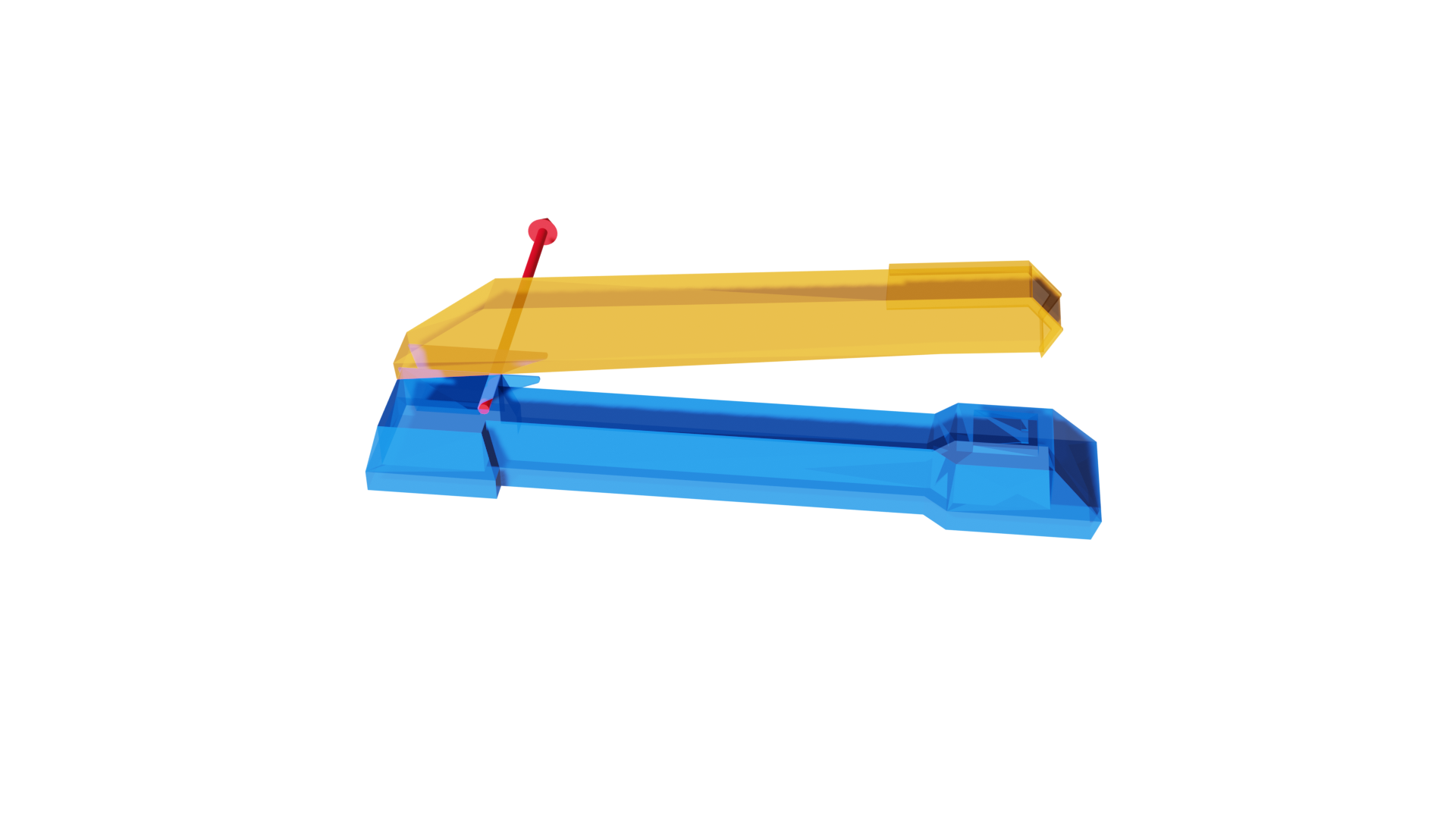} &
    \includegraphics[width=0.33\linewidth]{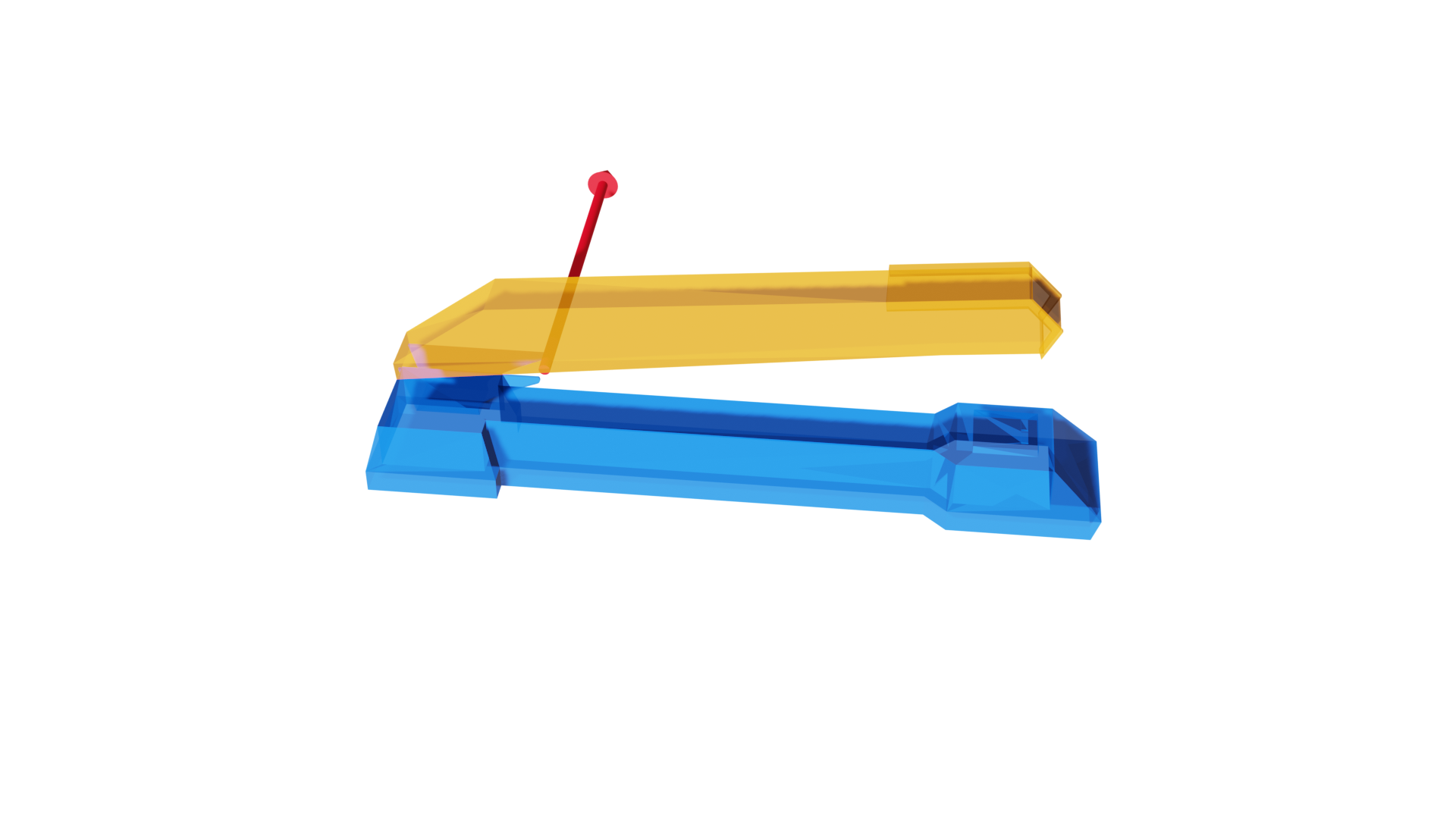} &
    \includegraphics[width=0.33\linewidth]{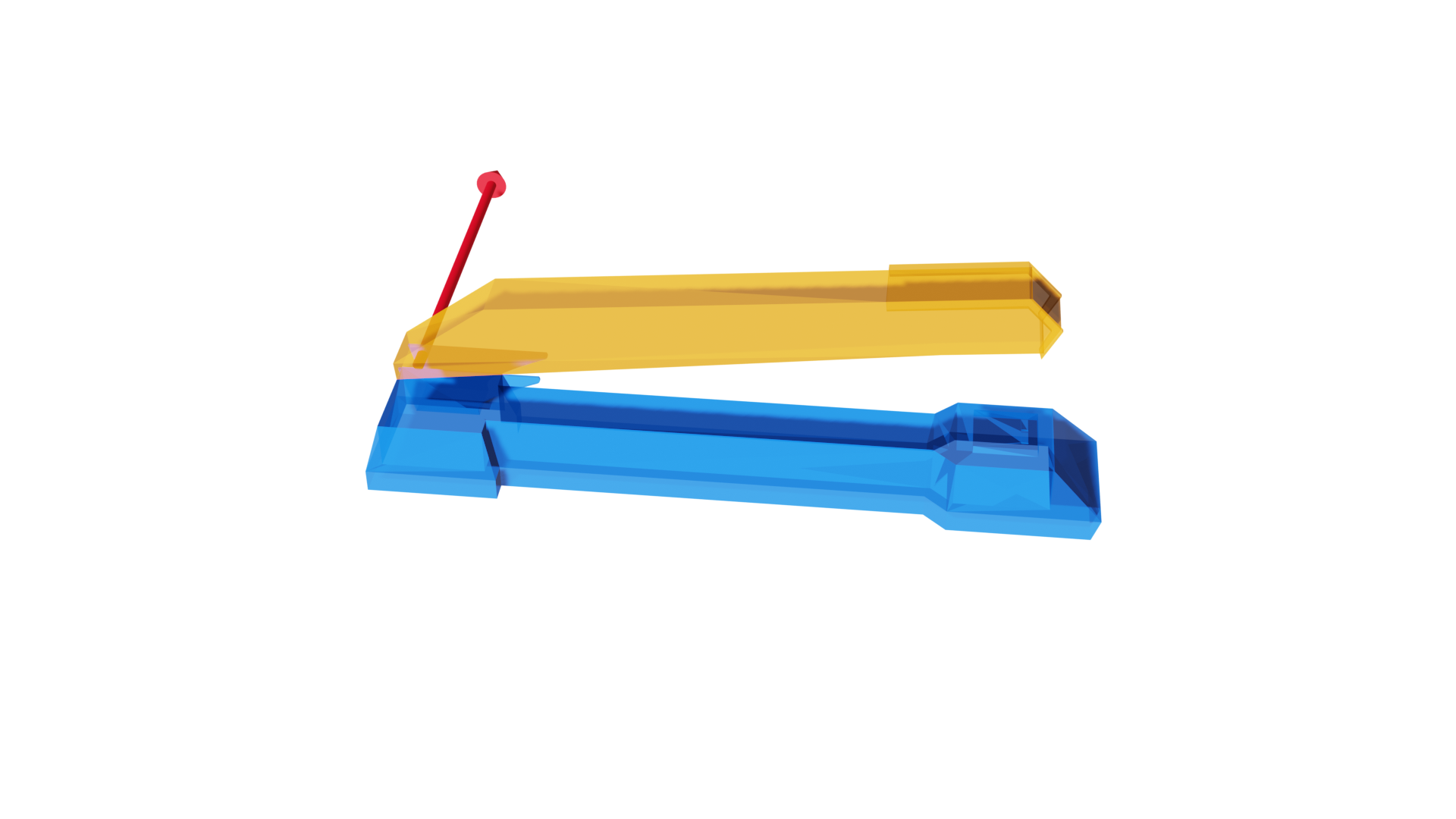}
    \\
    \includegraphics[width=0.33\linewidth]{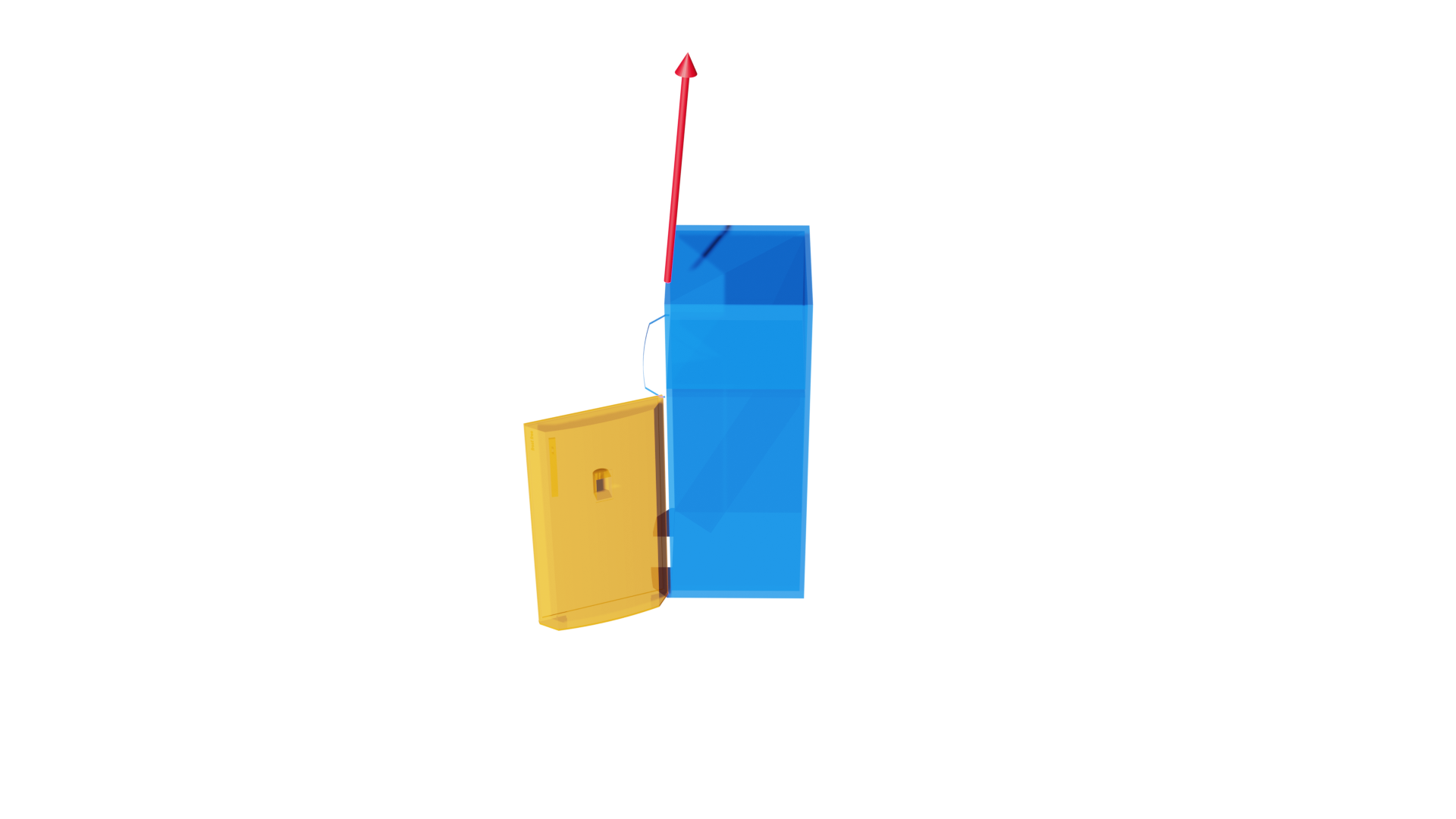} &
    \includegraphics[width=0.33\linewidth]{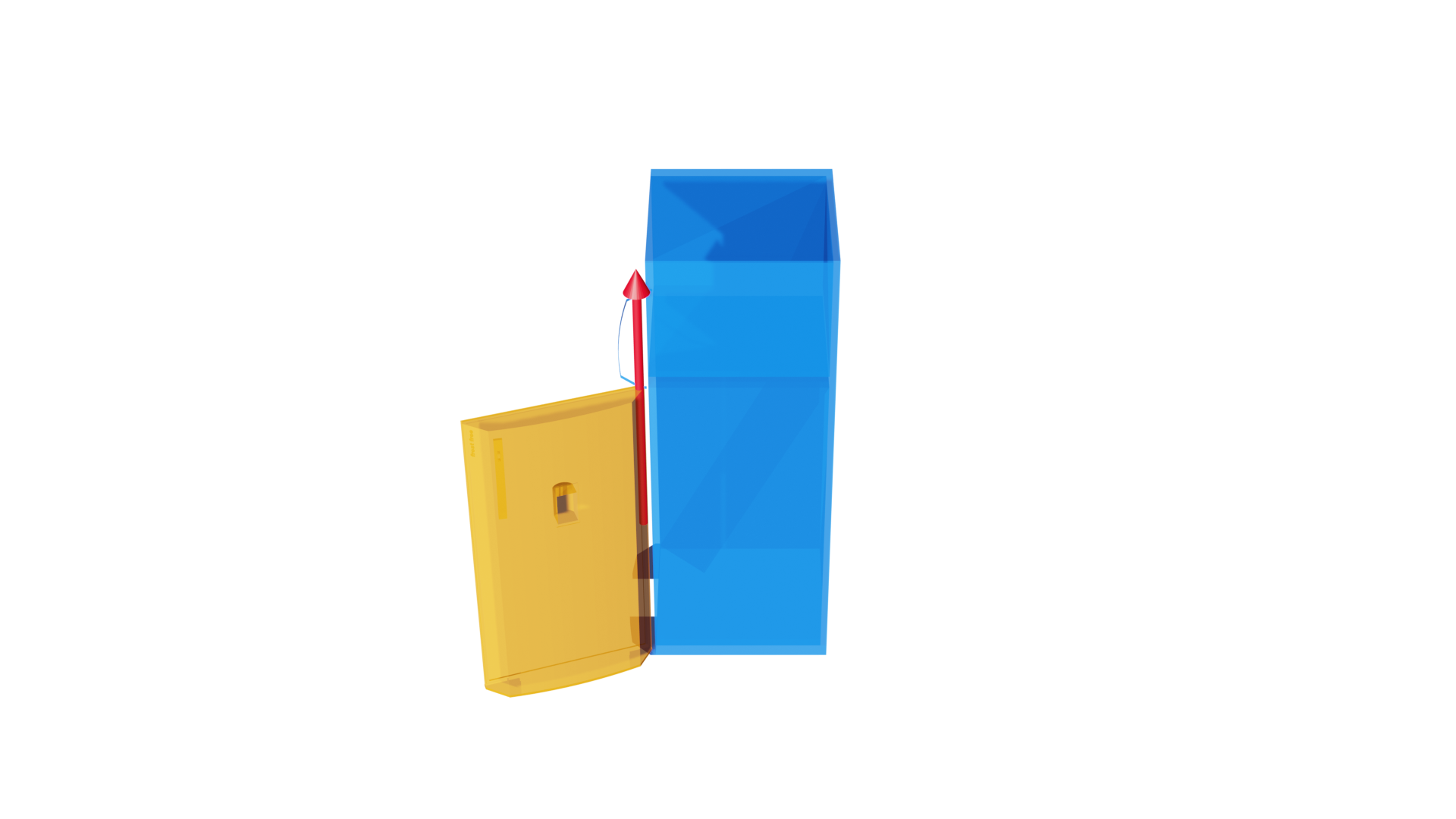} &
    \includegraphics[width=0.33\linewidth]{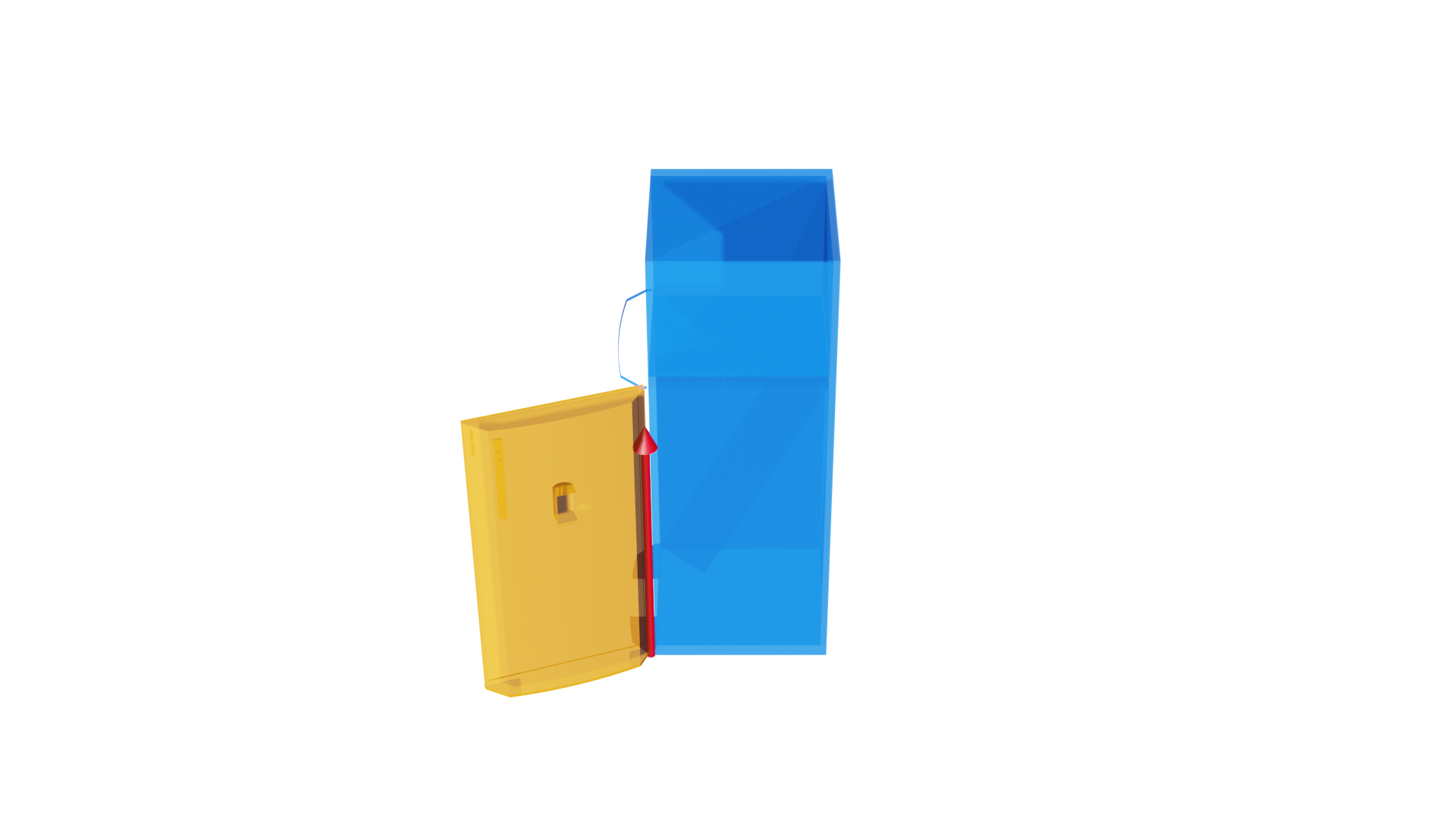}
    \\
    \includegraphics[width=0.33\linewidth]{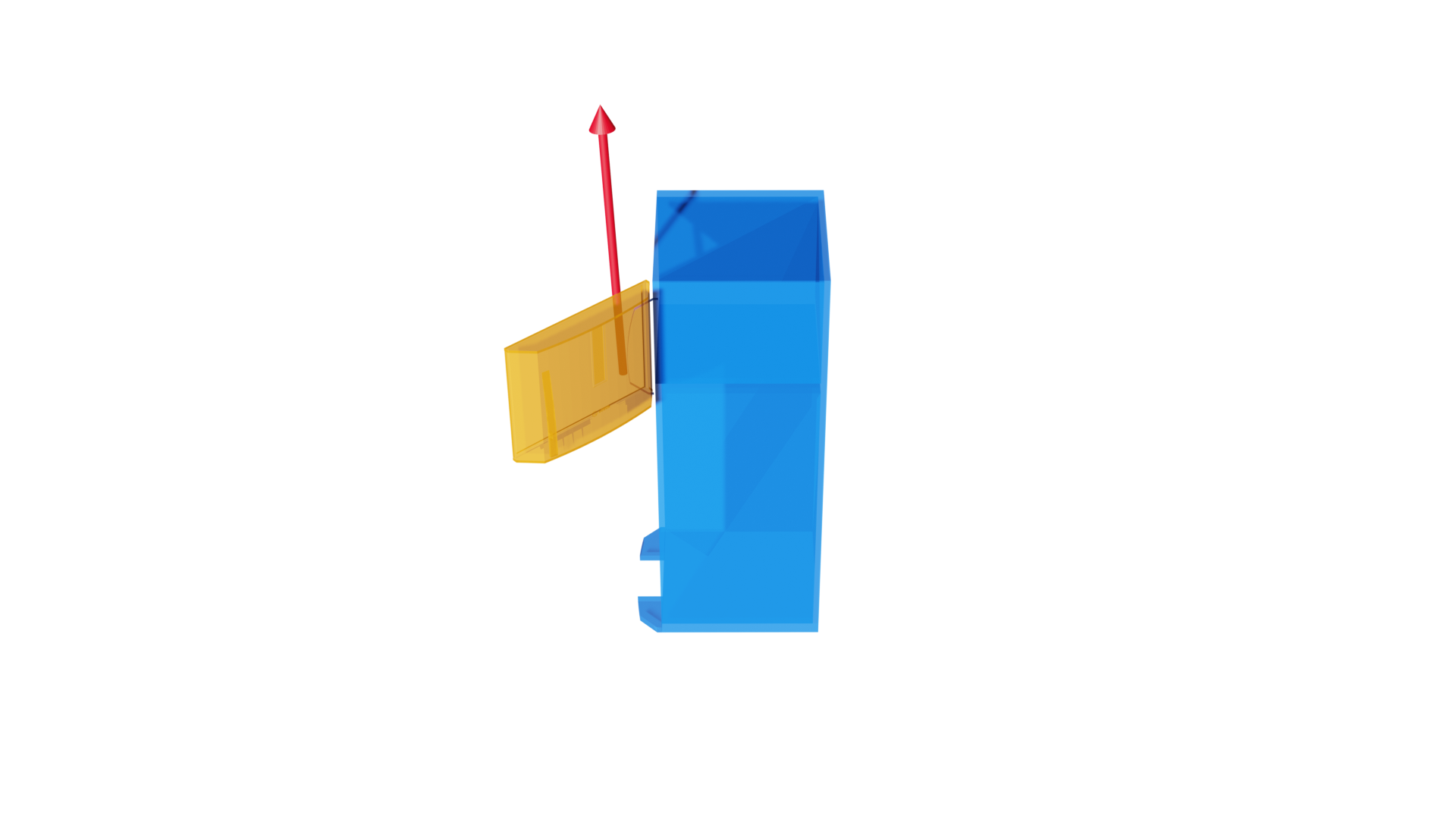} &
    \includegraphics[width=0.33\linewidth]{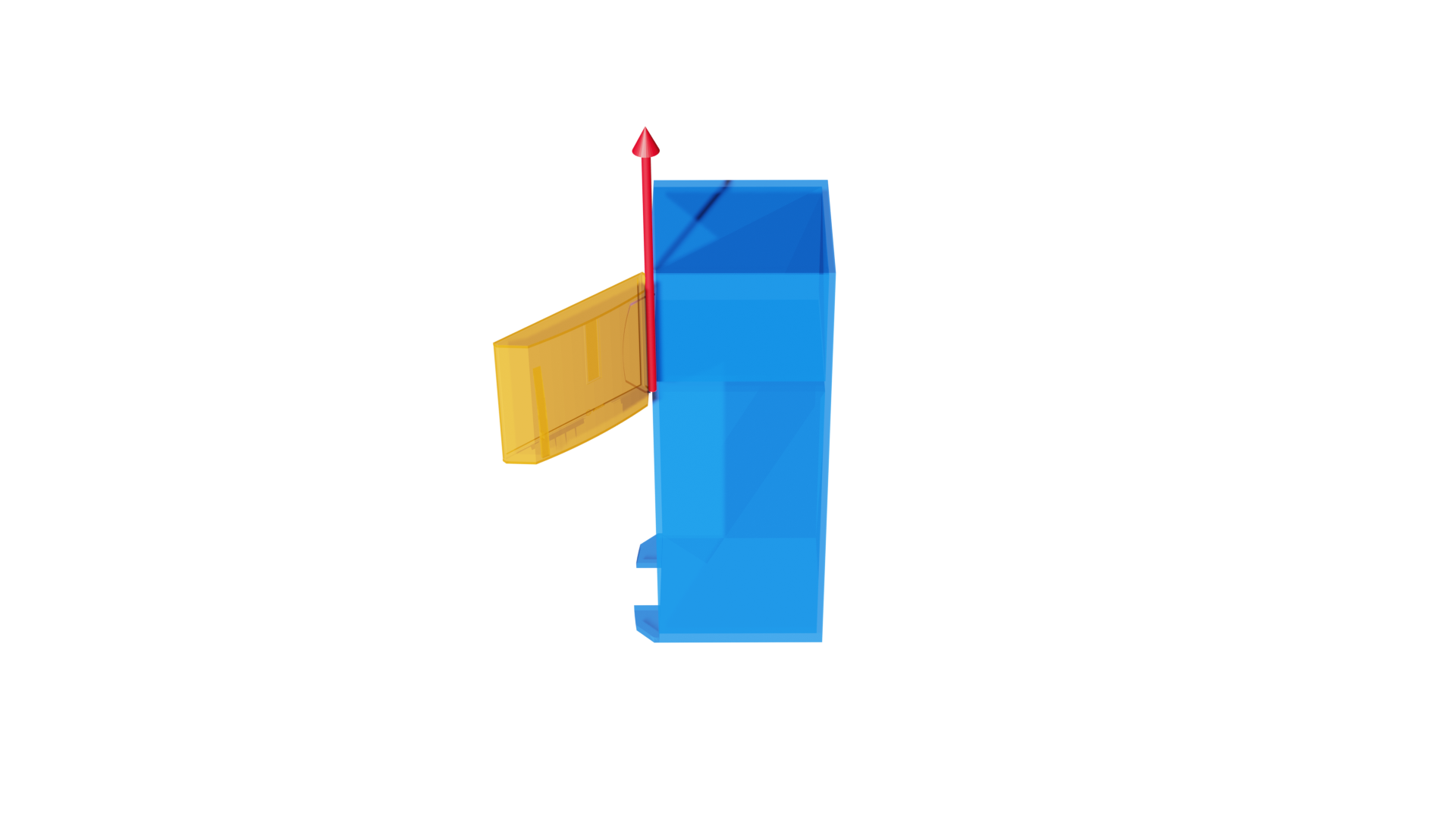} &
    \includegraphics[width=0.33\linewidth]{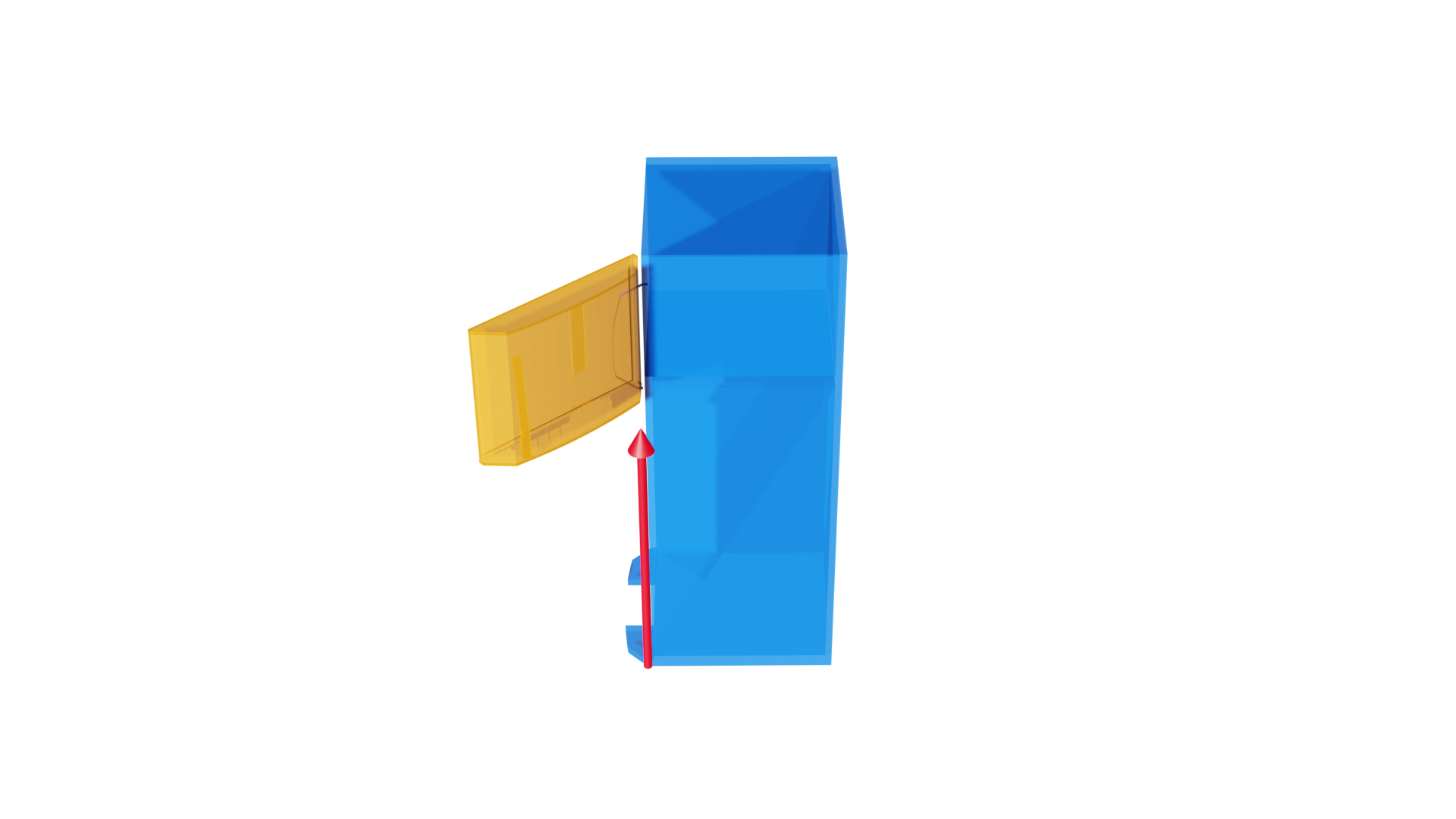}
    \\
    \includegraphics[width=0.33\linewidth]{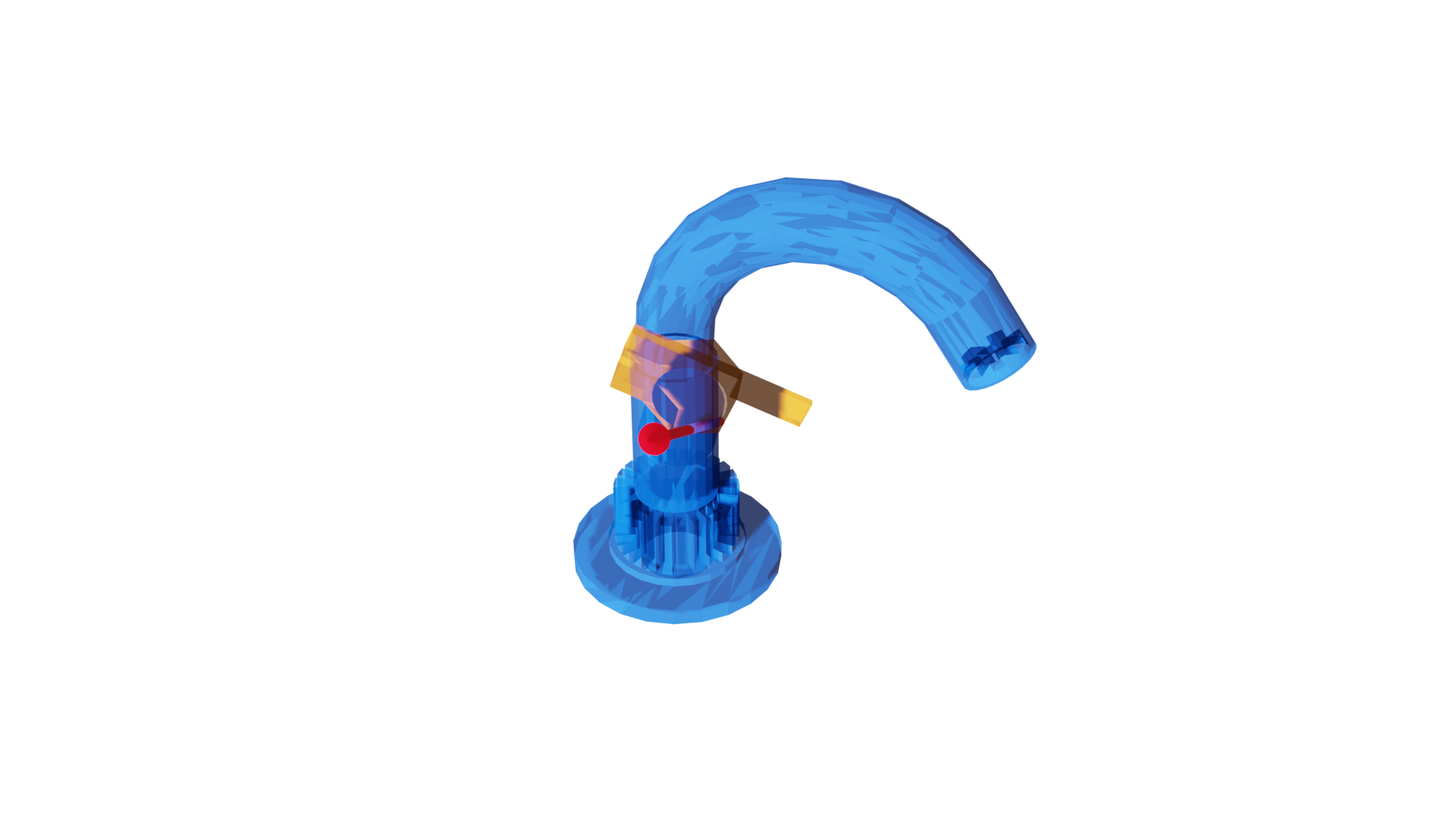} &
    \includegraphics[width=0.33\linewidth]{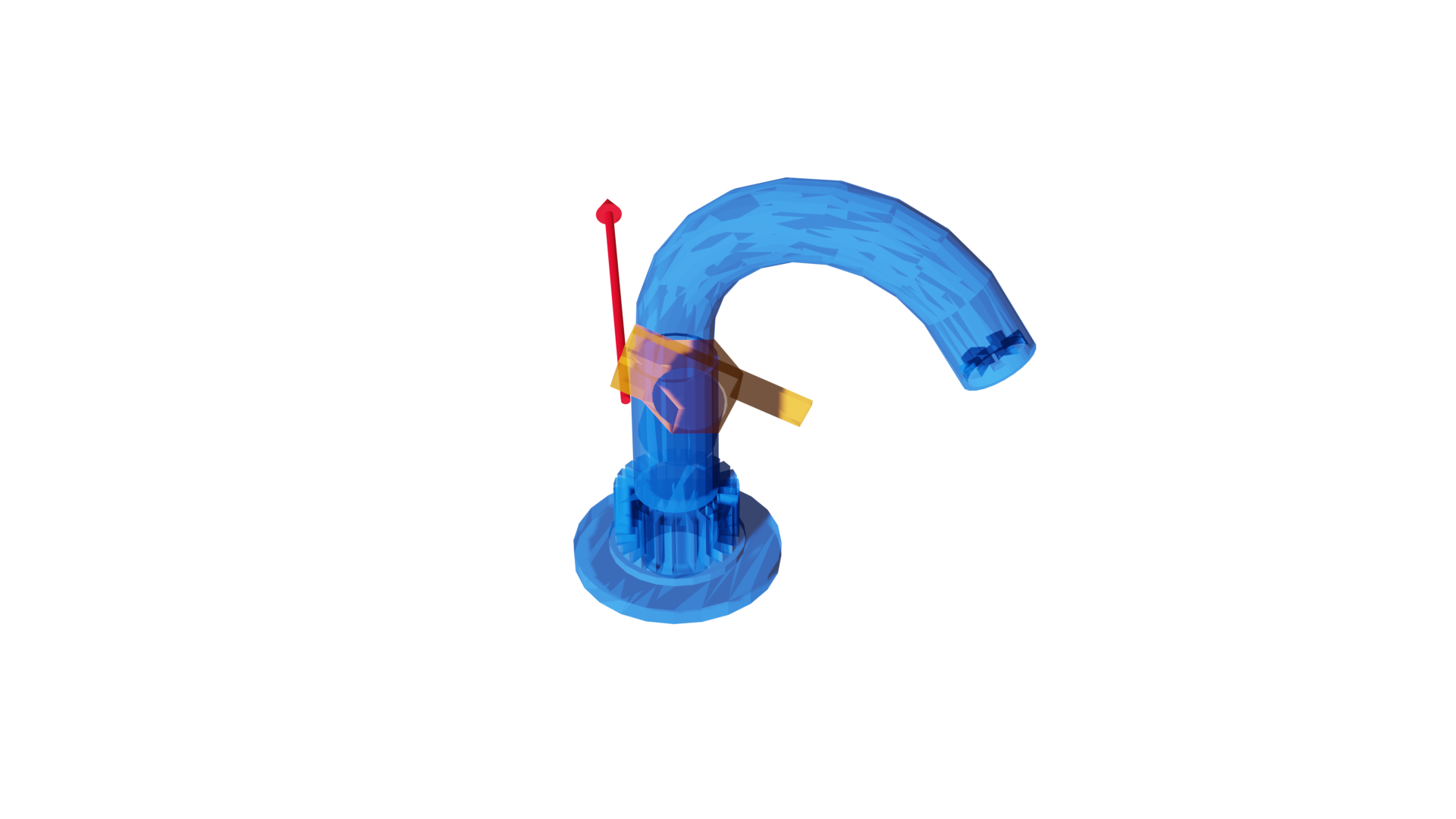} &
    \includegraphics[width=0.33\linewidth]{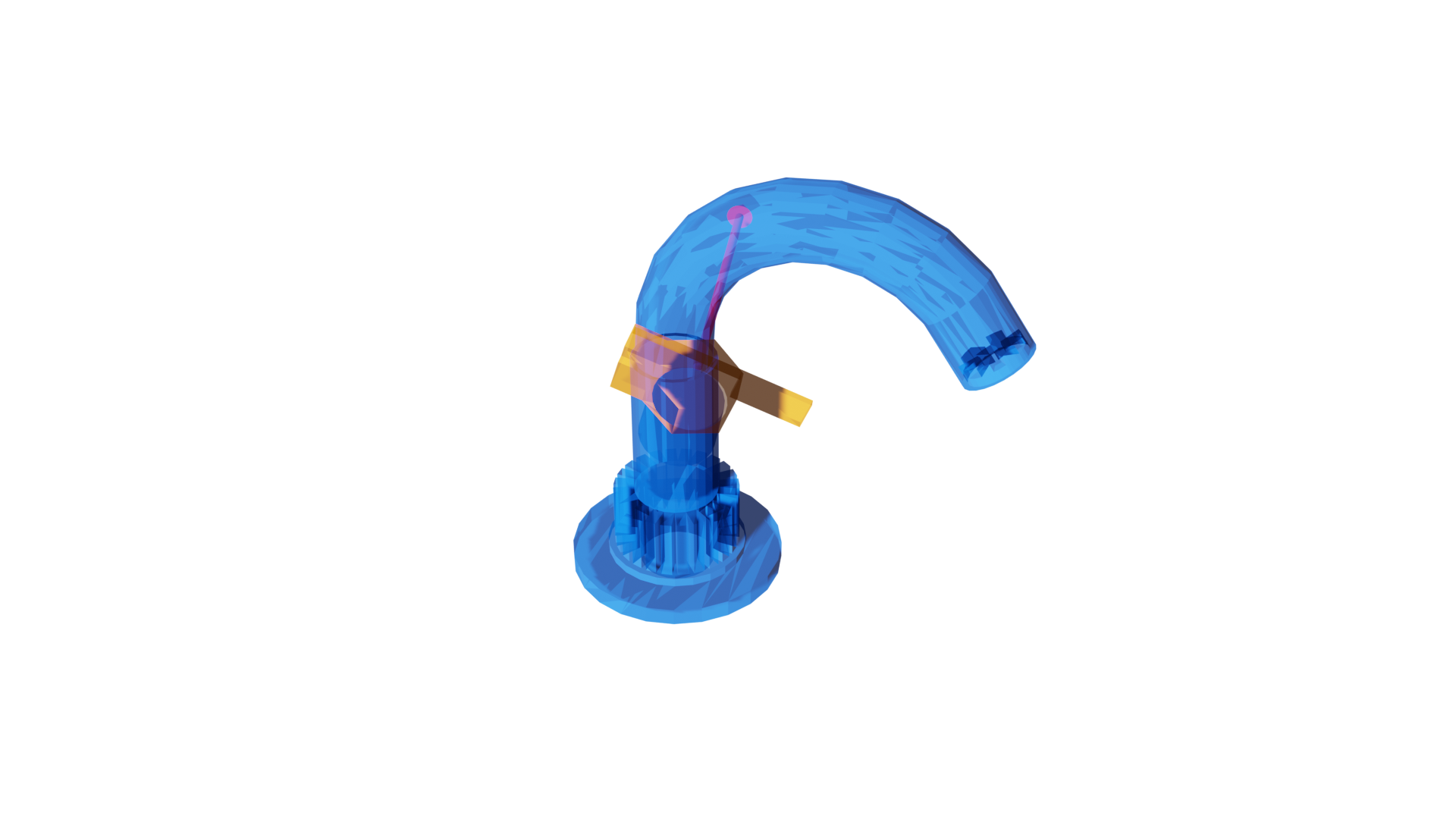}
    \\
    \end{tabular}
    \caption{
    Additional qualitative comparison of our method with the supervised BaseNet baseline
    }
    \label{figure:qualitative_comparison4}
\end{figure*}

\begin{figure*}[ht!]
    \centering
    \setlength{\tabcolsep}{1pt}
    \begin{tabular}{ccc}
    \textbf{BaseNet} & \textbf{Ours} & \textbf{GT}
        \\
    \includegraphics[width=0.33\linewidth]{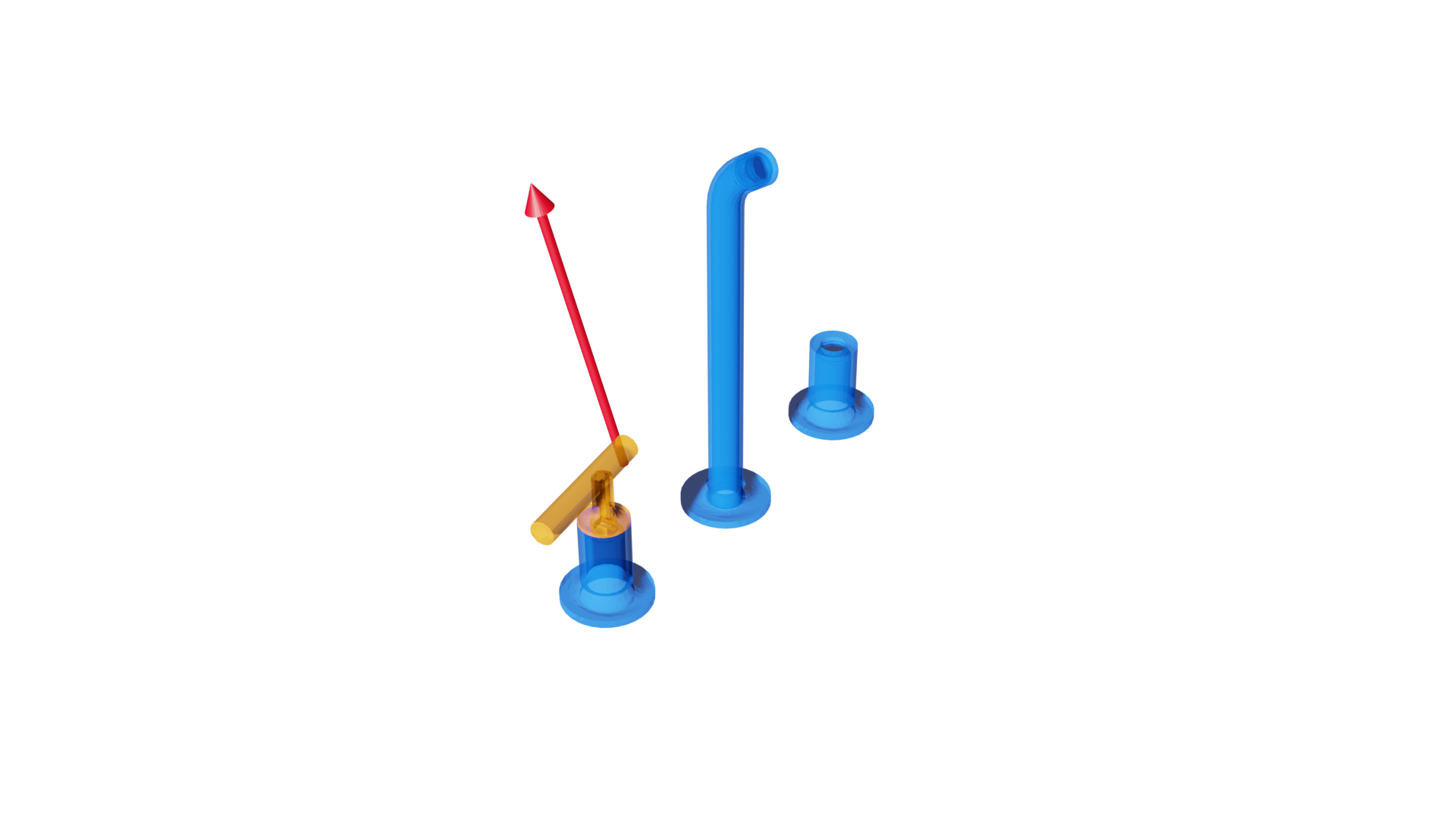} &
    \includegraphics[width=0.33\linewidth]{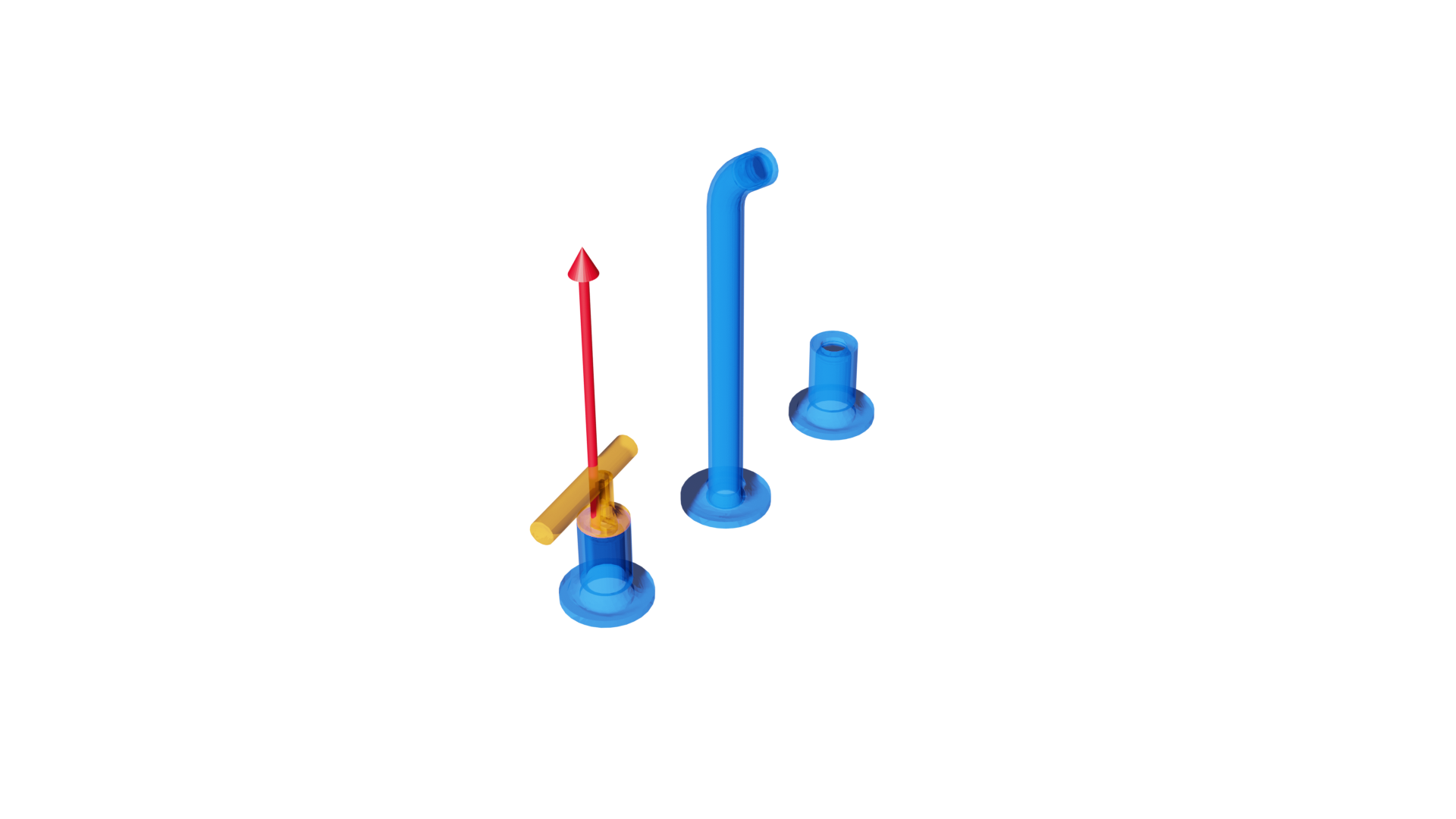} &
    \includegraphics[width=0.33\linewidth]{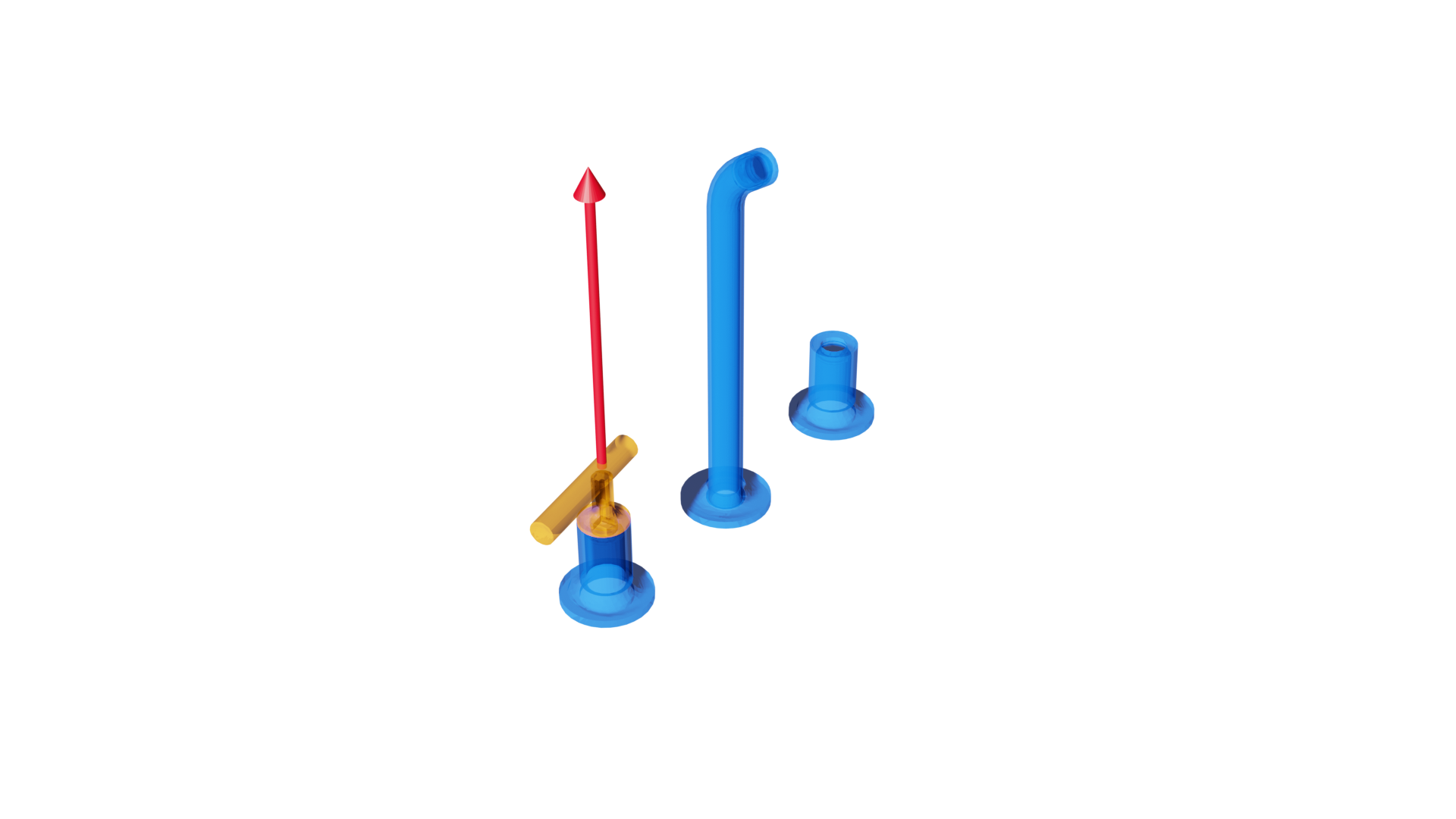}
    \\
    \includegraphics[width=0.33\linewidth]{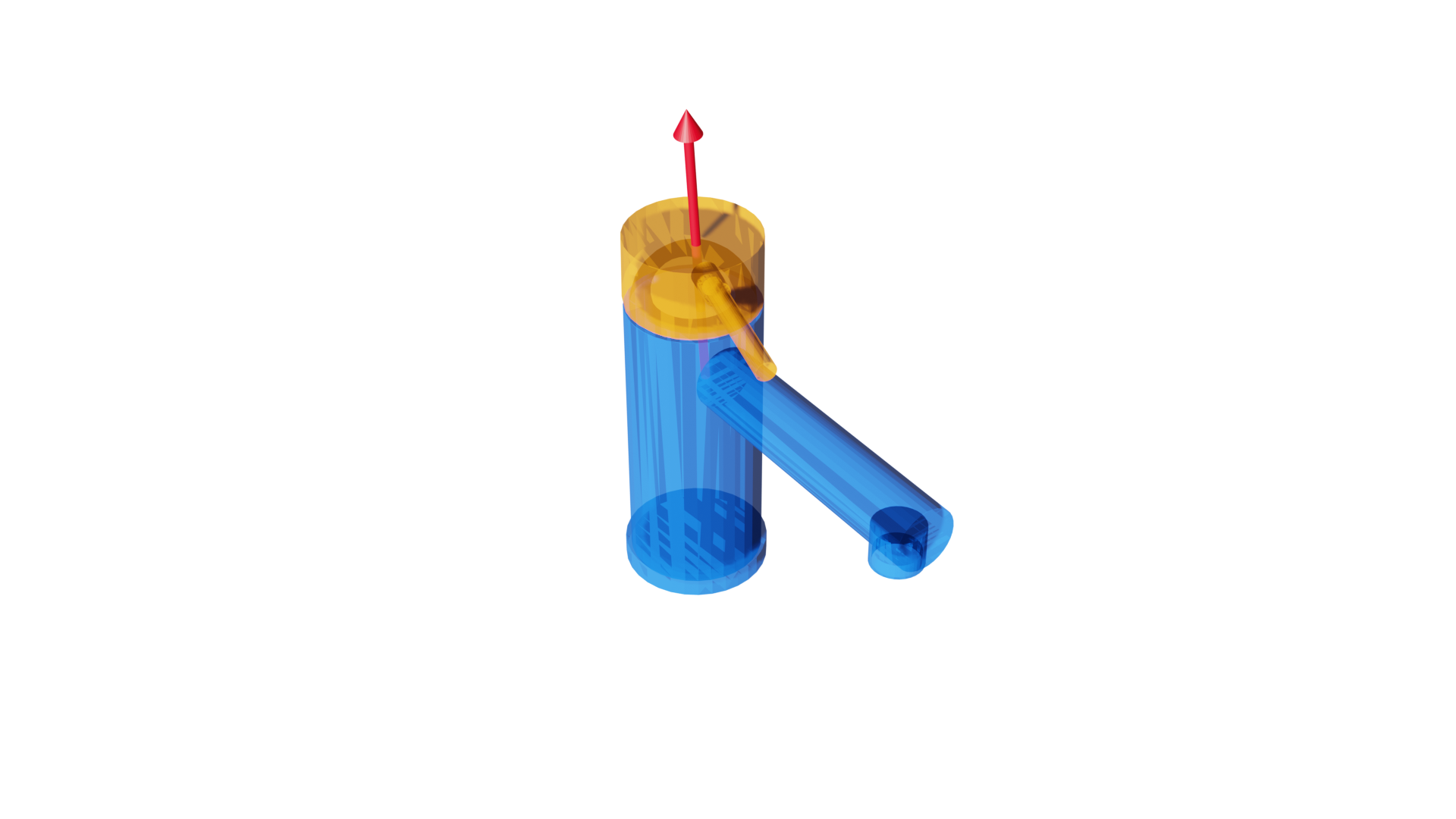} &
    \includegraphics[width=0.33\linewidth]{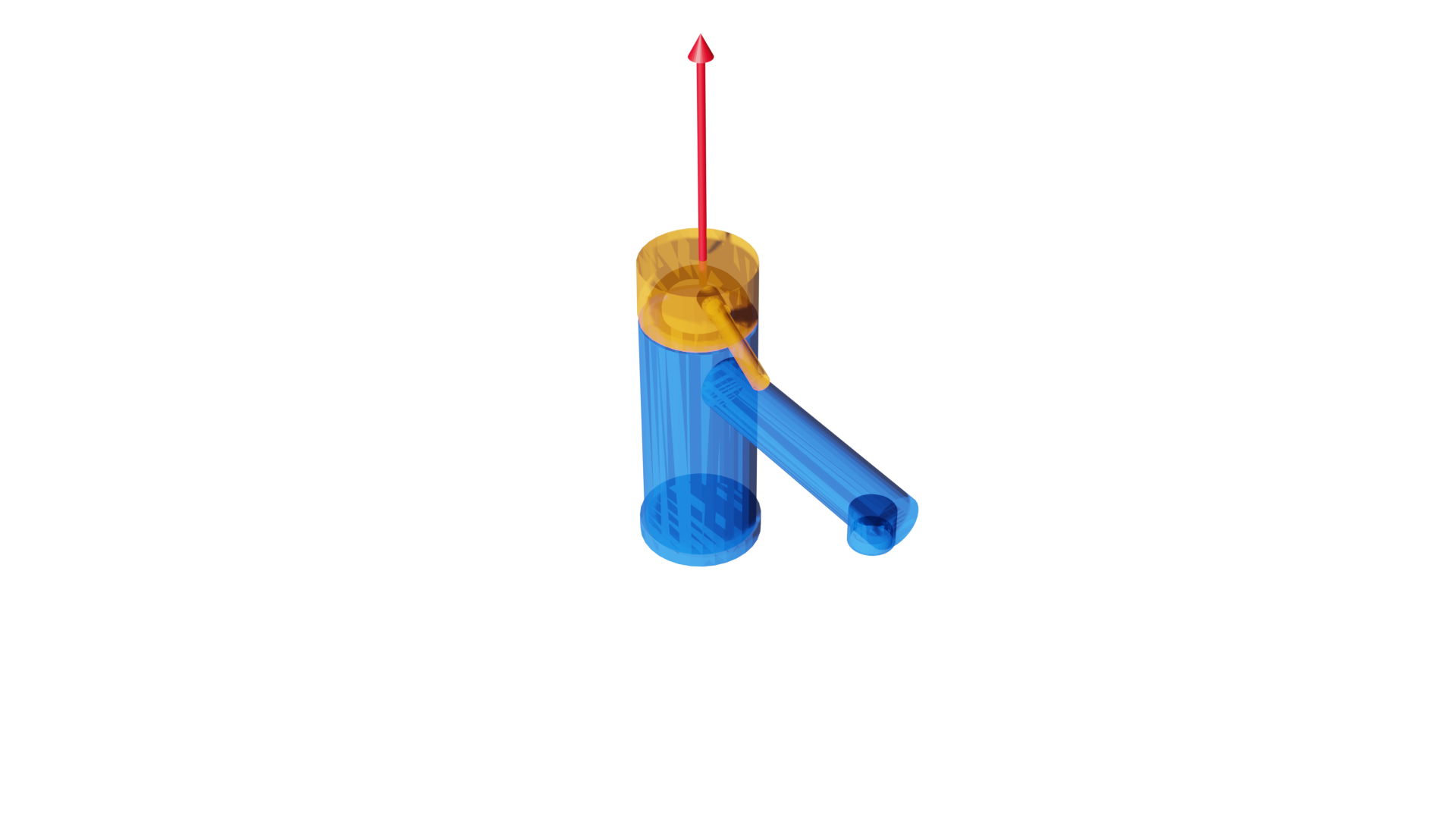} &
    \includegraphics[width=0.33\linewidth]{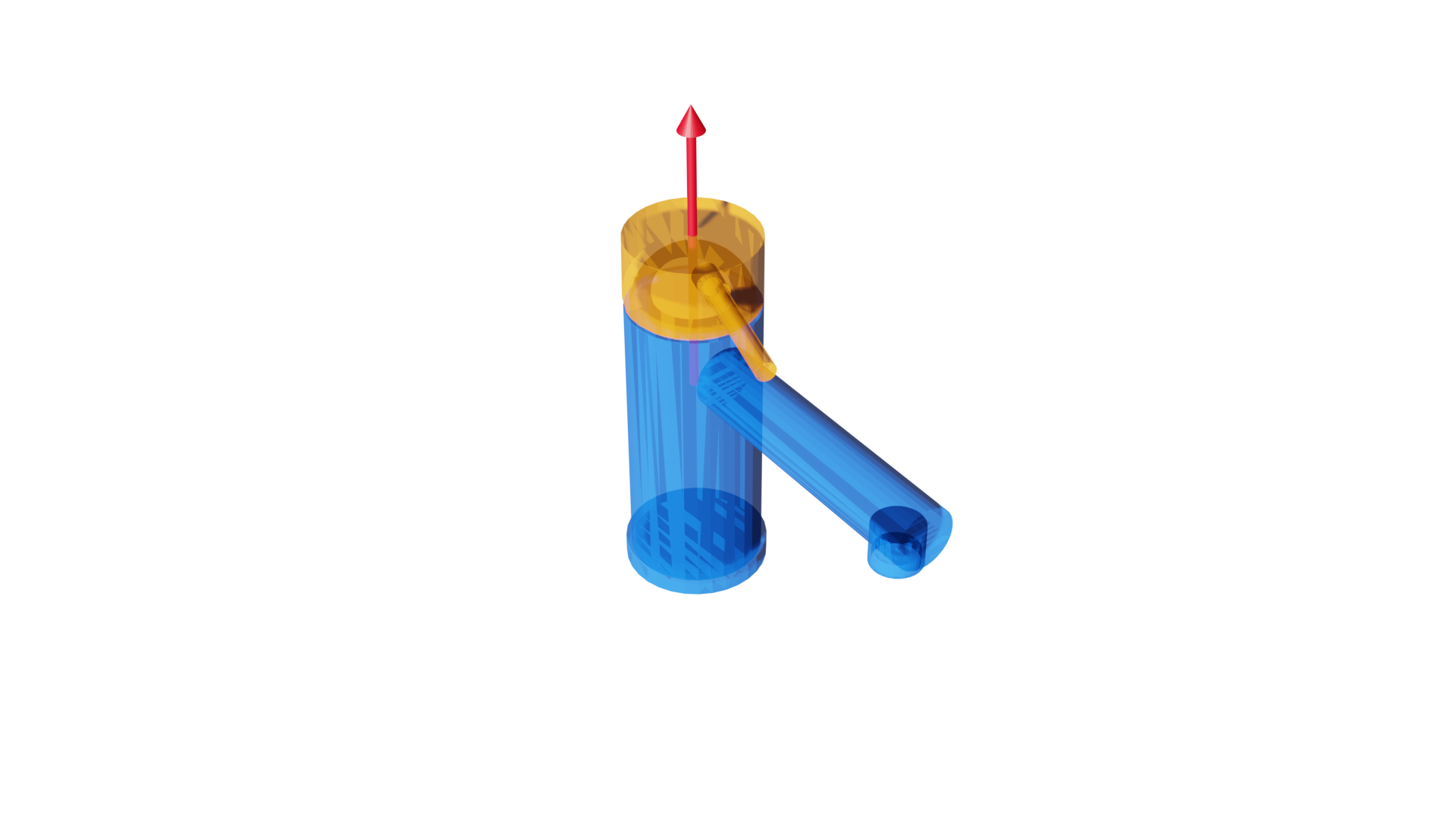}
    \\
    \includegraphics[width=0.33\linewidth]{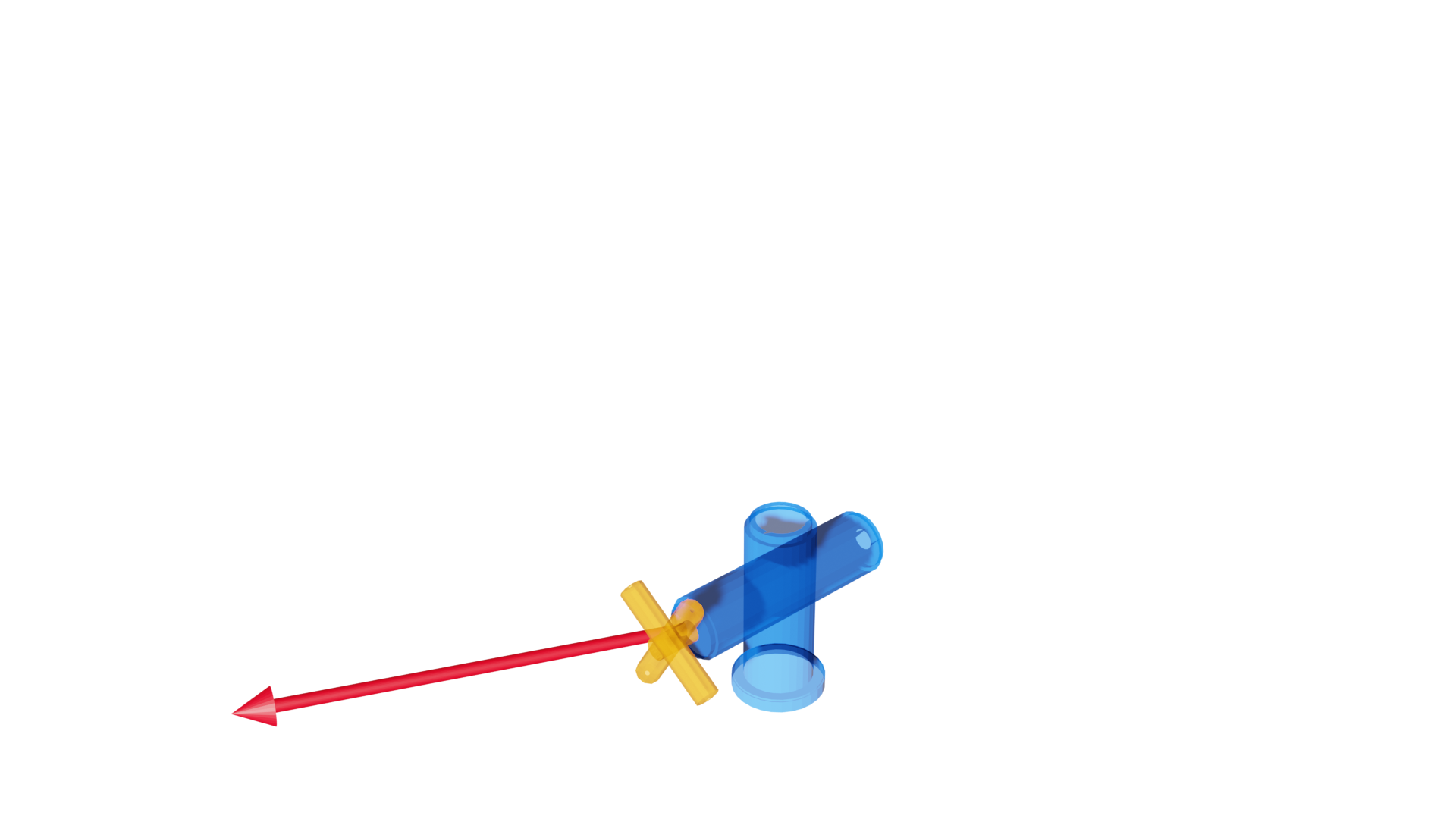} &
    \includegraphics[width=0.33\linewidth]{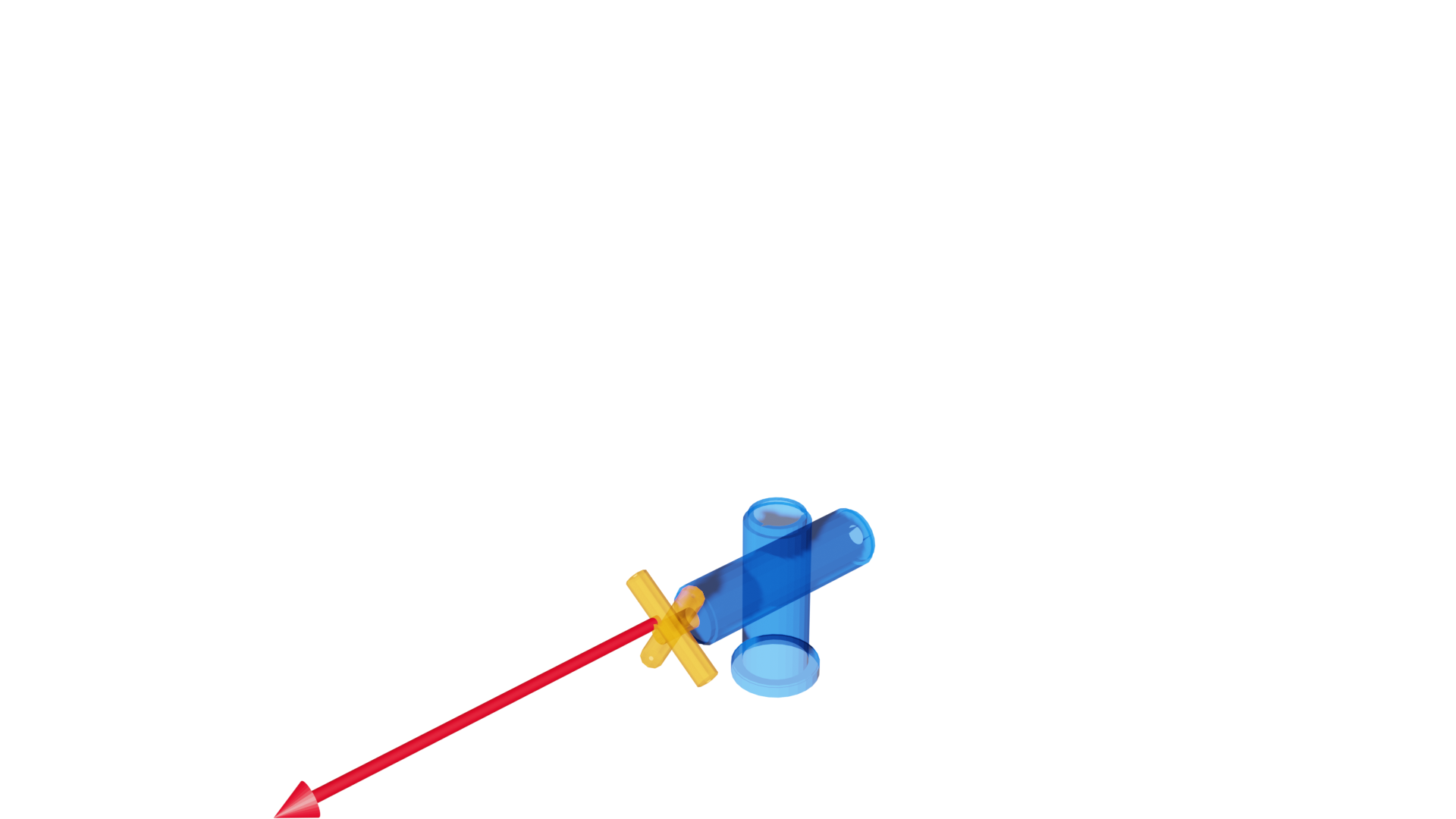} &
    \includegraphics[width=0.33\linewidth]{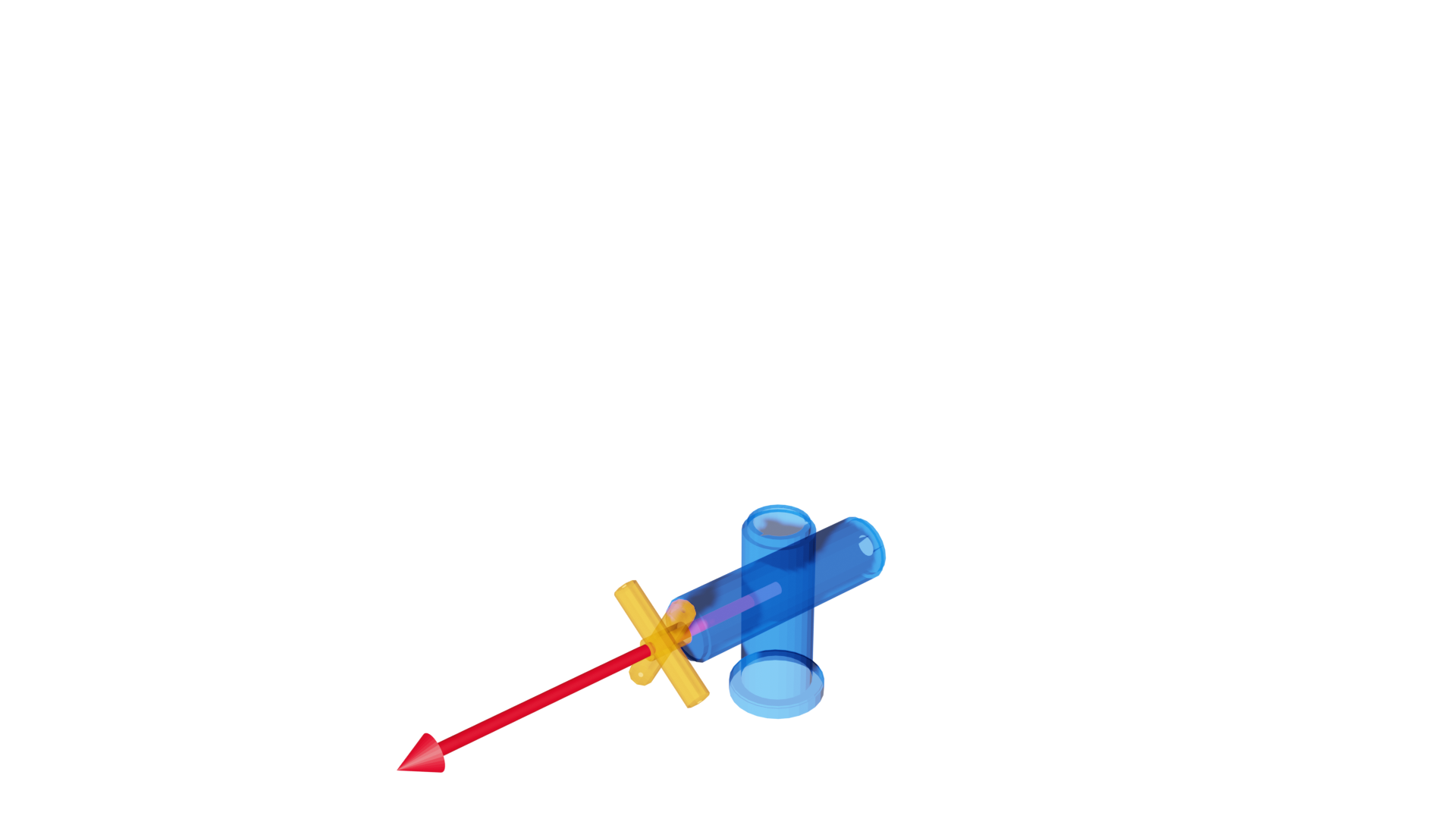}
    \\
    \includegraphics[width=0.33\linewidth]{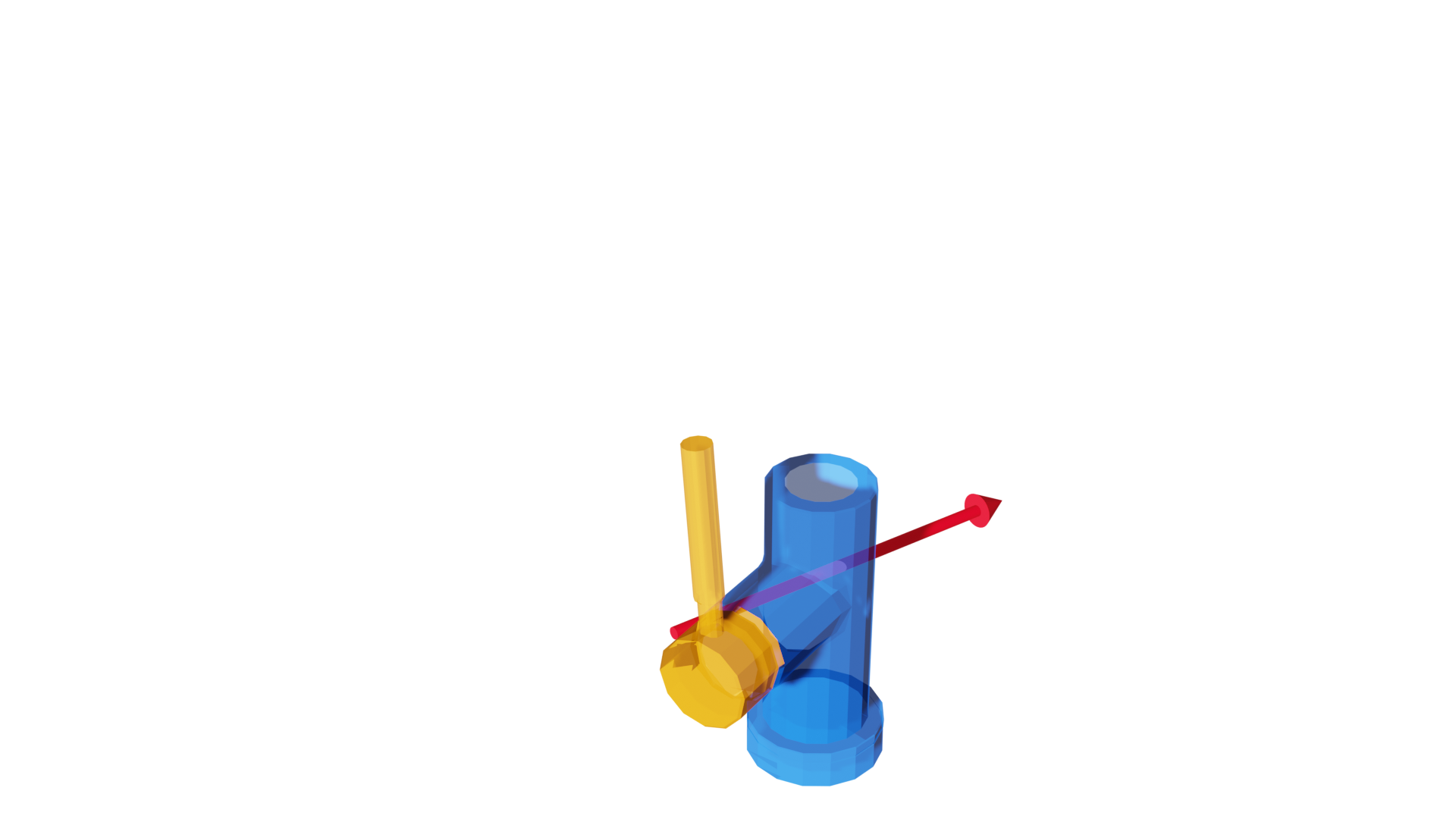} &
    \includegraphics[width=0.33\linewidth]{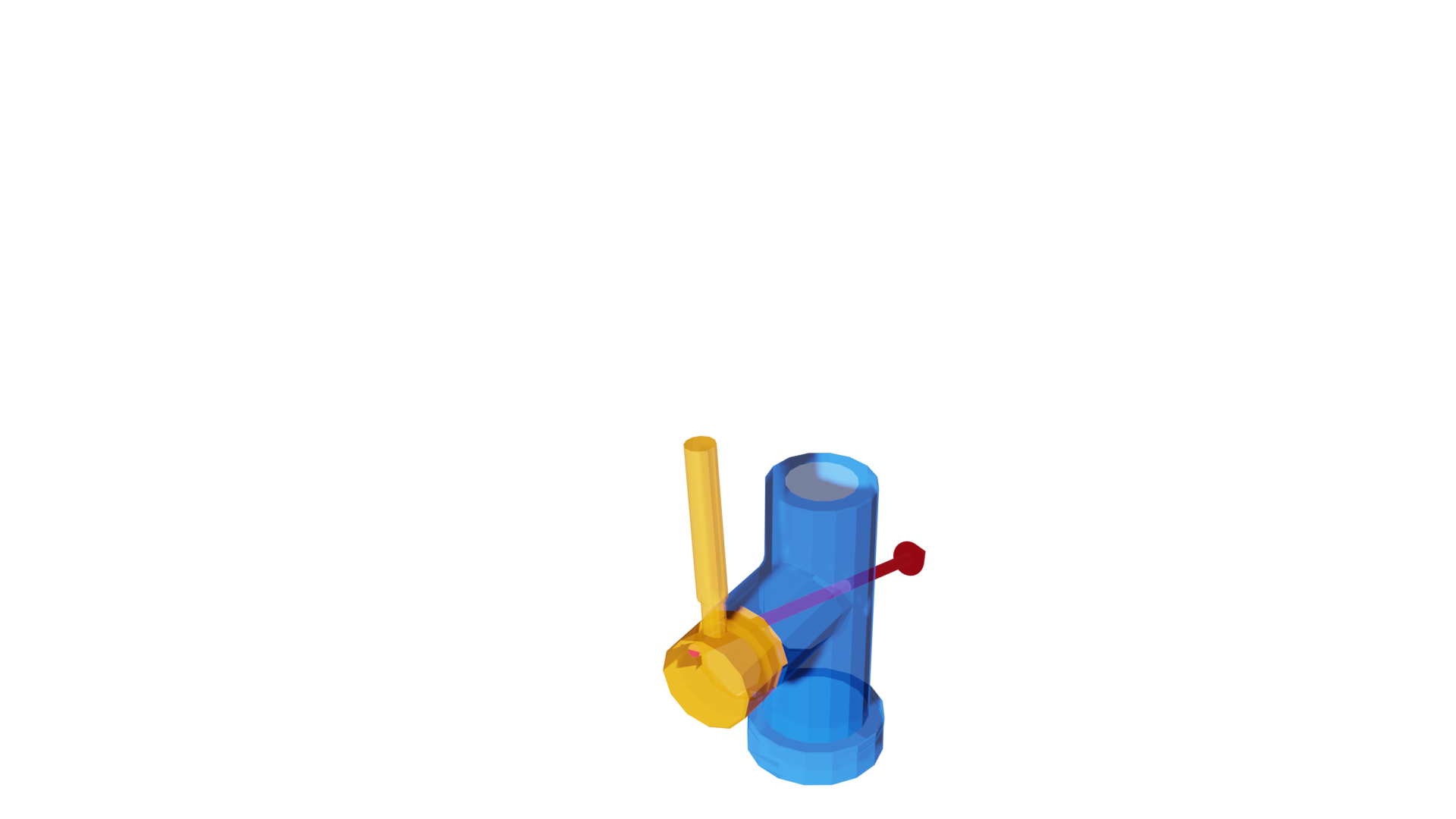} &
    \includegraphics[width=0.33\linewidth]{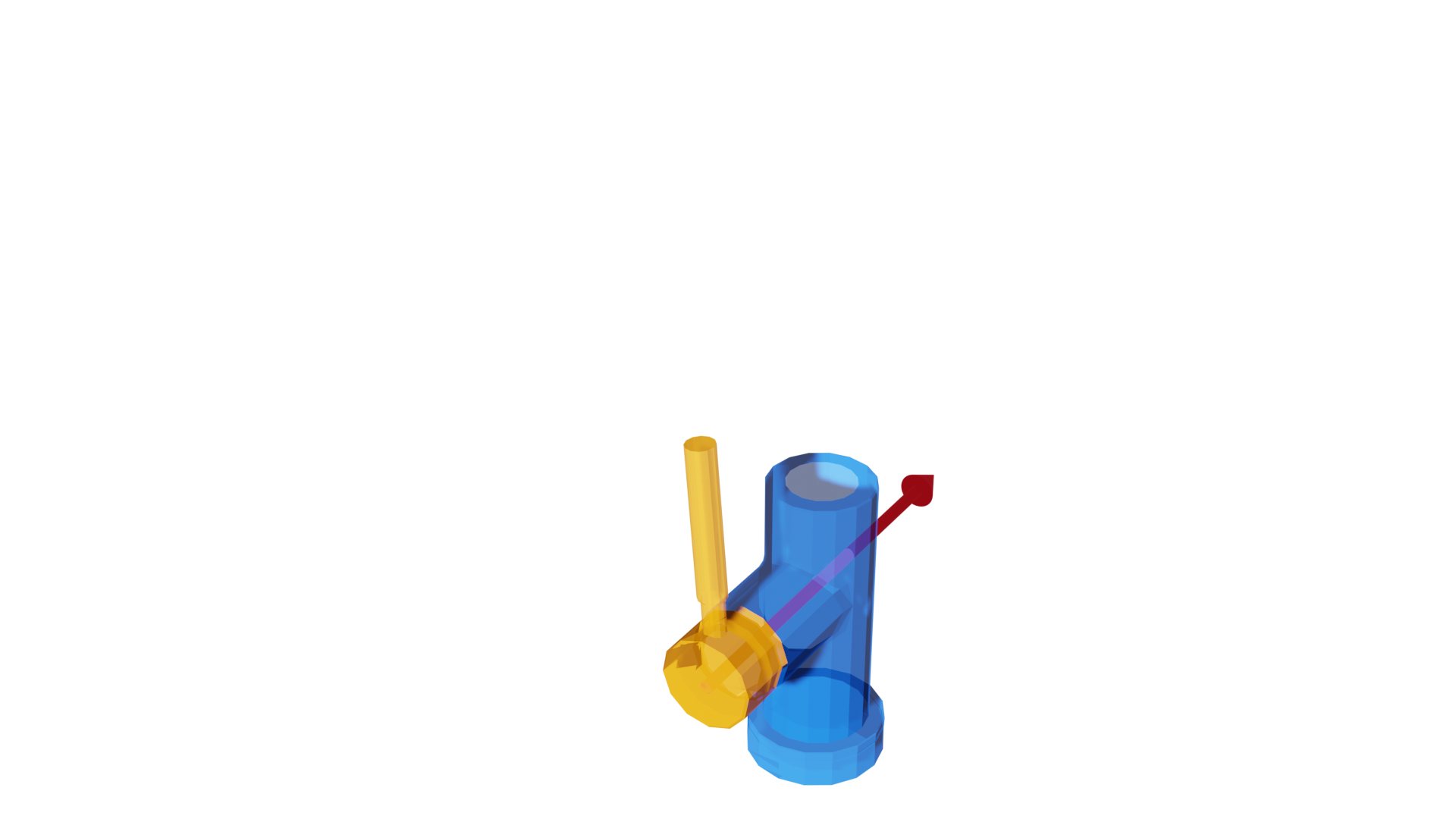}
    \\
    \includegraphics[width=0.33\linewidth]{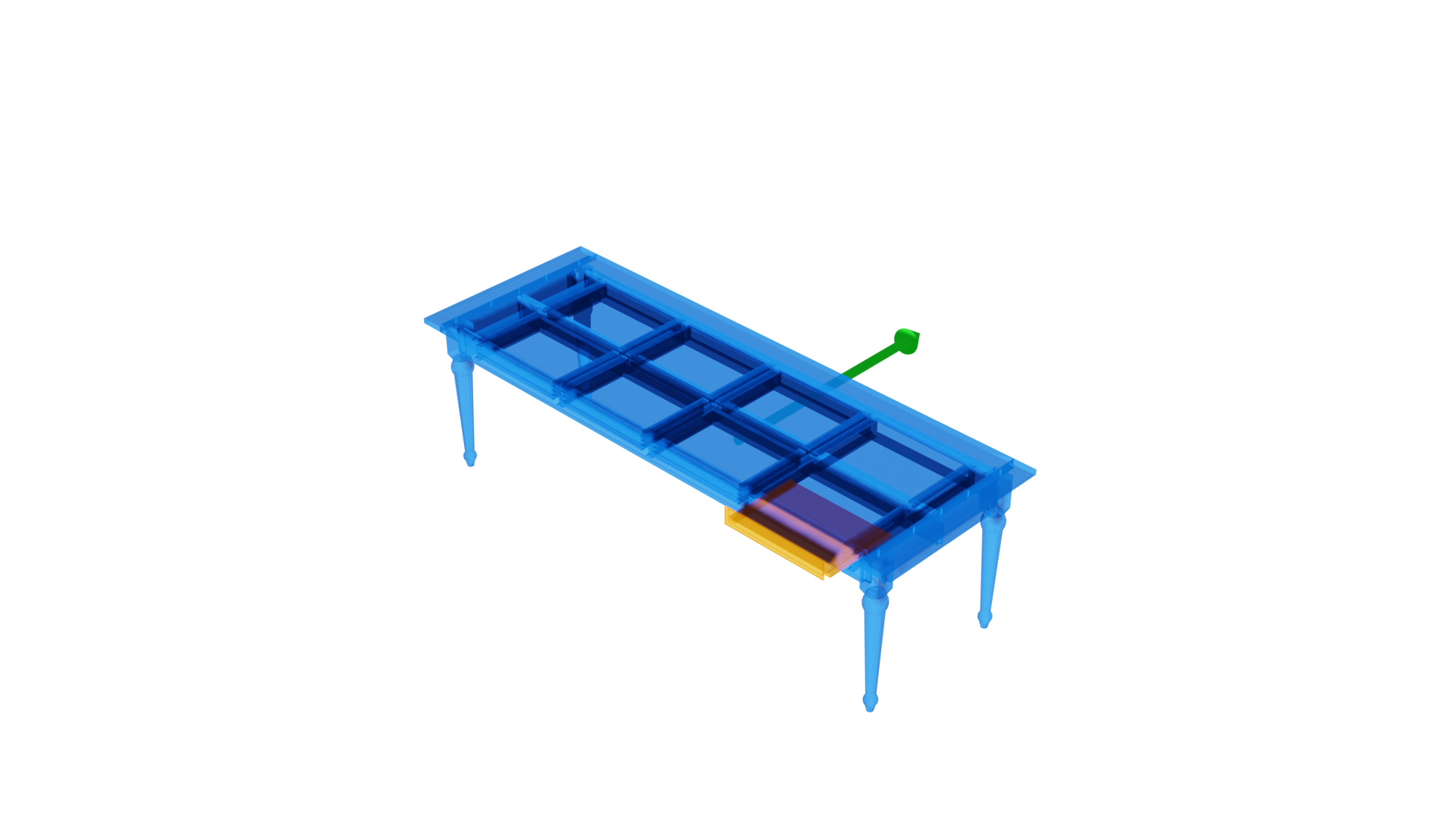} &
    \includegraphics[width=0.33\linewidth]{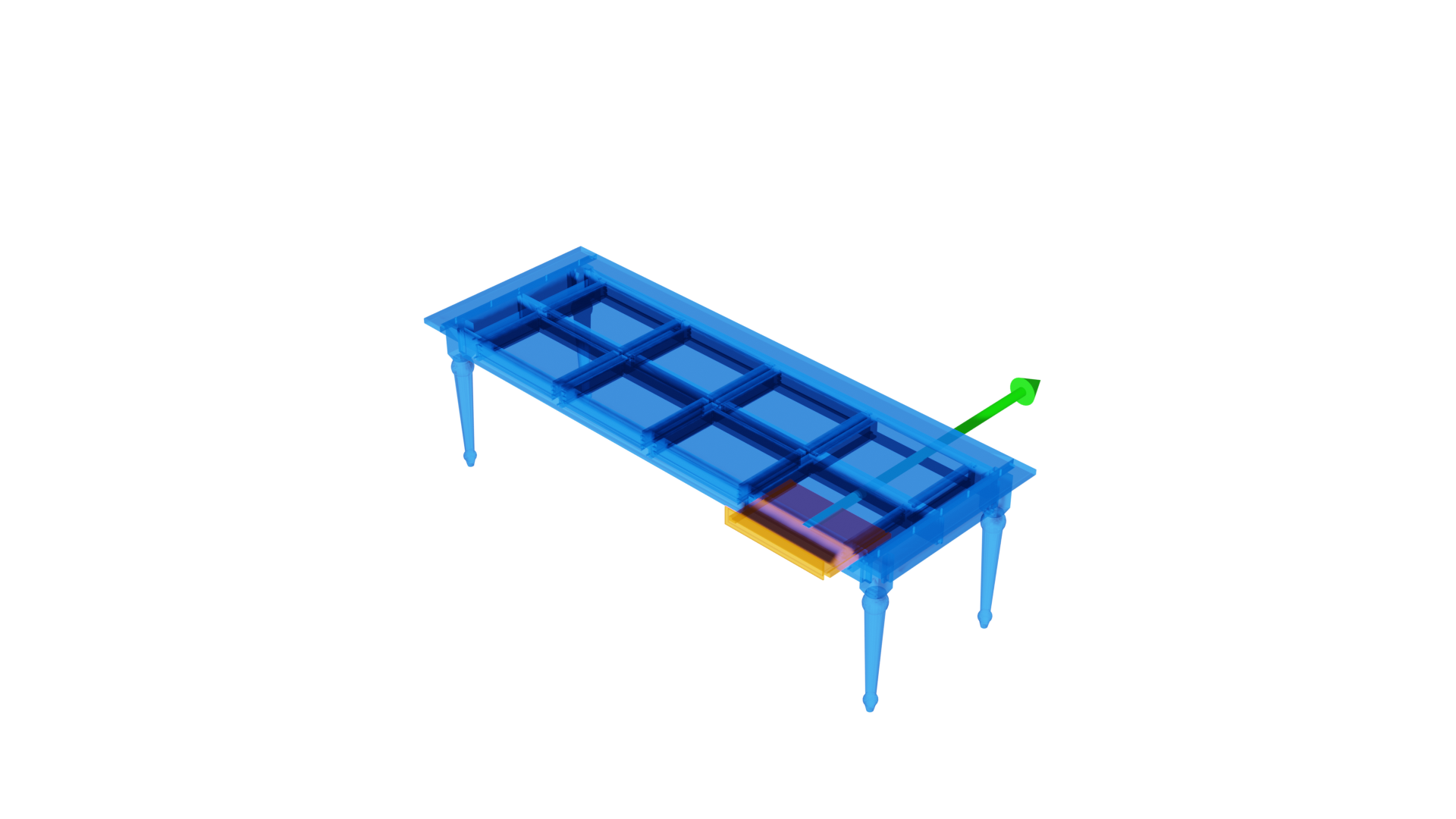} &
    \includegraphics[width=0.33\linewidth]{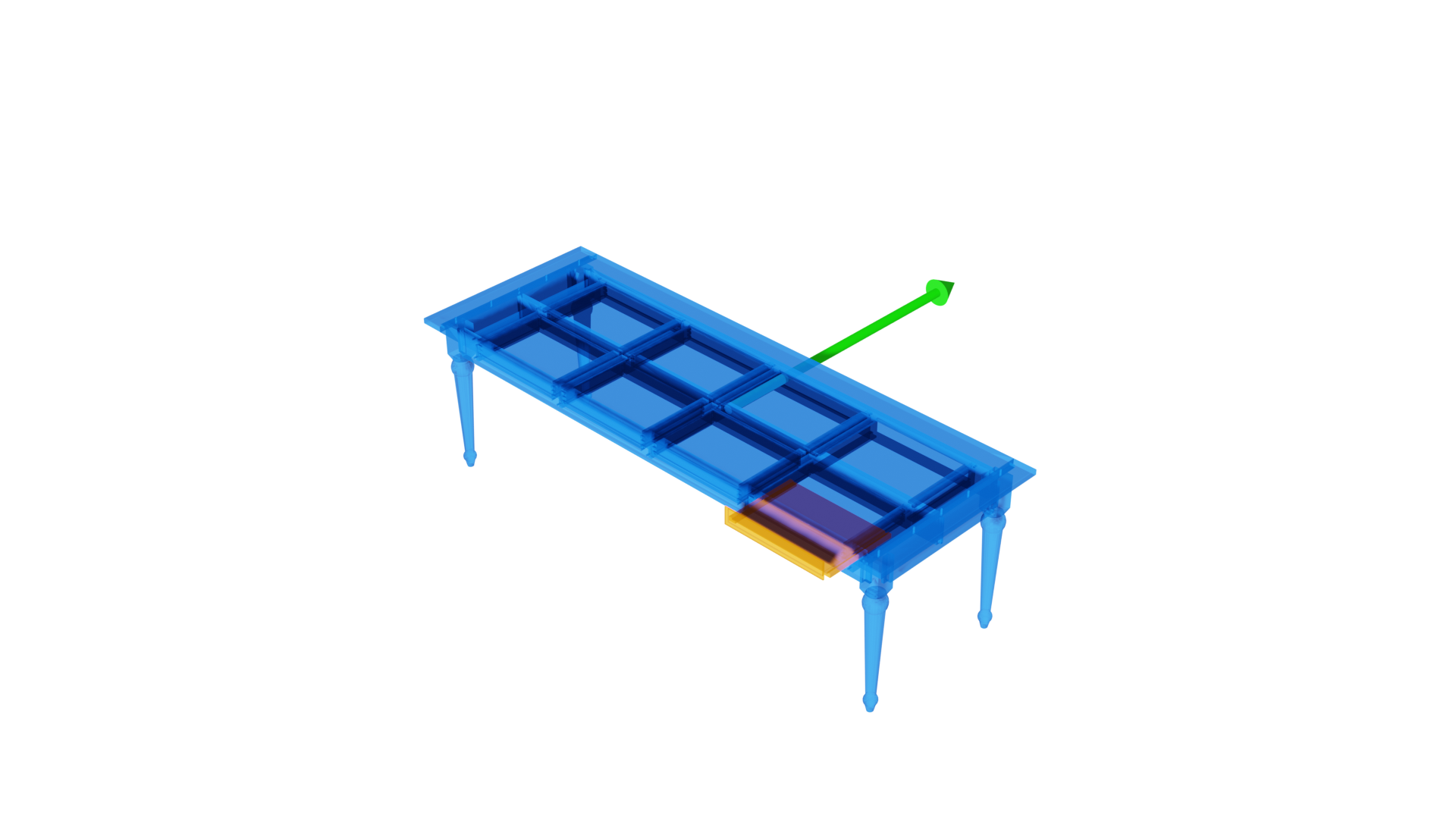}
    \\
    \end{tabular}
    \caption{
    Additional qualitative comparison of our method with the supervised BaseNet baseline
    }
    \label{figure:qualitative_comparison5}
\end{figure*}

Please see Figure~\ref{figure:qualitative_comparison1},~\ref{figure:qualitative_comparison2},~\ref{figure:qualitative_comparison3},~\ref{figure:qualitative_comparison4},~\ref{figure:qualitative_comparison5} for additional qualitative comparison of our method with the supervised baseNet baseline.

\clearpage
\newpage

\end{document}